\documentclass[11pt,a4paper,twoside]{book}
\usepackage{eepic}
\usepackage{makeidx}
\usepackage[]{epic,eepic}
\usepackage[xdvi]{graphicx}
\usepackage[pagebackref]{hyperref}
\usepackage{eufrak}
\usepackage{latexsym}

\makeatletter
\def\ps@myheadings{\let\@mkboth\markboth 
   \def\@evenhead{\vtop{\hbox to
\hsize{\noindent\normalsize\rm \thepage\hfil\slshape\rm\leftmark}%
              \vskip5pt \hbox to \hsize{\hrulefill}}}
   \def\@oddhead{\vtop{\hbox to
\hsize{\noindent{\slshape\normalsize \rm \rightmark}\hfil \rm \thepage}%
              \vskip5pt \hbox to \hsize{\hrulefill}}}

  \def\@evenfoot{}%
  \def\@oddfoot{}%
   \def\chaptermark##1{\markboth{##1}{##1}}%
 \def\sectionmark##1{\markright {\ifnum \c@secnumdepth >\z@
   \thesection \ \fi ##1}}
}
\makeatother

\makeatletter
\def\@makechapterhead#1{%
  \vspace*{50\p@}%
  {\parindent \z@ \raggedright \normalfont
    \interlinepenalty\@M
    \LARGE \bfseries #1\par\nobreak
    \vskip 40\p@ \thispagestyle{myheadings}  %% \thispagestyle von mir reingenommen   
  }}
\makeatother

\makeatletter
\def\@makeschapterhead#1{%
  \vspace*{50\p@}%
  {\parindent \z@ \raggedright
    \normalfont
    \interlinepenalty\@M
    \LARGE \bfseries  #1\par\nobreak
    \vskip 40\p@ \thispagestyle{myheadings}  %% \thispagestyle von mir reingenommen 
  }}
\makeatother

\makeatletter
\renewcommand\appendix{\par
\setcounter{chapter}{1}%
\setcounter{section}{0}%
\gdef\@chapapp{\appendixname}%
\gdef\thechapter{\@Alph\c@chapter}}
\makeatother

\makeatletter
\def\@dottedtocline#1#2#3#4#5{%
 \ifnum #1>\c@tocdepth
 \else
  \vskip \z@ plus .2pt
  {\leftskip #2\relax \rightskip \@tocrmarg plus2em% v.0.16
   \parfillskip -\rightskip
   \parindent #2\relax
   \@afterindenttrue
   \interlinepenalty\@M
   \leavevmode
   \@tempdima #3\relax \advance\leftskip \@tempdima \hbox{}\hskip -\leftskip
   #4\nobreak
   \hfill \nobreak
   \hbox to\@pnumwidth{\hfil\rm #5}\par}%
 \fi
}
\makeatother

\makeatletter
\renewcommand\tableofcontents{%
    \if@twocolumn
      \@restonecoltrue\onecolumn
    \else
      \@restonecolfalse
    \fi
    \chapter*{\contentsname
        \@mkboth{%
           \contentsname}{\contentsname}}%
    \@starttoc{toc}%
    \if@restonecol\twocolumn\fi
    }
\makeatother

\makeatletter
\renewcommand*\l@chapter[2]{%
  \ifnum \c@tocdepth >\m@ne
    \addpenalty{-\@highpenalty}%
    \vskip 1.0em \@plus\p@         % alter wert \vskip 1.0em
    \setlength\@tempdima{1.5em}%
    \begingroup
      \parindent \z@ \rightskip \@pnumwidth
      \parfillskip -\@pnumwidth
      \leavevmode %\bfseries         %%% Fettdruck rausgenommen
      \advance\leftskip\@tempdima
      \hskip -\leftskip
      #1\nobreak\hfil \nobreak\hb@xt@\@pnumwidth{\hss #2}\par
      \penalty\@highpenalty
    \endgroup
  \fi}
\makeatother

\makeatletter
\renewcommand*\l@section{\@dottedtocline{1}{1.5em}{3.0em}}    
\renewcommand*\l@subsection{\@dottedtocline{2}{1.5em}{3.0em}} 
\renewcommand*\l@subsubsection{\@dottedtocline{3}{7.0em}{4.1em}}
\renewcommand*\l@paragraph{\@dottedtocline{4}{10em}{5em}}
\renewcommand*\l@subparagraph{\@dottedtocline{5}{12em}{6em}} \makeatother

\renewcommand\bibname{References}


\makeatletter
\renewcommand\listoffigures{%
    \addcontentsline{toc}{chapter}{\listfigurename}  %% von mir damit List if Fig. in TOC
    \if@twocolumn
      \@restonecoltrue\onecolumn
    \else
      \@restonecolfalse
    \fi
    \chapter*{\listfigurename
      \@mkboth{\listfigurename}%
              {\listfigurename}}%
    \@starttoc{lof}%
    \if@restonecol\twocolumn\fi
    }
\makeatother

\makeatletter
\renewcommand\fnum@figure{\footnotesize \figurename~\thefigure}
\renewcommand\thefigure
      {\@arabic\c@figure}
\makeatother

\makeatletter
\renewcommand \thetable
     {\@arabic\c@table}
\makeatother

\let\ifuniqrefs\iftrue

\def\gammap{\gamma^{\prime}}
\def\betap{\beta^{\prime}}
\def\betaMF{\beta_{\ssstyle \mathrm{MF}}}

\def\sigmaMF{\sigma_{\ssstyle \mathrm{MF}}}
\def\thetaMF{\theta_{\ssstyle \mathrm{MF}}}
\def\deltaMF{\delta_{\ssstyle \mathrm{MF}}}
\def\gammaMF{\gamma_{\ssstyle \mathrm{MF}}}
\def\gammapMF{\gamma^{\prime}_{\ssstyle \mathrm{MF}}}
\def\nuperpMF{\nu_{\senk,{\ssstyle \mathrm{MF}}}}
\def\nuperpstar{\nu_{\senk,{\ssstyle \mathrm{MF}}}^{\ast}}
\def\nuparaMF{\nu_{\para,{\ssstyle \mathrm{MF}}}}
\def\zMF{z_{\ssstyle \mathrm{MF}}}
\def\deltarho{\delta \hspace{-0.3ex} \rho}
\def\deltap{\delta \hspace{-0.2ex} p}
\def\deltaT{\delta \hspace{-0.1ex} T}
\def\lambdac{\lambda_{\scriptscriptstyle \mathrm{c}}}
\def\Nc{N_{\scriptscriptstyle \mathrm{c}}}
\def\Dc{D_{\scriptscriptstyle \mathrm{c}}}
\def\Tc{T_{\scriptscriptstyle \mathrm{c}}}
\def\pc{p_{\scriptscriptstyle \mathrm{c}}}
\def\rhoc{\rho_{\scriptscriptstyle \mathrm{c}}}
\def\rhocR{\rho_{\scriptscriptstyle \mathrm{c},R}}
\def\rhoa{\rho_{\scriptscriptstyle \mathrm{a}}}
\def\rhoanull{\rho_{\scriptscriptstyle \mathrm{a},0}}
\def\rhoe{\rho_{\scriptscriptstyle \mathrm{e}}}
\def\rhoi{\rho_{\scriptscriptstyle \mathrm{i}}}
\def\chia{\chi_{\scriptscriptstyle \mathrm{a}}}

\def\Pa{P_{\scriptscriptstyle \mathrm{a}}}
\def\Na{N_{\scriptscriptstyle \mathrm{a}}}
\def\eff{\scriptscriptstyle \mathrm{eff}}
\def\Reff{R_{\scriptscriptstyle \mathrm{eff}}}
\def\Pperc{P_{\scriptscriptstyle \mathrm{perc}}}
\def\asympprop{\,{\tilde{\propto}}\,}
\def\ggl{\hspace{0.5ex}\raisebox{0.4ex}{$\scriptstyle\geq$}\hspace{0.5ex}}
\def\kgl{\hspace{0.5ex}\raisebox{0.4ex}{$\scriptstyle\leq$}\hspace{0.5ex}}
\def\senk{\hspace{-0.25ex}\raisebox{0.0ex}{$\scriptscriptstyle\perp$}\hspace{0.0ex}}
\def\para{\hspace{-0.15ex}\raisebox{0.0ex}{$\scriptscriptstyle\parallel$}\hspace{0.0ex}}

\def\plmi{\hspace{0.0ex}\raisebox{0.2ex}{$\scriptstyle\pm$}\hspace{0.0ex}}

\def\ssstyle{\scriptscriptstyle}
\def\reffigname#1{Figure\,\protect\ref{#1}}
\def\refFigname#1{Figure\,\protect\ref{#1}}
\def\reffigs2name#1#2{Figure\,\protect\ref{#1} and Figure\,\protect\ref{#2}}
\def\refeqname#1{Eq.\,(\protect\ref{#1})}
\def\refeqsname#1#2{Eqs.\,(\protect\ref{#1}-\protect\ref{#2})}
\def\refeqs2name#1#2{Eq.\,(\protect\ref{#1}) and Eq.\,(\protect\ref{#2})}
\def\reftablename#1{Table\,\protect\ref{#1}}
\def\reftables2name#1#2{Table\,\protect\ref{#1} and Table\,\protect\ref{#2}}

\begin{document}

\frontmatter

%%% aus einem komischen Grund, muss dass hier stehen
%\makeatletter
%\renewcommand\newcounter{figure}[part]
%\makeatother

\flushbottom 
  
\pagestyle{myheadings}

\title{{\huge \textbf{Universal scaling behavior\\[5mm] of non-equilibrium
phase transitions}}\\[2cm]
%HABILITATIONSSCHRIFT\\[1cm]
%{\Large der Fakult{\"a}t f{\"u}r Naturwissenschaften\\
%der Universit{\"a}t Duisburg-Essen\\
%vorgelegt von }
}

\author{{\huge Sven L{\"u}beck}\\[5mm]
Theoretische Physik, Univerit\"at Duisburg-Essen,\\
47048 Duisburg, Germany, \quad sven@thp.uni-duisburg.de\\[2cm]
December 2004}

\date{}
 
\maketitle

\thispagestyle{empty}
%\cleardoublepage
%\thispagestyle{empty}

%\setcounter{page}{1}
%\markright{Journal of Statistical Physics {\bf 102}, 1 (2001)}

%\cleardoublepage
%\thispagestyle{empty}
%\input{_sonstiges/_widmung.tex}
%\thispagestyle{empty}
%\cleardoublepage
%\thispagestyle{empty}
%\input{_sonstiges/_zusammen.tex}
\thispagestyle{empty}
\cleardoublepage
\thispagestyle{empty}
\begin{center}
\LARGE{Summary}
\end{center}

\vspace{0.5cm}

Non-equilibrium critical phenomena have attracted a lot of
research interest in the recent decades.
%Accurate numerical data became available
%in the recent years due to the continuing hardware improvement.
Similar to equilibrium critical phenomena, the concept of
universality remains the major tool to order the great variety 
of non-equilibrium phase transitions systematically.
All systems belonging to a given universality class
share the same set of critical exponents, and certain
scaling functions become identical near the critical point.
It is known that the scaling functions vary
more widely between different universality classes
than the exponents.
Thus, universal scaling functions offer a sensitive 
and accurate test for a system's universality class.
On the other hand, universal scaling functions 
demonstrate the robustness of a given universality class
impressively.
Unfortunately, most studies focus on the determination
of the critical exponents, neglecting the universal
scaling functions.

In this work a particular class of non-equilibrium
critical phenomena is considered, the so-called 
absorbing phase transitions.
Absorbing phase transitions are expected to occur
in physical, chemical as well as biological
systems, and a detailed introduction is presented.
The universal scaling behavior of two different universality 
classes is analyzed in detail, namely
the directed percolation and the Manna universality class.
Especially, directed percolation 
is the most common universality 
class of absorbing phase transitions.
The presented picture gallery of universal scaling functions 
includes steady state, dynamical as well as finite
size scaling functions.
In particular, the effect of an external field conjugated
to the order parameter is investigated.
Incorporating the conjugated field, it is possible
to determine the equation of state,
the susceptibility, and to perform a modified
finite-size scaling analysis appropriate for absorbing phase
transitions.
Focusing on these equations, the obtained results can be
applied to other non-equilibrium continuous
phase transitions observed in numerical simulations
or experiments.
Thus, we think that the presented picture
gallery of universal scaling functions is
valuable for future work.

Additionally to the manifestation of universality classes, universal
scaling functions are useful in order to check
renormalization group results quantitatively.
Since the renormalization group theory is the basis of
our understanding of critical phenomena, it is of fundamental
interest to examine the accuracy of the obtained results.
Due to the continuing improvement of computer hardware,
accurate numerical data have become available,
resulting in a fruitful interplay
between numerical investigations and renormalization
group analyzes.

%\thispagestyle{empty}
%\cleardoublepage
%\thispagestyle{empty}

\tableofcontents

\cleardoublepage
\mainmatter
\chapter{Introduction} 
\setcounter{figure}{0}

\label{chapter:introduction}

\section{Opening remarks and outline}
\label{sec:opening}

One of the most impressive features of continuous
phase transitions is the concept of universality,
that allows to group the great variety of different 
critical phenomena into a small number of universality
classes~(see~\cite{STANLEY_1} for a recent review).
All systems belonging to a given universality class
have the same critical exponents, and certain
scaling functions (equation of state, correlation 
functions, etc.) become identical near the critical point.
%Often one refers the universality also to certain 
%amplitude combinations but these amplitudes
%are just particular values of the scaling 
%functions.
The universality has its origin in the long range character
of the fluctuations.
Close to the transition point, the corresponding
correlation length becomes much larger than the 
typical range of molecular interactions.
Then the behavior of the cooperative phenomena 
becomes independent of the microscopic details
of the considered system.
The concept of universality is well established for 
equilibrium phase transitions
where a unifying theoretical framework exists.
In that case, the universal behavior of short-range 
interacting systems depends only on few fundamental
parameters, namely the dimensionality of space and
the symmetry of the order parameter~\cite{KADANOFF_2}.
Classical examples of such universal behavior 
in critical equilibrium systems are for instance 
the coexistence curve of liquid-vapor systems~\cite{GUGGENHEIM_1} 
and the equation of state in ferromagnetic 
systems~(e.g.~\cite{STANLEY_1,MILOSEVIC_2}).
A complete understanding of the critical behavior 
of a given system would require
to calculate the critical exponents and 
the universal scaling functions exactly.
In general, this is only possible above the 
upper critical dimension where mean field theories
apply.
But the universality ensures that even rather crude modeling
of complicated microscopic behavior provides 
quantitatively many essential features of the
critical behavior.
Therefore, highly accurate estimates of critical exponents 
of various universality classes are known (see e.g.~\cite{PELISSETTO_3}).

In contrast to equilibrium critical phenomena,
less is known
in case of non-equi\-li\-brium phase transitions.
Here, a generalized treatment is not known, 
lacking an analog to the equilibrium free energy.
Thus the rich and often surprising variety of
non-equilibrium phase transitions 
(see for example~\cite{HAKEN_2,HINRICHSEN_1,ODOR_1,WEIDLICH_1,HELBING_1})
observed in
physical, chemical, biological, as well as 
socioeconomic systems,
has to be studied for each system individually.
But similar to equilibrium systems it is believed that
non-equilibrium critical phenomena can be grouped into
universality classes and the concept of universality plays again
a central role in theoretical and numerical analysis.
%Like in equilibrium, the concept of universality is applied 
%and remains the major tool to group the great
%variety of non-equilibrium critical phenomena.
The universality implies in turn that
oversimplified representations or caricatures of
nature provide quantitatively correct results, if
the essential features are captured which are responsible for
non-equilibrium ordering.
Partial differential equations as well as interacting 
lattice models are two established approaches 
to study non-equilibrium systems.
In the first case a set of partial differential equations
is usually constructed on a mean field level 
by directly translating the reaction scheme 
(e.g.~of a chemical reaction) into 
equations for gain and loss of certain 
quantities.
The typically non-linear dynamics is described 
by deterministic equations, and 
phase transitions are related to bifurcations~\cite{HAKEN_2}.
Adding suitably chosen noise functions, 
improved results can be obtained within a
Langevin approach or a Fokker-Planck 
description (see e.g.~\cite{KAMPEN_1,RISKEN_1}).
In that case, field theoretical approaches assisted by
renormalization group techniques are  	successfully
applied to obtain results beyond the mean field level.
On the other hand, microscopic interacting particle systems 
like lattice-gas models or cellular automata~\cite{WOLFRAM_1}
provide another insight into non-equilibrium 
critical phenomena.
Although an exciting development has been seen in the last
decade leading to a series of exact solutions of
interacting particle systems (see e.g.~\cite{SCHUETZ_1}),
most models are not accessible to exact mathematical
treatment, in particular in higher than one dimension.
Thus numerical simulations on increasingly powerful
computers are widely used in order 
to obtain quantitative results.

As pointed out, a full classification of the universality 
classes of non-equilibrium phase transitions is still
lacking, i.e., neither the universality classes nor their 
defining fundamental parameters are known.
Therefore, numerous (mostly phenomenologically 
motivated) classifications schemes are discussed in the 
literature.
These universality hypotheses have to be checked 
model by model.
Due to a lack of analytical solutions,  
numerical simulations or renormalization group treatments are
used (often in a fruitful interplay) to identify 
a system's critical behavior, i.e., to specify the order 
parameter, predicting the order of the transition, and describing
the scaling behavior in the vicinity of the transition point
via critical exponents and scaling functions.
Unfortunately, most work focuses on the determination of the
critical exponents only, neglecting the determination
of the universal scaling functions.
It turns out that checking the universality class it is 
often a more exact test to consider scaling functions 
rather than the values of the 
critical exponents.
While for the latter ones the variations between 
different universality classes are often small, 
the scaling functions may differ significantly.
Thus the agreement of universal scaling functions
provides not only additional but also independent and more convincing 
evidence in favor of the conjecture that the phase
transitions of two models belong to the same universality
class.

It is the aim of this work to demonstrate the usefulness
of universal scaling functions for the analysis of non-equilibrium
phase transitions.
In order to limit the coverage of this article, we do not
present an overview of non-equilibrium phase transitions
as it was done for example in recent review 
articles~\cite{HINRICHSEN_1,ODOR_1}.
Instead we focus on a particular class of non-equilibrium 
critical phenomena, the so-called absorbing phase transitions.
These phase transitions arise from a competition of 
opposing processes, usually creation and annihilation processes.
The transition point separates 
an active phase and an absorbing phase in 
which the dynamics is frozen.
A systematic analysis of universal scaling
functions of absorbing phase transitions is presented, 
including static, dynamical, and finite-size
scaling measurements.
As a result a picture gallery of universal scaling functions 
is presented which allows to identify and to distinguish 
universality classes.

The outline of this work is as follows:~in the remaining
part of the introduction a transparent formulation of
scaling and universality is presented, since both notions 
are central to the understanding of critical phenomena.
Therefore, we follow the historic perspective and 
survey phenomenologically 
the basic ideas of equilibrium critical phenomena, and
discuss the concepts of critical exponents, 
generalized homogenous functions,
scaling forms, as well as universal amplitude combinations.
A foundation for an understanding of scaling
and universality is provided by Wilson's
renormalization group theory~\cite{WILSON_1,WILSON_2}, 
which is a topic on its own
and not presented in this work.
Instead we focus on the implications on universal 
scaling 
%In order to limit the coverage of this article we 
and just illustrate the main results, e.g.~we sketch 
how the renormalization group
theory allows to decide on the relevant
system parameters determining the universality class.
The renormalization group also provides a tool for computing critical 
exponents as well as the universal scaling functions,
and it explains the existence of an upper critical dimension.
For a rigorous substantiation of scaling and universality
the interested reader is referred to established
textbooks~(e.g.~\cite{PFEUTY_1,PLISCHKE_1,YEOMANS_1})  
and review articles~\cite{WILSON_3,FISHER_4,WEGNER_2}.
While we have attempted to present the introduction as general as
possible, we use for the sake of concreteness the language
of ferromagnetic phase transitions.
In conclusion, the subsequent sections can be used as an 
introduction to the theory of phase transitions for readers
unfamiliar with the concepts of scaling and universality.
Readers familiar with equilibrium critical phenomena 
and interested in non-equilibrium systems
are recommended to skip the introduction.

In chapter~\ref{chapter:apt} we introduce the definitions and notations of
absorbing phase transitions.
To be specific, we present a mean field 
treatment of the contact process providing a qualitative
but instructive insight.
Various simulation techniques are discussed for 
both steady state and dynamical measurements.
In particular, the mean field results are 
compared briefly to those obtained from simulations of the 
high-dimensional contact process.
The chapters~\ref{chapter:dp} and~\ref{chapter:aptcf}
are devoted to the investigation
of two different universality classes:~the 
directed percolation universality class 
and the so-called Manna universality class.
According to its robustness and ubiquity
the universality class of\index{directed percolation} 
directed percolation\index{DP} (DP) is recognized 
as the paradigm of the critical behavior
of several non-equilibrium systems exhibiting a 
continuous phase transition from an active 
to an absorbing state.
The widespread occurrence of such models describing critical
phenomena in physics, biology, as well as 
catalytic chemical reactions is reflected by the 
universality hypothesis of Janssen and Grassberger:
Short-range interacting models, which
exhibit a continuous phase transition into a unique 
absorbing state, belong to the directed percolation 
universality class,
provided they are characterized by a one-component order 
parameter~\cite{JANSSEN_1,GRASSBERGER_2}.
Different universality classes are expected to occur 
e.g.~in the presence of additional symmetries or
quenched disorder.
The universality class of directed percolation is well
understood. 
In particular, a field theoretical description is well
established, and renormalization group treatments 
provide within an $\epsilon$-expansion useful
estimates of both the critical exponents and 
the scaling functions.
But it should be remembered that the renormalization group
rarely provides exact results, i.e., it is essential
to have an independent (usually numerical or experimental) check.
These checks are of fundamental interest
since the renormalization group theory is the basis of
our understanding of critical phenomena.
Remarkably, a detailed analysis shows that renormalization
group approximations reveals more accurate results 
for directed percolation than for standard
equilibrium models.
In contrast to directed percolation, less is known in
case of the Manna universality class.
For example, field theoretical approaches using 
renormalization group techniques run into difficulties.
Thus a systematic $\epsilon$-expansion is still lacking, 
and most quantitative results are obtained 
from numerical simulations.
The Manna universality class is of particular interest since
it is related to the concept of self-organized criticality.

For both universality classes considered
several lattice models are investigated.
A systematic analysis of certain scaling functions
below, above as well as at the upper critical
dimension is presented.
At the upper critical dimension, the usual power-law behavior 
is modified by logarithmic corrections.
These logarithmic corrections are well known for
equilibrium critical phenomena, but they have been 
largely ignored for non-equilibrium phase transitions.
Due to the considerably high numerical effort, sufficiently
accurate simulation data have become available recently,
triggering further renormalization group
calculations and vice versa.
The results indicate that the scaling behavior
can be described in terms of universal scaling functions
even at the upper critical dimension.
Independent of the dimension, we investigate
steady state scaling functions, 
dynamical scaling functions, as well as 
finite-size scaling functions.
Focusing on common scaling functions 
such as the equation of state our method of analysis can be applied to 
other non-equilibrium continuous phase
transitions observed numerically or experimentally.
Thus we hope that the presented gallery 
of universal scaling functions will be useful for future
work where the scaling behavior of a given system
has to be identified.
Additionally, certain universal amplitude combinations are considered.
In general, universal amplitude combinations are related
to particular values of the scaling functions.
Systematic approximations of these amplitude combinations
are often provided by renormalization group treatments.
%Since the renormalization group theory is the basis of
%our understanding of critical phenomena it is of fundamental
%interest to check the obtained results.{\bf vielleicht weiter oben}
Furthermore, the validity of certain scaling laws connecting
the critical exponents is tested.

Crossover phenomena between different universality classes
are considered in chapter~\ref{chapter:crossover}.
Although crossover phenomena are well understood 
in terms of competing fixed points, 
several aspects are still open and are discussed in 
the literature.
For example, the question whether the full crossover
region, that spans usually several decades 
in temperature or conjugated field, 
can be described in terms of universal scaling 
functions attracted a lot of research activity.
Here, we consider several models belonging to the Manna universality
class with various interaction range.
Our analysis reveals that the corresponding
crossover from mean field to non-mean field
scaling behavior can be described in terms of universal 
scaling functions. 
This result can be applied to continuous phase
transitions in general, including equilibrium 
crossover phenomena.

Concluding remarks are presented 
in Chapter~\ref{chapter:outlook}.
Furthermore, we direct the reader's attention to areas
where further research is desirable.
The work ends with appendices containing tables
of critical exponents as well as  universal amplitude combinations.

A comment is worth making in order to
avoid confusion about the mathematical notation
of asymptotical scaling behavior.
Throughout this work the\index{Landau notation}
Landau notation \cite{LANDAU_E_1,ABRAMOWITZ_1} 
is used, i.e., the symbols $\,\propto\,$, $\,\sim\,$, and 
$\mathcal{O}$ denote that two functions $f(x)$ and $g(x)$
are 
\begin{displaymath}
\begin{array}{llll}
f(x) \propto g(x) & \Longleftrightarrow &
\frac{\displaystyle f(x)}
{\,\displaystyle g(x)^{\vphantom{X}}\,} 
= c \, , \quad \forall \, x 
& \quad \mathrm{(proportional)}\, ,\\[ 4mm]
f(x) = g(x) + \mathcal{O}(x^n) & \Longleftrightarrow &
\left |\frac{\displaystyle \,f(x)-g(x)\,}
{\displaystyle x^{n \vphantom{X}}} \right |
< c \, ,  \forall \, x > x_{\ssstyle 0}
& \quad \mathrm{(of\;the\;order\;of)}, \\[ 4mm]
f(x) \sim g(x) & \Longleftrightarrow &
\lim\limits_{\ssstyle x\to x_{\mathrm{c}}}
\frac{\displaystyle f(x)}{\displaystyle \,g(x)^{\vphantom{X}}\,} = 1
& \quad \mathrm{(asymptotically\; equal)}\, . 
\end{array}
\end{displaymath}
In the following, the mathematical limit $x\to x_{\ssstyle {\mathrm c}}$ 
corresponds to the physical 
situation that a phase transition is approached 
(usually $x\to 0$ or $x \to \infty$).
Beyond this standard notation the symbol $\,\asympprop\,$ is 
used to denote that the functions are asymptotically
proportional, i.e., 
\begin{eqnarray}
f(x) \asympprop g(x) & \Longleftrightarrow &
\lim_{\ssstyle x\to x_{\mathrm{c}}}
\frac{f(x)}{\,g(x)^{\vphantom{X}}\,} = c 
\quad\quad\quad\;\;\,\mathrm{(asymptotically\; proportional)}. \nonumber
\end{eqnarray}\\

%%%%%%%%%%%%%%%%%%%%%%%%%%%%%%%%%%%%%%%%%%%%%%%%%%%%%%%%%%%%%%%%%%%%%%%%%%%%%%%%%
%%%%%%%%%%%%%%%%%%%%%%%%%%%%%%%%%%%%%%%%%%%%%%%%%%%%%%%%%%%%%%%%%%%%%%%%%%%%%%%%%
%%%%%%%%%%%%%%%%%%%%%%%%%%%%%%%%%%%%%%%%%%%%%%%%%%%%%%%%%%%%%%%%%%%%%%%%%%%%%%%%%

\section{Scaling theory}
\label{sec:scaling_theory}

Phase\index{scaling theory} 
transitions in equilibrium systems are characterized
by a singularity in the free energy and its
derivatives~\cite{ONSAGER_1,YANG_1,LEE_2}.
This singularity causes a discontinuous behavior of 
various physical quantities when the transition point
%or critical point 
is approached.
Phenomenologically the phase transition is described
by an order parameter\index{order parameter}, 
having non-zero value in the ordered
phase and zero value in the disordered phase~\cite{LANDAU_5}. 
Prototype systems for equilibrium phase transitions
are simple ferromagnets\index{ferromagnets}, 
superconductors\index{superconductors}, 
liquid-gas systems\index{liquid-gas},
ferroelectrics\index{ferroelectrics},
as well as systems exhibiting superfluidity\index{superfluidity}.

\begin{figure}[t] 
\centering
%\leavevmode 
\includegraphics[clip,width=13cm,angle=0]{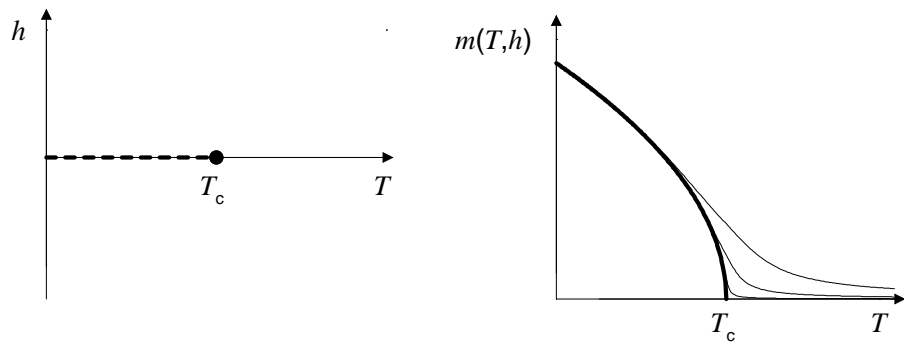}
\caption[Phase diagram and order parameter of a simple ferromagnet]
{Sketch of the
ferromagnetic phase diagram (left) and the corresponding order parameter behavior (right).
The phase diagram contains a line of first-order transitions 
at zero field (dashed line) that ends in a critical point
at temperature $\Tc$.
The temperature dependence of the magnetization is sketched
for zero-field (thick line) and three different values
of the conjugated field.
\label{fig:ferro_mag_01}} 
\end{figure}

In case of ferromagnetic systems the transition point
separates the ferromagnetic phase with non-zero magnetization 
from the paramagnetic phase with zero magnetization.
The well-known phase diagram is 
shown in \reffigname{fig:ferro_mag_01}.
Due to the reversal symmetry of ferromagnets 
%%$S(\underline r) \to -S(\underline r)$
all transitions occur at zero external field~$h$.
The phase diagram exhibits a boundary line along the
temperature axis, terminating at the critical point~$\Tc$.
Crossing the boundary 
for $T< \Tc$ the magnetization
changes discontinuously, 
i.e.~the systems undergoes a first order 
phase transition.
The discontinuity decreases if one approaches the critical
temperature.
At $\Tc$ the magnetization is continuous but its derivatives
are discontinuous.
Here the system undergoes a continuous or so-called second order 
phase transition.
For $T>\Tc$ no singularities of the free energy occur and
the systems changes continuously from a state
of positive magnetization to a state of negative magnetization.

In zero field the high temperature or paramagnetic phase 
is characterized by a vanishing magnetization.
Decreasing the temperature, a phase transition takes place
at the critical temperature and for $T<\Tc$ one
observes an ordered phase which is spontaneously
magnetized (see \reffigname{fig:ferro_mag_01}).
Following Landau, the magnetization~$m$ is the order 
parameter\index{order parameter}
of the ferromagnetic phase transition~\cite{LANDAU_5}. 
Furthermore the temperature~$T$ is the control parameter 
of the phase transition and the external field~$h$ is conjugated
to the order parameter\index{conjugated field}.
As well known, critical systems are characterized by 
power-laws sufficiently close to the critical point,
e.g.~the behavior of the order parameter can be 
described by 
\begin{equation}
m(T,h=0) \; \asympprop \; (-\deltaT)^{\beta}
\label{eq:mag_temp_beta}
\end{equation}   
with the reduced temperature $\deltaT=(T-\Tc)/\Tc$
and the exponent~$\beta$.

For non-zero field the magnetization increases smoothly
with decreasing temperature.
At the critical isotherm ($\deltaT=0$) the magnetization
obeys another power-law for $h\to 0$
\begin{equation}
m(T=\Tc,h) \; \asympprop \; | h |^{1/\delta} \, .
\label{eq:mag_field_delta}
\end{equation}   
Further exponents are introduced to describe the 
singularities of the
order parameter susceptibility~$\chi$,
the specific heat~$C$, as well as the correlation length~$\xi$
\begin{eqnarray}
\label{eq:mag_sus_gamma}
\chi (T,h=0) & \asympprop &  |\deltaT |^{-\gamma} \, , \\
\label{eq:mag_spech_alpha}
C(T,h=0)     & \asympprop &  |\deltaT |^{-\alpha} \, , \\
\label{eq:mag_corrl_nu}
\xi (T,h=0)  & \asympprop &  |\deltaT |^{-\nu} \, .
\end{eqnarray}  
Another exponent~$\eta$ describes the spatial 
decay of the correlation function at criticality
\begin{equation}
\Gamma(r)  \; \asympprop  \; r^{-D+2-\eta} \, ,
\label{eq:corr_func_eta} 
\end{equation} 
where $D$ denotes the dimensionality of the system.
%Thus, the exponent $\eta$ describes the deviations from the
%Ornstein-Zernicke behavior (see, e.g.~\cite{YEOMANS_1,STANLEY_2}).
%{\bf vielleicht diesem Satz ganz wegnehmen, ist sonst nicht
%klar, was das ist}

A comment about vanishing exponents is worth making.
A zero exponent corresponds either to a jump 
of the corresponding quantity at the critical point or 
to a logarithmic singularity since 
\begin{equation}
-\ln{|\deltaT|} \; = \; \lim_{s\to 0}
{\frac{\,|\deltaT|^{-s}-1\,}{s}} \, .
\label{eq:log_sing_exp_zero}
\end{equation}
Often, it is notoriously difficult to distinguish
from experimental or numerical data 
a logarithmic singularity from a small positive value
of the exponent.

The exponents $\alpha$, $\beta$, $\gamma$, $\delta$,
$\eta$, and $\nu$ are called critical exponents\index{critical exponents}.
Notice in \refeqsname{eq:mag_sus_gamma}{eq:mag_corrl_nu}
the equality of the critical exponents below and
above the critical point, which is an
assumption of the scaling theory. 
%verified by experiments
%and simulations.
The phenomenological scaling theory was developed by 
several authors in the 1960s
(e.g.~\cite{ESSAM_2,DOMB_2,WIDOM_1,KADANOFF_3,PATASHINSKII_1,GRIFFITHS_2})
and has been well verified by experiments as well as simulations.
In particular, the scaling theory predicts that 
the\index{scaling law, Rushbrooke}\index{scaling law, Widom} 
critical\index{scaling law, Fisher}\index{scaling law, Josephson} 
exponents mentioned above\index{Rushbrooke scaling law}\index{Widom scaling law}
are\index{Fisher scaling law}\index{Josephson scaling law} 
connected by the 
scaling laws\index{scaling law}
\begin{eqnarray}
\label{eq:rushbrooke}
\alpha+2\beta + \gamma  \; = \; 2  \quad & ~ & (\mathrm{Rushbrooke}), \\
\label{eq:widom}
\gamma \; = \; \beta\,(\delta-1)   \quad & ~ & (\mathrm{Widom}), \\
\label{eq:fisher}
\gamma \; = \; (2-\eta) \, \nu     \quad & ~ & (\mathrm{Fisher}), \\
\label{eq:josephson}
2-\alpha  \; = \;  \nu \, D        \quad & ~ & (\mathrm{Josephson}) \, .
\end{eqnarray}  
The Josephson law includes the spatial dimension~$D$ of the 
system and is often termed the hyperscaling 
relation\index{hyperscaling}.
In this way the critical behavior of an equilibrium
system is determined by two independent critical exponents.

The scaling theory rests on the assumption that 
close to the critical point the singular part of 
a particular thermodynamic potential is 
asymptotically a generalized homogeneous\index{generalized homogeneous function} 
function~(see e.g.~\cite{HANKEY_1}).
A function $f(x_{\ssstyle 1},x_{\ssstyle 2},\ldots)$ is a 
generalized homogeneous function if it satisfies
the following equation for all positive values of $\lambda$ 
\begin{equation}
\lambda \; f(x_{\ssstyle 1},x_{\ssstyle 2}\ldots)
\; = \;
f(\lambda^{s_{\ssstyle 1}} x_{\ssstyle 1},
\lambda^{s_{\ssstyle 2}} x_{\ssstyle 2},\ldots) \, .
\label{eq:def_gen_hom_func}
\end{equation}
The exponents $s_{\ssstyle 1}, s_{\ssstyle 2},\ldots$ are
usually termed scaling powers\index{scaling powers} and the variables 
$x_{\ssstyle 1}, x_{\ssstyle 2},\ldots$ are termed 
the scaling fields\index{scaling field}.
In case of ferromagnetism the singular part of the
Gibbs\index{Gibbs potential} potential per spin
$g_{\ssstyle \mathrm {sing}}(\deltaT,h)$ 
is assumed to scale asymptotically as 
\begin{equation}
g_{\ssstyle \mathrm {sing}}(\deltaT,h) \; \sim \; \lambda^{\alpha-2}
\; \; {\tilde g}(\deltaT \, \lambda, h \, \lambda^{\beta \delta} ) \, ,
\label{eq:gibbs_pot_hom_func}
\end{equation}
with the scaling function ${\tilde g}(x,y)$.
The scaling power of the conjugated field is often
denoted as the gap-exponent~$\Delta=\beta\, \delta$.\index{gap exponent}
Strictly speaking, \refeqname{eq:gibbs_pot_hom_func} is 
only asymptotically valid, i.e., only when $\deltaT$
and $h$ tend to zero.
Corrections occur away from this limit.
It turns out that all Legendre transformations of
generalized homogenous functions and all partial derivatives
of generalized homogenous functions are also 
generalized homogenous functions.
Thus with the Gibbs potential all thermodynamic
potentials and all thermodynamic functions that are
expressible as derivatives of thermodynamic potentials,
like the magnetization, specific heat, etc.,
are generalized homogeneous functions.

%Consider, for example, the specific heat.
%Differentiating 
%\refeqname{eq:gibbs_pot_hom_func}
%with respect to the temperature we find
%\begin{equation}
%c_{\ssstyle h, \mathrm {sing}}(\deltaT,h) 
%\; = \; - \, T \, 
%\left ( \frac{\partial^2 g_{\ssstyle \mathrm {sing}}}{\partial T^2}
%\right )_h
%\; \sim \;
%\frac{T}{\Tc^2} \; \lambda^{\alpha}
%\; {\tilde c}(\deltaT \, \lambda, h \, \lambda^{\beta \delta} ) \, ,
%\label{eq:spec_heat_hom_func_01}
%\end{equation}
%where the scaling function is given by 
%\begin{equation}
%{\tilde c}(x,y) \; = \; - \,
%\left (
%\frac{\partial^2 {\tilde g}(x,y)}{\partial x^2} 
%\right )_y
%\end{equation}
%Since this equation is valid for all positive values of $\lambda$
%it holds for $\lambda = 1/ |\deltaT |$, 
%hence one obtains at zero-field
%\begin{equation}
%c_{\ssstyle h, \mathrm {sing}}(\deltaT,0) 
%\; = \; |\deltaT |^{-\alpha} \;
%\frac{1}{\Tc} \; 
%{\tilde c}(\plmi 1, 0) \, .
%\label{eq:spec_heat_hom_func_02}
%\end{equation}
%We can expect that 
%${\tilde c}(+1,0) \neq {\tilde c}(-1,0)$ in general, i.e.,
%the amplitude of the power-law 
%behavior \refeqname{eq:spec_heat_hom_func_02} is different
%below ($T<\Tc$) and above ($T>\Tc$) the transition. 

Consider, for example, the magnetization
and the corresponding susceptibility.
Differentiating 
\refeqname{eq:gibbs_pot_hom_func}
with respect to the conjugated field we find\index{conjugated field}
\begin{eqnarray}
\label{eq:mag_hom_func_01}
m(\deltaT,h) 
& = & - 
\left ( \frac{\partial \, g_{\ssstyle \mathrm {sing}}}{\partial \, h}
\right )_T
\; \sim \;
\lambda^{\alpha-2+\beta \delta}
\; \; {\tilde m}(\deltaT \, \lambda, h \, \lambda^{\beta \delta} ) \, , \\
\label{eq:sus_hom_func_01}
\chi(\deltaT,h) 
& = & - 
\left ( \frac{\partial^2 \, g_{\ssstyle \mathrm {sing}}}{\partial \, h^2}
\right )_T
\; \sim \;
\lambda^{\alpha-2+2\beta \delta}
\; \; {\tilde \chi}(\deltaT \, \lambda, h \, \lambda^{\beta \delta} ) \, ,
\end{eqnarray}
where the scaling functions are given by 
\begin{equation}
\label{eq:mag_hom_func_02}
{\tilde m(x,y)} 
 \; = \;  -  
\left (
\frac{\partial \, {\tilde g}(x,y)}{\partial \, y} 
\right )_x \, , \quad \quad \quad 
%\label{eq:sus_hom_func_02}
{\tilde \chi(x,y)} 
 \; = \;  -  
\left (
\frac{\partial^2 \, {\tilde g}(x,y)}{\partial \, y^2} 
\right )_x \, .
\end{equation}
Since the \refeqsname{eq:mag_hom_func_01}{eq:sus_hom_func_01}
are valid for all positive
values of $\lambda$ they hold for $\lambda = 1/ |\deltaT |$, 
hence one obtains at zero-field
\begin{eqnarray}
\label{eq:mag_hom_func_03}
m(\deltaT,0) 
& \sim & 
(-\deltaT )^{-(\alpha-2+\beta \delta)}
\; \; {\tilde m}(-1,0) \, , \\
\label{eq:sus_hom_func_03}
\chi(\deltaT,0) 
& \sim &  
|\deltaT |^{-(\alpha-2+2\beta \delta)}
\; \; {\tilde \chi}(\plmi 1,0) \, ,
\end{eqnarray}
where the magnetization is defined for $\deltaT <0$ only.
Notice that one expects 
${\tilde \chi}(+1,0) \neq {\tilde \chi}(-1,0)$ in general, i.e.,
the amplitudes of the susceptibility are different 
below ($T<\Tc$) and above ($T>\Tc$) the transition. 
Comparing with \refeqs2name{eq:mag_temp_beta}{eq:mag_sus_gamma}
we find 
\begin{equation}
\beta \; = \; -\alpha+2-\beta\,\delta \, ,
\quad \quad \quad
\gamma \; = \; \alpha-2+2\, \beta\, \delta  \, ,
\label{eq:rushbrook_widom_herleitung}
\end{equation}
which leads directly to the Rushbrook [\refeqname{eq:rushbrooke}] 
and Widom [\refeqname{eq:widom}] 
scaling law.
The Fisher and Josephson scaling laws can be obtained
in a similar way from the scaling form of the
correlation function (see e.g.~\cite{YEOMANS_1}).
In order to obtain the Josephson law both thermodynamic scaling forms
and correlation scaling forms have to be combined.
Scaling relations obtained in this way are usually
termed\index{hyperscaling} hyperscaling laws  
and do not hold above the upper-critical dimension.

\begin{figure}[t] 
\centering
%\leavevmode 
\includegraphics[clip,width=13cm,angle=0]{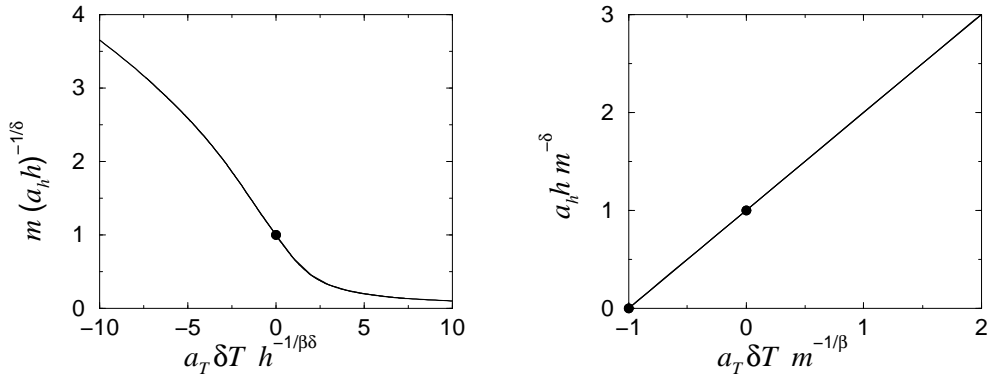}
\caption[Scaling forms of the equation of state of a simple ferromagnet]
{The data-collapsing of the $M$-$h$-$t$ equation of state. 
The data correspond to the three different magnetization curves
of \reffigname{fig:ferro_mag_01}.
All data points collapse onto the curves 
${\tilde m}(x,1)$ (left, Hankey-Stanley scaling 
form~[\refeqname{eq:rescaled_mag_01}])
and ${\tilde h}(x,1)$ (right, Widom-Griffiths scaling 
form~[\refeqname{eq:rescaled_mag_01}]),
respectively.
The so-called metric factors are defined in 
section \protect\ref{sec:equil_uni_hyp}.
The circles mark the normalization 
conditions \refeqs2name{eq:normalization_M}{eq:normalization_H}.
\label{fig:ferro_mag_02}} 
\end{figure}

The scaling theory implies still more. 
Consider for instance the $M$-$h$-$t$ equation of 
state [\refeqname{eq:mag_hom_func_01}].\index{equation of state}
Choosing $h \lambda^{\beta\delta}=1$
in \refeqname{eq:mag_hom_func_01}
we find 
\begin{equation}
m(\deltaT,h) 
\; \sim \;
\left .
\lambda^{-\beta} \; \; 
{\tilde m}(\deltaT \, \lambda, h \, \lambda^{\beta \delta} ) 
\right |_{\lambda=h^{-1\beta\delta}}
\; = \;
h^{1/ \delta}
\; \; {\tilde m}(\deltaT \, h^{-1/\beta\delta},1 ) \, .
\label{eq:mag_field_scal_01}
\end{equation}
At the critical isotherm ($\deltaT=0$) we 
recover \refeqname{eq:mag_field_delta}.
Furthermore, the equation of state may be written in the
rescaled form 
\begin{equation}
m_{\ssstyle h} \; \sim \; 
{\tilde m}(\deltaT_{\ssstyle h},1)
\quad \quad {\mathrm {with}} \quad \quad 
m_{\ssstyle h} = m \, h^{-1/ \delta} \, , \quad 
\deltaT_{\ssstyle h} = \deltaT \, h^{-1/ \beta\delta} \, .
\label{eq:rescaled_mag_01}
\end{equation}
In this way the equation of state is described by the 
single curve ${\tilde m}(x,1)$ and all $M$-$h$-$t$ data 
points will collapse onto the single curve ${\tilde m}(x,1)$ if one plots
the rescaled order parameter $m_{\ssstyle h}$ as a function
of the rescaled control parameter $\deltaT_{\ssstyle h}$.
This data-collapsing is shown in \reffigname{fig:ferro_mag_02}
for the magnetization curves of \reffigname{fig:ferro_mag_01}.

A different scaling form is obtained if one considers
instead of the Gibbs\index{Gibbs potential} potential the 
Helmholtz potential\index{Helmholtz potential} $a = g + m \,h$.
Since Legendre transforms of generalized homogeneous
functions are also generalized homogeneous functions
the singular part of the Helmholtz potential obeys
\begin{equation}
a_{\ssstyle \mathrm {sing}}(\deltaT,m) \; \sim \; 
\lambda^{\alpha-2}
\; \; {\tilde a}(\deltaT \, \lambda, m \, \lambda^{\beta} ) \, .
\label{eq:helmholtz_pot_hom_func}
\end{equation}
This equation leads to the scaling form of the conjugated field
\begin{equation}
h(\deltaT,m) 
\; = \;  
\left ( \frac{\partial\, a_{\ssstyle \mathrm {sing}}}{\partial\, m}
\right )_T
\; \sim \;
\lambda^{-\beta\delta}
\; \; {\tilde h}(\deltaT \, \lambda, m \, \lambda^{\beta} ) \, . 
\end{equation}
Choosing $m \, \lambda^{\beta}=1$ we find 
\begin{equation}
h_{\ssstyle m} \; \sim \; 
{\tilde h}(\deltaT_{\ssstyle m},1)
\quad \quad {\mathrm {with}} \quad \quad 
h_{\ssstyle m} = h \, m^{-\delta} \, , \quad 
\deltaT_{\ssstyle m} = \deltaT \, m^{-1/ \beta} \, .
\label{eq:rescaled_conj_field_01}
\end{equation}
Both, ${\tilde h}(x,1)$ and ${\tilde m}(x,1)$ are analytic
in the neighborhood of $x=0$, i.e., at the critical temperature.
The function ${\tilde h}(x,1)$ is often called the 
Widom-Griffiths scaling\index{Widom-Griffiths scaling form}\index{scaling forms}
function~\cite{WIDOM_1,GRIFFITHS_2}
whereas we term in the following ${\tilde m}(x,1)$
as the Hankey-Stanley\index{Hankey-Stanley scaling form}
scaling function~\cite{HANKEY_1}.
%% Widom fuer liquid gas
%% Griffiths fuer magnetismus
The corresponding data-collapses for both scaling forms 
are presented in \reffigname{fig:ferro_mag_02}.
The Hankey-Stanley scaling form is just the order parameter
curve as a function of the control parameter in a fixed 
conjugated field.
It is therefore the natural and perhaps more elegant way to present 
data-collapses of the equation of state.\index{equation of state}
%%Pfeuty bezeichnet es so
But often the mathematical forms of the Hankey-Stanley 
functions are rather complicated whereas the 
Widom-Griffiths scaling forms are analytically tractable.
Therefore ${\tilde h}(x,1)$ is often calculated within
certain approximation schemes, e.g.~$\epsilon$- or $1/n$-expansions
within a renormalization group approach.

Notice that the definition of a generalized homogeneous 
function [\refeqname{eq:def_gen_hom_func}]
and the data-collapse representations\index{generalized homogeneous function} 
\begin{eqnarray}
\label{eq:def_gen_hom_func_data_coll_x1}   
|x_{\ssstyle 1}|^{-1/s_{\ssstyle 1}} \; f(x_{\ssstyle 1},x_{\ssstyle 2})
& = &
f( \plmi1, x_{\ssstyle 2} \, |x_{\ssstyle 1}|^{-s_{\ssstyle 2}/s_{\ssstyle 1}} ) \, ,\\
\label{eq:def_gen_hom_func_data_coll_x2}
|x_{\ssstyle 2}|^{-1/s_{\ssstyle 2}} \; f(x_{\ssstyle 1},x_{\ssstyle 2}) 
& = &
f( x_{\ssstyle 1} \, |x_{\ssstyle 2}|^{-s_{\ssstyle 1}/s_{\ssstyle 2}},  \plmi1)
\end{eqnarray}
are mathematically equivalent, i.e., a function $f(x_{\ssstyle 1},x_{\ssstyle 2})$
is a generalized homogeneous function if and only if 
\refeqname{eq:def_gen_hom_func_data_coll_x1}, 
or equally \refeqname{eq:def_gen_hom_func_data_coll_x2},
is fulfilled~\cite{HANKEY_1}.
This was important for the phenomenological formulation of the 
scaling theory since it has been borne 
and has been confirmed
by numerous data-collapses of experimental and numerical data.

%
%where Widom used a notation appropiate for fluids
%rather than for magnets as used by Griffiths.
%

%Notice that the scaling function of the Gibbs potential
%${\tilde g}$ and therefore ${\tilde a}$, ${\tilde m}$, 
%${\tilde h}$, etc.~depend on the details  

%As well known all equilibrium thermodynamic variables
%are determined if a thermodynamic potential, e.g.~the
%Helmholtz free energy is given.
%However, a knowlegde of the equation of state  together
%with the phase boundary curve is almost as good in the 
%vicinity of the critical point if we are interested
%in the scaling behavior in this region (nochmal drueber nachdenken). 
% aus Widom paper

\section{Universality}
\label{sec:equil_uni_hyp}

According to the scaling laws \refeqsname{eq:rushbrooke}{eq:josephson}
equilibrium phase transitions are characterized by two 
independent critical exponents.
In the 1950s and 1960s it was experimentally recognized
that quantities like~$\Tc$
depend sensitively on the details of the interactions
whereas the critical exponents are universal, i.e., 
they depend only on a small number of general features.
This led to the concept of universality which\index{universality hypothesis}
was first clearly phrased by Kadanoff~\cite{KADANOFF_2},
but based on earlier works 
including e.g.~\cite{FISHER_3,JASNOW_1,WATSON_1,GRIFFITHS_1,BETTS_1}.
The universality hypothesis reduces the great
variety of critical phenomena to a small number of
equivalence classes, so-called universality\index{universality classes} classes,
which depend only on few fundamental parameters.
All systems belonging to a given universality class
have the same critical exponents\index{critical exponents} 
and the corresponding scaling functions become identical near 
the critical point.
For short range interacting equilibrium systems the fundamental 
parameters determining the universality class
are the symmetry of the order parameter and the dimensionality
of space~\cite{KADANOFF_2,GRIFFITHS_1}.
The specific nature of the transition, i.e.~the
details of the interactions, such as the lattice
structure and the range of interactions 
(as long as it remains finite)
do not affect the scaling behavior.
%This allows to group the great variety of different models showing
%critical phenomena into a small number of 
%universality classes\index{universality classes}.
For instance, ferromagnetic systems with one axis of easy magnetization
are characterized by a one component order 
parameter ($n=1$) and belong to the universality
class of the Ising\index{universality class, Ising}  ferromagnet.
Furthermore, liquid-gas 
transitions~\cite{HUBBARD_1,BRILLIANTOV_1,HOCKEN_1,LEVELT_1,SENGERS_1,LEVELT_2}, 
binary mixtures of liquids~\cite{BALZARINI_1},
and systems exhibiting an order-disorder 
transition in alloys such as beta-brass
are described by a scalar order parameter, too, 
and belong therefore to the same universality 
class (see e.g.~\cite{ALSNIELSEN_2}).
Even phase transitions occurring in high-energy physics
are expected to belong to the Ising universality class.
For example, within the electroweak theory the early universe 
exhibits a transition from a symmetric (high temperature) phase
to a broken so-called Higgs phase~\cite{KAJANTIE_1}.
The occurring line of first order phase transitions
terminates at a second order point which is argued to belong 
to the Ising class.

Ferromagnetic systems with a plane of easy magnetization
are characterized by a two-component\index{universality class, XY-model} 
order parameter ($n=2$).
%Ferromagnetic ordering breaks the rotational symmetry below
%the critical temperature.
Representatives of this universality class are the XY-model,
superconductors,
as well as liquid crystals that undergo a phase transition
from a smectic-A to a nematic phase~\cite{HALPERIN_1,HALPERIN_2,CHAIKIN_1}.
But the most impressive prototype is the superfluid
transition of $^4{\mathrm{He}}$ along the $\lambda$-line.
Due to its characteristic features like the 
purity of the samples, the weakness of the singularity 
in the compressibility of the fluid, as well as the 
reasonably short thermal relaxation times, superfluid
$^4{\mathrm{He}}$ is more suitable for high-precision 
experiments than any other system~\cite{AHLERS_1}.
For example, orbital heat capacity measurements of liquid
% im Orbit, da Tc vom Druck abhaengt, daher Druckgradienten in Probe auf Erde
helium to within $2\,\mathrm{nK}$ from the $\lambda$-transition
provide the estimate $\alpha=-0.01285 \pm 0.00038$~\cite{LIPA_1,LIPA_2}.
Thus, the $\lambda$-transition of $^4\mathrm{He}$ 
offers an exceptional opportunity to test theoretical predictions, 
obtained in particular from renormalization group 
theory (see e.g.~\cite{DOHM_1} and references therein).
%A particular feature of this universality class in two-dimensional
%systems, like thin films of superfluid Helium-4, 
%is the 
%%absence of conventional long-range~\cite{MERMIN_1}
%%in two-dimensional systems.
%Kosterlitz-Thouless phase \index{Kosterlitz-Thouless phase transition}
%transition~\cite{KOSTERLITZ_1,KOSTERLITZ_2}(Plischke Bergersen).

The well known Heisenberg universality class describes
isotropic ferromagnetic systems that are characterized
by a three component 
order parameter ($n=3$).\index{universality class, Heisenberg}
Despite of the Ising, XY, and Heisenberg 
universality classes other classes with $n\ggl 4$
are discussed in the literature.
For instance, the $n=5$ universality class is expected
to be relevant for the description of high-$\Tc$ 
superconductors~\cite{ZHANG_2,DEMLER_1}
whereas $n=18$ is reported in the context of 
superfluid $^3{\mathrm{He}}$~\cite{JONES_1}.
Furthermore, the limit $n\to 0$ corresponds
to the critical behavior of polymers and self-avoiding 
random walks~\cite{GENNES_1}.
\index{polymers}\index{random walk, self-avoiding}
The other limiting case $n\to \infty$ corresponds
to the exactly\index{spherical model} solvable spherical
model~\cite{BERLIN_1,STANLEY_3,JOYCE_1}.

%Performing an analytical continuation of $n$ one can
%identify further universality classes, for example 
%$n=0$ (e.g.~polymers, self-avoiding random walks),

%Phenomenologically the universality is explained 
%by the singular behavior of the correlation length.
%Approaching the transition point the correlation
%length diverges and is therefore significant larger
%one all other lengths scales of the system that
%are introduced by the interaction details (nicht schoen,
%noch besser formulieren).

In the scaling theory presented above, the exponents
are already universal but the scaling functions
are non-universal.
Thus two systems are characterized by e.g.~two different 
scaling functions ${\tilde m}$ although they are in
the same universality class.
The non-universal features can be absorbed into two
non-universal parameters leading to the universal
scaling function ${\tilde M}$
%{\bf oder doch den anderen Satz}
%The corresponding universal scaling function
%${\tilde M}$
%can be easily determined from ${\tilde m}$ by introducing
%non-universal metric factors
\index{universal scaling functions}
\begin{equation}
c_{\ssstyle m} \, m(\deltaT,h) 
\; \sim \;
\lambda^{-\beta} \; \; 
{\tilde M}(c_{\ssstyle T} \deltaT \, \lambda, 
c_{\ssstyle h} h \, \lambda^{\beta \delta} ) 
  \, .
\label{eq:mag_field_uni_scal_01}
\end{equation}
In this way, the universal scaling function ${\tilde M}$
is the same for all models belonging to a given universality
class whereas all non-universal system-dependent features
are contained in the metric factors $c_{\ssstyle m}$, $c_{\ssstyle T}$, 
and $c_{\ssstyle h}$~\cite{PRIVMAN_3}.
Since this scaling form is valid for all positive values 
of~$\lambda$, the number of metric factors can be reduced
by a simple transformation.
For the sake of convenience we choose 
$c_{\ssstyle m}^{1/\beta} \lambda \mapsto \lambda$, yielding
\begin{equation}
m(\deltaT,h) 
\; \sim \;
\lambda^{-\beta} \; \; 
{\tilde M}(a_{\ssstyle T} \deltaT \, \lambda, 
a_{\ssstyle h} h \, \lambda^{\beta \delta} ) \, ,
\label{eq:universal_HS_form}
\end{equation}
with $a_{\ssstyle T}=c_{\ssstyle T} c_{\ssstyle m}^{-1/\beta}$ 
and $a_{\ssstyle h}=c_{\ssstyle h} c_{\ssstyle m}^{-\delta}$, 
respectively.
The universal scaling function ${\tilde M}$ is normed by the conditions
\begin{equation}
{\tilde M}(-1,0) \; = \; 1 \, , \quad \quad \quad 
{\tilde M}(0,1) \; = \; 1 \, ,
\label{eq:normalization_M}
\end{equation}
and the non-universal metric factors are determined by the
amplitudes of 
\begin{eqnarray}
\label{eq:deter_aT_01}
m(\deltaT,h=0) 
& \sim &
\left .
\lambda^{-\beta} \; \; 
{\tilde M}(a_{\ssstyle T} \deltaT \, \lambda, 0 )
\right |_{a_{\ssstyle T} \deltaT \, \lambda=-1} 
\; = \; (-a_{\ssstyle T} \deltaT)^{\beta} \, ,\\
m(\deltaT=0,h) 
& \sim &
\left .
\lambda^{-\beta} \; \; 
{\tilde M}(0,a_{\ssstyle h} h \, \lambda^{\beta \delta}  )
\right |_{a_{\ssstyle h} h \, \lambda^{\beta \delta}=1} 
\; = \; (a_{\ssstyle h} h)^{1/\delta} \, .
\end{eqnarray}
Analogously, the universal Widom-Griffiths scaling form
is given by
\begin{equation}
a_{\ssstyle h}\, 
h(\deltaT,m) 
\; \sim \;
\lambda^{-\beta \delta} \; \; 
{\tilde H}(a_{\ssstyle T} \deltaT \, \lambda, 
 m\, \lambda^{\beta} ) \, .
\label{eq:universal_WG_form}
\end{equation}
The universal scaling function ${\tilde H}$ is usually normed
by the conditions
\begin{equation}
{\tilde H}(-1,1) \; = \; 0  \, , \quad \quad \quad 
{\tilde H}(0,1) \; = \; 1 \, , 
\label{eq:normalization_H}
\end{equation}
which correspond to \refeqname{eq:normalization_M}, i.e.,
the metric factors are again determined by the amplitudes
of the power-laws $m(\deltaT,h=0) \sim (-a_{\ssstyle T} \deltaT)^{\beta}$
and $m(\deltaT=0,h) \sim (a_{\ssstyle h} h)^{1/\delta}$,
respectively.

For example, let us consider a mean field theory of a 
simple\index{ferromagnets} 
ferromagnet.
\index{ferromagnets}
Following Landau, the\index{mean field theory, ferromagnets} 
free-energy\index{Landau theory}
is given by~\cite{LANDAU_5}
\begin{equation}
F(\deltaT,h,m) - F_{\ssstyle 0}\; = \; 
\frac{1}{2} \, b_{\ssstyle 2} \, \deltaT \, m^2 
\, + \,
\frac{1}{4} \, b_{\ssstyle 4} \, m^4 
\, - \, h \, m \, ,
\label{eq:Landau_freeenergy_fm}
\end{equation}
where the positive factors $b_{\ssstyle 2}$ and $b_{\ssstyle 4}$
are system dependent non-universal parameters.
Variation of the free energy with respect to the 
magnetization yields the equation of state\index{equation of state}
\begin{equation}
h \; = \; 
b_{\ssstyle 2} \, \deltaT \, m 
\, + \,
b_{\ssstyle 4} \, m^3 \, .
\label{eq:Landau_mht_fm}
\end{equation}
At zero-field we find
\begin{equation}
m \, = \, 0 \quad\quad\quad \mathrm{or}\quad\quad\quad 
m \, = \, \sqrt{ - \frac{b_{\ssstyle 2}}{b_{\ssstyle 4}} \, \deltaT \, }
\quad \Longrightarrow \quad 
\beta=\frac{1}{2} \, , \; \; 
a_{\ssstyle T} = \frac{b_{\ssstyle 2}}{b_{\ssstyle 4}} \, ,
\label{eq:Landau_mt_fm}
\end{equation}
whereas at the critical isotherm
\begin{equation}
m \, = \, \sqrt[3]{ \frac{1}{b_{\ssstyle 4}} \, h \,}
\quad \Longrightarrow \quad 
\delta=3 \, , \; \; 
a_{\ssstyle h} = \frac{1}{b_{\ssstyle 4}} \, .
\label{eq:Landau_mh_fm}
\end{equation}
From \refeqname{eq:Landau_mht_fm} it follows directly
\index{universal scaling functions}
\begin{equation}
{\tilde H}(x,y) \; = \; x \, y \, + \, y^3
\label{eq:Landau_H_fm}
\end{equation}
and in particular the universal Widom-Griffiths scaling form 
\index{Widom-Griffiths scaling form}
\begin{equation}
a_{\ssstyle h} h \, m^{-\delta} 
\; = \;
{\tilde H}(a_{\ssstyle T} \deltaT \, m^{-1/\beta},1)
\quad\quad \mathrm{with} \quad\quad
{\tilde H}(x,1) \; = \; 1\, +\, x \,.
\label{eq:Landau_WG_fm}
\end{equation}
On the other hand the form  
${\tilde h}(x,1) = b_{\ssstyle 4} + b_{\ssstyle 2} x$
is obviously non-universal.
Since the magnetization is a cube root 
function ($\delta=3$)
the universal Hankey-Stanley scaling form
is more complicated than the corresponding Widom-Griffiths
form
\index{universal scaling functions}
\index{Hankey-Stanley scaling form}
\begin{equation}
{\tilde M}(x,y>0) \; = \; 
\frac{\left ( 27 y + \sqrt{108 x^3+729 y^2} \right )^{1/3}}
{3 \, 2^{1/3}}
\, - \,
\frac{2^{1/3} \, x}
{\left ( 27 y + \sqrt{108 x^3+729 y^2}\right )^{1/3}}
\, .
\label{eq:Landau_HS_fm}
\end{equation}
The data presented in \reffigs2name{fig:ferro_mag_01}{fig:ferro_mag_02}
correspond to the mean field treatment presented above.
In addition to the universal scaling functions 
${\tilde M}(x,1)$ and ${\tilde H}(x,1)$ the normalizations
\refeqs2name{eq:normalization_M}{eq:normalization_H} 
are marked in \reffigname{fig:ferro_mag_02}.

\begin{figure}[t] 
\unitlength1.0cm   
\begin{picture}(14,7)
\put(0,0){\includegraphics[clip,width=7cm,angle=-0.55]{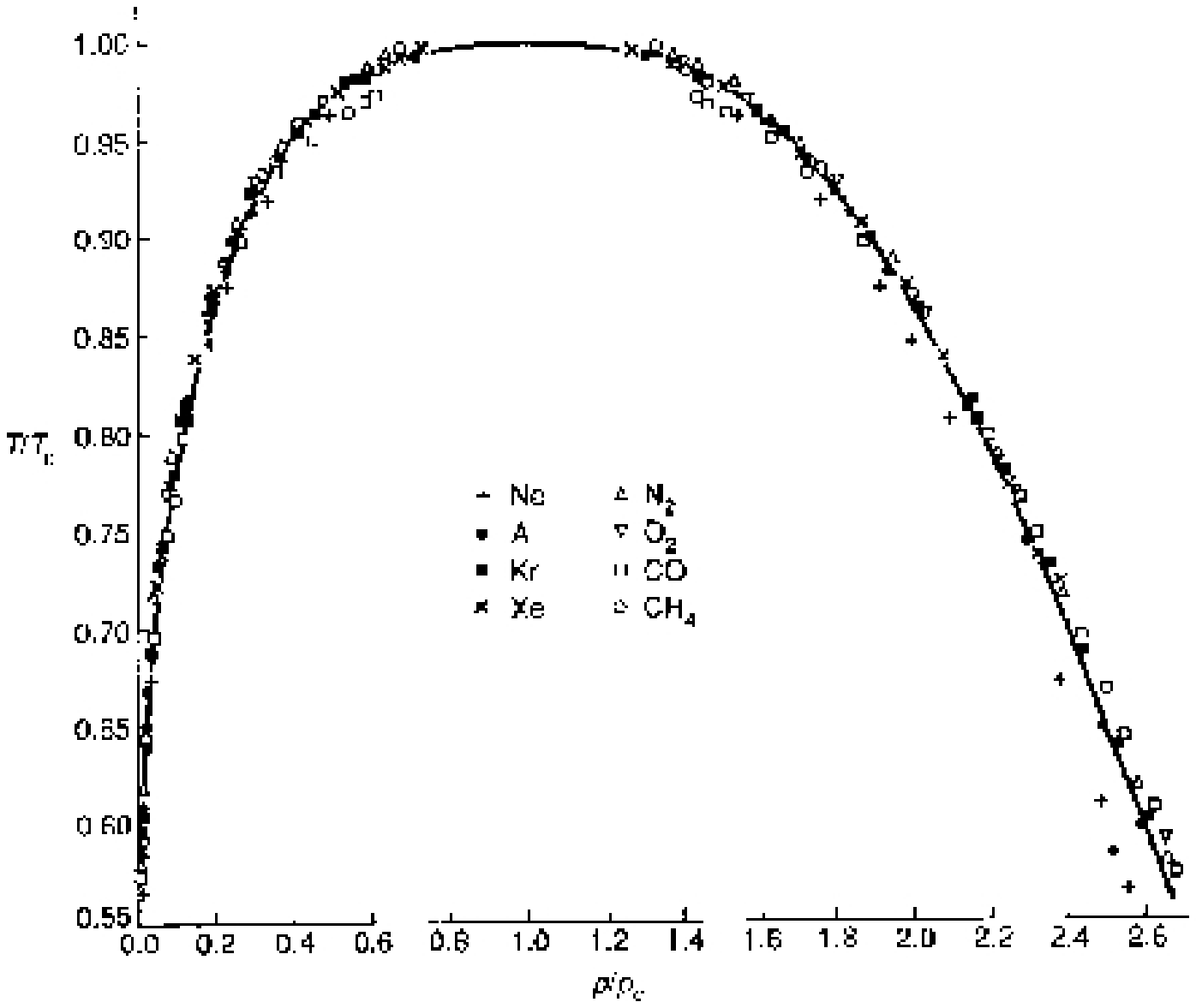}}
\put(8.3,5.6){\includegraphics[clip,width=6cm,angle=-89.8]{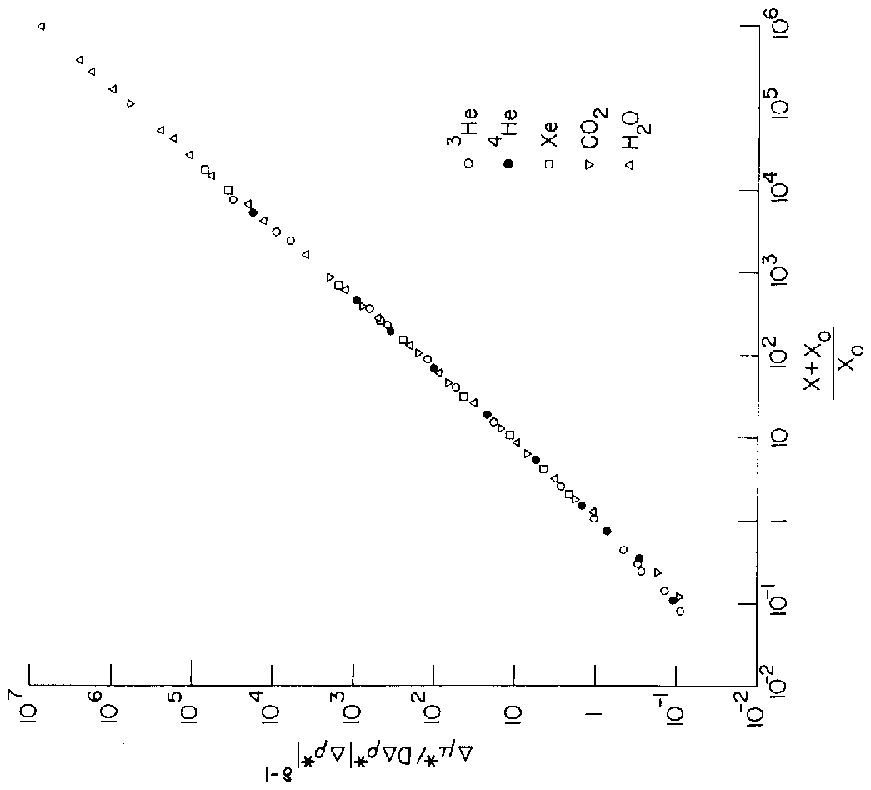}}
\end{picture}
\caption[Universal scaling plot of liquid-gas transitions ($n=1$) 
(from~\protect\cite{GUGGENHEIM_1,SENGERS_1})]
{The universal scaling plot of the coexistence curve 
of a eight different fluids, undergoing a liquid-gas
phase transition (left).
The solid line corresponds to a fit assuming the 
exponent $\beta=1/3$.
The figure is taken from
{E.\,A.~Guggenheim}, J.\,Chem.\,Phys.\,{\bf 13}, 253 (1945).
The right figure displays the corresponding universal 
Widom-Griffiths scaling function ${\tilde H}(x,1)$.
%of the chemical potential.
The scaling variable is defined 
as $x=\Delta T |\Delta \rho|^{-1/\beta}$
and $x_{\ssstyle 0}$ is related to the amplitude~$B$
of the power-law for the coexistence curve 
$\Delta\rho=B\,  \Delta T^{\beta}$.
The figure is taken from
{J.\,V.~Sengers} and {J.\,M.\,H.~Levelt\,Sengers}, 
{\textit{Critical phenomena in classical fluids}}
in {\textit{Progress in liquid physics}},
edited by {C.\,A.~Croxton}, 
(John Wiley\protect\&Sons, New York, 1978),
\label{fig:uni_plot_n1}} 
\end{figure}

Celebrated examples of universal scaling plots, providing
striking experimental evidences for the concept of 
universality at all, are shown in
\reffigs2name{fig:uni_plot_n1}{fig:uni_plot_n3}.
First, \reffigname{fig:uni_plot_n1} presents the coexistence curves of
eight different fluids characterized by different 
interatomic\index{universality class, Ising} interactions.
This plot was published by Guggenheim in 1945 and is
one of the oldest scaling plots in history~\cite{GUGGENHEIM_1}.
The corresponding Widom-Griffiths scaling function
is also shown in \reffigname{fig:uni_plot_n1}.
Here, the rescaled data of the chemical potential of 
five different gases collapse onto
a single universal curve.
This figure was published by Sengers and Levelt\,Sengers~\cite{SENGERS_1}.

%%\clearpage
%\begin{figure}[t] 
%\centering
%%\leavevmode 
%\includegraphics[clip,width=8cm,angle=-0.55]{_figures/guggenheim.eps}
%\caption[Universal scaling plot of liquid-gas transition ($n=1$) 
%(from~\protect\cite{GUGGENHEIM_1})]
%{The universal scaling plot of the coexistence curve 
%of a eight different fluids, undergoing a liquid-gas
%phase transition.
%The solid line corresponds to a fit assuming the 
%exponent $\beta=1/3$.
%The figure is taken from
%{E.\,A.~Guggenheim}, J.\,Chem.\,Phys.\,{\bf 13}, 253 (1945).
%\label{fig:uni_plot_n1}} 
%\end{figure} 

\unitlength1.0cm

\begin{figure}[t] 
\centering 
%\leavevmode 
\includegraphics[clip,width=8cm,angle=0]{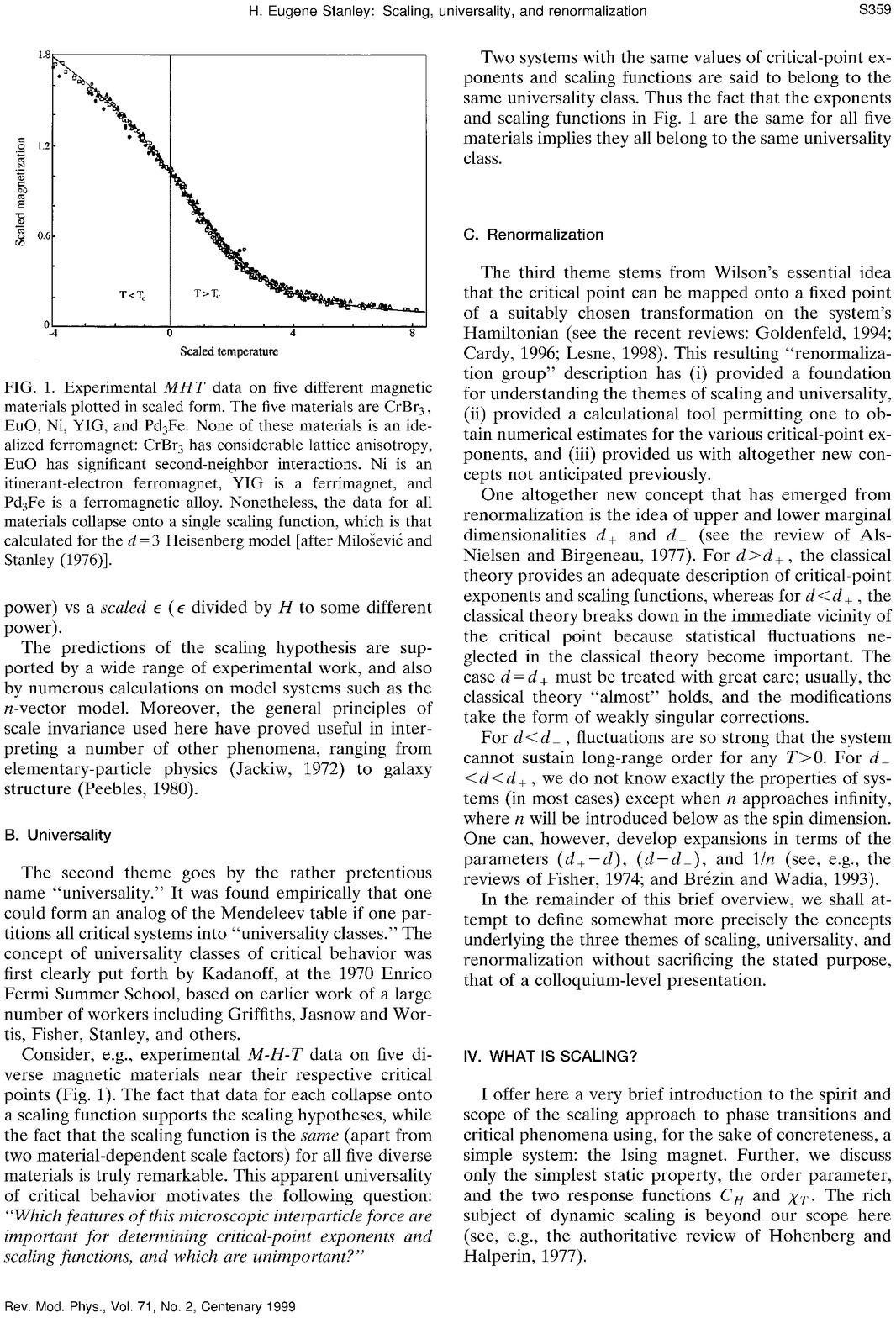}
\caption[Universal scaling plot of Heisenberg ferromagnets ($n=3$)
(from~\protect\cite{STANLEY_1})]
{The universal scaling plot of the equation of state of the
three-dimensional Heisenberg universality class ($n=3$).
The experimental data of five different magnetic materials
are shown:~$\mathrm{CrBr}_{\ssstyle 3}$ (characterized by a 
lattice anisotropy),
$\mathrm{EuO}$ (second neighbor interactions),
$\mathrm{Ni}$ (itinerant-electron ferromagnet),
$\mathrm{YIG}$ (Yttrium Iron Garnet, ferrimagnet),
$\mathrm{Pd}_{\ssstyle 3}\mathrm{Fe}$ (ferromagnetic alloy).
Despite of the different interaction details all 
data collapse onto the universal scaling function.
Furthermore the solid line was obtained from a series
expansion  of the Heisenberg model 
(see~\protect\cite{MILOSEVIC_2,MILOSEVIC_1}).
The figure is taken from
{H.\,E.~Stanley}, Rev.\,Mod.\,Phys.\,{\bf 71}, S358 (1999).
\label{fig:uni_plot_n3}} 
\end{figure} 
%\clearpage

Second, \reffigname{fig:uni_plot_n3} displays the universal scaling
function ${\tilde M}(x,1)$ of the three-dimen\-sional
Heisenberg universality\index{universality class, Heisenberg} 
class.
Here the data of five different magnetic materials (note that none
of these materials is an idealized ferromagnet~\cite{STANLEY_1}) agree
with the calculated curve of the Heisenberg model, obtained 
from a series expansion~\cite{MILOSEVIC_1,MILOSEVIC_2}.
Both universal scaling plots demonstrate perfectly the 
robustness of each universality class with respect to 
variations of the microscopic interactions.

In addition to the critical exponents and scaling functions
one often refers the universality also to certain 
amplitude\index{amplitude combinations}\index{universal amplitude combinations}
combinations (see~\cite{PRIVMAN_2} 
for an excellent review).
These amplitude combinations are very useful in order
to identify the universality class of a phase transition
since the amplitude combinations vary more widely than
the corresponding exponents.
Furthermore, the measurement of amplitude combinations
in experiments or numerical simulations yields a reliable
test for theoretical predictions, obtained e.g.~from
renormalization group approximations.
Consider the singular behavior of the susceptibility
by approaching the transition point from above and below
\begin{eqnarray}
\label{eq:sus_sing_ampl_01_p}   
\chi(T>\Tc,h=0) & \sim & a_{\ssstyle \chi, +} \, \deltaT^{-\gamma} \, , \\ 
\label{eq:sus_sing_ampl_01_m}
\chi(T<\Tc,h=0) & \sim & a_{\ssstyle \chi, -} \, (-\deltaT)^{-\gamma} \, .
\end{eqnarray}
The amplitudes $a_{\ssstyle \chi, +}$ and $a_{\ssstyle \chi, -}$ 
are related to the non-universal metric factors
and to particular values of universal scaling functions.
The universal scaling form of the susceptibility 
is obtained from \refeqname{eq:universal_HS_form}
\begin{equation}
\chi(\deltaT,h) \; = \; 
\left (
\frac{\partial\, m(\deltaT,h)}{\partial\, h}
\right )_{T}
\; \sim \;
a_{\ssstyle h} \;
\lambda^{\gamma} \; \; 
{\tilde \mathrm{X}}(a_{\ssstyle T} \deltaT \, \lambda, 
a_{\ssstyle h} h \, \lambda^{\beta \delta} ) \, ,
\label{eq:universal_HS_susc_form}
\end{equation}
with ${\tilde \mathrm{X}}(x,y)=\partial_{\ssstyle y} {\tilde M}(x,y)$.
Setting $a_{\ssstyle T} |\deltaT| \lambda=1$ we obtain 
for the amplitudes
\begin{eqnarray}
\label{eq:sus_sing_ampl_02_p}   
a_{\ssstyle \chi, +} & = & a_{\ssstyle h} 
\, a_{\ssstyle T}^{-\gamma} \, {\tilde \mathrm{X}}(+1,0) \, \\
\label{eq:sus_sing_ampl_02_m}
a_{\ssstyle \chi, -} & = & a_{\ssstyle h} 
\, a_{\ssstyle T}^{-\gamma} \, {\tilde \mathrm{X}}(-1,0) \, .
\end{eqnarray}
Obviously, the amplitudes $a_{\ssstyle \chi, +}$ and 
$a_{\ssstyle \chi, -}$ are non-universal, but the ratio
\begin{equation}
\frac{a_{\ssstyle \chi, +}}{\,a_{\ssstyle \chi, -}\,}
\; = \; \frac{\,{\tilde \mathrm{X}}(+1,0)\,}{{\tilde \mathrm{X}}(-1,0)}
\label{eq:uni_sus_ratio_01}
\end{equation}
is a universal quantity.
For example, the mean field behavior of the susceptibility takes 
the form
\begin{equation}
\chi(\deltaT,h=0) \; = \; \frac{1}
{\,b_{\ssstyle 2} \, \deltaT \, + \, 3 \, b_{\ssstyle 4} \, m^2\,} \, ,
\label{eq:sus_mean_field_fm}
\end{equation}
leading to [see \refeqname{eq:Landau_mt_fm}]
\begin{equation}
a_{\ssstyle \chi, +} \, = \, \frac{1}{b_{\ssstyle 2}} \,,
\quad \quad
a_{\ssstyle \chi, -} \, = \, \frac{1}{2 \, b_{\ssstyle 2}} \,,
\quad \quad \Longrightarrow \quad \quad
\frac{a_{\ssstyle \chi, +}}{\,a_{\ssstyle \chi, -}\,}
\; = \; 2 \, .
\label{eq:sus_mean_field_ampl_fm}
\end{equation}

Similar to the amplitude ratio of the susceptibility other
universal combinations can be defined.
Well known and experimentally important is the 
quantity (see~\cite{PRIVMAN_2})
\begin{equation}
R_{\chi} \; = \; \Gamma \, D_{\ssstyle \mathrm c} \, B^{\delta -1} \, . 
\label{eq:def_R_chi_fm}
\end{equation}
Here, $\Gamma$, $D_{\ssstyle  \mathrm c}$, $B$ are the traditional, but unfortunately
unsystematical, notations for the amplitudes of
\begin{equation}
\left . \chi  \sim \Gamma \, \deltaT^{-\gamma {\vphantom \beta}} \right |_{T>\Tc}  \, , 
\quad \quad
\left . h \sim D_{\ssstyle  \mathrm c} \, m^{\delta}  \right |_{T=\Tc} \, ,
\quad \quad
\left . m    \sim B  \, (-\deltaT)^{\beta}  \right |_{T<\Tc} \, . 
\label{eq:def_R_chi_fm_def_ampl_01}
\end{equation}
These amplitudes correspond to the values [see \refeqname{eq:universal_HS_form}]
\begin{equation}
\Gamma = a_{\ssstyle \chi,+} = a_{\ssstyle h} \,
a_{\ssstyle T}^{-\gamma} \, {\tilde \mathrm{X}}(1,0) \, ,  \quad \quad
D_{\ssstyle  \mathrm c} = a_{\ssstyle h}^{-1} \, {\tilde M}(0,1)^{-\delta} \, ,
\quad \quad
B = a_{\ssstyle T}^\beta \, {\tilde M}(-1,0) \, .
\label{eq:def_R_chi_fm_def_ampl_02}
\end{equation}
Using the normalizations ${\tilde M}(-1,0)={\tilde M}(0,1)=1$
we find for the amplitude combination
\begin{equation}
R_{\chi} \; = \; {\tilde \mathrm{X}}(1,0)
\label{eq:def_R_chi_fm_uni}
\end{equation}
which is obviously a universal quantity.
In case of the mean field theory we find $R_{\chi} =1$.
These two examples show how the universality of amplitude
combinations emerges naturally from the universality of 
the scaling functions, i.e., universal amplitude combinations
are just particular values of the universal 
scaling functions.

The above presented phenomenological concepts of scaling and universality
have been tested in a variety of systems with remarkable
success.
Nevertheless, they have certain shortcomings.
For example, there is no way of determining 
the critical exponents and scaling functions explicitly.
Furthermore, no mathematical substantiation for the 
underlying scaling form of thermodynamic potentials
is provided.
It requires Wilson renormalization group theory
to remedy these shortcomings.
This is sketched in the next section.

\section{Remarks on renormalization group theory}
\label{sec:rg_theory}

A foundation\index{renormalization group}\index{RG} 
for the understanding of the scaling
theory and the concept of universality has been provided
by Wilson's renormalization group (RG) theory~\cite{WILSON_1,WILSON_2}.
\index{renormalization group}
In equilibrium systems the RG theory maps the 
critical point onto a fixed point of a certain 
transformation of the system's Hamiltonian 
(introductions are presented in e.g.~\cite{FISHER_4,PFEUTY_1,YEOMANS_1,PLISCHKE_1}).
In case of the instructive real space RG~\cite{KADANOFF_3}
the transformation contains a rescaling of a microscopic
length scale, e.g.~the lattice spacing $a$,
by a factor $b$ ($a \mapsto b\, a$) and
the elimination of those degrees of freedom
that correspond to the range between $a$ and $b\, a$. 
This rescaling will change the system's properties
away from the critical point where the system 
exhibits only finite characteristic length scales.
But at criticality the infinite correlation length 
[\refeqname{eq:mag_corrl_nu}] determines
the physical behavior and the properties of the system
remain unaffected by the rescaling procedure.
In this way, criticality corresponds to a
fixed point of the renormalization transformation.

Denoting the system's Hamiltonian by ${\mathcal H}$
and the rescaled Hamiltonian by ${\mathcal H}^{\prime}$
the renormalization transformation is described by an
appropriate operator ${\mathcal R}$, 
\begin{equation}
{\mathcal H}^{\prime} \; = \; {\mathcal R} \circ {\mathcal H} \, .
\label{eq:rg_trans_H_op}
\end{equation}
Fixed point Hamiltonians ${\mathcal H}^{\ast}$ satisfy
\begin{equation}
{\mathcal H}^{\ast} \; = \; {\mathcal R} \circ {\mathcal H}^{\ast} \,
\label{eq:rg_trans_H_op_fp}
\end{equation}
and it turns out that different fixed point Hamiltonians
are related to different universality classes~\cite{FISHER_4}.
For the sake of concreteness consider the 
reduced Ising Hamiltonian
\begin{equation}
{\tilde {\mathcal H}} \; = \; - \, \frac{\mathcal H }{\, k_{\ssstyle \mathrm B} T \,} 
\; = \; 
\frac{J}{\, k_{\ssstyle \mathrm B} T \,}
\sum_{\ssstyle \langle i,j \rangle } 
\, S_{\ssstyle i}S_{\ssstyle j} \, + \, 
\frac{h}{\, k_{\ssstyle \mathrm B} T \,} 
 \sum_{\ssstyle i} \, S_{\ssstyle i}, 
\label{eq:Ising_H_op}
\end{equation}
with the spin variables $S= \plmi 1$, the nearest neighbor
interaction coupling $J$, and the homogeneous external field~$h$,
and where the first sum is taken over all pairs of
neighboring spins on a given $D$-dimensional lattice.
The partition function for a system of $N$ spins is 
\begin{equation}
Z_{\ssstyle N}(K_{\ssstyle 1},K_{\ssstyle 2})  \; = \;
\sum_{\ssstyle \{S \}} \, 
{\mathrm e}^{{\tilde {\mathcal H}}(K_{\ssstyle 1},K_{\ssstyle 2})} \, ,
\label{eq:partition_Z}
\end{equation}
where the sum is taken over all possible spin configurations
and where we introduce the couplings
\begin{equation}
K_{\ssstyle 1} \; = \; \frac{J}{\, k_{\ssstyle \mathrm B} T \,} \, ,
\quad \quad
K_{\ssstyle 2} \; = \; \frac{h}{\, k_{\ssstyle \mathrm B} T \,}  \, .
\label{eq:couplings_K1_K2}
\end{equation}
In general the Hamiltonian can be written as a function
of the couplings
\begin{equation}
{\tilde {\mathcal H}}(\underline K) \; = \; 
\sum_{\ssstyle i} \, K_{\ssstyle i} \, {\mathcal O}_{\ssstyle i} \, ,
\label{eq:gen_Hamiltonian}
\end{equation}
where ${\mathcal O}_{\ssstyle i}$ are the operators appearing in
the Hamiltonian.
Performing a renormalization transformation reduces the
spin numbers ($N \mapsto N^{\prime}=N/b^{D}$) and leads to a rescaled
Hamiltonian, characterized by the couplings
\begin{equation}
K^{\prime}_{\ssstyle 1} \; = \; 
K^{\prime}_{\ssstyle 1}(K_{\ssstyle 1},K_{\ssstyle 2}) \, ,
\quad \quad
K^{\prime}_{\ssstyle 2} \; = \; 
K^{\prime}_{\ssstyle 2}(K_{\ssstyle 1},K_{\ssstyle 2}) \, ,
\quad \quad
K^{\prime}_{\ssstyle 3} \; = \; 
K^{\prime}_{\ssstyle 3}(K_{\ssstyle 1},K_{\ssstyle 2}) \, ,
\ldots \; .
\label{eq:couplings_K1_K2_new}
\end{equation}
Here, $K^{\prime}_{\ssstyle i> 2}$ account for additional coupling terms
of the renormalized Hamiltonian which may appear as a result of 
the renormalization transformation even if they are not present
in the initial Hamiltonian.
These so-called RG recursion relations generate trajectories
in the space of couplings, i.e., the couplings $\underline K$ 
flow under successive renormalizations
\begin{equation}
{\underline K}  
\; \mathop{\longrightarrow}\limits_{\mathcal R} \; 
{\underline K}^{\ssstyle (1)} 
\; \mathop{\longrightarrow}\limits_{\mathcal R}   \;
{\underline K}^{\ssstyle (2)} 
\; \mathop{\longrightarrow}\limits_{\mathcal R}   \; 
\ldots 
\; \mathop{\longrightarrow}\limits_{\mathcal R}  \; 
{\underline K}^{\ssstyle (n)} 
\; \mathop{\longrightarrow}\limits_{\mathcal R}   \; 
{\underline K}^{\ssstyle (n+1)}    
\; \mathop{\longrightarrow}\limits_{\mathcal R}   \; 
\ldots \,
\end{equation}
along the RG trajectories towards a certain fixed 
point ${\underline K}^{\ast}$.
If the system is not initially at criticality the couplings
will flow towards a trivial fixed point, e.g., 
a fixed points that corresponds to zero or infinite
temperature.

Linearizing the problem close to a fixed point yields
\begin{eqnarray}
{\mathcal R} \circ {\tilde{\mathcal H}}({\underline K}^{\ast}+\delta{\underline K}) 
&= & {\tilde {\mathcal H}}^{\ast} \, + \, 
{\mathcal R} \circ \sum_{\ssstyle i} \, \delta K_{\ssstyle i} \, {\mathcal O}_{\ssstyle i}
\nonumber \\
&\approx  & {\tilde {\mathcal H}}^{\ast} \, + \, 
\sum_{\ssstyle i} \, \delta K_{\ssstyle i} \, 
\sum_{\ssstyle j} {\mathcal L}_{\ssstyle i,j} \, {\mathcal O}_{\ssstyle j} 
\nonumber \\
&=  & {\tilde {\mathcal H}}^{\ast} \, + \, 
\sum_{\ssstyle j}  \left ( 
\sum_{\ssstyle i} \, \delta K_{\ssstyle i} \, {\mathcal L}_{\ssstyle i,j} \right ) 
\, {\mathcal O}_{\ssstyle j} 
\nonumber \\
& =  & {\tilde {\mathcal H}}^{\ast} \, + \, 
\sum_{\ssstyle j}  \delta K_{\ssstyle j}^{\prime} \, {\mathcal O}_{\ssstyle j} \, .
\label{eq:rg_trans_H_op_fp_cal}
\end{eqnarray}
Assuming that the diagonalized operator ${\mathcal L}$ has the eigenoperators
$\Omega_{\ssstyle i}$ and eigenvalues $\Lambda_{\ssstyle i} $
such that ${\mathcal L} \Omega_{\ssstyle i} = \lambda_{\ssstyle i} \Omega_{\ssstyle i}$
we find that the couplings transform in the diagonal
representation ($\delta\kappa$) according to 
$\delta\kappa_{\ssstyle i}^{\prime} = \Lambda_{\ssstyle i} \delta\kappa_{\ssstyle i}$,
thus
\begin{equation}
{\mathcal R} \circ {\tilde{\mathcal H}}({\underline K}^{\ast}+\delta{\underline K}) 
\; = \; {\tilde {\mathcal H}}^{\ast} \, + \, 
\sum_{\ssstyle i} \, \Lambda_{\ssstyle i} \,
\delta\kappa_{\ssstyle i} \,
{\Omega}_{\ssstyle i} \, . \label{eq:rg_trans_H_op_fp_lambda}
\end{equation}
The couplings $\delta\kappa_{\ssstyle i}$ are termed
scaling fields \index{scaling field}
and their recursion relation can be expressed
by the rescaling factor~$b$ 
\begin{equation}
\delta\kappa_{\ssstyle i}^{\prime} 
\; = \; b^{y_{\ssstyle i}} \; \delta\kappa_{\ssstyle i}  
\quad\quad {\mathrm {with}} \quad \quad
\Lambda_{\ssstyle i} \; = \; b^{y_{\ssstyle i}} \, .
\label{eq:recursion_relation}
\end{equation}
Successive renormalizations correspond to 
\begin{equation}
\delta\kappa_{\ssstyle i}
\; \mathop{\longrightarrow}\limits_{\mathcal R} \; 
b^{y_{\ssstyle i}} \,
\delta\kappa_{\ssstyle i}
\; \mathop{\longrightarrow}\limits_{\mathcal R}   \;
b^{2 y_{\ssstyle i}} \, 
\delta\kappa_{\ssstyle i}
\; \mathop{\longrightarrow}\limits_{\mathcal R}   \; 
b^{3 y_{\ssstyle i}} \, 
\delta\kappa_{\ssstyle i}
\; \mathop{\longrightarrow}\limits_{\mathcal R}   \; 
\ldots \; .
\end{equation}
Thus, the renormalization flow depends in the vicinity
of a given fixed point ${\tilde{\mathcal H}}^{\ast}$ on the
exponents $y_{\ssstyle i}$.
For $y_{\ssstyle i}>0$ ($\Lambda_{\ssstyle i}>1$) 
the corresponding scaling 
field $\delta\kappa_{i}$ is called relevant since successive 
renormalization transformations drive the system away
from ${\tilde{\mathcal H}}^{\ast}$.
In case of a negative exponent $y_{\ssstyle i}<0$
($\Lambda_{\ssstyle i}<1$)
the system approaches the fixed point under repeated
transformations and the scaling field is termed irrelevant.
Marginal scaling fields correspond to $y_{\ssstyle i}=0$
($\Lambda_{\ssstyle i}=1$)
and require higher than linear order in the expansion.
%An effect of marginal scaling fields is a continuous
%dependence of critical exponents on the corresponding
%interaction parameters 
%{\bf (Quelle: Nienhuis, Domb Green 11 und andere)}.
In this way each fixed point is characterized by its 
associated scaling fields and by a domain of attraction
which corresponds to the set of points flowing
eventually to the fixed point.
This set forms a hypersurface in the space of couplings 
and is termed the critical surface.\index{critical surface}

In summary, a fixed point ${\tilde{\mathcal H}}^{\ast}$ 
is approached if all associated relevant
scaling fields are zero, otherwise the system flows away
from ${\tilde{\mathcal H}}^{\ast}$.
Examples for relevant scaling fields in ferromagnetism 
are the temperature $\deltaT$ and the conjugated field $h$.
Criticality is only achieved for $\deltaT \to 0$ and $h \to 0$,
therefore we may identify
\begin{eqnarray}
\label{eq:scaling_fields_identify_T}   
\delta\kappa_{\ssstyle 1} & = & \delta\kappa_{\ssstyle T} \; = \;
a_{\ssstyle T} \deltaT \, ,
\quad\quad 
y_{\ssstyle 1} \; = \; y_{\ssstyle T} \, > \, 0\, , \\
\label{eq:scaling_fields_identify_h}   
\delta\kappa_{\ssstyle 2} & = & \delta\kappa_{\ssstyle h} \; = \;
a_{\ssstyle h} h \, , \; \; \,
\quad\quad 
y_{\ssstyle 2} \; = \; y_{\ssstyle h} \, > \, 0\, ,
\end{eqnarray}
and $y_{\ssstyle i>2}<0$.
On the other hand, all Hamiltonians that differ from the 
fixed point ${\tilde{\mathcal H}}^{\ast}$ only by irrelevant 
scaling fields flow towards ${\tilde{\mathcal H}}^{\ast}$.
For example the five magnetic materials presented
in \reffigname{fig:uni_plot_n3} differ only by 
irrelevant scaling fields.

\begin{figure}[t] 
\centering
%\leavevmode 
\includegraphics[clip,width=7cm,angle=0]{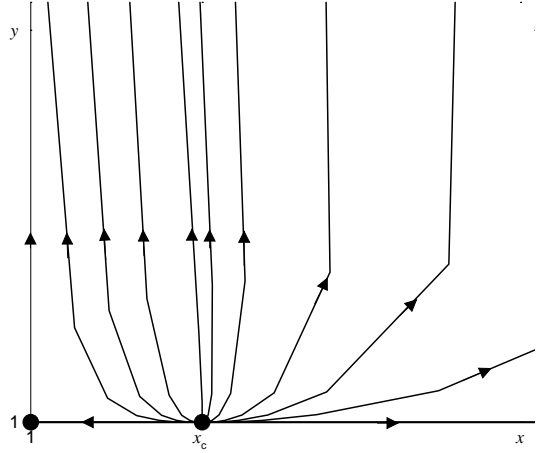}
\caption[Sketched renormalization group flow] {Sketch of the
renormalization group flow of an Ising ferromagnet 
(see~[\refeqname{eq:Ising_H_op}]) on a hierarchical lattice.
Hierarchical lattices are iteratively constructed
lattices~\cite{BERKER_1,KAUFMANN_1} where the so-called
Migdal-Kadanoff scheme of renormalization is 
exact~\cite{MIGDAL_1,KADANOFF_4}.
The flowlines show the motion of the coupling
constants $x=\exp{(2 K_{\ssstyle 1})}$ (temperature-like variable) 
and $y=\exp{(K_{\ssstyle 2})}$ (field-like variable) under 
successive iterations of the recursion relations
$x^{\prime}= (x^2+y^2)(x^{-2}+y^2)(1+y^2)^{-2}$
and $y^{\prime}=y^2(1+x^2 y^2)(x^2+y^2)^{-1}$.
The flowlines are attracted and repelled by the 
fixed points $(1,1)$, $(1,\infty)$, $(\infty,1)$, and 
$(x_{\ssstyle c},1)$ with $x_{\ssstyle c}=3.38298\ldots$.
The phase transition corresponds to the non-trivial fixed 
point $(x_{\ssstyle c},1)$ with the eigenvalues
$\Lambda_{\ssstyle 1}\approx 1.6785$
and $\Lambda_{\ssstyle 2}\approx 3.6785$, leading 
to $\beta\approx 0.16173$ and $\delta\approx 15.549$.
\label{fig:rg_flow_dia_02}} 
\end{figure}

Although the above linear recursion relations 
[\refeqname{eq:recursion_relation}] describe the 
RG trajectories only in the vicinities of fixed points they
provide some insight in the topology of the entire RG flow
(see \reffigname{fig:rg_flow_dia_02}).
These RG flow diagrams are useful in order to illustrate
the RG transformations schematically and 
present a classification scheme in terms of  
fixed point stability.
The stability of a given fixed point is 
determined by the number of relevant and irrelevant scaling fields.
Instable fixed points are characterized by at least one
relevant scaling field since Hamiltonians arbitrarily close
to the fixed point will flow away under successive RG 
iterations.
Ordinary critical points correspond to singly
unstable fixed points, i.e., unstable with respect to
the control parameter (e.g.~temperature) of the phase 
transition.
Tricritical points are characterized by a second
instability.
An applied external field conjugated to the order parameter
implies an additional instability of the fixed point.

Furthermore, the stability of fixed points depends on 
the spatial dimensionality~$D$ of a system.
Above a certain dimension~$\Dc$ the scaling behavior
is determined by a trivial fixed point with classical
mean field exponents,
whereas a different fixed point with non-classical
exponents determines the scaling behavior below~$\Dc$. 
This exchange of the scaling behavior is caused 
by an exchange of the stability of the corresponding
fixed points below and above~$\Dc$~\cite{PFEUTY_1,WEGNER_2,BREZIN_3}.
At the so-called upper critical dimension~$\Dc$ 
both fixed points are identical and marginally stable.
The scaling behavior is characterized by 
mean field exponents modified by logarithmic
corrections~\cite{WEGNER_1,WILSON_3}. 
\index{upper critical dimension} \index{critical dimension}
We will discuss the scaling behavior of certain
non-equilibrium phase transitions at the upper critical
dimension in detail in the following chapters.

Let us now consider how scaling emerges from the renormalization
transformation.
It is essential that the partition function is invariant under
the renormalization operation~${\mathcal R}$~\cite{FISHER_4}
\begin{equation}
Z_{\ssstyle N}(\delta\kappa_{\ssstyle T},
\delta\kappa_{\ssstyle h},\delta\kappa_{\ssstyle 3}, \ldots) 
\; = \;
Z_{\ssstyle N^{\prime}}(\delta\kappa^{\prime}_{\ssstyle T},
\delta\kappa_{\ssstyle h}^{\prime},
\delta\kappa_{\ssstyle 3}^{\prime},\ldots) \, .
\label{eq:partition_Z_inv}
\end{equation}
Therefore, the free energy per degree of freedom transforms according
to 
% da F=N f = kT Sp Z   und  F'=N' f' = kT Sp Z und F=F' 
\begin{equation}
f(\delta\kappa_{\ssstyle T},
\delta\kappa_{\ssstyle h},
\delta\kappa_{\ssstyle 3},\ldots)  
\; = \;   b^{-D} \, 
f(\delta\kappa^{\prime}_{\ssstyle T},
\delta\kappa_{\ssstyle h}^{\prime},
\delta\kappa_{\ssstyle 3}^{\prime},\ldots) \, .
\label{eq:free_energy_inv}
\end{equation}
Combining this equation with \refeqname{eq:recursion_relation}
and using the identifications 
\refeqs2name{eq:scaling_fields_identify_T}{eq:scaling_fields_identify_h}
yields the scaling form 
\begin{equation}
f(a_{\ssstyle T} \deltaT,
a_{\ssstyle h} h,
\delta\kappa_{\ssstyle 3},\ldots)  
\; \sim \;   b^{-D} \, 
f(b^{y_{\ssstyle T}}\, a_{\ssstyle T} \deltaT,
b^{y_{\ssstyle h}}\, a_{\ssstyle h} h,
b^{y_{\ssstyle 3}}\, \delta\kappa_{\ssstyle 3},\ldots) \, .
\label{eq:free_energy_scal_rg}
\end{equation}
Introducing $\lambda=b^{y_{\ssstyle T}}$ we obtain the
scaling form of the free energy [see \refeqname{eq:gibbs_pot_hom_func}]
\begin{equation}
f(a_{\ssstyle T} \deltaT,a_{\ssstyle h} h,
\delta\kappa_{\ssstyle 3},\ldots) 
\; \sim \;
\lambda^{\alpha-2} \; \; 
{f}(a_{\ssstyle T}\deltaT \, \lambda, 
a_{\ssstyle h} h \, \lambda^{\beta \delta},
\delta\kappa_{\ssstyle 3} \, \lambda^{\phi_{\ssstyle 3}}, 
\ldots) \, ,  
\label{eq:free_energy_hom_func}
\end{equation}
where we have identified the exponents
\begin{equation}
y_{\ssstyle T} \, = \, \frac{1}{\nu} \, ,  \quad
y_{\ssstyle h} \, = \, \frac{\beta\, \delta}{\nu} \, ,  \quad
y_{\ssstyle i>2} \, = \, \frac{\phi_{\ssstyle i>2}}{\nu} \, .
\label{eq:exponents_rg}
\end{equation}
The possible additional scaling fields $\delta\kappa_{i > 2}$
deserve comments.
Irrelevant scaling fields ($\phi_{i > 2}< 0$) may cause
corrections \index{corrections to scaling}
to the asymptotic scaling behavior~\cite{WEGNER_3}.
For example, choosing $\lambda = 1/|a_{\ssstyle T}\deltaT|$ we obtain at
zero field
\begin{eqnarray}
\label{eq:corr_to_scaling_irrel}   
f(a_{\ssstyle T} \deltaT,0,\delta\kappa_{\ssstyle 3}) 
& \sim & | a_{\ssstyle T} \deltaT|^{2-\alpha} \;
f(\plmi 1,0,\delta\kappa_{\ssstyle 3}\, 
|a_{\ssstyle T}\deltaT |^{|\phi_{\ssstyle 3}|})  \\
& \approx &
|a_{\ssstyle T}\deltaT|^{2-\alpha} \, f(\plmi 1,0,0) + 
|a_{\ssstyle T}\deltaT|^{2-\alpha+|\phi_{\ssstyle 3}|} \,
\delta\kappa_{\ssstyle 3} \, 
\left . 
\partial_{\ssstyle x}^{\vphantom X} f(\plmi 1,0,x) \right |_{\ssstyle x = 0} 
 +  \ldots \, . \nonumber 
\end{eqnarray}
The non-universal corrections to the leading order $|\deltaT|^{2-\alpha}$
are termed confluent singularities\index{confluent singularities}
and they determine the size of the\index{critical region} 
critical region.
Often confluent singularities have to be taken into
account in order to analyze high precision data.
Impressive examples of confluent singularity effects
of superfluid Helium are reviewed in~\cite{AHLERS_1}.
The above expansion of $f(\plmi 1,0,x)$ implies that the free energy is
an analytic function in $\delta\kappa_{\ssstyle 3}$.
If the free energy is non-analytic the scaling 
field~$\delta\kappa_{\ssstyle 3}$ is termed a dangerous
irrelevant\index{dangerous irrelevant variable}
variable~\cite{FISHER_5,PRIVMAN_4}.
In that case, the free energy exhibits e.g.~a power-law
divergence
\begin{equation}
f(x,y,z) \; = \; z^{-\mu} \, {\hat{f}}(x,y)
\label{eq:free_energy_dangerous}
\end{equation}  
characterized by the exponent $\mu>0$.
Singularities of this type occur for example in the
mean field regime of the well-known Landau-Ginzburg-Wilson 
Hamiltonian for short range interacting
ferromagnets (see e.g.~\cite{BRANKOV_2}).
There, the dangerous irrelevant variable corresponds
to the coupling constant of the $\phi^4$~interactions.
The nonanalytic behavior leads to the modified scaling
form of the free energy 
\begin{eqnarray}
f(a_{\ssstyle T} \deltaT,a_{\ssstyle h} h,
\delta\kappa_{\ssstyle 3}) 
& \sim & 
\lambda^{-\nu D} \; \; 
{f}(a_{\ssstyle T}\deltaT \, \lambda, 
a_{\ssstyle h} h \, \lambda^{\beta \delta},
\delta\kappa_{\ssstyle 3} \, \lambda^{\phi_{\ssstyle 3}}) \nonumber \\[2mm]
& = & \left .
\lambda^{-\nu D-\mu\phi_{\ssstyle 3}} \; \delta\kappa_{\ssstyle 3}^{-\mu}\; 
{\hat{f}}(a_{\ssstyle T}\deltaT \, \lambda, 
a_{\ssstyle h} h \, \lambda^{\beta \delta}) 
\right |_{\ssstyle h=0} \nonumber \\[2mm]
& = &
|a_{\ssstyle T}\deltaT|^{\nu D + \mu\phi_{\ssstyle 3}} \; 
\delta\kappa_{\ssstyle 3}^{-\mu} \;
{\hat{f}}(\plmi 1, 0) \, .
\label{eq:free_energy_dangerous_mod}
\end{eqnarray}  
Compared to the standard behavior $f \asympprop |\deltaT|^{2-\alpha}$
the above result reflects the breakdown of the hyperscaling
law $2-\alpha=\nu D$.
Additionally to the violation of scaling laws,
%Thus the non-analyticity of the free energy~$f$ 
%in the irrelevant variable~$\delta\kappa_{\ssstyle 3}$
%causes a violation of the scaling laws.
dangerous irrelevant variables also
cause the breakdown of common finite-size scaling 
within the mean field regime~\cite{BREZIN_5,PRIVMAN_4}.
This is well established in 
equilibrium e.g.~in the $n\to \infty$ limit.
We will address this point in detail in Chapter~\ref{chapter:dp}
where high-dimensional non-equilibrium phase transitions are considered.

%For example, the well-known Landau-Ginzburg-Wilson 
%Hamiltonian for short range interacting
%ferromagnets exhibits a dangerous
%irrelevant variable 
%(the coupling constant of the $\phi^4$~interactions) 
%which complicates the analysis
%and leads to an anomalous behavior
%(see e.g.~\cite{FISHER_5,PRIVMAN_4}).
%{\bf gilt dies nur fuer $D\ggl \Dc$}

The situation is different when the scaling 
field $\delta\kappa_{\ssstyle 3}$
is relevant, i.e., $\phi_{\ssstyle 3}>0$.
In that case the free energy at zero field is given by
\begin{equation}
f(a_{\ssstyle T} \deltaT,0,\delta\kappa_{\ssstyle 3}) 
\; \sim \; |a_{\ssstyle T} \deltaT|^{2-\alpha} \;
f(\plmi 1,0,\delta\kappa_{\ssstyle 3} \, 
| a_{\ssstyle T} \deltaT |^{-\phi_{\ssstyle 3}})  \, . 
\label{eq:corr_to_scaling_rel}
\end{equation}
For sufficient small arguments
($|\delta\kappa_{\ssstyle 3} |\deltaT |^{-\phi_{\ssstyle 3}}| \ll 1$)
the relevant scaling field leads again to corrections to
the asymptotic scaling behavior.
But approaching the transition point ($\deltaT \to 0$)
the scaling argument diverges and gives rise to a different 
critical behavior,
i.e., the system crosses over to a different universality class.
We will discuss crossover \index{crossover} phenomena in detail in 
Chapter\,\ref{chapter:crossover}.
Eventually a marginal scaling field causes logarithmic
corrections via
\begin{eqnarray}
\delta\kappa_{\ssstyle 3} \, 
|a_{\ssstyle T} \deltaT |^{-\phi_{\ssstyle 3}} 
& = & \delta\kappa_{\ssstyle 3} \,
\exp{(-\phi_{\ssstyle 3} \, \ln{|a_{\ssstyle T} \deltaT|})} \nonumber \\
& \mathop{\longrightarrow}\limits_{\phi_{\ssstyle 3} \to 0 }  &
\delta\kappa_{\ssstyle 3} \;
(\, 1-\phi_{\ssstyle 3} \, \ln{|a_{\ssstyle T} \deltaT|} 
\, + \, \ldots )\, .
\label{eq:corr_to_scaling_marg}
\end{eqnarray}
Often, these logarithmic contributions mask the power
law singularities and make the analysis of experimental 
or numerical data notoriously difficult.

Analogous to the free energy, the renormalization group
implies the scaling form of the correlation length~$\xi$.
Performing a renormalization transformation, the 
correlation length $\xi$, like all length scales,
is decreased by the factor~$b$,
\begin{equation}
\xi^{\prime} \; = \; b^{-1} \, \xi \, .
\label{eq:corr_length_inv}
\end{equation}
It is essential for the understanding of phase transitions
that fixed points are characterized by an infinite (or 
trivial zero) correlation length since $\xi$ satisfies
at a fixed point 
\begin{equation}
\xi^{\prime}  = \xi \, .
\label{eq:corr_length_fixed_point}
\end{equation}
In this way, a singular correlation length, or in other words,
scale invariance is the hallmark of criticality.
The scaling form of the correlation length is obtained 
from \refeqname{eq:corr_length_inv} 
\begin{equation}
\xi(\delta\kappa_{\ssstyle T},
\delta\kappa_{\ssstyle h},
\delta\kappa_{\ssstyle 3},\ldots)  
\; = \;   b \;
\xi(\delta\kappa^{\prime}_{\ssstyle T},
\delta\kappa_{\ssstyle h}^{\prime},
\delta\kappa_{\ssstyle 3}^{\prime},\ldots) \, ,
\label{eq:corr_length_trans}
\end{equation}
yielding with $\lambda=b^{y_{\ssstyle T}}$
and \refeqname{eq:recursion_relation} 
\begin{equation}
\xi(a_{\ssstyle T} \deltaT,a_{\ssstyle h} h,
\delta\kappa_{\ssstyle 3},\ldots) 
\; \sim \;
\lambda^{\nu} \; \; 
{\xi}(a_{\ssstyle T}\deltaT \, \lambda, 
a_{\ssstyle h} h \, \lambda^{\beta \delta},
\delta\kappa_{\ssstyle 3} \, \lambda^{\phi_{\ssstyle 3}}, 
\ldots) \, .
\label{eq:xi_hom_func}
\end{equation}

Instead of the real space renormalization
considered so far,
it is convenient to work in momentum space.
This can be achieved by reformulating the above derivations
in terms of Fourier transforms.
We refer the interested reader to the 
reviews of~\cite{WILSON_2,FISHER_3,PFEUTY_1}.
The momentum space formulation allows a perturbative 
definition of the RG which leads technically to a 
formulation in terms of Feyman graph expansions.
The appropriate small parameter for the perturbation expansion
is the dimensionality difference to the upper critical
dimension $\epsilon=\Dc-D$~\cite{WILSON_4}, 
i.e., the $\epsilon$-expansion\index{$\epsilon$-expansion} 
gives systematic corrections to mean field theory 
in powers of $\epsilon$.
The $\epsilon$-expansion provides a powerful tool
for calculating the critical exponents and the scaling functions.
For example, the exponent~$\gamma$ for $n$-component
magnetic systems with short range interactions is given 
in second order $\epsilon^2$
by~(see e.g.~\cite{BREZIN_3,WALLACE_1})
\begin{equation}
\gamma \; = \; 1 \, + \,
\epsilon \, \frac{n+2}{2(n+8)}
\, + \,
\epsilon^2 \, \frac{(n+2)}{4 (n+8)^3} \, (n^2+22n+52) 
\, + \, {\mathcal O}(\epsilon^3) \, .
\label{eq:gamma_eps_fm}
\end{equation}
Furthermore the Widom-Griffiths scaling function 
can be written as a power series in~$\epsilon$
\index{Widom-Griffiths scaling form}
\begin{equation}
{\tilde H}(x,1) \; = \;
1 \, + \, x \, + \, \epsilon \, {\tilde H}_{\ssstyle 1}(x,1)
\, + \, \epsilon^2 \, {\tilde H}_{\ssstyle 2}(x,1) 
\, +  \, {\mathcal O}(\epsilon^3) \, .
\label{eq:Widom_Griffiths_eps}
\end{equation}
Obviously the mean field scaling behavior 
[\refeqname{eq:Landau_WG_fm}] is obtained for $\epsilon=0$.
The functions become complicated with increasing
order and we just note~\cite{BREZIN_4}
\begin{eqnarray}
{\tilde H}_{\ssstyle 1}(x,1) & = & \frac{1}{2(n+8)} \;
 [ \, 3(x+3) \ln{(x+3)} \, + \, (n-1)(x+1) \ln{(x+1)} \nonumber \\
& & +\, 6 \, x \ln{2} \, - \, 9 (x+1) \ln{3} \,  ] \, .
\label{eq:WG_eps_eps1}
\end{eqnarray}
For ${\tilde H}_{\ssstyle 2}(x,1)$ we refer to the reviews
of~\cite{BREZIN_3,WALLACE_1}.
Thus the $\epsilon$-expansion provides estimates
of almost all quantities of interest as an asymptotic expansion
in powers of $\epsilon$ around the mean field values.
Unfortunately it is impossible to estimate within
this approximation scheme the corresponding error
bars since the extrapolation to larger values of 
$\epsilon$ is uncontrolled.
Detailed analyses turn out that the critical exponents
are more accurately estimated than the scaling functions
and therefore the amplitudes.
For example, the $\epsilon^2$-approximation [\refeqname{eq:gamma_eps_fm}]
for the susceptibility exponent of the two-dimensional Ising
model ($n=1$, $\epsilon=2$) yields $\gamma\approx 1.642$.
This value differs by $6\%$ from the exact 
value $\gamma=7/4$~\cite{BAROUCH_1}.
The amplitude ratio of the susceptibility
[\refeqname{eq:uni_sus_ratio_01}]
can be expanded as~\cite{NICOLL_1}
\begin{equation}
\frac{\,{\tilde \mathrm{X}}(+1,0)\,}{{\tilde \mathrm{X}}(-1,0)} \; = \;
2^{\gamma(\epsilon)-1} \; 
\frac{\gamma(\epsilon)}{\beta (\epsilon)}
\label{eq:uni_sus_ratio_fm_eps} 
\end{equation}
proposing the estimate $81.14\ldots$ for $\epsilon=2$. 
This result differs significantly ($115\%$) from the exact
value $37.69\ldots$~\cite{BAROUCH_1,DELFINO_2}.
The different accuracy reflects a conceptual difference
between the universality of critical exponents and 
the universality of scaling functions.
As pointed out clearly in~\cite{PRIVMAN_2}, the universality
of exponents arises from the linearized RG flow in the
vicinity of the fixed point,
whereas the scaling functions are obtained from the entire,
i.e., non-linear,\index{critical exponents}\index{universal scaling functions}
RG flow.\index{universal amplitude combinations}
More precisely, the relevant trajectories from the fixed
point of interest to other fixed points determine 
the universal scaling functions.
This illuminates also why exponents between different universality
classes may differ slightly while the scaling functions
and therefore the amplitude combinations differ
significantly.
Thus deciding on a system's universality class
by considering the scaling functions and amplitude
combinations instead of critical exponents appears
to be more sensitive.
Hence, nothing demonstrates universality
more convincing than the universal data-collapse 
of various systems (as shown 
in \reffigs2name{fig:uni_plot_n1}{fig:uni_plot_n3}).

%$ $\\
%In summary, we have sketched how Wilson's renormalization group 
%theory lays the foundation for an understanding of 
%scaling and universality.
%The RG explains the existence of an 
%upper critical dimension~$\Dc$ and turns out how 
%confluent singularities emerge from irrelevant scaling
%fields.
%Furthermore, the RG theory allows to decide on the relevant
%system parameters determining the universality class.
%It also yields a tool for computing critical exponents 
%as well as the universal scaling functions. 
%While critical exponents emerge from local properties
%near a given fixed point, scaling functions require
%the knowledge of the full RG flow along the trajectories between 
%neighboring fixed points.
%This illustrates why universal scaling functions and therefore 
%universal amplitude combinations are more
%{\it sensitive} than the corresponding critical exponents. 

\chapter{Absorbing phase transitions} 
\setcounter{figure}{5}
\label{chapter:apt}

The minimization\index{absorbing phase transitions}\index{APT} 
of the free energy is the governing 
principle of equilibrium statistical physics. 
In non-equilibrium situations no such framework exists. 
But nevertheless, the concepts and techniques which were developed 
for equilibrium critical phenomena can also be applied to non-equilibrium 
phase transitions.
In particular, the order parameter concept, 
scaling, universality as well as renormalization
group analyses are as powerful as in equilibrium, i.e., although
they were developed in equilibrium theories 
their applicability is far beyond.
Therefore, numerous non-equilibrium critical phenomena were 
successfully investigated by such methods, including
the geometrical problem of percolation~\cite{STAUFFER_3,ESSAM_3,STAUFFER_2}, 
interface motion and depinning 
phenomena~\cite{KARDAR_1,KARDAR_2,BARABASI_1},
irreversible growth processes like diffusion-limited
aggregation~\cite{WITTEN_1},
as well as turbulence~\cite{KOLMOGOROV_1,FRISCH_1}.
In the following we focus our attention on a certain class
of non-equilibrium phase transitions, the so-called absorbing
phase transitions (APT)~\cite{GRINSTEIN_2}.
These transitions were investigated intensively in the last
two decades (see~\cite{HINRICHSEN_1,ODOR_1} for recent reviews),
since they are related to a great variety of 
non-equilibrium critical phenomena such as 
forest fires~\cite{ALBANO_2,CLAR_2}, 
epidemic spreadings in biology~\cite{MOLLISON_1}, 
catalytic chemical reactions in chemistry~\cite{ZIFF_1},
spatio-temporal intermittency at the onset of 
turbulence~\cite{POMEAU_1,MENON_1,RUPP_1}, 
%%%%%intermittency was observed as a precursor to turbulence
scattering of elementary particles at high energies
and low-momentum transfer~\cite{CARDY_1},
interface growth~\cite{NARAYAN_1},
self-organized criticality~\cite{BAK_1,BAK_2,BAK_3,TURCOTTE_1},
damage spreading~\cite{HINRICHSEN_1,KAUFFMAN_1,HERRMANN_1},
as well as wetting transitions~\cite{HINRICHSEN_1,HINRICHSEN_9}.
But like in equilibrium critical phenomena, most of the work
focuses onto the determination of the critical exponents,
neglecting the determination of the universal scaling functions.
%{\bf Hier noch mehr. Cellular automata Vielleicht Satz aus Einleitung,
%at a first glance nach at first glance}
%{\bf Liggett zitieren fuer analytisches zum CP}

\section{Definitions and the contact process at first glance}
\label{sec:apt_defs}

Absorbing phase transitions occur in dynamical systems
that are characterized by at least one 
absorbing state.
Any configuration in which a system becomes trapped
forever is an absorbing state.\index{absorbing state}
The essential physics of absorbing phase transitions
is the competition between the proliferation and the
annihilation of a quantity of interest~$A$, for example
particles, energy units, molecules in catalytic reactions,
viruses, etc.
Often the proliferation and the annihilation processes
are described in terms of reaction-diffusion schemes, e.g.,
\begin{eqnarray}
\label{eq:reac_diff_proli}
{\mathrm {proliferation:}} & \quad &
A \, \longrightarrow \, 2\,A\, , \quad 2\, A \, \longrightarrow \, 3\,A\, , \quad
\hspace{1.5eX}  A \, \longrightarrow \, 3\,A\,, \ldots\\
\label{eq:reac_diff_annih}
{\mathrm {annihilation:}} & \quad &
A \, \longrightarrow \, 0\,, \hspace{2.0eX}  
\quad 2\, A \, \longrightarrow \, 0\, ,\hspace{2.0eX} \,
\quad 3\, A \, \longrightarrow \, 2\,A\,, \ldots \; .
\end{eqnarray}
Obviously any configuration with zero 
density ($\rho_{\ssstyle A}=0$) is an absorbing state if 
spontaneous particle generation processes
\begin{equation}
0 \, \longrightarrow \, A\, , \quad 0 \, \longrightarrow \, 2\,A\, , \quad
0 \, \longrightarrow \, 3\,A\,, \ldots 
\label{eq:reac_diff_creat}
\end{equation}
do not take place.
When the annihilation processes prevail 
the proliferation processes, the system will eventually reach 
the absorbing state after a transient regime.
On the other hand, the system will be characterized by a
non-zero steady-state density $\rho_{\ssstyle A}$ in the thermodynamic limit
if the proliferation
processes outweigh the annihilation processes.
In the latter case the system is said to be in the 
active phase whereas the absorbing phase contains
all absorbing configurations. 
Let us assume that the competition between the proliferation
and annihilation is described by a single rate~$r$.
An absorbing phase transition, i.e., a transition from
the active phase to the absorbing phase takes place
at the critical rate~$r_{\ssstyle \mathrm c}$ if the
steady state density $\rho_{\ssstyle A}$ 
vanishes below $r_{\ssstyle \mathrm c}$
\begin{equation}
\begin{array}{rcl}
\rho_{\ssstyle A}  = 0   & {\mathrm {for}} &  r < r_{\ssstyle \mathrm c} \, , \\[2mm]
\rho_{\ssstyle A}  > 0   & {\mathrm {for}} &  r > r_{\ssstyle \mathrm c} \, .
\end{array}
\label{eq:density_A_op}
\end{equation}
In other words, the density~$\rho_{\ssstyle A}$ is the order parameter
and the rate~$r$ is the control parameter of the absorbing
phase transition.
Analogous to equilibrium phase transitions, 
an order parameter exponent~$\beta$
can be defined if the order parameter vanishes continuously 
\begin{equation}
\rho_{\ssstyle A}  \; \asympprop \; (r-r_{\ssstyle \mathrm c})^{\beta} \, .
\label{eq:density_A_op_beta}
\end{equation}
It is worth mentioning that absorbing phase transitions
have no equilibrium counterparts since they are far from equilibrium
per definition.
In equilibrium, the associated transition rates
between two states
satisfy detailed\index{detailed balance}
balance~\cite{BINDER_1}.
In case of absorbing phase transitions the
rate out of an absorbing state is zero.
Therefore, an absorbing state can not obey detailed
balance with any other active state~\cite{GRINSTEIN_2}.

For the sake of concreteness,
let us consider the so-called
contact process (CP) on a $D$-dimensional\index{contact process}\index{CP}
lattice.
The contact process was introduced by Harris~\cite{HARRIS_2}
in order to model the spreading of epidemics 
(see for a review~\cite{MARRO_1}).
A lattice site  may be empty ($n=0$) 
or occupied ($n=1$) by a particle
representing a healthy or an infected site.
Infected sites will recover with probability~$1-p$ 
corresponding to a process of particle annihilation.
With probability $p$ an infected site will create 
an offspring at a randomly chosen vacant nearest neighbor site
reflecting particle propagation.
Thus the dynamics of the contact process is described by
the reaction scheme
\begin{equation}
A %\mathop{\longrightarrow}\limits_{p} 
\longrightarrow \, 0\, , \quad\quad\quad 
A %\mathop{\longrightarrow}\limits_{1-p} 
\longrightarrow \, 2\,A\, , 
\label{eq:reac_diff_cp}
\end{equation}
where the quantity~$A$ represents a particle.
Particles are considered as active in the sense that they
proliferate with rate $p$ and annihilate with rate $1-p$.
\label{page:cd_dynamic_rules}
Empty lattice sites remain inactive and the empty lattice
($\rho(A)=0$) is the unique absorbing state.
In the following we denote the density of active sites,
i.e., the order parameter of the absorbing phase 
transition with~$\rhoa$.

\begin{figure}[t] 
\centering
%\leavevmode 
\includegraphics[clip,width=13cm,angle=0]{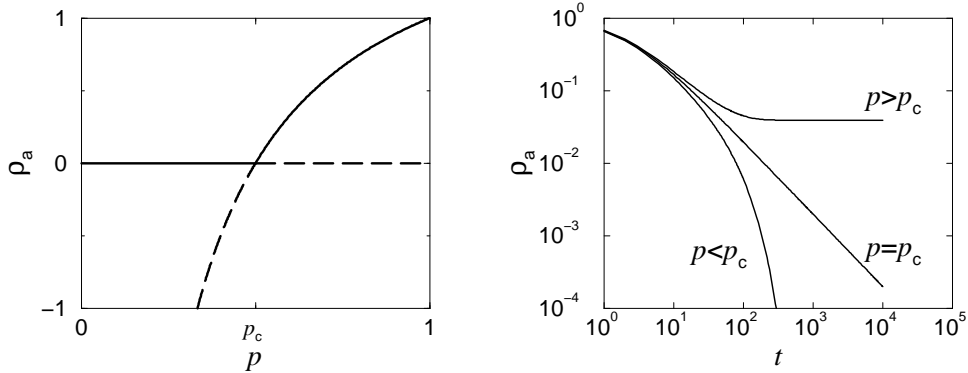}
\caption[Order parameter behavior of the mean field contact process, zero field] 
{Sketch of the steady state behavior (left) 
of the order parameter of the contact process within the mean 
field approximation [\refeqname{eq:mf_cp_one_site_steady_state}].
The dashed lines mark unphysical or unstable solutions.
%The upper right figure displays simulation data of the five-dimensional
%contact process on cubic lattices of linear size $L=16$ and $L=32$.
%In agreement with the mean field solution the order parameter 
%vanishes linearly at the transition point $\pc \approx 0.53238$.
The right figure displays the dynamical behavior of the order parameter.
The parameters of the presented mean field curves  
are $\rho_{\ssstyle {\mathrm a},t=0}=1$, $\deltap=0$ 
and $\deltap=\plmi0.02$, respectively.
%The lower right figure shows corresponding simulation data 
%of the five-dimensional contact process for $L=32$.
\label{fig:cp_mf_one_site_ord_zf}} 
\end{figure}

In order to provide a deeper insight into absorbing
phase transitions, we present a simple but 
intriguing\index{mean field theory, CP}
mean field\label{pg:mf_cp} treatment.
The probability that a given lattice is occupied is~$\rhoa$.
This particle will be annihilated with probability $1-p$.
Thus in an elementary update step the number of particles~$n$
is decreased with the probability $p(\Delta n=-1)=\rhoa (1-p)$.
On the other hand, a new particle is created with rate~$p$ at a 
randomly chosen vacant neighbor.
The associate probability is $p(\Delta n=1)=\rhoa \, p (1-\rhoa)$.
The number of particles remains unchanged with
probability $p(\Delta n=0)=(1-\rhoa)+ \rhoa^2 p$.
Since we have neglected spatial fluctuations and correlations
of $\rhoa$, the above reaction scheme describes the simplest mean field 
theory of the contact process.
This reaction scheme leads to the differential equation 
\begin{equation}
\partial_{\ssstyle t} \rhoa\; = \; 
\, p \, \rhoa(1-\rhoa) \,  - \, \rhoa \, (1-p)
\; = \; (2p-1)\rhoa \, - \, p \, \rhoa^2\, ,
\label{eq:mf_cp_one_site}
\end{equation}
with the steady state solutions ($\partial_{\ssstyle t}\rhoa=0$)
\begin{equation}
\rhoa \; = \; 0 \quad\quad \vee
\quad\quad
\rhoa \; = \; \frac{2p-1}{p} 
\quad \mathrm{for} \quad p>0 \, .
\label{eq:mf_cp_one_site_steady_state}
\end{equation}
The first equation corresponds to the absorbing state 
and is unstable for $p>1/2$.
The second equation yields unphysical results ($\rhoa<0$)
for $p<1/2$ but describes the order parameter for $p>1/2$.
The behavior of the order parameter is sketched 
in \reffigname{fig:cp_mf_one_site_ord_zf}.
In the vicinity of the critical value $\pc=1/2$ we find 
\begin{equation}
\rhoa \; = \; 2\, \deltap \, + \, {\mathcal O}(\deltap^2)
\label{eq:mf_cp_one_site_op}
\end{equation}
with the reduced control parameter $\deltap=(p-\pc)/\pc$.
Thus the order parameter exponent is $\beta=1$.

Consider now the dynamical behavior of the order parameter.
Solving \refeqname{eq:mf_cp_one_site} we obtain
\begin{equation}
\rhoa(t) \; = \; 
\frac{\deltap}
{\; p\,+\,({\deltap}/{\rho_{\scriptscriptstyle{\mathrm a},t=0}}-p)\;  
\exp{(-\deltap \; t)}\;} \, . 
\label{eq:mf_cp_one_site_dyn_01}
\end{equation}
Asymptotically ($t\to\infty$) the order parameter behaves as
\begin{eqnarray}
\label{eq:mf_cp_one_site_dyn_below}
\left . 
\rhoa(t)^{\vphantom X}  
\right |_{\scriptscriptstyle \deltap < 0}
& \sim &  - \, \deltap \,
\left (p- \frac{\deltap} { \rho_{\ssstyle {\mathrm a},t=0}} \right )^{-1}
%\frac{\, \delta\rho\,}{4} 
\; {\mathrm e}^{\deltap \, t}  \, , \\[2mm]
%\; \; \, \quad \quad \quad \quad {\rm for} \quad \delta\rho<0 \, , 
\label{eq:mf_cp_one_site_dyn_above}
\left . 
\rhoa(t)^{\vphantom X} 
\right |_{\scriptscriptstyle \deltap > 0}
& \sim & \frac{\,\deltap\,}{p} \; + \;
\frac{\deltap}{p^2}
\left (p- \frac{\deltap} { \rho_{\ssstyle {\mathrm a},t=0}} \right )
{\mathrm e}^{-\deltap \, t} \, .
%\quad \quad {\rm for} \quad \delta\rho>0 \, .
\end{eqnarray}
Apart from criticality,
the steady state solutions ($\rhoa=\deltap/p$ and 
$\rhoa=0$) are approached exponentially, independent of the
initial value $\rho_{\ssstyle {\mathrm a},t=0}$.
The associated correlation time diverges as
%$\tau_{\ssstyle {\mathrm {corr}}}=|\deltap|^{-1}$.  
$|\deltap|^{-1}$.  
%with $\nu_{\scriptscriptstyle \parallel}=1$.
At criticality ($\deltap=0$) 
the order parameter exhibits an algebraic decay
\begin{equation}
\rhoa(t) \; = \; 
\frac{1}{\, \rho_{\ssstyle \rm a,t=0}^{-1} \, + \, p\, t \, }
\; \mathop{\longrightarrow}\limits_{t\to \infty}
\; \frac{1}{\, p\, t \, } \, .
\label{eq:mf_cp_one_site_dyn_at_crit}
\end{equation}
The dynamical behavior of the order parameter is sketched 
in \reffigname{fig:cp_mf_one_site_ord_zf}.

\begin{figure}[t] 
\centering
%\leavevmode 
\includegraphics[clip,width=13cm,angle=0]{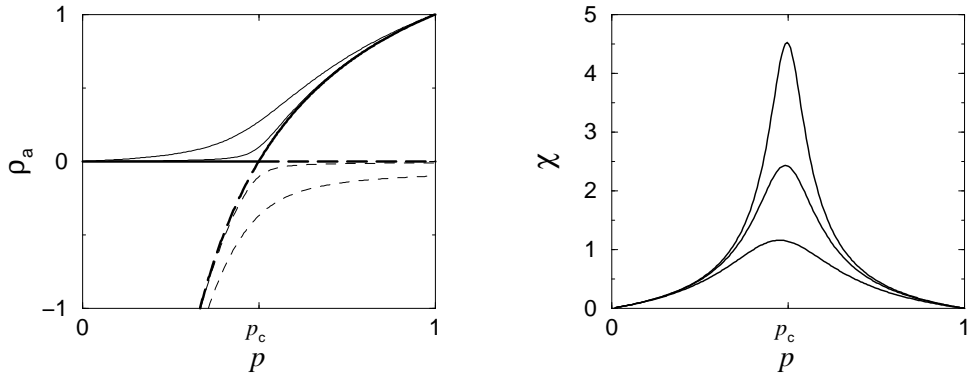}
\caption[Order parameter behavior of the mean field contact process, non-zero field] 
{The mean field order parameter (left) and the mean field order 
parameter susceptibility (right) of the contact process.
The dashed lines mark unphysical or unstable solutions.
Thick lines correspond to the zero-field behavior
whereas the thin lines sketch the smooth non-zero field 
behavior.
The susceptibility displays a finite peak for non-zero field.
In the limit $h\to 0$ this peak diverges
signalling the critical point.
%The right figures show simulation data of the five-dimensional
%contact process on cubic lattices of linear sizes $L=8,16,32$
%for various field values.
%In agreement with the mean field solution the order parameter
%exhibits a smooth behavior for finite field and the susceptibility
%displays the characteristic peak in the vicinity of the 
%critical point $\pc\approx0.53238$.
\label{fig:cp_mf_one_site_ord_f}} 
\end{figure}

Similar to equilibrium phase transitions, it is often 
possible for absorbing phase transitions
to apply an 
external field that is conjugated\index{conjugated field}
to the order parameter.
Being a conjugated field it has to destroy the absorbing
phase, it has to be independent of the control parameter,
and the corresponding linear response function
has to diverge at the critical point
\begin{equation}
\chi \; = \; \frac{\partial \, \rhoa}{\partial \, h} \; 
\longrightarrow \; \infty \, .
\label{eq:lin_resp_apt}
\end{equation}
For the contact process the conjugated field causes 
a spontaneous creation of particles ($0 \rightarrow A$).
Clearly spontaneous particle generation destroys 
the absorbing state and therefore the absorbing phase
transition at all.
In the above mean field scheme the conjugated field
creates a particle at an empty lattice site with 
probability $(1-\rhoa)h$.
This particle will survive the next update with probability~$p$
or will be annihilated with probability $1-p$, respectively.
The modified differential equation [\refeqname{eq:mf_cp_one_site}]
is given by
\begin{equation}
\partial_{\ssstyle t} \rhoa\; = \; 
\deltap \, \rhoa \, - \, p \, \rhoa^2\, 
+ \, p \, (1-\rhoa) \, h \, .
\label{eq:mf_cp_one_site_field}
\end{equation}
The steady state order parameter is a function of both
the control parameter and the conjugated field
\begin{equation}
\rhoa (\deltap , h)\; = \; \frac{\deltap - p\, h }{2\,p}\, \pm \,
\sqrt{\left ( \frac{\deltap-p\,h}{2\,p} \right )^2
\, + \, {h} \; } \, .
\label{eq:mf_cp_one_site_ord_field_gen}
\end{equation}
The solution with the $+$\,sign describes the 
order parameter as a function of the 
control parameter and of the conjugated field whereas
the $-$\,sign solution yields unphysical results $\rhoa<0$.
Both solutions are sketched in \reffigname{fig:cp_mf_one_site_ord_f}
for various field values.
As requested, the absorbing phase is no longer a 
steady state solution for $h>0$.
At criticality ($\deltap=0$) the order parameter behaves as
\begin{equation}
\rhoa (\deltap=0 , h)\; = \; \sqrt{{h}\,} \; + \; {\mathcal O}(h) \, .
\label{eq:mf_cp_one_site_ord_field_asymp}
\end{equation}

Now we examine the order parameter behavior close to the
critical point.
Therefore we perform in \refeqname{eq:mf_cp_one_site_field}
the limits $\rhoa\to 0$, $\deltap \to 0$, and
$h\to 0$ with the constraints that $\rhoa/\sqrt{h}$ and $\rhoa/\deltap$
are finite.
The remaining leading order yields
\begin{equation}
\rhoa (\deltap , h )\; \sim \; \frac{\deltap}{2\,p}\, + \,
\sqrt{\left ( \frac{\deltap}{2\,p} \right )^2
\, + \, {h} \; } \, .
\label{eq:mf_cp_one_site_ord_field_scal_asymp}
\end{equation}
The \refeqs2name{eq:mf_cp_one_site_op}{eq:mf_cp_one_site_ord_field_asymp}
are recovered from this result by setting \mbox{$h=0$} and 
\mbox{$\deltap=0$}, respectively.
It is straight forward to derive the dynamical behavior
of the order parameter $\rhoa(\deltap,h,t)$ close to the
critical point ($\rhoa \ll 1$). 
The differential equation \refeqname{eq:mf_cp_one_site_field} 
yields that the steady state solution is approached from
above as 
\begin{equation}
\rhoa (\deltap , h , t)\; \sim \; \frac{\deltap}{2\,p}\, + \, 
\sqrt{\left ( \frac{\deltap}{2\,p} \right )^2
\, + \, {h} \; } \;
\; \left ( 1 \, \plmi \, 2 \, c_{\ssstyle 0} 
{\mathrm e}^{-t/\xi_{\para}}
\, + \, {\mathcal O}({\mathrm e}^{-2t/\xi_{\para}}) \, \right  ) \, ,
\label{eq:mf_cp_one_site_ord_field_dyn}
\end{equation}
where the constant $c_{\ssstyle 0}$ contains the 
initial conditions.
The $+$\,sign corresponds to initial conditions with 
$\rho_{\ssstyle {\mathrm a},t=0 }>\rho_{\ssstyle {\mathrm a},t\to\infty }$ 
whereas the $-$\,sign is valid for 
$\rho_{\ssstyle {\mathrm a},t=0 }<\rho_{\ssstyle {\mathrm a},t\to\infty }$.
Again, the steady state value is exponentially
approached independent of the initial condition.
The corresponding temporal correlation length is usually
denoted as $\xi_{\para}$ for absorbing phase transitions,
and we find~\cite{GARDINER_1}
\begin{equation}
\xi_{\para}(\deltap ,h ) \; = \; \frac{1}
{2 \, p \, \sqrt{\left (\frac{\deltap}{2p} \right )^2+h\,} \,} \, .
\label{eq:temp_charac_length_mf}
\end{equation}
Thus the correlation time diverges at the critical point
as 
\begin{equation}
\xi_{\para}(\deltap ,h=0 ) \; = \; |\deltap|^{-1}
\quad\quad {\mathrm{ and}} \quad\quad
\xi_{\para}(\deltap=0 ,h ) \; = \; \frac{1}{2 \, p} \,
h^{-1/2} \, .
\label{eq:temp_charac_length_sing_mf}
\end{equation}
The zero-field singularity is usually associated with
the temporal correlation length exponent $\nu_{\para}=1$.

Furthermore, the derivative of $\rhoa$ with respect to the
conjugated field yields the order parameter susceptibility 
\begin{equation}
\chi (\deltap , h ) \; = \; \frac{\partial \, \rhoa}{\partial \, h} 
\;  \sim \; \frac{1}{2} \; 
\left [ \left ( \frac{\deltap}{2\,p} \right )^2
\, + \, {h} \; \right ]^{\,-1/2} \, .
\label{eq:mf_cp_one_site_sus_scal_asymp}
\end{equation}
As required, the order parameter susceptibility 
becomes infinite at the transition point
\begin{eqnarray}
\left . \chi(\deltap, h)^{\vphantom X} \right |_{h\to 0\,} & \propto &
|\deltap |^{-1} \; \; \, \mathop{\longrightarrow}\limits_{\deltap \to 0} \; \infty
\, , \\[2mm]
\left . \chi(\deltap, h)^{\vphantom X} \right |_{\deltap\to 0} & \propto &
h^{-1/2} \; \; \; \mathop{\longrightarrow}\limits_{h \to 0} \; \infty
\, .
\label{eq:mf_cp_one_site_sus_asymp}
\end{eqnarray}
The mean field behavior of the susceptibility is displayed in 
\reffigname{fig:cp_mf_one_site_ord_f}.

It is possible to incorporate spatial variations of the order parameter 
in the mean field theory.
Therefore \refeqname{eq:mf_cp_one_site_field} has to be modified
and we assume that the order parameter obeys close to
criticality ($\rhoa \ll 1$)  
\begin{equation}
\partial_{\ssstyle t} \rhoa (\underline x,t)\; = \; 
\deltap \, \rhoa(\underline x,t) \, - \, p \, \rhoa(\underline x,t)^2\, 
+ \, p \, h \,
+ \, \Gamma \, \nabla^2 \rhoa (\underline x, t) \, ,
\label{eq:mf_cp_field_equation}
\end{equation}
where $\rhoa(\underline x,t)$ represents the local
particle density and where we add the diffusive coupling
with $\Gamma>0$.
Usually, continuous equations
like \refeqname{eq:mf_cp_field_equation} are obtained
from microscopic models by a coarse graining procedure
or they are phenomenologically motivated by a Landau
expansion~\cite{LANDAU_5}.
According to Landau, a given system is described by an appropriate  
functional containing all analytic terms that are consistent
with the symmetries of the microscopic problem.
%Close to the critical point ($\rhoa \ll 1$) 
Consider small
spatial deviations from the homogeneous steady state 
value $\rhoa (\deltap,h)$ [\refeqname{eq:mf_cp_one_site_ord_field_scal_asymp}]
by introducing 
$\delta\rhoa(\underline x,t) = \rhoa(\underline x,t)-\rhoa(\deltap,h)$.
The order parameter variations $\delta\rhoa(\underline x, t)$ obey the differential
equation
\begin{equation}
\partial_{\ssstyle t} \, \delta\rhoa (\underline x,t)\; = \; 
[ \deltap \, - \, 2 \,p \, \rhoa(\deltap,h) ] \;
\delta\rhoa(\underline x,t) \, 
+ \, \Gamma \, \nabla^2 \, \delta\rhoa (\underline x, t) \, 
+ \, {\mathcal O}(\delta\rhoa^2) \, ,
\label{eq:mf_cp_spatial_variations}
\end{equation}
where nonlinear terms are neglected.
Working in Fourier space and 
using \refeqname{eq:mf_cp_one_site_ord_field_scal_asymp}
we find that small deviations with wavevector $\underline k$
decay as
\begin{equation}
\partial_{\ssstyle t} \, \delta\rhoa (\underline k,t)\; = \; 
- \, \Gamma \, \left  [ k^2 \, + \, 
%2 \,p \, \sqrt{ ( {\deltap}/{2\,p}  )^2 \, + \, {h} \; } 
\xi_{\senk}^{- 2}
\, \right ] \;
\delta\rhoa(\underline k,t) \, .
\label{eq:mf_cp_spatial_variations_decay}
\end{equation}
Here, the characteristic length 
\begin{equation}
\xi_{\senk}(\deltap,h) \; = \; 
\sqrt{\, \frac{\Gamma}{2\, p} \,} \;
\left [ \left ( \frac{\deltap}{2\,p} \right  )^2
\, + \, {h} \; \right ]^{\,-1/4}
\label{eq:spatial_charac_length_mf}
\end{equation}
is introduced that describes spatial correlations of the order
parameter variations~\cite{MORI_1,ELDERFIELD_1}.
Approaching the transition point, the spatial correlation length 
becomes infinite
\begin{equation}
\xi_{\senk}(\deltap ,h=0 ) \; = \; \sqrt{\Gamma\,} \, |\deltap|^{-1/2}
\quad\quad {\mathrm{ and}} \quad\quad
\xi_{\senk}(\deltap=0 ,h ) \; = \; \sqrt{ \frac{\Gamma}{2 \, p} \,} \,
h^{-1/4} \, .   
\label{eq:spatial_charac_length_sing_mf}
\end{equation}
Thus the mean field value of the exponent of the correlation
length at zero field is $\nu_{\senk} =1/2$.
Note that the temporal and spatial correlation length
are related via~\cite{GARDINER_1}
\begin{equation}
\xi_{\senk}^2 \; = \; \Gamma \, \xi_{\para} \, .
\label{eq:corr_spatial_temp_mf}
\end{equation}
Solving \refeqname{eq:mf_cp_spatial_variations_decay}
and taking the inverse transform we obtain the 
order parameter variations
\begin{equation}
\delta\rhoa (\underline x , t)\; = \; 
\rho_{\ssstyle {\mathrm a}, 0} \,
\left ( \frac{1}{4\,\pi\,\Gamma \,t}\, \right )^{D/2} \;
{\mathrm e}^{-t/\xi_{\para}} \;
{\mathrm e}^{-(\underline x - {\underline x}_{\ssstyle 0})^2/4\Gamma t} \, ,
\label{eq:mf_cp_spatial_variations_op_xt}
\end{equation}
where we have assumed a seed-like initial density variation
$\delta\rhoa(\underline x, t=0)=\rho_{\ssstyle {\mathrm a}, 0}
\delta(\underline x - {\underline x}_{\ssstyle 0})$.
Similar to the homogeneous case the lifetime of the 
localized density fluctuation is determined by the temporal
correlation time $\xi_{\para}$.
Furthermore the activity diffuses over regions of 
distance~$\sqrt{2 \Gamma t\,}$ from the origin~${\underline x}_{\ssstyle 0}$.

Fluctuations within the steady state can be investigated
by adding an appropriate noise term $\eta({\underline x},t)$ 
in \refeqname{eq:mf_cp_field_equation}
representing rapidly-varying degrees of 
freedom (see e.g.~\cite{GARDINER_1,MORI_1}).
The noise has zero mean and the correlator is assumed to be
given by 
\begin{equation}
\langle \, \eta({\underline x},t) \, 
\eta({\underline x}^{\prime},t^{\prime}) \, \rangle
\; = \; \kappa \; \rhoa \;
\delta({\underline x}-{\underline x}^{\prime}) \;
\delta(t-t^{\prime}) \, .
\label{eq:langevin_dp_corr_mf_01}   
\end{equation}
In that case, the mean field steady state fluctuations are 
given by~\cite{MORI_1} 
\begin{equation}
\langle \, \delta\rhoa^2 
%\delta\rhoa (\underline x^{\prime}) 
\, \rangle
\; \propto \; \frac{\kappa}{\,4 p\,} \;
\frac{\, 
\frac{\deltap}{2\,p}\, + \,
\sqrt{\left ( \frac{\deltap}{2\,p} \right )^2
\, + \, {h} \; } \, }
{\sqrt{\left ( \frac{\deltap}{2\,p} \right )^2
\, + \, {h} \; } \, } \, . 
\label{eq:langevin_dp_corr_mf_02}   
\end{equation}
For zero-field the fluctuations reduce to
\begin{equation}
\langle \, \delta\rhoa^2 \rangle \, \
\; \propto \; 
\left \{
\begin{array}{ll}
0 & \mathrm{if} \quad \deltap<0\\[2mm]
\kappa/2 p & \mathrm{if} \quad \deltap>0 \, ,
\end{array}
\right . 
\label{eq:langevin_dp_corr_mf_03}   
\end{equation}
i.e., the fluctuations do not diverge at the critical
point but exhibit a finite jump.

In summary, the mean field theory presented above describes the 
absorbing phase transition of the contact process.
The steady state and dynamical order
parameter behavior have been derived and certain 
critical exponents are determined.
In contrast to the one characteristic length scale of 
ordinary critical equilibrium systems,
absorbing phase transitions are characterized by two
length scales, $\xi_{\senk}$ and $\xi_{\para}$,
with different critical exponents 
$\nu_{\senk}$ and $\nu_{\para}$.
%As we will see this has consequences for the scaling laws.
This scaling behavior of directed percolation can be 
interpreted in terms of so-called strongly anisotropic 
scaling~\cite{HENKEL_5},
where a different steady state scaling\index{strongly anisotropic scaling} 
behavior occurs\index{scaling, strongly anisotropic} 
along different directions (indicated by different critical exponents, 
$\nu_{\para}/\nu_{\senk}\neq 1$).
Paradigms of strongly anisotropic scaling
are\index{Lifshitz points} 
Lifshitz points~\cite{HORNREICH_1}, e.g.~in Ising models
with competing interactions~\cite{SELKE_1,HUCHT_1}.

Unfortunately, the notations of some critical exponents
for absorbing phase transitions differ from those of
equilibrium systems.
In particular the field dependence of the order parameter
at the critical isotherm is written as~\cite{HINRICHSEN_1} 
\begin{equation}
\rhoa (\deltap=0,h) \; \asympprop \; h^{\beta/\sigma} \, ,
\label{eq:ord_par_crit_isotherm}
\end{equation}
in contrast to the usual equilibrium notation $M\asympprop h^{1/\delta}$.
Thus the field exponent $\sigma$ corresponds  
in equilibrium to $\beta \,\delta$, i.e., the exponent 
$\sigma$ of absorbing phase transitions 
is identical to the gap exponent $\Delta=\beta\, \delta$
of equilibrium systems.\index{gap exponent}
In order to avoid friction with the literature of absorbing 
phase transitions, where $\delta$ is reserved to describe the 
dynamical scaling behavior, we will 
use the notation of \refeqname{eq:ord_par_crit_isotherm} 
in the following.
Eventually we summarize that the mean field theory 
is characterized by the exponents
$\beta=1$, $\sigma=2$, $\nu_{\para}=1$, 
$\nu_{\senk}=1/2$, as well
as $\gamma=1$, where the latter one describes the divergence 
of the zero-field susceptibility 
($\chi \asympprop |\deltap|^{-\gamma}$).

\section{Numerical simulation methods}
\label{sec:num_sim}

Despite the simplicity of the contact process, a
rigorous solution is still lacking.
Like in equilibrium the mean field theory provides
correct results only above the upper critical
dimension~$\Dc$.
It fails below $\Dc$ since different sites are strongly
correlated whereas they are considered as independent
within the mean field approach.
Beside of approximation schemes like\index{series expansion}
series 
expansions~\cite{ESSAM_4,ADLER_1,ADLER_2,ESSAM_5,JENSEN_11,JENSEN_12,JENSEN_13,JENSEN_10,JENSEN_14,JENSEN_1,JENSEN_5,JENSEN_16}
and $\epsilon$-expansion
within renormalization group approaches,
numerical simulations were intensively used in order to determine 
the critical behavior of the
contact process and related models.
There are two primary ways to perform numerical
simulations of lattice models exhibiting absorbing phase transitions.
The first method is to study the steady state
behavior of certain quantities of interest, 
e.g., the order parameter and its fluctuations as 
a function of various parameters like
the control parameter and the external 
field~(e.g.~\cite{JENSEN_2,JENSEN_3,MENDES_1,LUEB_19}).
Especially the equation of state as well as the susceptibility
can be obtained in that way~\cite{LUEB_22,LUEB_24,LUEB_27,LUEB_26,LUEB_28,LUEB_32}.
But approaching the critical point, the 
simulations are affected by diverging fluctuations and 
correlations causing
finite-size and critical slowing down effects
(see e.g.~\cite{BINDER_1}).
Techniques such as finite-size scaling analysis have to be
applied in order to handle these effects.

In contrast to steady state measurements, the 
second method is based on the dynamical 
scaling behavior.
Often the evolution of a prepared state close
to the absorbing phase is investigated.
For example the spreading of activity started from a
single seed is studied in the vicinity of the
critical point~\cite{GRASSBERGER_4}.
This method is known to be very efficient and the critical
exponents can be determined with high accuracy
(see for
instance~\cite{GRASSBERGER_2,GRASSBERGER_8,JENSEN_8,DICKMAN_9,JENSEN_9,JENSEN_6,ODOR_2,VOIGT_1}). 
A drawback of this technique is that it is restricted  to the vicinity 
of the absorbing state. 
For example the conjugated field drives the systems
away from the absorbing phase and can therefore not be
incorporated.
In the following both methods are discussed
and the associated exponents are defined.
For the sake of concreteness corresponding data 
of the five-dimensional contact process are presented
as an exemplification and are compared to the mean field results. 
%The universal scaling forms as well as the scaling laws
%which connects the critical exponents are derived in the
%next chapter.

\subsection{Steady state scaling behavior}
\label{subsec:apt_steady_state_scal}

\begin{figure}[b] 
\centering
%\leavevmode 
\includegraphics[clip,width=13cm,angle=0]{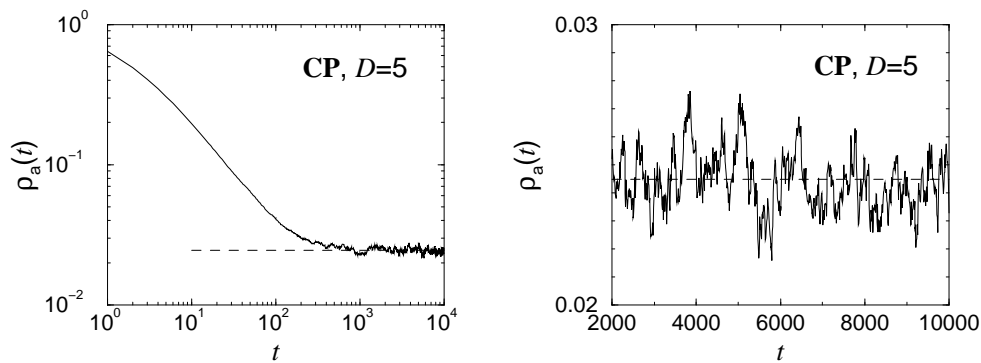}
\caption[Order parameter behavior of the mean field contact process, relaxation] 
{The density of active sites $\rhoa$ as a function
of time for the five-dimensional contact process 
($L=16$, $\deltap=0.02$, and $\rho_{\ssstyle {\mathrm a},t=0}=1$).
After a transient regime, which depends on the initial
configuration, the density of active sites
fluctuates around the steady state value $\rhoa$ (dashed line).
\label{fig:cp_5d_rho_t}} 
\end{figure} 

In order to perform simulations of 
a system exhibiting an absorbing phase transition
we have to specify the lattice type
and the boundary conditions.
In case of the contact process it is customary
to consider $D$-dimensional simple cubic lattices of linear size~$L$
and periodic boundary conditions.
Steady state simulations usually start far away from
the absorbing state, e.g.~with a fully
occupied lattice.
The system is updated according to the microscopic
rules presented on page~\pageref{page:cd_dynamic_rules}
using a randomly sequential update
scheme.\index{random sequential update}
Therefore all occupied sites are listed and one
active site after the other is updated, selected at random.
After a sufficient number of update steps the system
reaches a steady state where the number of active
sites fluctuates around the average value 
(see \reffigname{fig:cp_5d_rho_t}).
It is customary to interpret one complete lattice
update with $n_{\ssstyle \mathrm a}$ active sites 
as one time step ($t \to t+1$).
Thus an elementary update of one lattice site corresponds
to a time increment $t \to t + 1/n_{\ssstyle \mathrm a}$.
Monitoring the density of active sites $\rhoa(t)$ 
in the steady state
one obtains an estimate of the order 
parameter~$\rhoa=\langle \rhoa(t) \rangle$
as well as of its fluctuations
\begin{equation}
\Delta\rhoa \; = \; L^D \, 
\left ( \,\langle \rhoa(t)^2 \rangle  \, - \,
\langle \rhoa(t) \rangle^2 \, \right ) \, ,
\label{eq:def_fluc}
\end{equation}
where $\langle \ldots \rangle$ corresponds to the temporal
average
\begin{equation}
\langle \rhoa(t)^k \rangle \; = \;
\frac{1}{T} \, \sum_{\ssstyle t=1}^{\ssstyle T}
\, \rhoa(t)^k \, .
\label{eq:temporal_average}
\end{equation}
Of course reliable results are only obtained
if the number of update steps significantly exceeds 
the correlation time $T \gg \xi_{\para}$.
This becomes notoriously difficult close to the
critical point since $\xi_{\para}\to \infty$.
Furthermore the accuracy can be improved if an additional
averaging over different initial configurations is 
performed.

\begin{figure}[t]
\centering
%\leavevmode
\includegraphics[clip,width=13cm,angle=0]{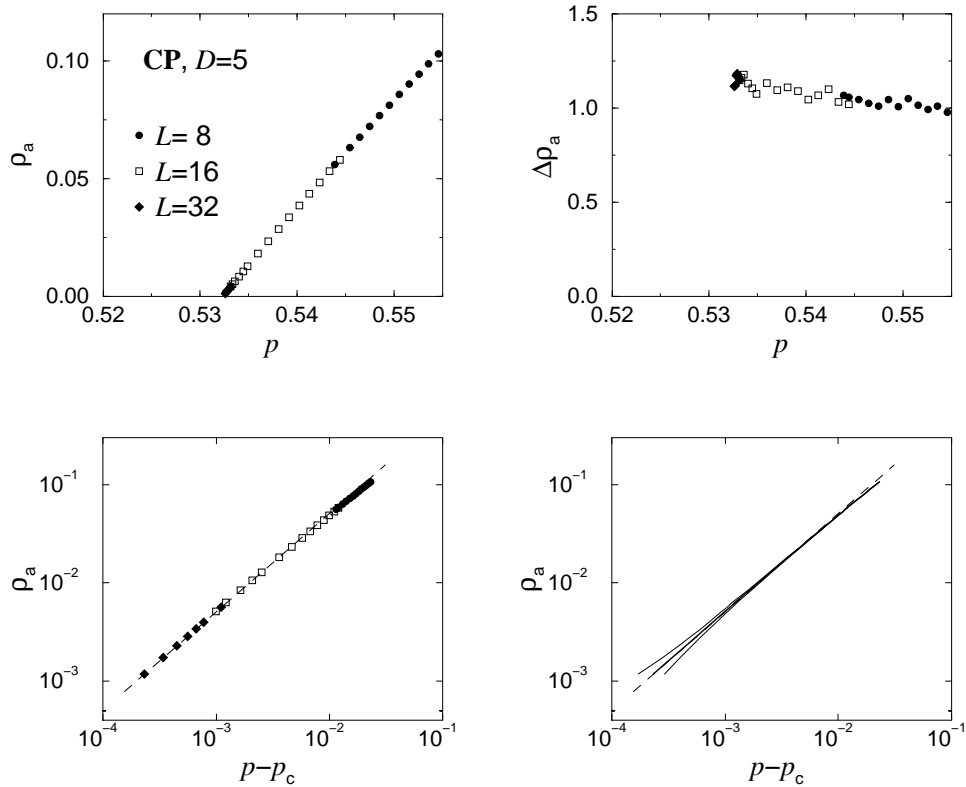}
\caption[Order parameter behavior of the contact process, $D=5$ at $h=0$]
{The order parameter (upper left) and its fluctuations (upper right)
of the five-dimensional contact process for various system sizes~$L$.
The data are averaged over at least $T=10^{7}$ lattice updates.
The lower figures show the order parameter as a function of
$p-\pc$.
The dashed line corresponds to the mean field behavior with $\beta=1$.
A straight line is obtained for $\pc=0.53237$, whereas
$\pc=0.53243$ and $\pc=0.53231$ lead to significant curvatures
(see lower right figure).
For the sake of simplicity lines are
plotted instead of symbols.
\label{fig:cp_5d_ord_fluc_01}}
\end{figure}

The average density of active sites as well as its fluctuations 
are plotted 
in \reffigname{fig:cp_5d_ord_fluc_01} as a function of the control
parameter~$p$ for various system sizes~$L$.
As can be seen the order parameter tends to zero in the
vicinity of $p \approx 0.532$.
Assuming that the scaling behavior of $\rhoa$ obeys
asymptotically the power law 
\begin{equation}
\rhoa(p) \; \asympprop \; (p-\pc)^\beta \, ,
\label{eq:ord_par_def_beta_01}
\end{equation}
the critical value~$\pc$ 
is varied until a straight line in a log-log plot is obtained
(see \reffigname{fig:cp_5d_ord_fluc_01}).
Convincing results are observed for $\pc=0.53237 \plmi 0.00006$
and the corresponding curve is shown in \reffigname{fig:cp_5d_ord_fluc_01}.
For $\pc=0.53231$ and $\pc=0.53243$ significant curvatures
in the log-log plot occur.
In this way it is possible to estimate the critical
value $\pc$ as well as its error-bars.
Then a regression analysis yields the value of the order 
parameter exponent $\beta = 0.991\plmi 0.015$ in agreement with 
the mean field value $\beta=1$.

Usually the order parameter fluctuations 
diverge at the transition point
\begin{equation}
\Delta\rhoa \; \asympprop \; (p- \pc)^{-\gammap} \, .
\label{eq:def_fluc_gammap}
\end{equation}
But as can be seen the fluctuations are characterized
by a finite jump at~$\pc$
corresponding to the value $\gammap=0$.
This agrees well with the mean field 
result~\refeqname{eq:langevin_dp_corr_mf_03}.
We will see in the next chapter that low-dimensional
systems are characterized by a non-zero exponent.

The effects of finite system sizes\index{finite-size effects} 
deserve comment.
Of course a finite system reaches the absorbing state 
with a finite probability even for $p > \pc$.
But this probability tends to zero for $L\to \infty$.
It turns out that sufficiently large systems 
will never reach the absorbing phase within reasonable 
simulation times for $p > \pc$.
The situation changes close to the transition point
since the finite system size~$L$ prevents the correlation
length~$\xi_{\senk}$ from becoming infinite.
As a result so-called finite-size effects occur, i.e.,
the corresponding singularities become
rounded and shifted (see for example~\cite{BINDER_1}).
A typical feature of finite-size effects in equilibrium 
systems is that a given system may pass within the 
simulations from one phase to the other.
This behavior is caused by critical fluctuations that increases
if one approaches the transition point.
In case of absorbing phase transitions
the scenario is different.
Approaching the transition point the correlation 
length~$\xi_{\senk}$ increases and as soon as~$\xi_{\senk}$ 
is of the order of~$L$ the system may pass to an absorbing 
state and is trapped forever.
For example finite-size effects are expected to occur 
for $\deltap < \Gamma L^{-2}$ according to the mean field 
result \refeqname{eq:spatial_charac_length_sing_mf}.
In that case the absorbing state is reached by
finite fluctuations even for $p>\pc$.
A simple way to handle that problem is to increase the
system size before finite-size effects occur 
(see~\reffigname{fig:cp_5d_ord_fluc_01}).
Considering overlapping data regions for different system
sizes ensures that
the obtained results are not affected by finite-size effects.
Additionally it is suggested (see, for instance~\cite{JENSEN_3})
to consider metastable or quasisteady-state values of the
order parameter.
But as pointed out in~\cite{LUEB_23} this method is 
inefficient and can be seriously questioned since
the metastable values are not well defined.

\begin{figure}[t] 
\centering
%\leavevmode 
\includegraphics[clip,width=13cm,angle=0]{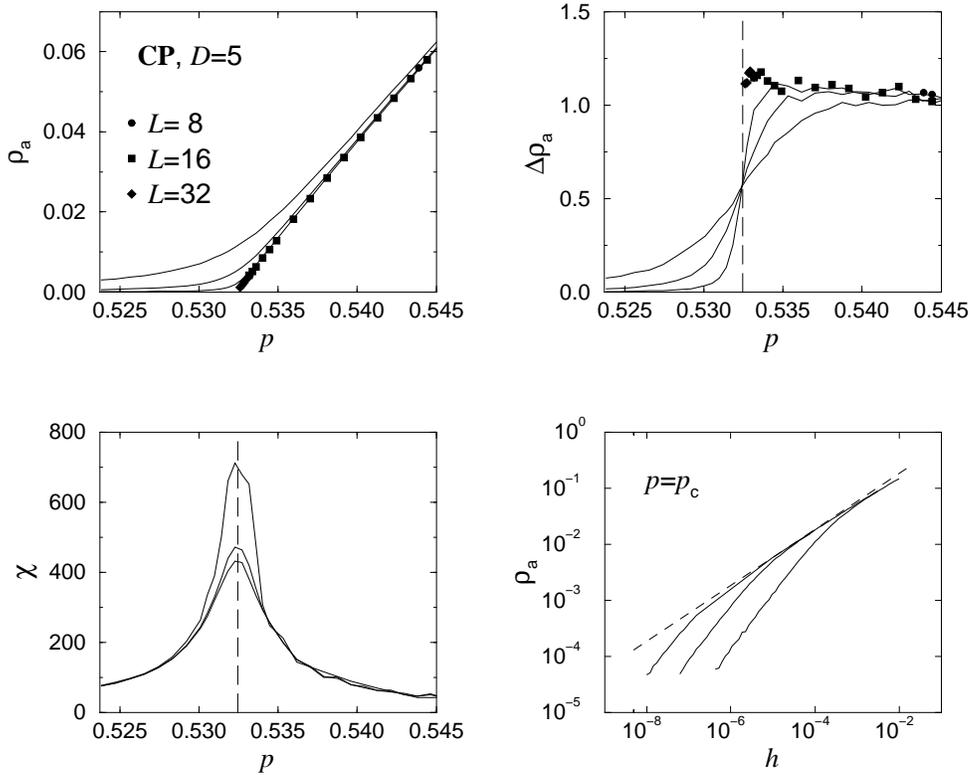}
\caption[Order parameter behavior of the contact process, $D=5$ for $h>0$] 
{The order parameter (upper left) and its fluctuations (upper right)
of the five-dimensional contact process for various field 
values ($h=3\,10^{-5},8\,10^{-6},10^{-6}$). 
The order parameter and the fluctuations
exhibit an analytic behavior for non-zero field.
The order parameter susceptibility (lower left)
displays the characteristic peak at the 
critical point (dashed line).
The lower right figure shows the order parameter 
as a function of the conjugated field at $p=\pc$
for various system sizes~$L=4,8,16$ (from right to left).
The dashed line corresponds to the mean field behavior
with $\beta/\sigma=1/2$.
\label{fig:cp_5d_ord_fluc_02}} 
\end{figure}

A better way to analyze finite-size effects
for absorbing phase transitions is to incorporate
the conjugated field.
Due to the conjugated field the system cannot be
trapped forever in the absorbing phase.
Therefore steady state quantities are available for all
values of the control parameter.
Analogous to equilibrium phase transitions, the\index{conjugated field}
conjugated field results in a rounding of the zero-field
curves, i.e., the order parameter and its fluctuations
behave smoothly as a function of the control parameter~$p$
for finite field values (see \reffigname{fig:cp_5d_ord_fluc_02}).
For $h\to 0$ we recover the non-analytical behavior
of $\rhoa$ and $\Delta\rhoa$, respectively.
Furthermore, the order parameter susceptibility~$\chi$
can be obtained by performing the numerical derivative
of the order parameter $\rhoa(\deltap,h)$ with respect to the 
conjugated field [\refeqname{eq:lin_resp_apt}].
Similar to the mean field solution the susceptibility
displays the characteristic peak at the 
critical point $\pc$.
In the limit $h\to 0$ this peak diverges,
signalling the critical point.

The order parameter and fluctuation data presented 
in \reffigname{fig:cp_5d_ord_fluc_02}
are obtained from simulations 
where the correlation length is small compared to the
systems size.
Thus these data do not suffer from finite-size effects.
But approaching the transition point finite-size
effects emerge.
\refFigname{fig:cp_5d_ord_fluc_02} shows the 
order parameter at $\pc$ as a function of the
conjugated field.
The behavior of the infinite system ($L \gg \xi_{\senk}$)
agrees with the mean field result $\rhoa \asympprop h^{1/2}$.
Approaching the transition point ($h\to 0$), 
finite-size effects, i.e., deviations from the
behavior of the infinite system occur. 
The larger the system size~$L$ the larger
is the scaling regime where the power-law 
behavior is observed.
In this way it is possible to study finite-size effects
of steady state measurements in absorbing phase transitions.
A detailed analysis of the order parameter, the fluctuations,
as well as of a fourth order cumulant is presented
in the next chapter.
Similar to equilibrium phase transitions, we formulate
scaling forms which incorporate 
the system size as an additional scaling field.

\subsection{Dynamical scaling behavior}
\label{subsec:apt_dynamical_scal}

\begin{figure}[b] 
\centering
%\leavevmode 
\includegraphics[clip,width=8cm,angle=0]{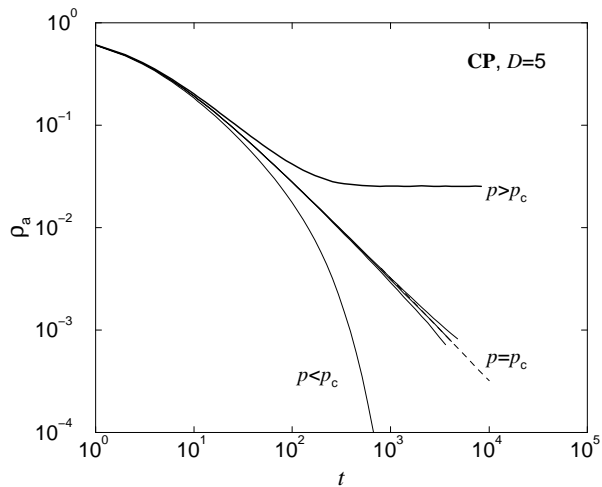}
\caption[Dynamical behavior of the order parameter for $D=5$] 
{The dynamical behavior of the order parameter 
of the five-dimensional contact process for \mbox{$L=32$}.
The parameters of the presented curves are $\deltap=0$,
$\deltap = \plmi 0.02$ (thick lines), 
$\deltap = \plmi 0.00006$ (thin lines) respectively.
The data are averaged over at least $N=10^{3}$ different runs.
Started from a fully occupied lattice 
($\rho_{\ssstyle {\mathrm a},t=0}=1$) the order parameter
saturates for $p> \pc$ and decays exponentially for $p< \pc$.
At the critical point the order parameter displays an
algebraic decay. 
The dashed line corresponds to the mean field exponent
$\alpha=1$ and is shown to guide the eye.
\label{fig:cp_5d_rho_a_t_01}} 
\end{figure}

In the following we investigate the dynamical scaling
behavior at zero field close to the critical point.
First, we consider how the order parameter decays
starting from a fully occupied lattice, i.e.,
starting from a so-called homogenous particle source.
Studying the temporal evolution of the system
one has to average over different runs
$\rhoa(t) = [\rhoa(t,n_{\ssstyle {\mathrm {rnd}}})]$
with 
\begin{equation}
[ \rhoa(t) ] \; = \;
\frac{1}{N} \, \sum_{\ssstyle k=1}^{\ssstyle N}
\, \rhoa(t, n_{\ssstyle {\mathrm {rnd}},k}) \, .
\label{eq:random_average}
\end{equation}
Here, $n_{\ssstyle {\mathrm {rnd}}}$ denotes the seed number 
of the random number generator.
Simulation results for various values of the control parameter 
are shown in \reffigname{fig:cp_5d_rho_a_t_01}.

Similar to the mean field behavior
[\refeqsname{eq:mf_cp_one_site_dyn_below}{eq:mf_cp_one_site_dyn_at_crit}],
the order parameter tends exponentially
to the steady state value above~$\pc$
(positive curvature in the log-log plot). 
Below the transition point
the absorbing state is 
approached exponentially (negative curvature).
At the critical point the order parameter decays asymptotically as
\begin{equation}
\rhoa(t) \; \asympprop \; t^{-\alpha} \, .
\label{eq:ord_decay_def_alpha}   
\end{equation}
As can be seen in \reffigname{fig:cp_5d_rho_a_t_01}
the decay exponent agrees with the mean field
value $\alpha=1$.
Again, deviations from the power-law behavior allow 
to estimate the critical value~$\pc$.
The obtained result and its uncertainty is in agreement 
with the steady state measurement~$\pc=0.53237\pm0.00006$.

\begin{figure}[t] 
\centering
%\leavevmode 
\includegraphics[clip,width=12cm,angle=0]{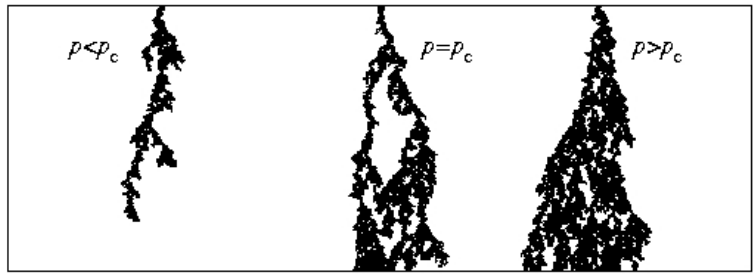}
\caption[Snapshots of activity spreading in the contact process] 
{Snapshots of activity spreading in the one-dimensional 
contact process (time increases downward from $t=0$ to $10^{3}$). 
The parameters of the presented clusters are $\deltap=0$
and $\deltap= \plmi 0.02$.
\label{fig:cp_act_spread_snapshot}} 
\end{figure}

More accurate estimates for the critical value~$\pc$
and for the critical exponents can be obtained by\index{activity spreading}
considering the activity spreading generated from a single
active seed~\cite{GRASSBERGER_4}, a so-called localized particle source.
Three typical snapshots of activity spreading of the 
one-dimensional contact process are shown 
in \reffigname{fig:cp_act_spread_snapshot}.
Below the critical point the activity ceases after a 
certain transient regime whereas above~$\pc$ the activity 
spreading continues infinitely with a finite probability.
At the transition point a cluster is generated by the 
temporal evolution that exhibits fractal properties~\cite{GRASSBERGER_8}.
Interpreting active sites, e.g., as viruses this figure illustrates
the spreading of epidemics through a population.

\begin{figure}[t] 
\centering
%\leavevmode 
\includegraphics[clip,width=13cm,angle=0]{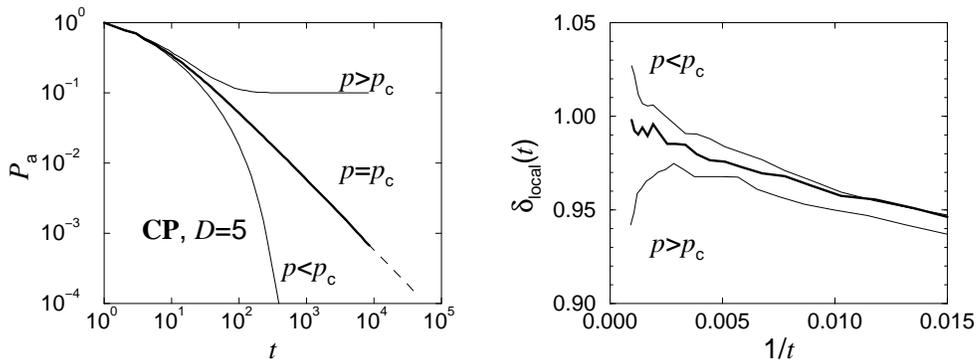}
\caption[Survival probability of activity spreading of the contact process, $D=5$] 
{The survival probability of the five-dimensional contact
process close to criticality (left figure).
The parameters of the presented data are $\deltap=0$ (straight asymptotic line)
and $\deltap= \plmi 0.02$ (curved lines).
The dashed line corresponds to the asymptotic mean field 
behavior ($\delta=1$).
The right figure shows the local slope 
$\delta_{\ssstyle {\mathrm {local}}}$
[\refeqname{eq:def_delta_local}]. 
The parameters of the presented data are $\deltap=0$
(thick line) and $\deltap= \plmi 0.00002$ 
(upper and lower thin line).
The data are averaged over $N=10^{7}$ independent runs
and $c=8$ is used to calculate the local slope.
As can be seen the mean field value $\delta=1$ is 
approached for $\deltap=0$ and $t\to \infty$, whereas
the local slopes tend to $0$ or $\infty$ for $\deltap\neq 0$. 
\label{fig:cp_5d_sur_prob}} 
\end{figure}

In order to analyze the scaling behavior of 
activity spreading it is useful to measure the 
survival probability $\Pa(t)$, 
the number of active sites $\Na(t)$
as well as the average mean square distance of the
spreading $R_{\ssstyle \mathrm a}(t)^2$~\cite{GRASSBERGER_4}.
Here, $\Na(t)$ is averaged over all clusters
whereas $R_{\ssstyle \mathrm a}(t)^2$ is averaged 
only over surviving clusters.
At criticality, the following power-law behavior 
are expected to occur asymptotically
\begin{equation}
\Pa(t) \, \asympprop \, t^{-\delta} \, , \quad\quad
\Na(t) \, \asympprop \, t^{\theta}   \, , \quad\quad
R_{\ssstyle \mathrm a}(t)^2 \, \asympprop \, t^{2/z}  \, ,
\label{eq:def_delta_theta_z}   
\end{equation}
where we have used the notation of~\cite{HINRICHSEN_1}.
The so-called dynamical exponent~$z$ describes the relationship
between spatial and temporal correlations
and is therefore identified as 
\begin{equation}
z \; = \; \frac{\, \nu_{\para} \,}{\nu_{\senk}} \, .
\label{eq:def_dyn_exp}
\end{equation}
According to \refeqname{eq:corr_spatial_temp_mf} the mean field value
is given by $z=2$.\index{spreading exponents}\index{critical exponents}
\index{dynamical exponent}
The survival probability~$\Pa(t)$ is shown in \refFigname{fig:cp_5d_sur_prob}.
Again, a positive curvature for $t\to \infty$ 
indicates in a log-log plot the active phase
whereas a negative curvature indicates the absorbing phase.
Similar to the steady state [\refeqname{eq:ord_par_def_beta_01}]
and dynamical [\refeqname{eq:ord_decay_def_alpha}]
order parameter behavior, 
the survival probability obeys power-law 
behavior \refeqname{eq:def_delta_theta_z}  
only asymptotically, i.e., confluent \index{confluent singularities}
singularities have to be taken into consideration.
Often, local slopes are considered 
\begin{equation}
\delta_{\ssstyle {\mathrm{local}}}(t) \, = \, - \,
\frac{\, \ln{[\, \Pa(t)\, / \, \Pa(t/c)\, ]}\,}{\ln{c}} 
\label{eq:def_delta_local}   
\end{equation}
to handle this problem.
The parameter~$c$ defines the distance of the
local slopes and typical values used in previous studies
are for example $c=4$~\cite{ODOR_6}, 
$c=5$~\cite{JENSEN_6,JENSEN_15}, $c=8$~\cite{GRASSBERGER_8}
as well as $c=10$~\cite{NOH_1}.
Of course, for $c\to 1$ the local slopes equals the
so-called effective\index{effective exponent}  
exponent~\cite{RIEDEL_1}
\begin{equation}
\delta_{\eff}(t) \; = \; - \, 
\frac{\partial\, \ln{\Pa(t)}}{ \partial\, \ln{t}} \, .
\label{eq:def_delta_eff}
\end{equation}
The analysis of the local slope is illustrated 
in \reffigname{fig:cp_5d_sur_prob}.
Above criticality the local slope tends to
zero whereas $\delta_{\mathrm{local}}(t)\to \infty$ 
for $\deltap<0$.
Both limiting cases are separated by 
the local slope at criticality 
and the exponent~$\delta$ can be estimated from an extrapolation
to $t \to \infty$.

%%%%%%%%%%%%%%%%%%%%%%%%%%%%%%%%%%%%%%%%%%%%%%%%%%%%%%%%%%%%%%%%%%%%%%%%%%%%%%%%%%%%%%%%
%%%%%%%%%%%%%%%%%%%%%%%%%%%%%%%%%%%%%%%%%%%%%%%%%%%%%%%%%%%%%%%%%%%%%%%%%%%%%%%%%%%%%%%%%%

The exponents $\theta$ and $z$ can be determined in
an analogous way and the mean field values $\delta=1$,
$\theta=0$, and $z=2$ are obtained in case of the five
dimensional contact process.
The dynamical scaling behavior at $\pc$ equals a 
critical\index{branching process} branching
process (see appendix)
and the value $\theta=0$ corresponds to 
\mbox{$\Na(t) \sim {\mathrm{const}}$}.
The accuracy of the exponents can be 
checked using the hyperscaling relation~\cite{GRASSBERGER_4}
\index{hyperscaling}
\begin{equation}
\theta \, + \, 2 \, \delta \; = \; \frac{\, D \, }{z} \, ,
\label{eq:hyperscal_dp_theta_delta_z}
\end{equation}
which is valid for $D\kgl \Dc$.
A derivation of this scaling law is presented in the 
next chapter.
Notice that the mean field values $\delta=1$,
$\theta=0$, and $z=2$ fulfill the scaling law 
\refeqname{eq:hyperscal_dp_theta_delta_z} for $D=4$, 
indicating that four is the upper critical dimension
of the contact process.\index{upper critical dimension}

Other exponents can be derived from $\delta$, $\theta$, and $z$
by additional scaling laws.
For example, the average number of active sites 
per surviving cluster scales as  
$N_{\ssstyle {\mathrm {a, sur}}}(t)=\Na(t)/\Pa(t) \asympprop t^{\theta+\delta}$.
As usually, the fractal dimension~$D_{\ssstyle {\mathrm f}}$
is defined via 
$N_{\ssstyle {\mathrm {a, sur}}}(t) \asympprop R(t)^{D_{\ssstyle {\mathrm f}}}$
yielding~\cite{GRASSBERGER_8}
\begin{equation}
D_{\ssstyle {\mathrm f}} \; = \; z \, (\theta+\delta) \, .
\label{eq:def_frac_dim_cluster}   
\end{equation}   
In case of the one-dimensional contact process, the critical clusters are
characterized by the fractal \index{fractal dimension}
dimension \mbox{$D_{\ssstyle {\mathrm f}} \approx 0.75$},
whereas the mean field values yield $D_{\ssstyle {\mathrm f}}=2$.

In summary, activity spreading measurements
provide very accurate estimates for the critical value~$\pc$
as well as for the exponents $\delta$, $\theta$, and $z$
of systems exhibiting absorbing phase transitions
(see 
e.g.~\cite{GRASSBERGER_2,GRASSBERGER_8,JENSEN_8,DICKMAN_9,JENSEN_9,JENSEN_6,JENSEN_15,VOIGT_1}). 
The efficiency of this technique can be increased significantly
by performing off-lattice simulations
using the fact that clusters of active sites are fractal at the
critical point, i.e., huge parts of the lattice remains empty.
Storing just the coordinates of active sites
in dynamically generated lists makes the algorithm
much more efficient compared to steady state exponents.
Another advantage of such off-lattice simulations
is that no finite-size effects occur.
But a drawback of activity spreading measurements is that the determination of
the exponents $\delta$, $\theta$, and $z$ via 
\refeqname{eq:def_delta_theta_z} is not
sufficient to describe the scaling behavior of
absorbing phase transition.
Due to the hyperscaling relation [\refeqname{eq:hyperscal_dp_theta_delta_z}]
only two independent exponents can be determined.
But we will see in the next chapter that the
scaling behavior of the contact process is characterized
by three independent exponents.
Thus additional measurements have to be performed in 
order to complete the set of critical exponents.

Additionally to the consideration of lattice models,
direct numerical integrations of the corresponding 
Langevin equations\index{Langevin equation}
are used to investigate the behavior of 
the contact process or other 
related models~\cite{DICKMAN_17,LOPEZ_1,RAMASCO_1}.
But this technique runs into difficulties, for example
negative densities occur if the absorbing
phase is approached.
To bypass this problem, a discretization of the 
density variable was proposed in~\cite{DICKMAN_17}.
This approach has been applied to various absorbing 
phase transitions.
The obtained estimates of the exponents 
agree with those of simulations of lattice models but are
of less accuracy.

\chapter{Directed percolation} 
\setcounter{figure}{13}
\label{chapter:dp}

\begin{figure}[t] 
\centering
%\leavevmode 
\includegraphics[clip,width=13cm,angle=0]{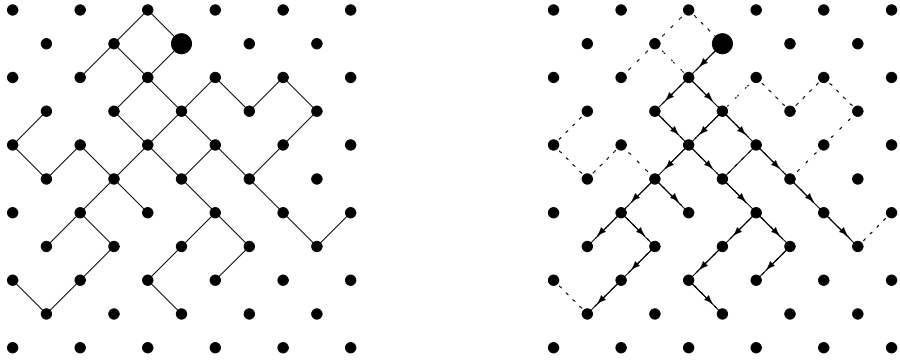}
\caption[Isotropic and directed percolation] 
{Isotropic bond percolation (left) and directed bond percolation (right) on a 
rotated square lattice.
Only the cluster is shown that is connected to the marked
lattice site (bold circle).
The same bond configuration leads to different clusters
for isotropic and directed percolation.
In case of directed percolation the cluster propagates
only along the preferred direction (arrows).
Therefore, each site generates an individual cluster 
for a given bond configuration.
\label{fig:iso_direc_perc}} 
\end{figure}

The\index{directed percolation}\index{percolation}
problem of directed percolation was introduced 
in the mathematical literature in 1957 by
Broadbent and Hammersley
to mimic various processes like 
epidemic spreading, wetting in porous media
as well as wandering in mazes~\cite{BROADBENT_1}.
In a lattice formulation, directed percolation
is an anisotropic modification of isotropic percolation.
Consider for instance the problem of bond directed
percolation as shown in \reffigname{fig:iso_direc_perc}.
Here, two lattice sites are connected by a bond with
probability~$p$~\cite{STAUFFER_3,ESSAM_3,STAUFFER_2}.
If the bond probability is sufficiently large, a cluster
of connected sites will propagate through the system.
The probability that a given site belongs to a
percolating cluster is the order parameter of the 
percolation transition and obeys
\begin{equation}
\Pperc(p) \; \asympprop \;   (p-\pc)^{\betap} 
\label{eq:perc_prob_beta_prime}
\end{equation}
for $p > \pc$, whereas it is zero below the critical
value $\pc$.
Due to a duality symmetry the critical value of bond
percolation on a square lattice is 
$\pc=1/2$.
%and the critical exponent is known exactly 
%$\betap=5/36$ and $\nu=4/3$.

In case of directed\index{DP} 
percolation (DP) a preferred spatial direction 
is introduced.
For example, only branching of the cluster in the 
downward direction in \reffigname{fig:iso_direc_perc} is 
allowed.
Similar to isotropic percolation one considers
the probability~$\Pperc(p)$ that a path of connected sites percolates 
through the entire system.
Again, depending on the probability~$p$, a continuous 
phase transition takes place at a certain threshold~$\pc$
and the percolation probability~$\Pperc(p)$ vanishes
algebraically [\refeqname{eq:perc_prob_beta_prime}].
Thus both percolation transitions are characterized by the
order parameter~$\Pperc(p)$, but as expected the critical
value for directed percolation ($\pc=0.644700185(5)$~\cite{JENSEN_16}) is
larger than the isotropic value~$\pc=1/2$.
Another important difference is that isotropic percolation
can be mapped 
exactly to the equilibrium $q$-state Potts\index{Potts model} 
model~\cite{POTTS_1,WU_1}
in the limit \mbox{$q\to 1$}~\cite{FORTUIN_1,FORTUIN_2} 
and the critical exponents are known exactly,
$\betap=5/36$ and $\nu=4/3$~\cite{NIJS_1,NIENHUIS_1}.
On the other hand, directed percolation is a non-equilibrium
system and is not exactly solved so far.
Furthermore, the critical exponents do not seem to be given
by rational numbers in contrast to isotropic percolation.
And finally the upper critical dimension of isotropic 
percolation is \mbox{$\Dc=6$}~\cite{STAUFFER_2} 
whereas \mbox{$\Dc=4$}~\cite{CARDY_1,OBUKHOV_1}
in case of directed percolation.
In other words, despite their similarities isotropic and
directed percolation belong to different universality classes.

Although, directed percolation is a geometrical problem 
according to the original definition,
it appears in different contexts as a dynamic process.
In those cases the preferred direction is interpreted as
a representation of time instead of space.
Therefore, directed percolation describes the 
propagation or spreading of a non-conserved quantity (particles, viruses etc.)
through a one-dimensional cellular automata.
Each row represents the state of the system at time step~$t$
and particles are associated to lattice sites that 
belong to a cluster at that time step.
Obviously, an empty row is an absorbing state if
no spontaneous particle creation takes place.
In this way the geometrical percolation transition of 
two-dimensional directed percolation is mapped to an 
absorbing phase transition of a one-dimensional 
cellular\index{cellular automaton} automaton.  
Furthermore, the percolation order 
parameter [\refeqname{eq:perc_prob_beta_prime}]
corresponds to the ultimate survival probability
\begin{equation}
\Pperc(p) \; = \; \lim_{t \to \infty} \, \Pa(t,p) 
\label{eq:ulti_sur_prob}
\end{equation}
that a cluster generated by an active single seed propagates
through the entire system.
Thus the critical behavior of directed percolation
can be described by both the percolation probability
and the steady state density order parameter
\begin{equation}
\Pperc(p) \; \asympprop \;   (p-\pc)^{\betap} \, ,
\quad\quad\quad 
\rhoa(p) \; \asympprop \; (p-\pc)^{\beta} \, .
\label{eq:both_ord_par}
\end{equation}

Today, directed percolation is considered as 
the most important universality class of absorbing
phase transitions.
Directed percolation transitions occur, for example,
in systems describing the spreading of epidemics
like the contact process,
catalytic reactions,
the percolation in porous media, etc.~(see for recent 
reviews~\cite{HINRICHSEN_1,ODOR_1}).
The widespread occurrence of directed percolation
is reflected by the universality hypothesis of Janssen
and Grassberger~\cite{JANSSEN_1,GRASSBERGER_2}:~Short-range
interacting systems, exhibiting a continuous phase transition
into a unique absorbing state generically belong to the
universality class of directed percolation, provided they
have no additional symmetries.\index{universality class, DP}
\index{universality hypothesis}
Different universality classes are expected to occur
in the presence of additional symmetries,
e.g.~particle-hole symmetry, or relevant disorder effects.

Similar to equilibrium critical phenomena, the universality
of directed percolation is understood by renormalization
group treatments of an associated continuous field theory.
Amazingly the field theoretic description of directed percolation
is well established in high energy physics, termed Reggeon field
theory~(see e.g.~\cite{CARDY_1,ABARBANEL_1}).\index{Reggeon field theory}
Reggeon field theory describes hadron scattering 
at high energies and low-momentum transfer.
The time direction of directed percolation is related 
to the longitudinal momentum, the spatial direction equals the
impact parameter and the total cross section corresponds
to the square of the percolation probability, i.e.,
to the order parameter~\cite{CARDY_1}.

The process of directed percolation might be 
represented by the Langevin\index{Langevin equation}
equation~\cite{JANSSEN_1} 
%(in so-called Ito sense, see e.g.~\cite{KAMPEN_1}) 
\begin{equation}
\partial_{\ssstyle t} \, \rhoa({\underline x},t)
\; = \; r \, \rhoa({\underline x},t) 
\, - \, u \, \rhoa^2({\underline x},t) 
\, + \, \Gamma \, \nabla^2 \, \rhoa({\underline x},t) 
\, + \, 
\eta({\underline x},t) 
\label{eq:langevin_dp_01}
\end{equation}
which describes the order parameter $\rhoa({\underline x},t)$
on a mesoscopic scale and where $\eta$ denotes the multiplicative
noise which accounts for fluctuations of the particle
density~$\rhoa({\underline x},t)$.
According to the central limit theorem, 
$\eta({\underline x},t)$ 
is a Gaussian random variable with zero mean
and whose correlator is given by
\begin{equation}
\langle \, \eta({\underline x},t) \, 
\eta({\underline x}^{\prime},t^{\prime}) \, \rangle
\; = \; \kappa \; \rhoa({\underline x},t) \;
\delta({\underline x}-{\underline x}^{\prime}) \;
\delta(t-t^{\prime}) \, .
\label{eq:langevin_dp_corr_01}
\end{equation}
Notice, that the multiplicative noise ensures that the 
systems is trapped in the absorbing state \mbox{$\rhoa({\underline x},t)=0$}.
Furthermore, higher order terms as 
$\rhoa({\underline x},t)^3,\rhoa({\underline x},t)^4, \ldots$ 
(or 
$\nabla^4\rhoa({\underline x},t),\nabla^6\rhoa({\underline x},t), \ldots$)
are irrelevant under renormalization group transformations
as long as $u > 0$.
Negative values of $u$ give rise to a first order
phase transition whereas $u=0$ is associated to
a tricritical\index{tricritical point} point~\cite{OHTSUKI_1,OHTSUKI_2} 
similar to the Landau functional of ferromagnets 
[\refeqname{eq:Landau_freeenergy_fm}]
with terms of order $m^6$~\cite{LANDAU_5}.

Obviously, the above Langevin equation corresponds to the
mean field equation of the contact 
process [\refeqname{eq:mf_cp_field_equation}]
if the noise can be neglected.
More precisely, the mean field equation describes the macroscopic
behavior correctly if the noise is irrelevant on large scales,
whereas it fails if the noise is relevant on large
scales.
The question whether the noise is relevant or irrelevant 
can be answered by the so-called dimensional analysis.
Similar to the renormalization group approach in equilibrium
(see~\ref{sec:rg_theory}) one considers how the 
coupling constants $r$, $u$, $\Gamma$, $\kappa$ of 
the Langevin equation behave under the 
rescaling transformations (see e.g.~\cite{HINRICHSEN_1,GRINSTEIN_2})
\begin{equation}
{\underline x} \, \mapsto \, b \, {\underline x} \, , \quad  \quad  \quad
t \, \mapsto \, b^{z} \, t \,  \, , \quad \quad  \quad
\rhoa \, \mapsto \, b^{-\beta/\nu_{\senk}} \, \rhoa \,  \, , \quad \quad  \quad
\label{eq:rescaling}
\end{equation}
with $b>1$.
Replacing ${\underline x}$, $t$, and $\rhoa$ 
in \refeqs2name{eq:langevin_dp_01}{eq:langevin_dp_corr_01}
by the rescaled quantities yields the recursion 
relations for the couplings constants
\begin{equation}
r\, \mapsto \, b^{z} \, r \, ,  \quad 
u\, \mapsto \, b^{z-\beta/\nu_{\senk}} \, u \, ,   \quad 
\Gamma \, \mapsto \, b^{z-2} \, \Gamma \, ,   \quad 
\kappa \, \mapsto \, b^{\beta/\nu_{\senk} + z - D} \, \kappa \, .
\label{eq:rescaling_noise_constant}
\end{equation}
Thus the mean field solution is consistent
for $D>4$ since the Langevin equation is 
at criticality ($r=0$) invariant under 
the transformations \refeqname{eq:rescaling} for the
mean field values $\beta=1$, $\nu_{\senk}=1/2$, $z=2$.
The situation is different below the upper critical
dimension $\Dc=4$ where the noise is relevant and 
the mean field solution does not describe the macroscopic 
behavior of the system.

\begin{figure}[t] 
\centering
%\leavevmode 
\includegraphics[clip,width=8cm,angle=0]{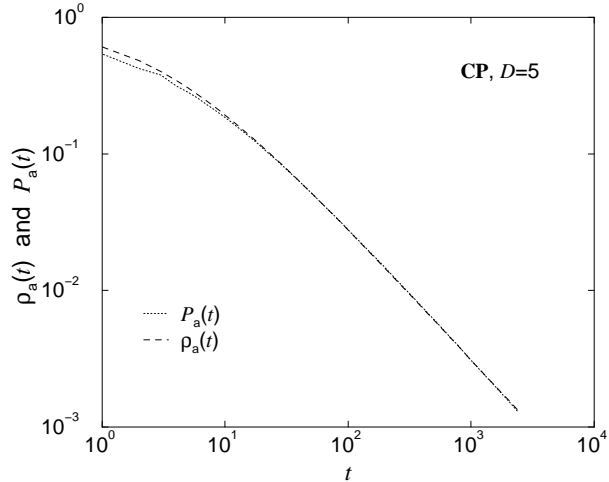}
\caption[Rapidity reversal, asymptotic equivalence of $\rhoa(t)$ and $\Pa(t)$ for DP] 
{The asymptotic equivalence of $\rhoa(t)$ and $\Pa(t)$ 
for the five dimensional contact process.
The survival probability $\Pa(t)$ (here multiplied by a factor) 
measures the probability 
that a cluster generated by a single seed is still active
after $t$ time steps.
On the other hand~$\rhoa(t)$
describes the decay of the particle density started from
a fully occupied lattice.
\label{fig:cp_5d_duality}} 
\end{figure}

Below the upper critical dimension, 
renormalization group techniques have to be applied
to determine the critical exponents and the
scaling functions.
In that case path integral formulations\index{path integral} are more
adequate than the Langevin equation~(see e.g.~\cite{CHAICHIAN_1}).
Stationary correlation functions as well as response
functions can be determined by calculating 
path integrals with weight $\exp{(-{\mathcal S})}$,
where the dynamic functional ${\mathcal S}$ describes
the considered stochastic process.
Up to higher irrelevant orders the dynamic functional 
associated to directed percolation is given 
by~\cite{JANSSEN_1,JANSSEN_7,DEDOMINICIS_1,JANSSEN_8} 
\begin{equation}
{\mathcal S}[{\tilde n},n] \; = \; \int
{\mathrm d}^D{\underline x} \,
{\mathrm d}t \;
{\tilde n}  \, \left [ \, 
\partial_{\ssstyle t} n  -  (r + \nabla^2 ) n
 -  \left (  \frac{\kappa}{2} \, {\tilde n}  - u \, n \right ) n \, 
\right ]
\label{eq:action_reggeon_field_theory}
\end{equation}
where the particle density equals $n({\underline x},t)$ 
and where ${\tilde n}({\underline x},t)$
denotes the response field conjugated to the 
Langevin noise field~\cite{MARTIN_1}.
This functional corresponds to the Lagrangian
of Reggeon field theory~\cite{ABARBANEL_1}.
Rescaling the fields 
\begin{equation}
{\tilde n}({\underline x},t) \; = \; \mu \, 
{\tilde s}({\underline x},t) \, , \quad\quad
{n}({\underline x},t) \; = \; \mu^{-1} \, 
{s}({\underline x},t) \, 
\label{eq:rescal_field}
\end{equation}
the functional $S$ is invariant under the duality
transformation (so-called rapidity reversal in Reggeon field 
theory)\index{rapidity reversal symmetry}\index{symmetry, rapidity reversal}    
\begin{equation}
{\tilde s}({\underline x},t) \, \longleftrightarrow
\, -\, {s}({\underline x},-t) \, 
\label{eq:rapidity_trans}
\end{equation}
for $\mu^2=2u/ \kappa$. Note that $\mu$ is a redundant variable
from the renormalization group point of view~\cite{JANSSEN_1,JANSSEN_4}.
As usual, the duality transformation defines a 
dual stochastic process that might differ from the original one~\cite{LIGGETT_1}.
%in general differs from the original one.
Close to criticality, the average density of particles 
of the dual process $\rhoa^{\ssstyle \mathrm{dual}}$ is 
connected to the survival probability~\cite{JANSSEN_4} via
\begin{equation}
\Pa(t) \; \sim \; \mu^2 \, \rhoa^{\ssstyle \mathrm{dual}}(t) \, .
\label{eq:p_sur_alpha_rho_dual}
\end{equation}
Here, $\Pa(t)$ denotes again the probability 
that a cluster generated by a single seed is still active
after $t$ time steps.
On the other hand, $\rhoa^{\ssstyle \mathrm{dual}}(t)$
describes the particle decay of the dual process started from
a fully occupied lattice.
The self-duality of directed percolation 
implies \mbox{$\rhoa(t)=\rhoa^{\ssstyle \mathrm{dual}}(t)$}
and therefore~\cite{GRASSBERGER_4,JANSSEN_4}  
\begin{equation}
\Pa(t) \; \sim \; \mu^2 \, \rhoa(t) \, .
\label{eq:p_sur_alpha_rho}
\end{equation}
Numerical simulations confirm this result, as can be
seen in \reffigname{fig:cp_5d_duality} for the five dimensional
contact process.
The asymptotic equivalence ensures that both quantities 
have the same exponents~\cite{GRASSBERGER_4},
including
\begin{equation}
\beta \; = \; \betap \, .
\label{eq:beta_betap_equal}
\end{equation}
It is worth mentioning that \refeqname{eq:p_sur_alpha_rho}
follows from the rapidity reversal symmetry of the dynamical
functional~${\mathcal{S}}[{\tilde n}, n]$, i.e., it is a specific
property of the directed percolation universality class.
Thus, compared to general absorbing phase transitions,
the number of independent critical exponents
for directed percolation is reduced.
Furthermore, the self-duality is restricted to the field theoretical
treatment (e.g.~after a course-graining procedure)
and does not necessarily represent a physical symmetry 
of microscopic models~\cite{JANSSEN_1}.

Using standard renormalization
group techniques such as $\epsilon$-expansion, the 
critical exponents as well as the scaling\index{$\epsilon$-expansion} 
functions were obtained.\label{page:dp_eps_results}
For example the exponents $\beta$, $z$, and $\nu_{\senk}$
are in two loop order~\cite{JANSSEN_1,BRONZAN_1,BRONZAN_2,BRONZAN_3}
\begin{eqnarray}
\label{eq:dp_exp_epsilon2_beta}   
\beta       & = & 1 \, - \, \frac{\epsilon}{6} \, 
\left [ 1- \left ( \frac{11}{288} - \frac{53}{144} \, \ln{\frac{4}{3}}
\right )  \, \epsilon \,  + \ldots \,
\right ] \, , \\[2mm]
\label{eq:dp_exp_epsilon2_nu_perp}   
\nu_{\senk} & = & \frac{1}{2} \, + \, \frac{\epsilon}{16} \, 
\left [ 1+ \left ( \frac{107}{288} - \frac{17}{144} \, \ln{\frac{4}{3}}
\right ) \, \epsilon \, + \ldots \,
\right ]\, , \\[2mm]
\label{eq:dp_exp_epsilon2_beta_z}   
z           & = &  2 \, - \, \frac{\epsilon}{12} \, 
\left [ 1+ \left ( \frac{67}{288} - \frac{59}{144} \, \ln{\frac{4}{3}}
\right )  \, \epsilon \,  + \ldots \,
\right ] \, ,
\end{eqnarray}
with $\epsilon=4-D$.
Furthermore, the Widom-Griffiths scaling form of the equation of
state is given close to the transition point (to be specific
for $x<0$ and $|x|\ll 1$) by~\cite{JANSSEN_2} 
\begin{eqnarray}
{\tilde H}(x,1) & = & 1 \, - \, x \, + \,
\frac{\epsilon}{6} \, K \,
\left [ (2-x) \ln{(2-x) - 2 (1-x) \ln{2}}
\right ] \nonumber \\
&  & + \, \frac{\epsilon^2}{72} \, 
\left [ (4-x) \ln^{2}{(2-x) - 4 (1-x) \ln^{2}{2}}
\right ]
\label{eq:dp_eqos_WG_epsilon2}
\end{eqnarray}
with 
\begin{equation}
K \; = \; 1 \, + \, \epsilon \,
\left ( \frac{85}{288} + \frac{29}{72} \, \ln{2}
- \frac{53}{144} \, \ln{3} \, 
\right ) \, .
\label{eq:K_aus_eqos_epsilon2}
\end{equation}
Below, these universal quantities will be compared to various
lattice models belonging to the directed percolation
universality class.

\section{Lattice models of directed percolation}
\label{sec:dp_lattice_models}

In the following we consider various lattice 
models that belong to the universality class of directed
percolation.
First, we revisit the contact process that is well
known in the mathematical literature.
%The master equation is presented and an important conclusion
%for the universality class of directed percolation 
%will be drawn from it.
Second we consider the Domany-Kinzel (DK) cellular 
automaton~\cite{DOMANY_1} which is very useful
in order to perform numerical investigations
of directed bond and directed site percolation.
Furthermore, the Domany-Kinzel automaton exhibits 
a non-trivial phase diagram and allows therefore
to study how non-directed percolation behavior
arises if an additional symmetry occurs.
Third we consider the pair contact process (PCP)~\cite{JENSEN_2}
that is characterized in contrast to the other models 
by infinitely many absorbing states. 
In contrast to the first two models the universal behavior of 
the pair contact process is still a matter of controversial 
discussions in the literature.
In particular the scaling behavior in higher 
dimension and the effects of additional particle 
diffusion are investigated.
Furthermore we briefly discuss the threshold transfer 
process~\cite{MENDES_1} as well as the 
Ziff-Gular{\'\i}-Barshad model~\cite{ZIFF_1}.
The latter one mimics the catalysis of the 
carbon monoxide oxidation.

\subsection{Contact process}
\label{subsec:contact_process}

\index{contact process}

The contact process (CP) is\index{contact process}\index{CP} 
a continuous-time Markov
process that is usually defined on a $D$-dimensional
simple cubic lattice.
A lattice site  may be empty ($n=0$) 
or occupied ($n=1$) by a particle and the 
dynamics is characterized by spontaneously occurring
processes, taking place with certain transition rates.
In numerical simulations the asynchronous 
update\index{synchronous update scheme}
is realized by a random sequential update 
scheme\index{random sequential update}
(see section~\ref{sec:num_sim} for details):~A particle 
on a randomly selected lattice site $n_{\ssstyle
i}$ is annihilated with rate one, 
whereas particle creation takes places on an 
empty neighboring 
site with rate $\lambda n / 2 D$, i.e.,
\begin{eqnarray}
\label{eq:trans_rate_cp_annih}
n_{\ssstyle i} = 1 \quad  
&  \mathop{\longrightarrow}\limits_{1} &
\quad n_{\ssstyle i} = 0 \, ,\\
\label{eq:trans_rate_cp_creat}
n_{\ssstyle i} = 0 \quad  
&  \mathop{\longrightarrow}\limits_{\lambda n / 2 D} & 
\quad n_{\ssstyle i} = 1 \, ,
\end{eqnarray}
where~$n$ denotes the number of occupied neighbors
of~$n_{\ssstyle i}$.
Notice that the rates are defined as transition probabilities
per time unit, i.e., they are not normalized probabilities 
and may be larger than one.
Thus rescaling the time will change the transition rates.
In simulations a discrete time formulation of the
contact process is performed.
In that case a particle creation takes place at a randomly 
chosen lattice site with probability $p=\lambda/(1+\lambda)$
whereas particle annihilation occurs with 
probability $1-p=1/(1+\lambda)$.
In dynamical simulations, the time increment $1/\Na$ is
associated to each attempted elementary update 
step,
where $\Na$ denotes the number of active sites
(see sections~\ref{sec:apt_defs} and \ref{subsec:apt_steady_state_scal}). 
It is usual to present the critical value in terms
of $\lambdac$ instead of $\pc$.
To avoid friction with the literature of absorbing 
phase transitions we list the corresponding values
of $\lambdac$ in \reftablename{table:pc_values_dp}.

For the sake of completeness, we mention that the 
master equation\index{master equation}
of the contact process may be written as~(see e.g.~\cite{HINRICHSEN_1})
\begin{eqnarray}
\partial_{\ssstyle t} \, 
P(n_{\ssstyle 1},\ldots,n_{\ssstyle N},t) \; = \;
\sum_{i=1}^{N} \, (2 n_{\ssstyle i}-1) \hspace{-1.2ex}
& \left [ \vphantom{X^X} \right .  &\lambda \, n_{\ssstyle i-1} \; 
 P(n_{\ssstyle 1},\ldots,n_{\ssstyle i-2},1,0,n_{\ssstyle i+1},\ldots,n_{\ssstyle N},t)  
\nonumber \\
& + & \lambda \, n_{\ssstyle i+1} \; 
P(n_{\ssstyle 1},\ldots,n_{\ssstyle i-1},0,1,n_{\ssstyle i+2},\ldots,n_{\ssstyle N},t)
\nonumber \\
& - &
 P(n_{\ssstyle 1},\ldots,n_{\ssstyle i-1},1,n_{\ssstyle i+1},\ldots,n_{\ssstyle N},t)  
\left . \vphantom{X^X} \right ] \, .
\label{eq:master_eq_cp}
\end{eqnarray}
Here, \mbox{$P(n_{\ssstyle 1},\ldots,n_{\ssstyle N},t)$} 
denotes the probability to find the system at time $t$ in the
configuration $(n_{\ssstyle 1},n_{\ssstyle 2} \ldots, n_{\ssstyle N})$.
Although a complete analytical solution of the contact
process is still lacking, a number of rigorous results 
are available.
For example, the transition is known to be continuous,
upper and lower bounds on~$\lambdac$
exist, as well as a complete convergence theorem has been 
proven~(see e.g.~\cite{LIGGETT_1,DURRETT_1,DURRETT_2,BEZUIDENHOUT_1}).

%%%%%%%%%%%%%%%%%%%%%%%%%%%%%%%%%%%%%%%%%%%%%%%%%%%%%%%%%%%%%%%%%%%%%%%%%%%%%%%%%%%%%%
%%%%%%%%%%%%%%%%%%%%%%%%%%%%%%%%%%%%%%%%%%%%%%%%%%%%%%%%%%%%%%%%%%%%%%%%%%%%%%%%%%%%%%
%%%%%%%%%%%%%%%%%%%%%%%%%%%%%%%%%%%%%%%%%%%%%%%%%%%%%%%%%%%%%%%%%%%%%%%%%%%%%%%%%%%%%%

\subsection{Domany-Kinzel automaton}
\label{subsec:domany_kinzel}

\index{Domany-Kinzel automaton}

\begin{figure}[t] 
\centering
%\leavevmode 
\includegraphics[clip,width=13cm,angle=0]{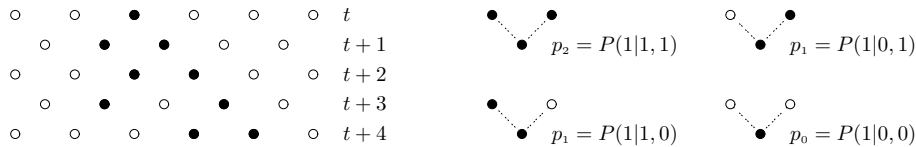}
\caption[Domany-Kinzel automaton] 
{The $1+1$-dimensional Domany-Kinzel automaton.
Occupied sites are marked by full circles.
A spreading of particles is sketched in the left part.
The state of the one dimensional system at time $t+1$ is
obtained by an iteration of the dynamics rules
that are embodied in the conditional probabilities
$P(n_{\ssstyle t+1}|n_{{\ssstyle \mathrm {l},t}},n_{{\ssstyle \mathrm {r},t}})$.
Thus the occupation $n_{\ssstyle t+1}$ of a given site at time~$t$
depends on the state of the left (l) and right (r) backward neighbors 
($n_{{\ssstyle \mathrm {l},t}},n_{{\ssstyle \mathrm {r},t}}$) 
at time $t$.
Spontaneous particle creation corresponds to the conjugated field
and is usually forbidden, i.e., $p_{\ssstyle 0}=h=0$.
\label{fig:dk_model_01}} 
\end{figure}

Another important $1+1$-dimensional stochastic cellular automaton 
exhibiting directed percolation scaling behavior is 
the\index{Domany-Kinzel automaton}\index{DK} 
Domany-Kinzel (DK) automaton~\cite{DOMANY_1}.
It is defined on a diagonal square lattice with a discrete time variable
and evolves by parallel update\index{parallel update} 
according to the following 
rules (see \reffigname{fig:dk_model_01}).
A site at time $t$ is occupied with 
probability~$p_{\ssstyle 2}$ 
($p_{\ssstyle 1}$) if both (only one) backward
sites (at time $t-1$) are occupied.
Otherwise the site remains empty.
If both backward sites are empty a 
spontaneous particle creation takes place with 
probability $p_{\ssstyle 0}$.
Similar to the contact process, the spontaneous 
particle creation destroys the absorbing phase (empty lattice) 
and corresponds therefore to the conjugated field~$h$.\index{conjugated field}

As pointed out above, the geometrical problem of $D+1$-dimensional 
directed percolation corresponds to a $D$-dimensional dynamic
process.
This mapping becomes evident for the Domany-Kinzel 
automaton~\cite{DOMANY_1} where the problems of bond 
and site directed percolation are recovered for a
particular choice of the conditional probabilities 
$p_{\ssstyle 1}$ and $p_{\ssstyle 2}$.
Suppose that the bonds of the square lattice 
of \reffigname{fig:dk_model_01} are present with probability~$q$
and that the sites of the lattice are present with~ probability~$p$.
This equals the 
site\index{directed percolation, site}\index{site directed percolation}\index{sDP}
directed percolation (sDP) problem for \mbox{$q=1$} whereas the 
bond\index{directed percolation, bond}\index{bond directed percolation}\index{bDP}
directed percolation (bDP) problem is recovered
for \mbox{$p=1$}.
These occupation rules correspond on the other hand to
the conditional probabilities $p_{\ssstyle 1}=p \, q$ 
and $p_{\ssstyle 2}=p \, q (2-q)$~\cite{DOMANY_1}, 
respectively. 
Thus bond directed percolation \mbox{($p=1$)} is recovered for 
\begin{equation}
p_{\ssstyle 2} \; = \; p_{\ssstyle 1}\, (2-p_{\ssstyle 1} ) \, , 
\label{eq:DK_bond_DP}
\end{equation}
whereas site directed percolation \mbox{($q=1$)} corresponds to 
\begin{equation}
p_{\ssstyle 2} \; = \; p_{\ssstyle 1} \, .
\label{eq:DK_site_DP}
\end{equation}

\begin{figure}[b] 
\centering
%\leavevmode 
\includegraphics[clip,width=3cm,angle=0]{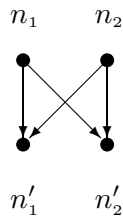}
\caption[Domany-Kinzel automaton, mean field approximation] 
{The mean field treatment of the $1+1$-dimensional Domany-Kinzel automaton.
In the simplest case one considers a lattice consisting of
two sites $i=1,2$.
The sites are updated ($n_{\ssstyle i} \to n_{\ssstyle i}^\prime$)
in parallel according to the conditional probabilities 
presented in \protect\reffigname{fig:dk_model_01}.
%Better mean field approximations of the behavior of the
%Domany-Kinzel automaton can be obtained by increasing the 
%number of lattice sites, e.g.~ $i=1,2,3$ (right).
\label{fig:dk_model_02}} 
\end{figure}

\begin{figure}[t] 
\centering
%\leavevmode 
\includegraphics[clip,width=13cm,angle=0]{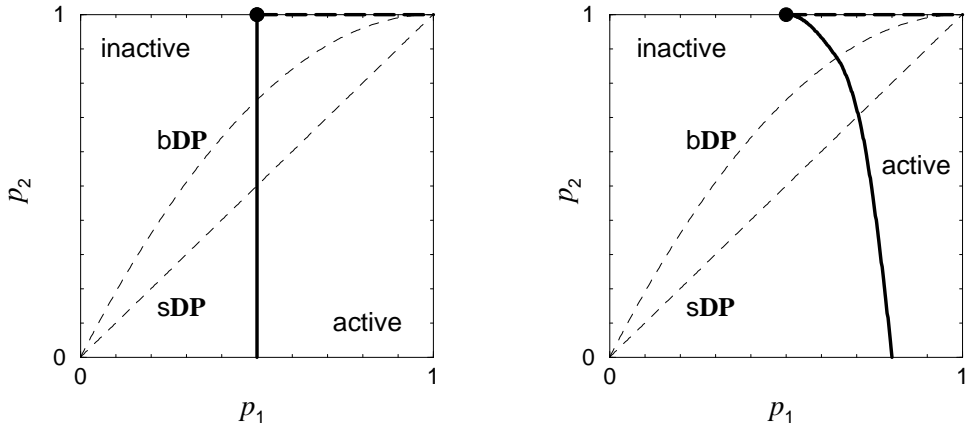}
\caption[Domany-Kinzel automaton, phase diagram] 
{The phase diagram of the Domany-Kinzel automaton.
The left figure shows the phase diagram of the mean field 
approximation.
%The simplest mean field solution is marked by the solid black
%line whereas the gray line corresponds to a more accurate
%approximation ({\bf Funtkion ncoh ausrechnen und angeben, ref hier hin}).
The $1+1$-dimensional behavior is
sketched in the right figure.
In both cases, the inactive phase \mbox{($\rhoa=0$)} is separated from the
active phase \mbox{($0<\rhoa<1$)} 
by a line of second-order transitions (solid line).
Bond directed percolation corresponds to 
$p_{\ssstyle 2} = p_{\ssstyle 1}(2-p_{\ssstyle 1})$
whereas site directed percolation is obtained for
$p_{\ssstyle 1} = p_{\ssstyle 2}$.
For $p_{\ssstyle 2}=0$ the Domany-Kinzel automaton corresponds
to the cellular automata rule "18" of Wolfram's classification
scheme~\protect\cite{WOLFRAM_1,ZEBENDE_1}.
Furthermore, the fully occupied lattice \mbox{($\rhoa=1$)} remains inactive 
for $p_{\ssstyle 1}>1/2$ and $p_{\ssstyle 2}=1$ (thick dashed line).
For $p_{\ssstyle 2}=1$ the system is characterized by a particle-hole
symmetry which gives rise to a different universal
scaling behavior (marked by the full circle), 
so-called compact directed percolation.
\label{fig:dk_model_phase_diagram}} 
\end{figure}

Analogous to the contact process (see page~\pageref{pg:mf_cp}), 
mean field treatments present some insight into 
the critical behavior\index{mean field theory, DK}
of the Domany-Kinzel 
automaton~(see e.g.~\cite{KINZEL_2,TOME_1,RIEGER_1,ATMAN_1}).
Here, we consider an automaton consisting of 
two-sites $i=1,2$. 
Depending on the state of the system 
$\{ n_{\ssstyle 1},n_{\ssstyle 2} \} $
a new configuration 
$\{ n_{\ssstyle 1}^{\prime},n_{\ssstyle 2}^{\prime} \}$
is obtained by applying the dynamical rules that
are embodied in the conditional probabilities 
\mbox{$p_{\ssstyle 0}=h$}, $p_{\ssstyle 1}$, and $p_{\ssstyle 2}$.
Neglecting correlations, the associated probabilities for a change
of the number of active sites by 
$\Delta n_{\ssstyle \mathrm a}$ are
\begin{eqnarray}
p(\Delta n_{\ssstyle \mathrm a}=-2) 
& = & \rhoa^2 \, (1-p_{\ssstyle 2})^2 \, , \nonumber \\
p(\Delta n_{\ssstyle \mathrm a}=-1) 
& = & 2 \, \rhoa^2 \, p_{\ssstyle 2} (1-p_{\ssstyle 2}) \,
+ \, 2 \, \rhoa \, \rhoe (1-p_{\ssstyle 1})^2 \, ,\nonumber \\
p(\Delta n_{\ssstyle \mathrm a}=\phantom{-}0) 
& = & 2 \, \rhoa^2 \, p_{\ssstyle 2}^2  \,
+ \, 4 \, \rhoa \, \rhoe p_{\ssstyle 1} \,(1-p_{\ssstyle 1}) \,
+ \, \rhoe^2 \, (1-h)^2 \, ,\nonumber \\
p(\Delta n_{\ssstyle \mathrm a}=\phantom{-}1) 
& = & 2 \, \rhoa \, \rhoe \, p_{\ssstyle 1}^2 \,
+ \, 2 \, \rhoe^2 \, h \, (1-h) \,  , \nonumber \\
p(\Delta n_{\ssstyle \mathrm a}=\phantom{-}2) 
& = & \rhoe^2 \, h^2 \, ,
\end{eqnarray}   
where $\rhoa$ denotes the density of (active) occupied sites
and the density of empty sites is $\rhoe=1-\rhoa$.
This reaction scheme leads to the differential
equation
\begin{equation}
\partial_{\ssstyle t} \, \rhoa(p_{\ssstyle 1}, p_{\ssstyle 2}, h) 
\; = \; 2 \, (2 p_{\ssstyle 1}-1) \, \rhoa 
\, - \, 2 \, (2 p_{\ssstyle 1} - p_{\ssstyle 2}) \,
\rhoa^2 \, + \, 2 (1-\rhoa)^2 \, h \, .
\label{eq:DK_mean_Langevin}
\end{equation}
Focusing on the steady state behavior ($\partial_{\ssstyle t}\rhoa=0$) 
we find that the
order parameter at zero field is given by
\begin{equation}
\rhoa(p_{\ssstyle 1}, p_{\ssstyle 2}, h=0)\; = \; 0
 \quad\quad \vee
\quad\quad
\rhoa(p_{\ssstyle 1}, p_{\ssstyle 2}, h=0) 
\; = \; \frac{2 p_{\ssstyle 1}-1}{\,2 p_{\ssstyle 1} - p_{\ssstyle 2}\,}
\label{eq:DK_mean_field_op_h0}
\end{equation}
The first solution is stable for $p_{\ssstyle 1}<1/2$ and
corresponds to the inactive phase.
The active phase is described by the second solution 
which is stable for $p_{\ssstyle 1}>1/2$.
Thus within this approximation, the phase diagram of the 
Domany-Kinzel automaton exhibits two phases separated by the
critical line $\pc =p_{\ssstyle 1} =1/2$
(see~\reffigname{fig:dk_model_phase_diagram}).
Along this critical line the order parameter vanishes
linearly ($\beta=1$) in the active phase 
\begin{equation}
\rhoa(p_{\ssstyle 1}, p_{\ssstyle 2}, h=0)
\; = \; \frac{1}{1-p_{\ssstyle 2}} \, \deltap
\, + \, {\mathcal O}(\deltap^2)
\label{eq:DK_mean_field_op_h0_scal}
\end{equation}
with $\deltap=(p_{\ssstyle} - \pc)/\pc$.
The field dependence of the order parameter at the critical line
is given by 
\begin{equation}
\rhoa(p_{\ssstyle 1}=p_{\ssstyle \mathrm c}, p_{\ssstyle 2}, h)\; = \; 
\sqrt{ \frac{h}{1-p_{\ssstyle 2}}\,}
\, + \, {\mathcal O}(h) \, ,
\label{eq:DK_mean_field_op_pc}
\end{equation}
yielding the field exponent~$\sigma=2$.
The asymptotic behavior of the equation of state
\begin{equation}
\rhoa(p_{\ssstyle 1}, p_{\ssstyle 2}, h)\; \sim \; 
\frac{a_{\ssstyle p}\deltap}{2} \, + \, 
\sqrt{ \left ( \frac{a_{\ssstyle p}\deltap}{2} \right )^2 \, +  \,
a_{\ssstyle h} h \;} \, ,
\label{eq:DK_mean_field_eqos}
\end{equation}
is obtained by performing the limits $\rhoa \to 0$, $\deltap\to 0$,
and $h\to 0$ with the constraints that $\rhoa / \sqrt{h}$ and
$\rhoa / \deltap$ remain finite.
The metric factors are given by 
$a_{\ssstyle r}=a_{\ssstyle p}=1/(1-p_{\ssstyle 2})$.
Thus the mean field behavior of the Domany-Kinzel
automaton is characterized by the same critical exponents and
despite of the metric factors $a_{\ssstyle p}$, $a_{\ssstyle h}$
by the same scaling function as the mean field contact 
process [see \refeqname{eq:mf_cp_one_site_ord_field_scal_asymp}].

It is possible to improve the mean field
approximations presented above
%by increasing the number of incorporates sites 
%(see \reffigname{fig:dk_model_02} and 
%\reffigname{fig:dk_model_phase_diagram}) 
%or equivalently 
by considering higher orders of correlations (see e.g.~\cite{TOME_1}).
In this way, better approximations of the 
phase diagram of the $1+1$-dimensional 
system can be obtained.
But the results are still mean field-like, i.e., the 
critical exponents as well as the asymptotic scaling functions
equal $\beta=1$, $\sigma=2$, and \refeqname{eq:DK_mean_field_eqos}.

\begin{figure}[t] 
\centering
%\leavevmode 
\includegraphics[clip,width=8cm,angle=0]{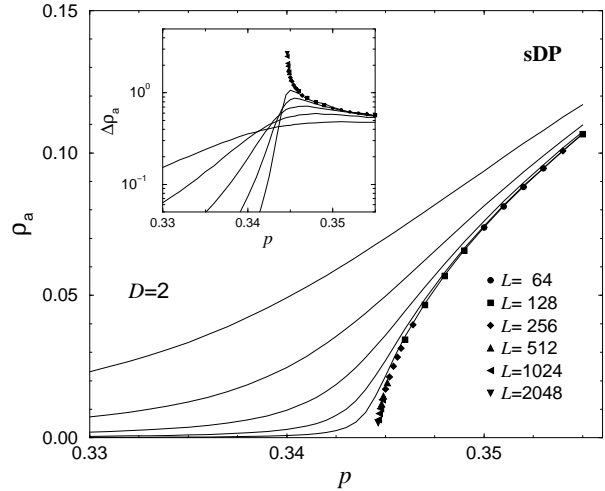}
\caption[Domany-Kinzel automaton, order parameter and its fluctuations] 
{The order parameter and its fluctuations of the $2+1$-dimensional
generalized Domany-Kinzel automaton on a bcc lattice for various
field values (from $h=10^{-6}$ to $h=2\,10^{-4}$).
The data corresponds to the site directed percolation (sDP)
process.
For non-zero field (lines) $\rhoa$ and $\Delta\rhoa$ exhibit an 
analytical behavior.
For zero field (symbols) the typical critical behavior of 
the order parameter
and its fluctuations are obtained.
The data are obtained from simulations on various system
sizes $L \kgl 2048$ with periodic boundary conditions.
\label{fig:sDP_2d_op}} 
\end{figure}

Clearly, the above derived mean field results are not valid
for $p_{\ssstyle 2}=1$. 
In that case no particle annihilation takes place
within a domain of occupied sites, i.e., the creation
and annihilation processes are bounded to the domain 
walls where empty and occupied sites are adjacent.
This corresponds to the particle-hole 
symmetry\index{particle-hole symmetry}\index{symmetry, particle-hole}
\begin{equation}
n \; \; \leftrightarrow \; \; 1-n
\label{eq:particle_hole_symmetry}
\end{equation}
which is also reflected in the mean field differential
equation [\refeqname{eq:DK_mean_Langevin}]
\begin{equation}
\partial_{\ssstyle t} \, \rhoa(p_{\ssstyle 1}, p_{\ssstyle 2}=1, h=0) 
\; = \; 4 \, (2 p_{\ssstyle 1}-1) \; \rhoa \, (1-\rhoa) \, .
\label{eq:DK_mean_Langevin_CDP}
\end{equation}
Obviously, the fully occupied lattice is another absorbing state.
The behavior of the domain wall boundaries can be mapped to 
a simple random walk~\cite{DOMANY_1} and the domains
of particles
grow on average for $p_{\ssstyle 1}>1/2$ whereas they
shrink for $p_{\ssstyle 1}<1/2$.
Thus the steady state density $\rhoa$ is zero 
below \mbox{$p_{\ssstyle \mathrm c}=1/2$} and \mbox{$\rhoa=1$} above
\mbox{$p_{\ssstyle \mathrm c}$}.
At the critical value the order parameter $\rhoa$ exhibits a jump. 
The associated critical exponent $\beta=0$
differs in all dimensions from the directed percolation 
values $0<\beta_{\ssstyle \mathrm {DP}}\kgl 1$.
Since domain branching does not take place  
the dynamical process for $p_{\ssstyle 2}=1$
is\index{compact directed percolation}\index{directed percolation, compact}\index{CDP} 
termed compact directed percolation (CDP).
The critical behavior equals that of the $1+1$-dimensional 
voter model~\cite{LIGGETT_1} and is analytically tractable 
due to the mapping to random walks.
Exact results are derived for the critical exponents~\cite{DOMANY_1,ESSAM_1} 
($\beta=0$, $\betap=1$, $\nu_{\para}=2$, $\nu_{\senk}=1$, 
see also \reftablename{table:cdp_exponents})
as well as for certain finite-size scaling functions~\cite{KEARNEY_1}.
In particular, the domain growths from a single seed exhibits a 
one-to-one correspondence to a pair of annihilating random walkers.
That correspondence allows to calculate the percolation 
probability~\cite{ESSAM_1} [\refeqname{eq:ulti_sur_prob}]
\begin{equation}
\Pperc(p_{\ssstyle 1}) \; = \;
\left \{
\begin{array}{lcl}
0 & \mathrm{if} & p_{\ssstyle 1}<1/2 \, ,\\[2mm]
\frac{2\, p_{\ssstyle 1}-1}{p_{\ssstyle 1}^2} & \mathrm{if} & p_{\ssstyle 1}>1/2 \, ,
\end{array}
\right .
\label{eq:CDP_Pperc}
\end{equation}
yielding the percolation exponent $\betap=1$.
In contrast to directed percolation the 
universality class of compact directed 
percolation\index{universality class, CDP} 
is\index{universality class, compact directed percolation} 
characterized by the inequality
\begin{equation}
\beta \; \neq \; \betap \, .
\label{eq:beta_betap_nequal}   
\end{equation}
The number of independent critical exponents is therefore
greater than for directed percolation. 
In summary, the universality class of compact directed
percolation is characterized by a continuously vanishing
percolation order parameter $\Pperc$ and algebraically
diverging correlations lengths~\cite{ESSAM_1},
indicating a second order phase transition.
But due to the misleading discontinuous behavior 
of the steady state order parameter~$\rhoa$ 
the phase transition was often considered as first order.

It is straight forward to generalize the $1+1$-dimensional
Domany-Kinzel automaton to higher 
dimensions (see e.g.~\cite{GRASSBERGER_8,GRASSBERGER_3,LUEB_28}).
In the following, we consider cellular automata on a 
$D+1$-dimensional body centered cubic (bcc) lattice where 
the time corresponds to the $[0,0,\ldots,0,1]$ direction.
A lattice site at time~$t$ is occupied with probability~$p$
if at least one of its $2^D$ backward neighboring 
sites (at time $t-1$) is occupied.
Otherwise the site remains empty.
This parameter choice corresponds to the condition probabilities
$p_{\ssstyle 1}=p_{\ssstyle 2}=\ldots=p_{\ssstyle 2^D}=p$,
i.e., site directed percolation is considered.
In \reffigname{fig:sDP_2d_op} we present simulation results
for the $2+1$-dimensional Domany-Kinzel automaton.
The data show the expected critical behavior of the 
order parameter and its fluctuations.

%%%%%%%%%%%%%%%%%%%%%%%%%%%%%%%%%%%%%%%%%%%%%%%%%%%%%%%%%%%%%%%%%%%%%%%%%%%%%%%%%%%%%%
%%%%%%%%%%%%%%%%%%%%%%%%%%%%%%%%%%%%%%%%%%%%%%%%%%%%%%%%%%%%%%%%%%%%%%%%%%%%%%%%%%%%%%
%%%%%%%%%%%%%%%%%%%%%%%%%%%%%%%%%%%%%%%%%%%%%%%%%%%%%%%%%%%%%%%%%%%%%%%%%%%%%%%%%%%%%%

\subsection{Pair contact process}
\label{subsec:pair_contact_process}

\index{pair contact process}

\begin{figure}[t] 
\centering
%\leavevmode 
\includegraphics[clip,width=10cm,angle=0]{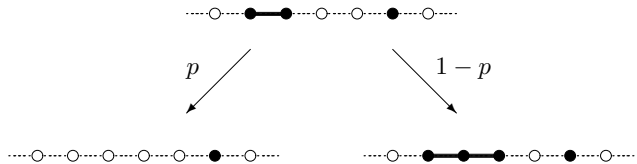}
\caption[Pair contact process] 
{The $1+1$-dimensional pair contact process~\protect\cite{JENSEN_1}.
Lattice sites may be either empty (open circles) or occupied by a particle 
(full circles).
Pairs of occupied sites (solid bonds) are considered as 
active whereas isolated particles remain inactive.
A pair is annihilated with rate~$p$, otherwise an 
offspring is created at an empty  neighboring site selected at random.
\label{fig:pcp_model_01}} 
\end{figure}

\begin{figure}[b] 
\centering
%\leavevmode 
\includegraphics[clip,width=8cm,angle=0]{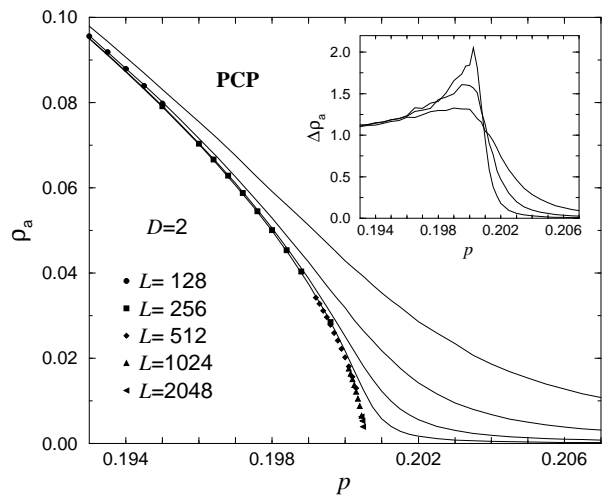}
\caption[Pair contact process, order parameter and its fluctuations] 
{The order parameter and its fluctuations (inset) 
of the $2+1$-dimensional
pair contact process on a square lattice for various
values of the field (from $h=10^{-6}$ to $h=6\,10^{-5}$).
The symbols mark zero-field behavior.
The data are obtained from simulations on various system
sizes $L \kgl 2048$ with periodic boundary conditions.
\label{fig:PCP_2d_op}} 
\end{figure}

The pair contact process (PCP) was introduced 
by\index{pair contact process}\index{PCP} 
Jensen~\cite{JENSEN_2} and is one of the 
simplest models with infinitely many absorbing states 
showing a continuous phase transition.
The process is defined on a $D$-dimensional cubic lattice
and an asynchronous (random sequential) update 
scheme is applied.
A lattice site may be 
either occupied ($n=1$) or empty ($n=0$).
Pairs of adjacent occupied sites, linked by an active bond, 
annihilate each other with rate $p$ otherwise an 
offspring is created at a neighboring site provided the 
target site is empty (see \reffigname{fig:pcp_model_01}).
The density of active bonds $\rho_{\rm{a}}$ is the order 
parameter of a continuous phase transition from an active 
state  
to an inactive absorbing state without particle pairs. 
Similar to the contact process and to the Domany-Kinzel
automaton a spontaneous particle creation acts as
a conjugated field~\cite{LUEB_27}.
The behavior of the PCP order parameter and its fluctuations
are plotted in \reffigname{fig:PCP_2d_op}.

Since isolated particles remain inactive, any
configuration containing only isolated particles is 
absorbing.
In case of the $1+1$-dimensional pair contact process
with $L$~sites and periodic boundary conditions
the number of absorbing states is asymptotically 
given by the golden mean~\cite{CARLON_1}
\begin{equation}
N_{\ssstyle \mathrm {absorb.~states}} \; \sim \,
\left (
\frac{\,1+\sqrt{5}}{2}
\right )^{L} \, .
\label{eq:pcp_1d_n_abs_states_pbc}
\end{equation}
In the thermodynamic limit the pair contact process
is characterized by infinitely many absorbing states.
%%%%%%%%%%%%%%%%%%%%%%%%%%%%%%%%%%%%%%%%%%%%%%%%%%%%%%%%%%%%%%%%%%%%%%%%%%%
%%%%%%%%%%%%%%%%%%%%%%%%%%%%%%%%%%%%%%%%%%%%%%%%%%%%%%%%%%%%%%%%%%%%%%%%%%% 
Due to that non-unique absorbing phase the universality
hypothesis of Janssen and Grassberger can not be applied.
Therefore, the critical behavior of the pair contact process
was intensively investigated by 
simulations (including~\cite{JENSEN_1,JENSEN_3,DICKMAN_11,DICKMAN_4,DICKMAN_5}).
It was shown numerically that the critical scaling behavior 
of the $1+1$-dimensional pair contact process 
is characterized by the same critical 
exponents~\cite{JENSEN_2,JENSEN_3} as well
as by the same universal scaling functions as directed 
percolation~\cite{LUEB_27}.
In particular the latter result provides a convincing
identification of the universal behavior.
Thus despite the different structure of the absorbing phase
the $1+1$-dimensional pair contact process belongs to the 
directed percolation universality class.
This numerical evidence confirms a corresponding 
renormalization group analysis
predicting DP universal behavior~\cite{MUNOZ_1}
in all dimensions.
But the scaling behavior of the PCP in higher dimension is still
a matter of controversial discussions.
A recently performed renormalization group analysis conjectures 
that the pair contact process exhibits a
dynamical\index{dynamical percolation} 
percolation-like\index{universality class, dynamical percolation}
scaling behavior~\cite{WIJLAND_1,WIJLAND_3}.
A dynamical percolation cluster at criticality equals
a fractal cluster of ordinary percolation\index{percolation} 
on the same lattice.
Thus, the dynamical percolation universality 
class~\cite{GRASSBERGER_9,CARDY_5,CARDY_3,JANSSEN_6} differs from the 
directed percolation universality class. 
In particular the upper 
critical dimension\index{upper critical dimension}\index{critical dimension}
equals \mbox{$\Dc=6$} instead of \mbox{$\Dc=4$} for DP.
So far the investigations of the PCP are limited to the
$1+1$- and $2+1$-dimensional~\cite{SILVA_4} systems.
Only recently performed numerical simulations~\cite{LUEB_32}
provide estimates of the scaling
behavior in higher dimensions, 
including the exponents as well as the
scaling functions.
The results are presented in this work.

A modification of the pair contact process, the so-called
pair contact process with diffusion (PCPD),\index{PCPD}
attracted a lot of attention in the last years
(e.g.~\cite{GRASSBERGER_2,HOWARD_2,ODOR_4,CARLON_1,HINRICHSEN_4,HINRICHSEN_5,ODOR_3,NOH_1,DICKMAN_10,ODOR_5,PARK_1,KOCKELKOREN_1,BARKEMA_1,HINRICHSEN_6,PAESSENS_1,JANSSEN_12}). 
In contrast to the original PCP isolated particles are allowed
to diffuse in case of the PCPD.
Clearly, the diffusion of isolated particles changes the
structure of the absorbing phase. 
It only contains the empty lattice and a state 
with a single endlessly wandering particle.
The question of the universal scaling behavior of the 
pair contact process with diffusion is still a matter 
of controversial discussions in the 
literature (see~\cite{HENKEL_1} for a recent review).
For example, non-universal continuously varying 
exponents~\cite{DICKMAN_10,NOH_1}, 
a new universality class~\cite{HINRICHSEN_4,PARK_1,KOCKELKOREN_1}
as well as a crossover to the universality class of directed 
percolation~\cite{BARKEMA_1} 
are reported.
The striking variety of the observed conflicting scaling 
scenarios reflects the non-trivial character of the 
pair contact process with diffusion which remains a
topic for future research.

%%%%%%%%%%%%%%%%%%%%%%%%%%%%%%%%%%%%%%%%%%%%%%%%%%%%%%%%%%%%%%%%%%%%%%%%%%%%%%%%%%%%%%
%%%%%%%%%%%%%%%%%%%%%%%%%%%%%%%%%%%%%%%%%%%%%%%%%%%%%%%%%%%%%%%%%%%%%%%%%%%%%%%%%%%%%%
%%%%%%%%%%%%%%%%%%%%%%%%%%%%%%%%%%%%%%%%%%%%%%%%%%%%%%%%%%%%%%%%%%%%%%%%%%%%%%%%%%%%%%

\subsection{Other models}
\label{subsec:dp_other_models}

In this section we briefly consider two lattice models
exhibiting a directed percolation-like phase transition.
The first one is the threshold transfer process (TTP)
which was\index{threshold transfer process}\index{TTP} 
introduced in~\cite{MENDES_1}.
Here, lattice sites may be empty ($n=0$), 
occupied by one particle ($n=1$),
or occupied by two particles ($n=2$).
Double occupied lattice sites are 
considered as active.
In that case both particles may move to the 
left ($\mathrm l$) and right ($\mathrm r$) 
neighbor if possible, i.e.,
\begin{eqnarray}
\label{eq:ttp_dynamic_rules}   
n_{\ssstyle \mathrm{l}} & \longrightarrow &
n_{\ssstyle \mathrm{l}} + 1 \quad\quad
{\mathrm{if}} \quad n_{\ssstyle \mathrm{l}}<2 \, ,\nonumber \\
n_{\ssstyle \mathrm{r}} & \longrightarrow &
n_{\ssstyle \mathrm{r}} + 1 \quad\quad
{\mathrm{if}} \quad n_{\ssstyle \mathrm{r}}<2 \, .
\end{eqnarray}
Additionally to the particle movement, creation and annihilation
processes are incorporated.
A particle is created at an empty lattice site ($0 \to 1$) 
with probability~$r$ whereas a particle annihilation
($1\to 0$) takes place with probability $1-r$.
In the absence of double occupied sites the dynamics 
is characterized by a fluctuating steady state with a 
density~$r$ of single occupied sites.
The density of double occupied sites is identified as the
order parameter of the process, and any configuration
devoid of double occupied sites is absorbing.
The probability~$r$ controls the particle density,
and a non-zero density of active sites occurs only
for sufficiently large values of~$r$.
In contrast to the infinitely many frozen absorbing 
configurations of the pair contact process, the
threshold transfer process is characterized by 
fluctuating absorbing states.
Nevertheless steady state numerical simulations of 
the $1+1$-dimensional threshold transfer process 
yield critical exponents that are in agreement 
with the corresponding DP values~\cite{MENDES_1}.
So far no systematic analysis of the TTP in higher
dimensions was performed.

In this work we focus on the $2+1$-dimensional model.
In that case one tries to transfer both particles
of a given active site to randomly chosen empty or single
occupied nearest neighbor sites.
Furthermore we apply an external field that is conjugated
to the order parameter.
In contrast to the models discussed above the conjugated field
can not be implemented by a particle creation.
A particle creation with rate $h$ would affect\index{conjugated field} 
the particle density, i.e., the control parameter of the
phase transition.
But the conjugated field has to be independent of the 
control parameter.
Therefore we implement the conjugated field by a diffusion-like
field that acts by particle movements.
A particle on a given lattice site moves to a randomly 
selected neighbor with probability~$h$, if $n<2$.
Thus the conjugated field of the TTP differs from the
conjugated field of the Domany-Kinzel automaton, the contact 
process, and the pair contact process.

%%%%%%%%%%%%%%%%  ZGB %%%%%%%%%%%%%%%%%%%%%%%%%%%%%%%%%%%

\begin{figure}[t] 
\centering
%\leavevmode 
\includegraphics[clip,width=13cm,angle=0]{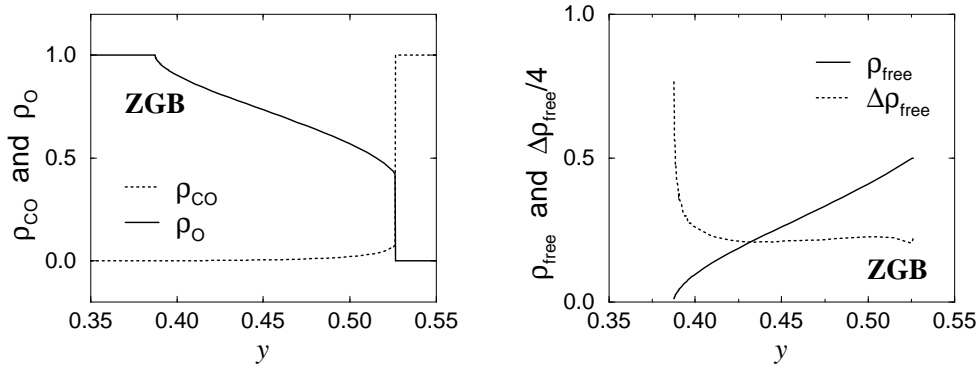}
\caption[Ziff-Gular{\'\i}-Barshad model for catalytic reaction] 
{The schematic phase diagram of the Ziff-Gular{\'\i}-Barshad
model for the catalytic carbon-monoxide oxidation.
The steady state concentrations of ${\mathrm {CO}}$ and 
${\mathrm {O}}_{\ssstyle 2}$ on the catalytic surface
are plotted as function
of the ${\mathrm {CO}}$ concentration $y$ in the gas phase.
The system undergoes a second order phase
transition to the oxygen passivated phase whereas
a first order phase transition takes place if the
${\mathrm {CO}}$ passivated phase is approached.
The right figure shows the density of vacant lattice sites
and its fluctuations.
\label{fig:zgb_phase_dia}} 
\end{figure}

Another model exhibiting a directed percolation-like
absorbing phase transition is 
the\index{Ziff-Gular{\'{\i}}-Barshad model}\index{ZGB} 
Ziff-Gular{\'\i}-Barshad
(ZGB) model~\cite{ZIFF_1}.
This model mimics the heterogeneous catalysis of the
carbon monoxide oxidation \index{catalytic reactions}
\begin{equation}
2\, {\mathrm {CO}} \; + \; {\mathrm O}_{\ssstyle 2} \;
\longrightarrow \; 2\, {\mathrm {CO}}_{\ssstyle 2} 
\label{eq:zgb_CO_O2_2CO2}
\end{equation}
on a catalytic material, e.g.~platinum.
The catalytic surface is represented by a square lattice
where ${\mathrm {CO}}$ or ${\mathrm {O}}_{\ssstyle 2}$ can
be adsorbed from a gas phase with concentration $y$ for 
carbon monoxide and $1-y$ for oxygen, respectively.
The concentration $y$ is the control parameter of the 
model determining the density of the components 
on the catalytic surface.
Adsorbed oxygen molecules  %~${\mathrm {O}}_{\ssstyle 2}$ 
dissociate at the catalytic surface into pairs
of ${\mathrm {O}}$~atoms.
It is assumed that the lattice sites are either
empty, occupied by an ${\mathrm {CO}}$ molecule,
or occupied by an ${\mathrm {O}}$ atom.
Adjacent ${\mathrm {CO}}$ and ${\mathrm {O}}$ react
instantaneously and the resulting ${\mathrm {CO}}_{\ssstyle 2}$
molecule leaves the system.
Obviously, the system is trapped in absorbing configurations
if the lattice is completely covered by carbon monoxide
or completely covered by oxygen.
The dynamics of the system is attracted by these absorbing
configurations for sufficiently large ${\mathrm {CO}}$ concentrations
and for sufficiently large ${\mathrm {O}}_{\ssstyle 2}$ concentrations.
Numerical simulations show that catalytic activity occurs 
in the range \mbox{$0.390 < y < 0.525$}~\cite{JENSEN_8} 
only (see \reffigname{fig:zgb_phase_dia}).
The system undergoes a second order phase
transition to the oxygen passivated phase whereas
a first order phase transition takes place if the
${\mathrm {CO}}$ passivated phase is approached.
In particular, the continuous phase transition  
belongs to the universality class of directed 
percolation~\cite{GRINSTEIN_3,JENSEN_8,VOIGT_1}. 
This might be surprising at a first glance
since the ZGB model is characterized by two
distinct chemical components, ${\mathrm {CO}}$
and ${\mathrm {O}}$.
But the catalytic activity is connected
to the density of vacant sites, i.e., to a
single component order parameter~\cite{GRINSTEIN_3}.
Thus the observed directed percolation behavior is in 
full agreement with the universality hypothesis of Janssen 
and Grassberger.

But one has to stress that the ZGB model is an 
oversimplified representation of the catalytic 
carbon monoxide oxidation.
A more realistic modeling has to incorporate for
example mobility and desorption processes
as well as a repulsive interaction between 
adsorbed oxygen molecules (see e.g.~\cite{ZIFF_1,LIU_1}).
These features affect the critical behavior and
drive the system out of the directed percolation
universality class.
%{\bf
%In particular, desorption processes increase the 
%density of vacant sites, i.e., they act like a 
%conjugated\index{conjugated field}
%field.}

Eventually we just mention that depinning
phenomena can be mapped to directed percolation.
Whereas certain interface depinning phenomena 
are related to $1+1$-dimensional DP, higher 
dimensional behavior is related to line pinning
phenomena.
Instead of going into the details we refer the
interested readers to the reviews~\cite{HINRICHSEN_1,ODOR_1,KARDAR_2}.

\section{Steady state scaling behavior}
\label{sec:dp_steady_state_scaling}

In the following we consider the steady state scaling 
behavior of different models in various dimensions.
We focus our attention on the order parameter, i.e., the 
equation of state,
its fluctuations as well as on 
the order parameter susceptibility.
First we consider the scaling behavior below the upper
critical dimension and compare the results to
those of an renormalization group approach.
Then we show that the scaling behavior above the upper
critical dimension is in full agreement with the
mean field predictions.
The scaling scenario at the upper critical dimension
is characterized by logarithmic corrections, and will be
discussed in detail.
Furthermore, certain universal amplitude combinations
are considered which are related to the order parameter
and its fluctuations.
Analogous to equilibrium phase transitions, these amplitude
combinations are directly related to particular values
of the universal scaling functions and are of great 
experimental interest.
A finite-size scaling analysis of the order parameter, its
fluctuations as well as a fourth order cumulant is
presented at the end of this section.

%%%%%%%%%%%%%%%%%%%%%%%%%%%%%%%%%%%%%%%%%%%%%%%%%%%%%%%%%%%%%%%%%%%%%%%%%%%%%%%%%%%%%%
%%%%%%%%%%%%%%%%%%%%%%%%%%%%%%%%%%%%%%%%%%%%%%%%%%%%%%%%%%%%%%%%%%%%%%%%%%%%%%%%%%%%%%
%%%%%%%%%%%%%%%%%%%%%%%%%%%%%%%%%%%%%%%%%%%%%%%%%%%%%%%%%%%%%%%%%%%%%%%%%%%%%%%%%%%%%%

\subsection{Below the upper critical dimension}
\label{subsec:dp_below_dc}

Similar to equilibrium phase transitions 
we assume that sufficiently close to the critical point 
the order parameter~$\rhoa$, its fluctuations~$\Delta\rhoa$,
and the order parameter susceptibility~$\chi$ can be
described by generalized homogeneous functions
(see \refeqname{eq:universal_HS_form})\index{generalized homogeneous function}
\begin{eqnarray}
\label{eq:scal_ansatz_EqoS_HS}
%a_{\ssstyle a} \,
\rhoa(\deltap, h) 
\; & \sim & \; 
\lambda^{-\beta}\, \, {\tilde R}
(a_{\ssstyle p}  
\deltap \; \lambda, a_{\ssstyle h} h \;
\lambda^{\sigma}) \, ,\\[2mm]
\label{eq:scal_ansatz_Fluc_HS}
a_{\ssstyle \Delta} \,
\Delta \rhoa(\deltap, h) 
\; & \sim & \; 
\lambda^{\gamma^{\prime}}\, \, {\tilde D}
(a_{\ssstyle p} \deltap \; \lambda, 
a_{\ssstyle h} h \, \lambda^{\sigma})  \, , \\[2mm]
\label{eq:scal_ansatz_Susc_HS}
a_{\ssstyle \chi} \,
\chi(\deltap, h) 
\; & \sim & \; 
\lambda^{\gamma}\, \, {\tilde {\mathrm X}}
(a_{\ssstyle p} \deltap \; \lambda, 
a_{\ssstyle h} h \, \lambda^{\sigma})  \, .
\end{eqnarray}
The fluctuation exponent fulfills the hyperscaling law
[see~\refeqname{eq:dim_analysis_fluc}]
\begin{equation}
\gammap \; = \; \nu_{\senk} \, D \, - \, 2 \, \beta \, .
\label{eq:gammap_nu_D_2beta}
\end{equation}   
Note that in contrast to the scaling behavior of 
equilibrium phase transitions 
the non-equilibrium absorbing phase transition 
of directed percolation is characterized 
by $\gamma\neq \gammap$.                    
Furthermore, the spatial and temporal correlation length 
are expected to scale as
\begin{eqnarray}
\label{eq:scal_ansatz_xi_senk_HS}
a_{\senk} \,
\xi_{\senk}(\deltap, h) 
\; & \sim & \; 
\lambda^{-\nu_{\senk}}\, \, {\tilde \Xi}_{\senk}
(a_{\ssstyle p}  
\deltap \; \lambda, a_{\ssstyle h} h \;
\lambda^{\sigma}) \, ,\\[2mm]
\label{eq:scal_ansatz_xi_para_HS}
a_{\para} \,
\xi_{\para}(\deltap, h) 
\; & \sim & \; 
\lambda^{-\nu_{\para}}\, \; {\tilde \Xi}_{\para}
(a_{\ssstyle p} \deltap \; \lambda, 
a_{\ssstyle h} h \, \lambda^{\sigma})  \, . 
\end{eqnarray}
As usual, $h$ denotes the conjugated field
and $\deltap$ denotes the distance to the critical point,
e.g.~$\deltap = (\lambda - \lambdac)/\lambdac$ 
for the contact process,
$\deltap = (p-\pc)/\pc$ for site directed percolation and for the
pair contact process, 
$\deltap = (r - r_{\ssstyle \mathrm c})/r_{\ssstyle \mathrm c}$ 
for the threshold transfer process, etc.
Additionally to the above Hankey-Stanley scaling 
forms\index{Hankey-Stanley scaling form}
one\index{Widom-Griffiths scaling form}
may\index{scaling forms}
consider the corresponding 
Widom-Griffiths scaling forms, e.g.
\begin{equation}
a_{\ssstyle h}\, 
h(\deltap,\rhoa) 
\; \sim \;
\lambda^{-\sigma} \; \; 
{\tilde H}(a_{\ssstyle p} \deltap \, \lambda, 
 \rhoa\, \lambda^{\beta} ) \, .
\label{eq:scal_ansatz_EqoS_WG}
\end{equation}

Once the non-universal metric factors are chosen in
a specified way, 
the universal scaling functions ${\tilde R}$, ${\tilde H}$, 
${\tilde D}$, ${\tilde {\mathrm X}}$, 
${\tilde \Xi}_{\senk}$ and 
${\tilde \Xi}_{\para}$ are the
same for all systems within a given universality
class.
All non-universal system-dependent features 
(e.g.~the lattice structure, the range of interaction,
as long as the interaction
decreases sufficient rapidly as function of separation, etc.)
are contained in the non-universal metric factors 
$a_{\ssstyle p}$, $a_{\ssstyle h}$, 
$a_{\ssstyle\Delta}$, $\ldots$~\cite{PRIVMAN_3}. 
The above scaling forms are valid 
for $D\neq \Dc$.
At the upper critical dimension $\Dc$
they have to be modified by logarithmic 
corrections~\cite{LUEB_26}.

Note that all universal functions 
are characterized by the same arguments.
Identical scaling arguments 
arise naturally in equilibrium
thermodynamics where in principle all functions can be 
derived from a single thermodynamic 
potential, e.g.~the free energy (see \ref{sec:scaling_theory}). 
In case of non-equilibrium phase transitions one
can argue that the functions can be derived from a
corresponding Langevin equation.
Furthermore, beyond the equation of state [\refeqname{eq:scal_ansatz_EqoS_HS}], 
one new metric factor is introduced for each 
additional quantity being under consideration, 
here $a_{\ssstyle \Delta}$, $a_{\ssstyle \chi}$, 
$a_{\senk}$, and $a_{\para}$.
But this phenomenological ansatz does not imply 
that the metric factors are independent of each other.
For example only two metric factors are needed to describe the
scaling behavior in equilibrium~\cite{STAUFFER_4},
often termed\index{two scale factor universality}\index{universality, two scale factor}  
two-scale-factor-universality.

\begin{figure}[t] 
\centering
%\leavevmode 
\includegraphics[clip,width=13cm,angle=0]{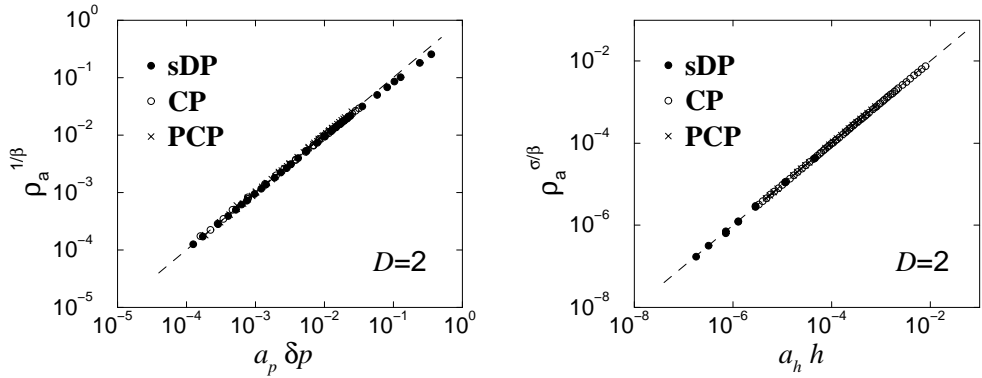}
\caption[Determination of the non-universal metric factors]
{The determination of the non-universal metric factors
$a_{\ssstyle p}$ and $a_{\ssstyle h}$ for site directed percolation (sDP), 
the contact process (CP), 
and the pair contact process (PCP) in $D=2$.
The left figure shows the order parameter at zero-field ($h=0$)
whereas the right figure displays the order parameter
at the critical value ($\deltap=0$).
Plotting $\rhoa^{1/\beta}$ vs. $a_{\ssstyle p}\deltap$ and
$\rhoa^{\sigma/\beta}$ vs. $a_{\ssstyle h} h$ the curves tend
asymptotically to the function $f(x)=x$ (dashed lines) when  
the transition point ($\deltap\to 0$ and $h\to 0$) is approached.
\label{fig:metric_factors_sDP_CP_PCP_D2}} 
\end{figure}

Similar to equilibrium [\refeqname{eq:normalization_M}] 
we norm the universal scaling functions
by 
\begin{equation}
{\tilde R}(1,0) \; = \; 1 \, , \quad \quad 
{\tilde R}(0,1) \; = \; 1 \, , \quad \quad 
{\tilde D}(0,1) \; = \; 1 \, .
\label{eq:uni_scal_norm_R_D}
\end{equation}
The first two normalizations imply
\begin{equation}
{\tilde H}(1,1)\, = \, 0 \, ,
\quad\quad\quad
{\tilde H}(0,1)\,=\,1 \, .
\end{equation}
In that way the non-universal metric factors 
$a_{\ssstyle p}$, $a_{\ssstyle h}$, and $a_{\ssstyle \Delta}$
are determined by the amplitudes of the power-laws
\begin{eqnarray}
\label{eq:metric_factors_a_rho}
\rho_{\ssstyle \mathrm a}(\deltap, h=0) \; & \sim & \; 
(a_{\ssstyle p} \, \deltap)^{\beta} \, ,\\
\label{eq:metric_factors_a_h}   
\rho_{\ssstyle \mathrm a}(\deltap =0, h) \; & \sim & \; 
(a_{\ssstyle h} \, h)^{\beta / \sigma} \, , \\
\label{eq:metric_factors_a_Delta}   
a_{\ssstyle \Delta} \,
\Delta\rho_{\ssstyle \mathrm a}(\delta p=0, h) \; & \sim &\; 
(a_{\ssstyle h} \, h)^{-\gamma^{\prime}/\sigma} \, .
\end{eqnarray}
The determination of the metric factors $a_{\ssstyle p}$
and $a_{\ssstyle h}$ for site directed percolation, the 
contact process, and the pair contact process is shown
in \reffigname{fig:metric_factors_sDP_CP_PCP_D2}.
For this purpose the best known estimates for the critical
exponents, as given in \reftablename{table:dp_exponents}, are
used.
The values of the non-universal metric factors
are listed in \cite{LUEB_28,LUEB_32}.

Taking into consideration that the susceptibility is defined as the
derivative of the order parameter with respect to the 
conjugated field [\refeqname{eq:lin_resp_apt}]
we find 
\begin{equation}
{\tilde {\mathrm X}}(x,y) \; = \; \partial_y \, {\tilde R}(x,y) \, ,
\quad \quad \quad
a_{\ssstyle \chi}\; = \; a_{\ssstyle h}^{-1} \, ,
\label{eq:sus_func_metric_factor}
\end{equation}
as well as the scaling law
\begin{equation}
\gamma \; = \; \sigma  \, - \, \beta \, .
\label{eq:widom_apt}
\end{equation}
This scaling law corresponds to the well 
known Widom\index{scaling law, Widom}\index{Widom scaling law}
law [\refeqname{eq:widom}] 
of equilibrium phase transitions.
Furthermore, 
comparing \refeqname{eq:scal_ansatz_Susc_HS}
for $\deltap=0$
to the definition of the susceptibility
\begin{equation}
a_{\ssstyle \chi} \,
\chi(\deltap, h) 
\; \sim  \; (a_{\ssstyle h} h)^{-\gamma/\sigma}\;
{\tilde {\mathrm X}}(0,1) \, ,
\quad\quad\quad\quad
\chi \; = \; \partial_{\ssstyle h} \, \rhoa 
\; = \; \partial_{\ssstyle h} \, (a_{\ssstyle h} h)^{\beta/\sigma}
\end{equation}   
leads to the condition
\begin{equation}
{\tilde {\mathrm X}}(0,1) \; = \; \frac{\,\beta\,}{\sigma} \, .
\label{eq:susc_Chi_01}
\end{equation}
The above result offers a useful consistency check of 
the numerical estimates of the susceptibility.
It is worth mentioning that the validity of the scaling
form \refeqname{eq:scal_ansatz_EqoS_HS} implies the required
singularity of the susceptibility [\refeqname{eq:lin_resp_apt}],
confirming that the applied external field is
conjugated to the order parameter.

\begin{figure}[t] 
\centering
%\leavevmode 
\includegraphics[clip,width=13cm,angle=0]{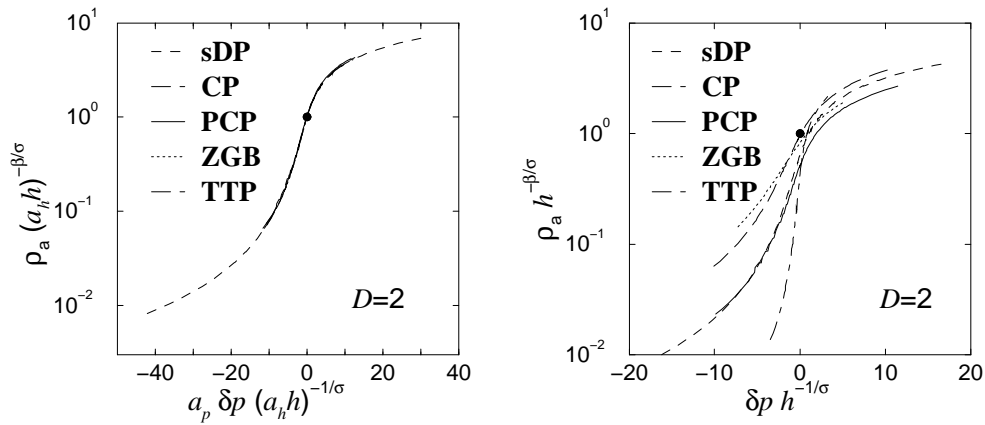}
\caption[Universal scaling function ${\tilde R}(x,1)$, DP class, $D=2$]
{The universal scaling function ${\tilde R}(x,1)$ of the
directed percolation universality class (left).
The data are plotted according 
to \refeqname{eq:scal_ansatz_EqoS_HS_collapse}.
All models considered are characterized by the same universal
scaling function, an impressive 
manifestation of the robustness of the directed 
percolation universality class with respect to 
variations of the microscopic interactions.
Neglecting the non-universal metric factors each model
is characterized by its own scaling function (right).
For all models considered the scaling plots contain at least four
different curves corresponding to four different
field values (see e.g.~\reffigname{fig:sDP_2d_op}
and \reffigname{fig:PCP_2d_op}).
The circles mark the condition ${\tilde R}(0,1)=1$.
\label{fig:uni_dp_eqos_HS_2d}} 
\end{figure}

Choosing $a_{\ssstyle h} h \, \lambda^{\sigma}=1$ in
\refeqsname{eq:scal_ansatz_EqoS_HS}{eq:scal_ansatz_Susc_HS} 
we obtain the scaling forms
\begin{eqnarray}
\label{eq:scal_ansatz_EqoS_HS_collapse}
%a_{\ssstyle a} \,
\rhoa(\deltap, h) 
\; & \sim & \; 
(a_{\ssstyle h}\, h)^{\beta/\sigma}\, \, {\tilde R}
(a_{\ssstyle p}  
\deltap \; (a_{\ssstyle h}\, h)^{-1/\sigma}, 1) \, ,\\[2mm]
\label{eq:scal_ansatz_Fluc_HS_collapse}
a_{\ssstyle \Delta} \,
\Delta \rhoa(\deltap, h) 
\; & \sim & \; 
(a_{\ssstyle h}\, h)^{-\gammap/\sigma}\, \, {\tilde D}
(a_{\ssstyle p}  
\deltap \; (a_{\ssstyle h}\, h)^{-1/\sigma}, 1) \, ,\\[2mm]
\label{eq:scal_ansatz_Susc_HS_collapse}
a_{\ssstyle \chi} \,
\chi(\deltap, h) 
\; & \sim & \; 
(a_{\ssstyle h}\, h)^{-\gamma/\sigma}\, \, {\tilde {\mathrm X}}
(a_{\ssstyle p}  
\deltap \; (a_{\ssstyle h}\, h)^{-1/\sigma}, 1) \, .
\end{eqnarray}
Thus plotting the rescaled quantities
\begin{equation}
\rhoa(\deltap, h) \; (a_{\ssstyle h}\, h)^{-\beta/\sigma} \, ,
\quad\quad
a_{\ssstyle \Delta} \, \Delta \rhoa(\deltap, h) 
\; (a_{\ssstyle h}\, h)^{\gammap/\sigma} \, ,
\quad\quad
a_{\ssstyle \chi} \, \chi(\deltap, h) \;
(a_{\ssstyle h}\, h)^{-\gamma/\sigma} \, 
\label{eq:rescal_quantities_steady_state}
\end{equation}
as a function of the rescaled control parameter
$a_{\ssstyle p} \deltap \, (a_{\ssstyle h} h)^{-1/\sigma}$
the corresponding data of all systems belonging to the
same universality class have to collapse onto the universal 
curves ${\tilde R}(x,1)$, ${\tilde D}(x,1)$, 
and ${\tilde {\mathrm X}}(x,1)$.
This is shown for ${\tilde R}(x,1)$
in \reffigname{fig:uni_dp_eqos_HS_2d}
where the rescaled order parameter is plotted
for various two-dimensional models.
As can be seen the data of all considered models
collapse onto the same scaling function, 
clearly supporting the assumption that all models
belong to the universality class of directed percolation.
Furthermore, the data-collapse confirms the accuracy of the
numerically estimated values~\cite{VOIGT_1} of the critical exponents.

\begin{figure}[t] 
\centering
%\leavevmode 
\includegraphics[clip,width=8cm,angle=0]{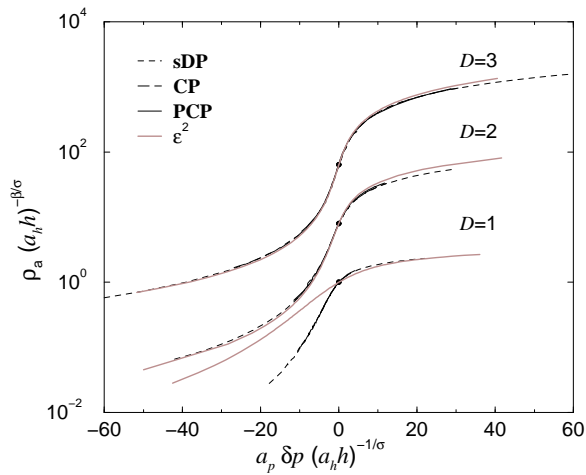}
\caption[Universal scaling function ${\tilde R}(x,1)$, DP class, $D=1,2,3$]
{The universal scaling function ${\tilde R}(x,1)$ of the
directed percolation universality class in various
dimensions.
The two- and three-dimensional data are vertically shifted
by a factor $8$ and $64$ in order to avoid overlaps.
The circles mark the condition ${\tilde R}(0,1)=1$.
The gray lines correspond to an $\epsilon$-expansion
of a renormalization group approach~\protect\cite{JANSSEN_2}.
\label{fig:uni_dp_eqos_HS_123d}} 
\end{figure}

The universal scaling function ${\tilde R}(x,1)$ is
displayed in \reffigname{fig:uni_dp_eqos_HS_123d} for $D=1,2,3$.
For each dimension, the data of the site directed percolation
process, of the contact process as well as of the pair contact
process collapse onto the unique scaling function.
As expected the scaling functions vary with 
the spatial dimensionality below~$\Dc$.

Similar to equilibrium
the powerful and versatile $\epsilon$-expansion\index{$\epsilon$-expansion} 
provides estimates for almost all quantities of
interest (see page\,\pageref{page:dp_eps_results}), 
in particular for the critical 
exponents~\cite{JANSSEN_1,BRONZAN_1,BRONZAN_2,BRONZAN_3} 
and for the equation of state~\cite{JANSSEN_2,JANSSEN_3}.
Using a parametric representation~\cite{SCHOFIELD_1,JOSEPHSON_1}
of the directed percolation phase transition 
Janssen {\it et al.}~showed 
that the equation of state is given within the 
$\epsilon^2$-approximation by the remarkable 
simple scaling function~\cite{JANSSEN_2} 
\begin{equation}
H(\theta) \; = \; \theta \, (2-\theta) + {\mathcal O}(\epsilon^3) \, .
\label{eq:para_repres_eqos}
\end{equation}
Here the scaling behavior of the 
quantities $\rhoa$, $\deltap$, and $h$ is
transformed to the variables $R$ and $\theta$
through the relations
\begin{equation}
b \, \deltap \; = \; R \, (1-\theta),
\quad\quad\quad
\rhoa \; = \; R^\beta \, 
\frac{\theta}{2} \, . 
\label{eq:para_transform}
\end{equation}
The equation of state is given by 
\begin{equation}
 a\, h\; = \; 
 \left ( \frac{R^\beta}{2} \right )^{\delta} 
 \; H(\theta) 
\label{eq:para_transform_equa_state} 
\end{equation}
with the metric factors $a$ and $b$.
The whole phase diagram is described by the parameter
range $R\ge 0$ and $\theta \in [ 0 , 2 ]$.
%%%%%%%%%%%%%%%%%%%%%%%%%%%%%%%%%%%%%%%%%%%%%%%%%%%%%%%%%%%%%%%%%%%%%%%%%%%%%%%%%%
\begin{figure}[t] 
\centering
%\leavevmode 
\includegraphics[clip,width=13cm,angle=0]{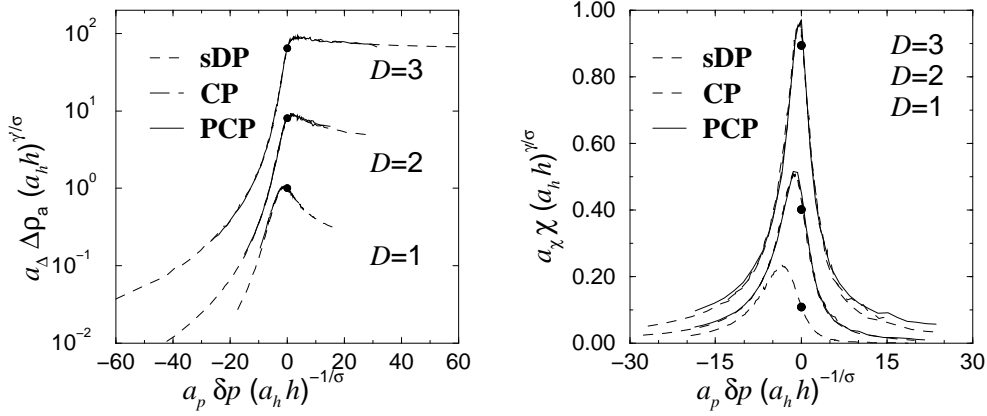}
\caption[Universal scaling functions ${\tilde D}(x,1)$ 
and ${\tilde {\mathrm X}}(x,1)$, DP class, $D=1,2,3$]
{The universal scaling functions of the fluctuations
${\tilde D}(x,1)$ (left) and of the susceptibility 
${\tilde {\mathrm X}}(x,1)$ (right) in various
dimensions.
The two- and three-dimensional fluctuation 
data are vertically shifted
by a factor $8$ and $64$ in order to avoid overlaps.
The circles mark the condition ${\tilde D}(0,1)=1$.
In case of the susceptibility the 
two- and three-dimensional data
are vertically shifted
by a factor $3/2$ and $9/4$ in order to avoid overlaps.
The circles mark the condition ${\tilde \mathrm{X}}(0,1)=\beta/\sigma$
and reflect the accuracy of the performed analysis.
\label{fig:uni_dp_fluc_susc_123d}} 
\end{figure} 
%%%%%%%%%%%%%%%%%%%%%%%%%%%%%%%%%%%%%%%%%%%%%%%%%%%%%%%%%%%%%%%%%%%%%%%%%%%%%%%%%%%%
Compared to the Widom-Griffiths scaling 
form [\refeqname{eq:dp_eqos_WG_epsilon2}]
the parametric representation is not restricted to 
sufficiently small scaling arguments in the absorbing phase.
The scaling function ${\tilde R}(x,1)$ can be 
calculated from the equation above, and the results 
are compared to the numerical data
in \reffigname{fig:uni_dp_eqos_HS_123d}.
As expected, the differences increase with 
increasing perturbation parameter~$\epsilon$, i.e., with decreasing 
dimension and are especially strong in $D=1$~\cite{LUEB_27}.
Thus higher orders than ${\mathcal O}(\epsilon^2)$ are
necessary to describe the scaling behavior of directed percolation
in low dimensions.
A more detailed discussion of the accuracy of the renormalization
group data will be presented in~\ref{subsec:dp_ampl_comb}.

The universal scaling functions of the order parameter
fluctuations and the order parameter susceptibility 
are shown in \reffigname{fig:uni_dp_fluc_susc_123d}.
The susceptibility is obtained by performing 
the numerical derivative
of the order parameter with respect to the 
conjugated field.
The perfect data-collapses confirm the scaling
forms \refeqsname{eq:scal_ansatz_Fluc_HS}{eq:scal_ansatz_Susc_HS}.
All scaling functions exhibit for $D=1,2,3$ a clear maximum 
signalling the divergence of $\Delta\rhoa$ and
$\chi$ at the critical point.
The susceptibility data fulfills \refeqname{eq:susc_Chi_01},
reflecting the accuracy of the performed analysis.
In summary all considered models are characterized
by the same critical exponents and the same steady state 
scaling functions ${\tilde R}$, ${\tilde D}$, and
${\tilde {\mathrm X}}$.
Thus the models are characterized by the same
scaling behavior for $D=1,2,3$.

%%%%%%%%%%%%%%%%%%%%%%%%%%%%%%%%%%%%%%%%%%%%%%%%%%%%%%%%%%%%%%%%%%%%%%%%%%%%%%%%%%%%%%
%%%%%%%%%%%%%%%%%%%%%%%%%%%%%%%%%%%%%%%%%%%%%%%%%%%%%%%%%%%%%%%%%%%%%%%%%%%%%%%%%%%%%%
%%%%%%%%%%%%%%%%%%%%%%%%%%%%%%%%%%%%%%%%%%%%%%%%%%%%%%%%%%%%%%%%%%%%%%%%%%%%%%%%%%%%%%

\subsection{Above the upper critical dimension}
\label{subsec:dp_above_dc}

According to the renormalization group scenario
the stable fixed point of the recursion relations for $D>\Dc$
is usually the trivial fixed point with classical,
i.e., mean field universal quantities.
Thus, in contrast to the situation below $\Dc$
the critical exponents as well as the universal
scaling functions are independent of the particular
value of the dimension for $D>\Dc$.
In most cases it is possible to derive these 
mean field exponents and even the scaling functions 
exactly since correlations and fluctuations can be 
neglected above~$\Dc$.
The mean field behavior of the 
contact process and the 
Domany-Kinzel automaton\index{Domany-Kinzel automaton}
was considered in section~\ref{sec:apt_defs}
and in section~\ref{subsec:domany_kinzel}, respectively.
A corresponding analysis of the pair contact 
process can be found in~\cite{LUEB_27}.
The obtained universal scaling functions
are given by (see e.g.~\cite{MORI_1})
\begin{eqnarray}
\label{eq:uni_scal_mf_EqoS_R}
{\tilde R}_{\ssstyle \mathrm {MF}}  
(x, y) & = &
\frac{x}{2} \, + \, \sqrt{y \, + \,
\left (\frac{x}{2} \right )^2 \;} \, ,   \\[2mm]
\label{eq:uni_scal_mf_Fluc_D}
{\tilde D}_{\ssstyle \mathrm {MF}}  
(x , y) & = &
\frac{{\tilde R}_{\ssstyle \mathrm {MF}}(x,y)}
{\,{\sqrt{y \, + \,\left ( {x}/{2} \right )^2 \;}}\,} \, , \\[2mm]
\label{eq:uni_scal_mf_Susc_X}
{\tilde \mathrm {X}}_{\ssstyle \mathrm {MF}}  
(x , y) & = & 
\frac{1}{\,2\, \,{\sqrt{y \, + \,\left ( {x}/{2} \right )^2 \;}}\,} \, \\[2mm]
\label{eq:uni_scal_mf_xi_senk_R}
{\tilde \Xi}_{\senk,\ssstyle \mathrm {MF}}  
(x, y) & = &  
\frac{1}{\,{\sqrt[4]{y \, + \,\left ( {x}/{2} \right )^2 \;}}\,} \, \\[2mm]
\label{eq:uni_scal_mf_xi_para_R}
{\tilde \Xi}_{\para,\ssstyle \mathrm {MF}}  
(x, y) & = &  
\frac{1}{\,{\sqrt{y \, + \,\left ( {x}/{2} \right )^2 \;}}\,} \, 
\end{eqnarray}
yieldling the mean field exponents 
$\betaMF=1$, 
$\sigmaMF=2$, 
$\gammaMF=1 $, 
$\gammapMF =0$,
$\nu_{\senk,{\ssstyle \mathrm{MF}}}=1/2$, and
$\nu_{\para,{\ssstyle \mathrm{MF}}}=1$.
These functions obey the scaling forms 
\refeqsname{eq:scal_ansatz_EqoS_HS_collapse}{eq:scal_ansatz_Susc_HS_collapse}.

\begin{figure}[t] 
\centering
%\leavevmode 
\includegraphics[clip,width=13cm,angle=0]{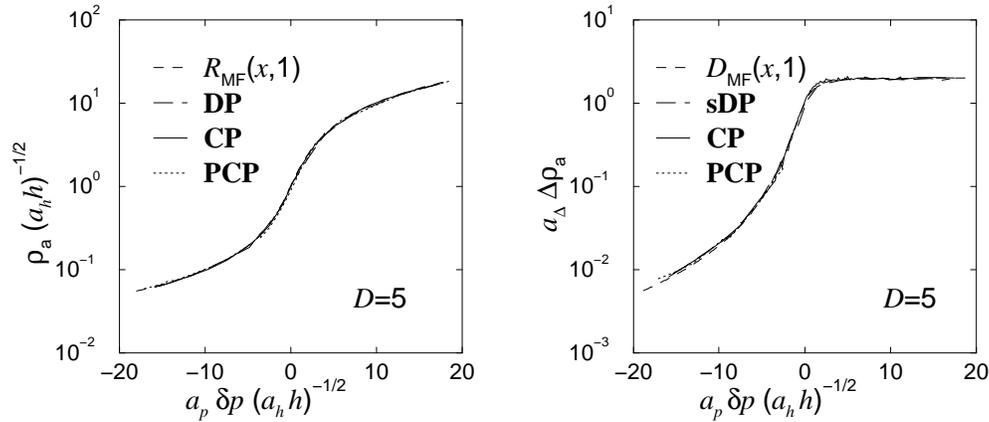}
\caption[Universal scaling functions ${\tilde R}(x,1)$ 
and ${\tilde D}(x,1)$, DP class, $D=5$]
{The universal scaling function of the order parameter
${\tilde R}(x,1)$ (left) and the fluctuations
${\tilde D}(x,1)$ (right) for $D=5$.
The five-dimensional data agree with the corresponding
mean field functions \refeqname{eq:uni_scal_mf_EqoS_Rx1}
and \refeqname{eq:uni_scal_mf_Fluc_Dx1}, respectively.
\label{fig:uni_dp_eqos_mf}} 
\end{figure}

The scaling behavior of the fluctuations deserves
comment.
The exponent $\gammapMF=0 $ corresponds to a jump of the 
fluctuations and the scaling 
form \refeqname{eq:scal_ansatz_Fluc_HS} reduces to
\begin{equation}
a_{\ssstyle \Delta} \; 
\Delta \rhoa(\deltap, h) 
\; \sim \; {\tilde D}_{\ssstyle \mathrm {MF}} 
(a_{\ssstyle p} \deltap \; \lambda, 
a_{\ssstyle h} h \, \lambda^{\sigma}) \, .
\label{eq:uni_fluc_scal_fluc_mf}
\end{equation}
Using again ${\tilde D}_{\ssstyle \mathrm {MF}} (0,1)=1$,
the non-universal metric factor 
$a_{\ssstyle \Delta}$ is determined
by 
\begin{equation}
a_{\ssstyle \Delta} \; = \;
\frac{1}{\, \Delta \rhoa (\deltap=0, h)\,} \, .
\label{eq:mf_fluc_d}
\end{equation}
The reduced scaling form \refeqname{eq:uni_fluc_scal_fluc_mf} 
implies
some interesting properties of the universal 
function ${\tilde D}_{\ssstyle \mathrm {MF}}$.
The value ${\tilde D}_{\ssstyle \mathrm {MF}}(1,0)$ is 
now given by 
\begin{equation}
{\tilde D}_{\ssstyle \mathrm {MF}} (1,0) \; = \;
\frac{\, \Delta \rhoa(\delta p, h=0) \, }
{\, \Delta \rhoa(\delta p=0, h) \, } \, ,
\label{eq:mf_fluc_D}
\end{equation}
and the mean field scaling function fulfills 
the symmetries 
\begin{eqnarray}
\label{eq:fluc_mf_symm_1}
{\tilde D}_{\ssstyle \mathrm {MF}} (1,x)
& = &{\tilde D}_{\ssstyle \mathrm {MF}} (x^{-1/\sigma},1) \, , \\[2mm]
\label{eq:fluc_mf_symm_2}
{\tilde D}_{\ssstyle \mathrm {MF}} (x,1) & = &
{\tilde D}_{\ssstyle \mathrm {MF}} (1,x^{-\sigma}) \, ,
\end{eqnarray}
for $x>0$.
In particular we obtain for $x \to 0$ 
${\tilde D}_{\ssstyle \mathrm {MF}} (1,0)
={\tilde D}_{\ssstyle \mathrm {MF}} (\infty,1)$
and ${\tilde D}_{\ssstyle \mathrm {MF}}(0,1)
={\tilde D}_{\ssstyle \mathrm {MF}}(1,\infty)$,
respectively.
This symmetry is an artefact of the zero fluctuation
exponent and does not occur below the upper critical 
dimension.
Therefore, it can be used to check whether a given 
$D$-dimensional system exhibits mean field behavior.
% gut falls man Funktion selbst nicht kennt
% kenne wir hier aber

Numerical data of the five-dimensional models are
presented in \reffigname{fig:uni_dp_eqos_mf}.
A good data-collapse of the rescaled data with the
universal mean field scaling functions
\begin{eqnarray}
\label{eq:uni_scal_mf_EqoS_Rx1}
{\tilde R}_{\ssstyle {\mathrm {MF}}} 
(x, 1) & = &
\frac{x}{2} 
\, + \, \sqrt{1 \, + \, \left ( \frac{x}{2} \right )^2 \;} \, \\[2mm]
\label{eq:uni_scal_mf_Fluc_Dx1}
{\tilde D}_{\ssstyle {\mathrm {MF}}} 
(x, 1) & = &  1 \, + \, 
\frac{x}
{\,2 \; \sqrt{1 \, + \, \left ( {x}/{2} \right )^2 \;}\, } \, 
\end{eqnarray}
is obtained.
The observed mean field behavior for $D=5$ agrees with
the renormalization group\index{upper critical dimension} 
result~$\Dc =4$~\cite{OBUKHOV_2,CARDY_1}.

\begin{figure}[t] 
\centering
%\leavevmode 
\includegraphics[clip,width=13cm,angle=0]{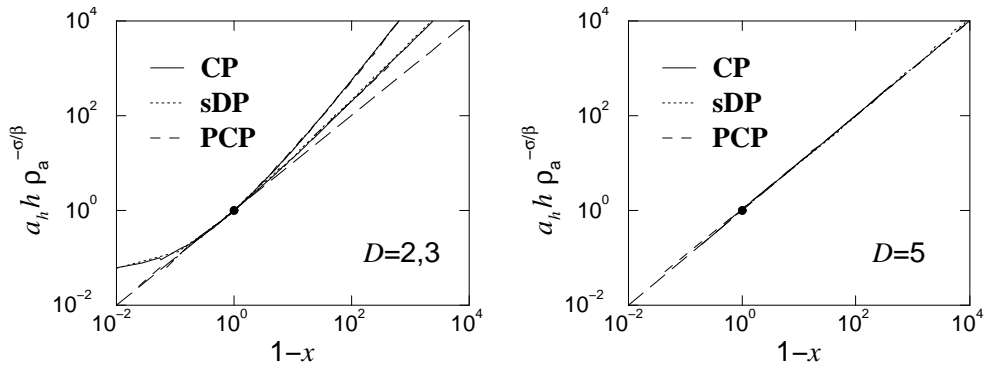}
\caption[Universal scaling functions ${\tilde H}(x,1)$, DP class, $D=1,2,3,5$]
{The universal Widom-Griffiths scaling function 
${\tilde H}(x,1)$ for various dimensions as a function
of $x=a_{\ssstyle p} \rhoa^{-1/\beta}$.
The numerically obtained curves ${\tilde H}(x,1)$
tend to the mean field result (long dashed lines) 
${\tilde H}_{\ssstyle \mathrm {MF}}(x,1)=1-x$
with increasing dimension. 
\label{fig:uni_dp_widom_scal_all_D}} 
\end{figure}

As already pointed out the scaling functions depend on the
spatial dimension below~$\Dc$. 
In particular the Widom-Griffiths scaling form 
[\refeqname{eq:scal_ansatz_EqoS_WG}] is 
well suited to examine how the universal 
functions approach the mean field solution with 
increasing dimension.\index{Widom-Griffiths scaling form}
The mean field Widom-Griffiths scaling form is given
by 
\begin{equation}
{\tilde H}_{\ssstyle \mathrm {MF}}  
(x, y) \; = \; y^2 \, - \, x \, y \, .
\label{eq:uni_scal_mf_EqoS_H}
\end{equation}
The rescaled field 
$a_{\ssstyle h} h \, \rhoa^{-\sigma/\beta}$ is plotted in 
\reffigname{fig:uni_dp_widom_scal_all_D} 
as a function of the rescaled control 
parameter $a_{\ssstyle p} \deltap \, \rhoa^{-1/\beta}$.
As can be seen the scaling functions ${\tilde H}(x,1)$ 
tend to the mean field curve ${\tilde H}_{\ssstyle \mathrm {MF}}(x,1)=1-x$
with increasing dimension.
The observed agreement of the data of the 
five-dimensional models with the mean field
scaling function ${\tilde R}_{\ssstyle \mathrm {MF}}$,
${\tilde H}_{\ssstyle \mathrm {MF}}$,
and ${\tilde D}_{\ssstyle \mathrm {MF}}$
leads to the result that the 
upper critical dimension is less than five.
This is a non-trivial result.
The value of $\Dc$ is often predicted by 
field theoretical treatments, and a reliable 
determination of the upper critical dimension 
is therefore of particular interest.
For example, two contrary field theories conjecture
for the pair contact process $\Dc=4$~\cite{MUNOZ_1}
and $\Dc=6$~\cite{WIJLAND_1,WIJLAND_3}, respectively.
The latter result is in clear contradiction to the
observed mean field behavior in $D=5$.
In this way, an upper bound on~$\Dc$
can be obtained just by checking whether the 
numerical or experimental data are consistent with the
usually known universal mean scaling functions.

%%%%%%%%%%%%%%%%%%%%%%%%%%%%%%%%%%%%%%%%%%%%%%%%%%%%%%%%%%%%%%%%%%%%%%%%%%%%%%%%%%%%%%
%%%%%%%%%%%%%%%%%%%%%%%%%%%%%%%%%%%%%%%%%%%%%%%%%%%%%%%%%%%%%%%%%%%%%%%%%%%%%%%%%%%%%%
%%%%%%%%%%%%%%%%%%%%%%%%%%%%%%%%%%%%%%%%%%%%%%%%%%%%%%%%%%%%%%%%%%%%%%%%%%%%%%%%%%%%%%

\subsection{At the upper critical dimension}
\label{subsec:dp_at_dc}

The scaling behavior around the upper critical
dimension $\Dc$ is
well understood within the renormalization 
group theory 
(see for instance~\cite{WEGNER_2,BREZIN_3,PFEUTY_1}).
For $D>\Dc$ the stable fixed point of the
corresponding renormalization equations is usually 
a trivial fixed point with mean field exponents.
This trivial fixed point is unstable for $D<\Dc$,
and a different stable fixed point exists
with nonclassical exponents.
%These exponents can be estimated by an
%$\epsilon$-expansion, for instance.
For $D=\Dc$ both fixed points are identical
and marginally stable.
In this case, the scaling behavior is 
characterized by mean field power-laws
modified by logarithmic corrections.
For instance the order parameter is expected to obey in 
leading order
\begin{eqnarray}
\label{eq:scal_OPzf_dc}
\rhoa(\deltap, h=0) & \asympprop &
\deltap \; | \ln{ \deltap}|^{\mathrm B}  \, , \\[2mm]
\label{eq:scal_OPcp_dc}
\rhoa (\deltap=0, h) & \asympprop &
\sqrt{ h} \; | \ln{h}|^{\Sigma} \, .
\end{eqnarray}
Greek capitals will be used in the following to denote the logarithmic 
correction exponents.
These logarithmic correction exponents 
are characteristic features of the whole
universality class similar to the usual critical exponents.
Numerous theoretical, numerical, as well as experimental
investigations of critical systems at the upper critical
dimension have been performed (see for 
example~\cite{LARKIN_1,AHARONY_1,GRIFFIN_1,BRINKMANN_1,AKTEKIN_1,GRASSBERGER_6,GRASSBERGER_7,OWCZAREK_1,LUEB_5,LUEB_10,LUEB_17,GRASSBERGER_5,FEDORENKO_1,LUEB_26,LUEB_28,JANSSEN_3,JANSSEN_5,STENULL_1,GRUENEBERG_1,KENNA_1}).
Logarithmic corrections make the data analysis
notoriously difficult.
Hence, most investigations focus
on the determination of the correction exponents only, 
lacking the determination of the scaling functions 
at~$\Dc$.
Recent numerical~\cite{LUEB_26,GRUENEBERG_1} 
as well as analytical results~\cite{JANSSEN_3}
reveal that the concept of universal scaling functions
can also be applied to the upper critical dimension.

A method of analysis to determine both the universal scaling 
functions and the correction exponents at the
upper critical dimension
was developed in~\cite{LUEB_26} 
and successfully applied to site directed percolation~\cite{LUEB_28}. 
In this work the following phenomenological scaling ansatz is used 
for the Hankel-Stanley form of the equation of state
(all terms in leading order)
\begin{equation}
a_{\ssstyle \mathrm a}  \, \rhoa(\deltap, h) 
\; \sim \; 
\lambda^{- \betaMF}\, | \ln{\lambda}|^{\Lambda} 
\; \; {\tilde R}
(a_{\ssstyle p}  
\deltap \; \lambda \, | \ln{\lambda}|^{\Pi} , 
a_{\ssstyle h} h \;
\lambda^{\sigmaMF}\, | \ln{\lambda}|^{\mathrm H}) \, .
\label{eq:uni_scal_EqoS_R_dc}
\end{equation}
%with directed percolation exponents $\betaMF=1$ 
%and $\sigmaMF=2$.
According to this scaling form, the order parameter at zero field 
($h=0$) and at the critical density ($\deltap=0$) is given by
\begin{eqnarray}
\label{eq:uni_scal_OPzf_dc}
a_{\ssstyle \mathrm a}  \, \rhoa(\deltap, h=0) 
& \sim & 
a_{\ssstyle p}  \,
\deltap \; \; | \ln{a_{\ssstyle p}  \deltap}|^{\mathrm B} 
\; \; {\tilde R}(1,0) , \\[2mm]
\label{eq:uni_scal_OPcp_dc}
a_{\ssstyle \mathrm a}  \, \rhoa (\deltap=0, h) 
& \sim & 
\sqrt{a_{\ssstyle h}  h} \;
\, | \ln{\sqrt{a_{\ssstyle h} h}}|^{\Sigma} 
\; \; {\tilde R}(0,1)
\end{eqnarray}
with ${\mathrm B}=\Pi+\Lambda$ and $\Sigma={\mathrm H}/2+\Lambda$.
Similar to $D \neq \Dc$ 
the normalization ${\tilde R}(0,1)={\tilde R}(1,0)=1$
is used. 
Furthermore, the scaling behavior of the equation of state 
is given in leading order by
\begin{equation}
a_{\ssstyle \mathrm a}  \, \rhoa(\deltap, h) 
\; \sim \; 
\sqrt{a_{\ssstyle h}  h}
\; | \ln{\sqrt{a_{\ssstyle h} h}}|^{\Sigma} 
\; \; {\tilde R}(x,1) 
\label{eq:uni_scal_EqoS_dc}
\end{equation}
where $x$ denotes the scaling argument 
\begin{equation}
x \; = \; 
a_{\ssstyle p} \deltap \,
\sqrt{a_{\ssstyle h} h\,}^{\, -1} \,
| \ln{\sqrt{a_{\ssstyle h}  h}\,}|^{\Psi} 
\label{eq:uni_scal_arg_x_dc}
\end{equation}
with $\Psi=\Pi-{\mathrm H}/2={\mathrm B}-\Sigma$.

In case of directed percolation it is possible to
confirm the scaling form \refeqname{eq:uni_scal_EqoS_R_dc}
by a renormalization group analysis~\cite{JANSSEN_3}, 
yielding in addition
%In particular the logarithmic correction exponents are 
%given by 
the values $\Lambda=7/12$, $\Pi=-1/4$, and ${\mathrm H}=-1/2$.
Thus the scaling behavior of the equation of state
is determined by
\begin{equation}
{\mathrm B} \; = \; \Sigma \; = \; 1/3 \, , 
\quad \quad  \quad 
\Psi=0 \, .
\label{eq:log_corr_exp_dp_dc}
\end{equation}
It is worth mentioning that in contrast to the 
renormalization group results
below the upper critical dimension, the logarithmic
correction exponents do not rely on approximation
schemes like $\epsilon$- or $1/n$-expansions.
Within the RG theory they are exact results.

\begin{figure}[t] 
\centering
%\leavevmode 
\includegraphics[clip,width=13cm,angle=0]{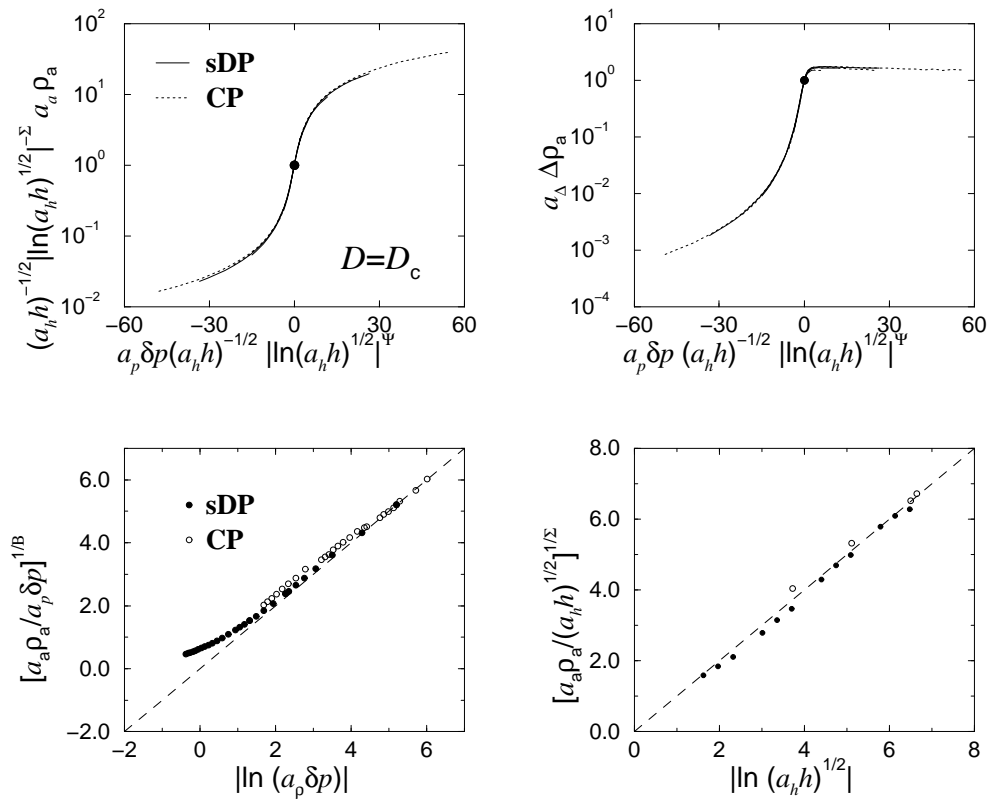}
\caption[Universal scaling functions ${\tilde R}(x,1)$
and ${\tilde D}(x,1)$, DP class, $D=\Dc$]
{The universal scaling functions
of the order parameter (upper left)
and its fluctuations (upper right) 
at the upper critical dimension $\Dc=4$.
The logarithmic correction exponents are given by
${\mathrm B} = \Sigma = 1/3$~\protect\cite{JANSSEN_3} and $\Psi=0$.
For both models considered, the scaling plots contain at least three
different curves corresponding to three different
field values.
The circles mark the condition ${\tilde R}(0,1)=1$ 
and  ${\tilde D}(0,1)=1$, respectively.
The lower figures show the order parameter 
for zero-field (left) and at the 
critical control parameter (right), respectively.
The order parameter is rescaled according to 
\refeqname{eq:uni_scal_OPzf_dc} and 
\refeqname{eq:uni_scal_OPcp_dc}.
Approaching the transition point 
($h\to 0$ and $\deltap \to 0$) the data tend to the function
$f(x)=x$ (dashed lines) as required (see text).
\label{fig:uni_dp_eqos_fluc_4d}} 
\end{figure}

Similarly to the order parameter the following 
scaling form is used for the fluctuations~\cite{LUEB_26}
\begin{equation}
 a_{\ssstyle \Delta}  \; \Delta\rhoa(\deltap, h) 
\; \sim \; 
\lambda^{\gammapMF}  \; | \ln{\lambda}|^{\mathrm K} 
\; \; {\tilde D}
(a_{\ssstyle p}  
\deltap \; \lambda \; | \ln{\lambda}|^{\Pi} , 
a_{\ssstyle h} h \;
\lambda^{-\sigma}\, | \ln{\lambda}|^{\mathrm H}) \, .
\label{eq:uni_scal_fluc_dc}
\end{equation}
Using the mean field value $\gammapMF=0$
and taking into account that numerical simulations
show~\cite{LUEB_28} that the fluctuations remain 
finite at the critical point
(i.e.~${\mathrm K}=0$) the scaling form
\begin{equation}
a_{\ssstyle \Delta}  \; \Delta\rhoa(\deltap, h) 
\; \sim \; 
\; {\tilde D}(x,1)
\label{eq:uni_scal_Fluc}
\end{equation}
is obtained, where the scaling argument~$x$ is given 
by \refeqname{eq:uni_scal_arg_x_dc} with $\Psi = 0$.
The non-universal metric factor $a_{\ssstyle \Delta}$
is determined again by the condition ${\tilde D}(0,1)=1$.

In that way, the scaling behavior of the order parameter and 
its fluctuations at $\Dc$
is characterized by two exponents (${\mathrm B}=1/3$
and $\Sigma=1/3$) and four unknown non-universal metric factors
($a_{\ssstyle \mathrm a},a_{\ssstyle p},a_{\ssstyle h},a_{\ssstyle \Delta}$).
Following~\cite{LUEB_26} these values are determined
by several conditions which are applied simultaneously:
first, both the rescaled order parameter and the rescaled
fluctuation data have to collapse to the 
universal functions ${\tilde R}(x,1)$ and ${\tilde D}(x,1)$.
Second, the order parameter behavior at zero field and at the 
critical density is asymptotically given by the simple
function $f(x)=x$ when plotting 
$[a_{\ssstyle \mathrm a} \rhoa(\deltap,0)
/a_{\ssstyle \rho} \deltap]^{1/{\mathrm B}}$
vs.~$|\ln{a_{\ssstyle p} \deltap}|$
and 
$[a_{\ssstyle \mathrm a} \rhoa(0,h)
/\sqrt{a_{\ssstyle h} h\,}]^{1/{\Sigma}}$
vs.~$|\ln{\sqrt{a_{\ssstyle h} h\,}}|$,
respectively.
Applying this analysis to data of both the contact process 
and site directed percolation,
convincing results are obtained
for ${\mathrm B}=\Sigma=1/3$, $\Psi=0$, and  
for the values of the non-universal metric factors listed 
in~\cite{LUEB_28,LUEB_32}.
The corresponding plots are presented in
\reffigname{fig:uni_dp_eqos_fluc_4d} and show
that the concept of universal scaling functions
can be applied to the upper critical dimension.

Notice that no data-collapse is obtained if logarithmic
corrections are neglected, i.e., for $\mathrm{B}=\Sigma=0$.
Thus, the leading logarithmic corrections have to be 
taken into account in order to study steady state
scaling functions.
It is therefore remarkable that
recently performed off-lattice simulations
of the dynamical scaling behavior at $\Dc=4$ reveals that
logarithmic corrections of higher orders
(e.g.~${\mathcal O}({\ln{\ln{t}}})$) 
are necessary to describe the numerical data~\cite{GRASSBERGER_12}.
Although the steady state results presented here are quite
convincing, we expect that even better results are obtained
by incorporation higher order corrections.
But in contrast to the dynamical scaling behavior, no
analytical results of higher logarithmic corrections for the
steady state scaling behavior are available so far.

%%%%%%%%%%%%%%%%%%%%%%%%%%%%%%%%%%%%%%%%%%%%%%%%%%%%%%%%%%%%%%%%%%%%%%%%%%%%%%%%%%%%%%
%%%%%%%%%%%%%%%%%%%%%%%%%%%%%%%%%%%%%%%%%%%%%%%%%%%%%%%%%%%%%%%%%%%%%%%%%%%%%%%%%%%%%%
%%%%%%%%%%%%%%%%%%%%%%%%%%%%%%%%%%%%%%%%%%%%%%%%%%%%%%%%%%%%%%%%%%%%%%%%%%%%%%%%%%%%%%

\subsection{Universal amplitude combinations}
\label{subsec:dp_ampl_comb}

Additionally to the critical exponents and universal
scaling functions, a universality class 
is also characterized by various
universal amplitude\index{universal amplitude combinations}
combinations~\cite{PRIVMAN_2,HENKEL_2}.
These amplitude combinations emerge from the universality
of the scaling functions since universal amplitude combinations
correspond to particular values of the scaling functions.
It is known from critical equilibrium systems that
universal amplitude combinations vary usually more widely than
the critical exponents.
Thus these amplitude combinations are very useful in order
to identify the universality class of a phase transition
or to provide additional and often convincing evidence for 
the universal behavior.
Furthermore, the measurement of amplitude combinations
in experiments or simulations yields a 
reliable test for theoretical predictions.
In particular, approximation schemes of the renormalization 
group, such as $\epsilon$- or $1/n$-expansions, are widely
used to obtain explicit and systematic estimates of the amplitude
combinations (see section~\ref{sec:rg_theory}).

Usually numerical investigations focus
on amplitude combinations arising from finite-size
scaling analyses (see e.g.~\cite{HENKEL_2}).
A well known example is the value of Binder's fourth
order cumulant at criticality~(see e.g.~\cite{BINDER_1}).
Instead of those finite-size properties we
focus our attention on bulk critical behavior
since bulk amplitude combinations are of great experimental 
interest.
Furthermore, they can be compared to the renormalization
group results of Janssen {\it et al.}~\cite{JANSSEN_2}.

\begin{figure}[t] 
\centering
%\leavevmode 
\includegraphics[clip,width=8cm,angle=0]{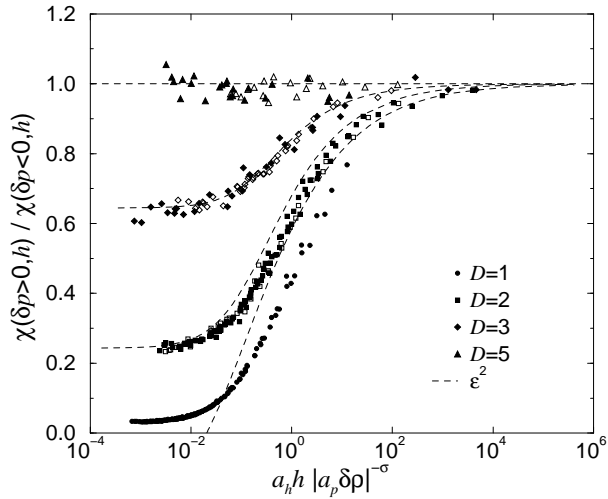}
\caption[Universal scaling function 
${\tilde \mathrm{X}}(1,x)/{\tilde \mathrm{X}}(-1,x)$, DP class]
{The universal scaling function 
${\tilde \mathrm{X}}(1,x)/{\tilde \mathrm{X}}(-1,x)$ 
for various dimensions and models.
Site directed percolation data are marked by closed
symbols.
Open symbols correspond to data
of the pair contact process for $D=2$ and to
data of the contact process for $D=3$ and $D=5$, respectively.
The dashed lines correspond to an $\epsilon$-expansion of a
renormalization group approach~\protect\cite{JANSSEN_2}.
The universal amplitude 
${\tilde \mathrm{X}}(1,0)/{\tilde \mathrm{X}}(-1,0)$ 
is obtained from the extrapolation 
$a_{\ssstyle h} h |a_{\ssstyle p} \, \delta p|^{-\sigma} \to 0$.
\label{fig:uni_dp_mom_susc}} 
\end{figure}

The amplitude combination considered first is related to
the susceptibility behavior below and above the 
transition.
In analogy to equilibrium (see \ref{sec:equil_uni_hyp}) 
the susceptibility diverges as
\begin{eqnarray}
\chi(\deltap >0, h=0) & \sim &
a_{\ssstyle \chi,+} \; \deltap^{-\gamma} \, , \\[2mm]
\chi(\deltap <0, h=0) & \sim &
a_{\ssstyle \chi,-} \; (-\deltap)^{-\gamma} \, ,
\end{eqnarray}
if the critical point is approached from above
and below, respectively.
Using the scaling form \refeqname{eq:scal_ansatz_Susc_HS}
the susceptibility ratio
\begin{equation}
\frac{\,\chi(\delta p >0, h)\,}{\chi(\delta p <0, h)}
\; = \;
\left .
\frac{\,{\tilde \mathrm{X}}
(\phantom{-}a_{\ssstyle p} \deltap \; \lambda, 
a_{\ssstyle h} h \, \lambda^{\sigma})\,}
{\,{\tilde \mathrm{X}}
(-a_{\ssstyle p} \deltap \; \lambda, 
a_{\ssstyle h} h \, \lambda^{\sigma})\,}
\, \right |_{a_{\ssstyle p} |\deltap| \lambda=1}
\; = \; 
\frac{\,{\tilde \mathrm{X}}
(+1,x)\,}
{\,{\tilde \mathrm{X}}(-1,x)\,}
\label{eq:uni_ampl_comb_sus_X_x1}
\end{equation}
is clearly a universal quantity for all values of the
scaling argument 
$x=a_{\ssstyle h} h |a_{\ssstyle p} \deltap|^{-\sigma}$.
In particular it equals the 
ratio ${a_{\ssstyle \chi,+}}/{a_{\ssstyle \chi,-}}$
for vanishing field
\begin{equation}
\frac{\, a_{\scriptscriptstyle \chi,+}\, }{a_{\scriptscriptstyle \chi,-}}
\; = \;
\lim_{h\to 0} \, 
\frac{\,{\tilde \mathrm{X}}(+1,x)\,}{\,{\tilde \mathrm{X}}(-1,x)\,}
\; = \;
\frac{\,{\tilde \mathrm{X}}(+1,0)\,}{\,{\tilde \mathrm{X}}(-1,0)\,} \, .
\label{eq:uni_ampl_comb_sus}
\end{equation}
In case of the mean field 
approximation \refeqname{eq:uni_scal_mf_Susc_X} leads to
\begin{equation}
\frac{{\, \tilde \mathrm{X}}_{\ssstyle {\mathrm {MF}}}(+1,x) \, }
{{\tilde \mathrm{X}}_{\ssstyle {\mathrm {MF}}}(-1,x)}
\; = \; 1 
\label{eq:uni_ampl_comb_sus_mf}
\end{equation}
for all $x$.
The susceptibility ratio ${\tilde \mathrm{X}}(+1,x)/{\tilde \mathrm{X}}(-1,x)$ 
is shown in \reffigname{fig:uni_dp_mom_susc} 
for various models and different dimensions. 
The five dimensional data agree well with the mean
field prediction.
For $D< D_{\scriptscriptstyle \mathrm c}$ the ratio
reflects the crossover from mean field to non-mean field 
behavior.
Far away from the transition point, the critical fluctuations
are suppressed and the behavior of the system is well 
described by the mean field solution.
Approaching criticality, the critical fluctuations
increase and a crossover to the $D$-dimensional behavior
takes place.

A parametric representation of the susceptibility
and therefore of the 
susceptibility ratio ${\tilde \mathrm{X}}(+1,x)/{\tilde \mathrm{X}}(-1,x)$ 
is derived in~\cite{JANSSEN_2}. 
It is instructive to compare these results 
to the numerical data (see Fig.\,\ref{fig:uni_dp_mom_susc})
since the theoretical curve reflects the accuracy
of the RG estimations of all involved quantities,
the exponent, the scaling functions, as well as the
non-universal metric factors~\cite{LUEB_28}.
The perfect agreement between the numerical data and
the RG curve for $D=3$ indicates that
all quantities are approximated well.
In the two-dimensional case a horizontal
shift is observed between the numerical data and the RG-estimates.
Thus the RG-approach yields good estimates for the exponents
and the scaling functions, but the metric factors are
of significantly less quality.
For $D=1$ the $\epsilon^2$-approximation does not 
provide appropriate estimates of the DP scaling
behavior similar to the equation of 
state (see \reffigname{fig:uni_dp_eqos_HS_123d}).

The consideration of the zero-field universal
amplitude 
ratio ${\tilde \mathrm{X}}(+1,0)/{\tilde \mathrm{X}}(-1,0)$
offers a more 
quantitative check of the renormalization group results.
Numerically this ratio is obtained by 
an extrapolation of the susceptibility ratio 
${\tilde \mathrm{X}}(+1,x)/{\tilde \mathrm{X}}(-1,x)$ 
to $x\to 0$.
The estimated values are listed in \reftablename{table:dp_exponents}.
Within the $\epsilon^2$-expansion\index{$\epsilon$-expansion}
this amplitude ratio is given by 
\begin{equation}
\frac{{\tilde \mathrm{X}}(+1,0)}{\,{\tilde \mathrm{X}}(-1,0)\,}
\; = \;
1 \, - \, \frac{\epsilon}{\, 3 \, } \,
\left [
\, 1 \, - \,
\left (
\frac{11}{288} \, - \, \frac{53}{144} \, \ln{\frac{4}{3}} \,
\right ) \, \epsilon 
\, + \, {\mathcal O}(\epsilon^2)
\right ] \, .
\label{eq:susc_ampl_ratio_epsilon}
\end{equation}
Despite of the negative and therefore unphysical 
results for $D=1$ the RG results agree well with the
numerical data (see also Fig.\,\ref{fig:uni_dp_mom_susc}).
For example, the two-dimensional values differ
only by $3\%$.
That has to be compared to the difference of the 
critical exponents.
For example the RG estimate of the
order parameter exponent [\refeqname{eq:dp_exp_epsilon2_beta}]
%%% der Wert der eps Entw. ist beta=0.622
differs for $D=2$ from the best known
numerical value by $6\%$ (see \reftablename{table:dp_exponents}).
% 0.6215 aus epsilon und 0.584 aus Simulationen
The similar accuracy of the critical exponent and of the
amplitude combination is contrary to an observation from
critical equilibrium systems where the exponents are usually 
calculated more accurately by $\epsilon$-expansions than 
universal amplitude combinations (see section~\ref{sec:rg_theory}
for an example, as well as~\cite{PRIVMAN_2}).
A possible explanation is that the $\epsilon^2$-approximation
yields for $\epsilon=1,2$ much better results for directed
percolation than e.g.~for the Ising model.
This is also reflected by the $\epsilon$-approximations of 
the equation of state. 
The various forms of the equation of state are less
complicated for directed percolation.
In particular the parametric representation 
of the equation of state [\refeqname{eq:para_repres_eqos}]
is remarkably simple.

\begin{figure}[t] 
\centering
%\leavevmode 
\includegraphics[clip,width=8cm,angle=0]{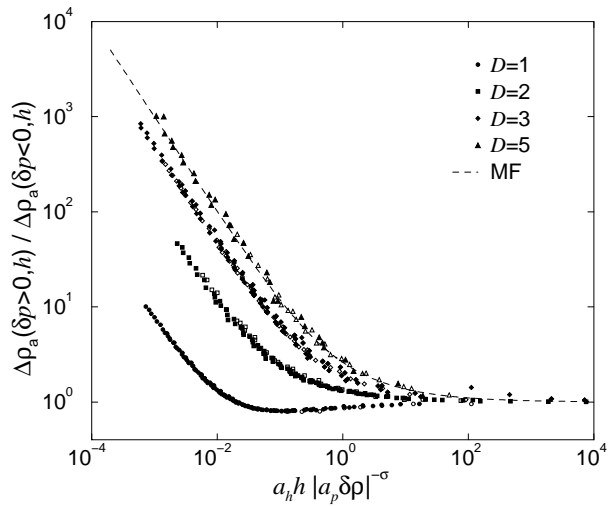}
\caption[Universal scaling function ${\tilde D}(1,x)/{\tilde D}(-1,x)$, DP class]
{The universal scaling function 
${\tilde D}(1,x)/{\tilde D}(-1,x)$ 
for various dimensions.
Site directed percolation data are marked by closed
symbols whereas open symbols correspond to data
of the contact process.
The dashed line corresponds to the mean field
scaling behavior.
\label{fig:uni_dp_mom_fluc}} 
\end{figure}

Analogous to the susceptibility the 
universal amplitude ratio of the fluctuations
is given by
\begin{equation}
\frac{\Delta\rhoa(\deltap >0, h)}
{\Delta\rhoa(\deltap <0, h)}
\; = \; \frac{\, {\tilde D}(+1,x)\, }{\, {\tilde D}(-1,x)\,}
\label{eq:uni_ampl_comb_fluc_D_x1}
\end{equation}
with $x=a_{\ssstyle h} h 
| a_{\ssstyle p} \deltap |^{-\sigma}$.
In case of absorbing phase transitions this 
ratio diverges for vanishing field.
For $\deltap <0$ the order parameter
fluctuations are zero for $h=0$ (absorbing state) 
whereas the fluctuations remain finite above the
transition ($\deltap >0$).
Thus absorbing phase transitions are generally 
characterized by 
\begin{equation}
\frac{\, {\tilde D}(+1,0)\, }{\, {\tilde D}(-1,0) \,} 
\; \to \; \infty .
\label{eq:uni_ampl_comb_fluc}
\end{equation}
In \reffigname{fig:uni_dp_mom_fluc} the fluctuation
ratio is plotted as a function
of the scaling variable~$x$
for various dimensions and models.
The fluctuation ratios diverge for $x \to 0$ in all dimensions.
The one-dimensional system exhibits a particular
behavior characterized by the minimum of the 
corresponding curve.
The origin of this behavior is that for $D=2,3$ 
the universal scaling function ${\tilde D}(x,1)$ exhibits 
a maximum for $x>0$, whereas for $D=1$ 
it is located at $x<0$ 
(see \reffigname{fig:uni_dp_fluc_susc_123d}). 
In case of the five-dimensional data a perfect 
agreement is observed with the mean field behavior
\begin{equation}
\frac{\, {\tilde D}_{\ssstyle {\mathrm {MF}}}(+1,x) \,}
{{\tilde D}_{\ssstyle {\mathrm {MF}}}(-1,x)}
\; = \; \frac{\,\phantom{-\,}1\,+\,\sqrt{1+4x\,}\,}
{\,-\,1\,+\,\sqrt{1+4x\,}\,} 
\; \; \mathop{\longrightarrow}\limits_{x\to 0} \; \; 
\frac{\,1+2x\, }{2 x} \, .
\label{eq:uni_ampl_comb_fluc_mf}
\end{equation}

% rausgenommen, nicht wichtig
%Surprisingly, the two- and three-dimensional data are also
%well approximated by this formula provided that 
%one performs a simple rescaling ($x  \mapsto a_{\scriptscriptstyle D} x$)
%which results in Fig.\,\ref{fig:moment_fluc}
%in a horizontal shift of the data. 
%We suppose that this behavior could be explained
%by a RG-analysis of the fluctuations.
%%% cite{JANSSEN_P} aus kommentiert

\begin{figure}[t] 
\centering
%\leavevmode 
\includegraphics[clip,width=8cm,angle=0]{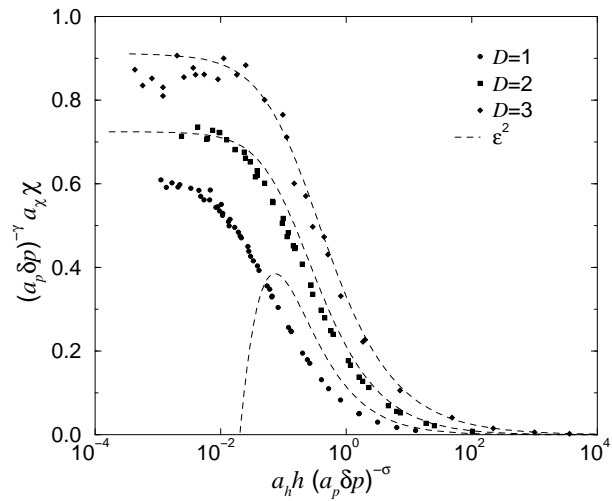}
\caption[Universal scaling functions ${\tilde \mathrm{X}}(x,1)$, DP class]
{The universal scaling function ${\tilde \mathrm {X}}(1,x)$ 
for various dimensions.
The dashed lines correspond to an $\epsilon$-expansion of a
renormalization group approach~\protect\cite{JANSSEN_2}.
The universal amplitude $R_{\scriptscriptstyle \chi}$ 
is obtained from the extrapolation 
$a_{\ssstyle h} h (a_{\ssstyle p} \, \deltap)^{-\sigma} \to 0$.
\label{fig:uni_dp_C_1x}} 
\end{figure}

Finally we consider the universal amplitude
combination $R_{\chi}$ (see \refeqname{eq:def_R_chi_fm})
which can be defined in the notation of absorbing
phase transitions by
\begin{equation}
R_{\chi} \; = \; 
a_{\ssstyle \chi,+}\; \,
a_{\ssstyle h}^{-1}\; \,
a_{\ssstyle p}^{\sigma- \beta} \, .
\label{eq:uni_ampl_comb_R_chi}
\end{equation}
The scaling forms 
\refeqs2name{eq:scal_ansatz_EqoS_HS}{eq:scal_ansatz_Susc_HS}
lead to 
\begin{equation}
R_{\chi} \; = \;
%\Gamma \, D_{\scriptscriptstyle \mathrm c} \, B^{\delta-1} \; = \; 
{\tilde \mathrm{X}}(1,0) 
\label{eq:uni_ampl_comb_R_chi_X_10}   
\end{equation}
which is obviously a universal quantity.
In \reffigname{fig:uni_dp_C_1x} 
the scaling function ${\tilde \mathrm{X} }(1,x)$ is 
plotted as a function of 
$x=a_{\scriptscriptstyle h} h\, (a_{\scriptscriptstyle p} \, \delta p)^{-\sigma}$
for various dimensions below $\Dc$.
The corresponding data saturates for $x\to 0$,
and the obtained estimates are listed in \reftablename{table:dp_exponents}.
Compared to the universal amplitude combination
${\tilde \mathrm{X}}(1,0)/{\tilde \mathrm{X}}(-1,0)$
the estimates of $R_{\chi}$ are of less accuracy
since the data are affected by the uncertainty
of the exponent $\gamma$ and the uncertainties of the
metric factors $a_{\ssstyle p}$, $a_{\ssstyle h}$~\cite{LUEB_28}.
These uncertainties increase the error-bars significantly.
But nevertheless the 
two- and three-dimensional data
agree quite well with the RG-results of~\cite{JANSSEN_2}
whereas unphysical results are obtained again for $D=1$ 
from the $\epsilon^2$-expansion.

In this way the analysis of combinations of the universal 
scaling functions provides a detailed and instructive check 
of the accuracy of corresponding renormalization group
results.
Compared to critical equilibrium systems the $\epsilon^2$-expansion
provides astonishingly accurate estimates of certain\index{$\epsilon$-expansion}
amplitude combinations for $\epsilon=1$ and $\epsilon=2$.
The $\epsilon^2$-approximation fails only in low dimensions 
($\epsilon=3$), i.e., here, higher orders
than ${\mathcal O}(\epsilon^2)$ are necessary to describe
the scaling behavior of directed percolation.

%%%%%%%%%%%%%%%%%%%%%%%%%%%%%%%%%%%%%%%%%%%%%%%%%%%%%%%%%%%%%%%%%%%%%%%%%%%%%%%%%%%%%%
%%%%%%%%%%%%%%%%%%%%%%%%%%%%%%%%%%%%%%%%%%%%%%%%%%%%%%%%%%%%%%%%%%%%%%%%%%%%%%%%%%%%%%
%%%%%%%%%%%%%%%%%%%%%%%%%%%%%%%%%%%%%%%%%%%%%%%%%%%%%%%%%%%%%%%%%%%%%%%%%%%%%%%%%%%%%%

\subsection{Finite-size scaling behavior}
\label{subsec:dp_fss_below_Dc}

So\index{finite-size scaling}\index{universal finite-size scaling}
far simulation data are considered where the 
correlation length~$\xi_{\senk}$ is small compared 
to the system size~$L$.
Thus, the data presented above do not suffer from finite-size
effects, such as rounding and shifting of the anomalies.
As pointed out in section \ref{subsec:apt_steady_state_scal},
an appropriate method to study finite-size effects
of absorbing phase transitions within the steady state 
is to examine the
field-dependence of the quantities of interest
at the critical value $\deltap=0$ for various system sizes
(see \reffigname{fig:cp_5d_ord_fluc_02}).
Similar to critical equilibrium scaling theory 
(see~\cite{CARDY_4} for a review), it is
assumed that the system size enters the scaling forms
[\refeqsname{eq:scal_ansatz_EqoS_HS}{eq:scal_ansatz_Fluc_HS}]
as an additional scaling field 
%with the scaling power~$\nu_{\senk}$
\begin{eqnarray}
\label{eq:scal_ansatz_EqoS_HS_fss}
%a_{\ssstyle a} \,
\rhoa(\deltap, h, L) 
\; & \sim & \; 
\lambda^{-\beta}\, \, {\tilde R}_{\ssstyle \mathrm{pbc},\Box}
(a_{\ssstyle p}  
\deltap \; \lambda, 
a_{\ssstyle h} h \;\lambda^{\sigma}, 
a_{\ssstyle L} L \;\lambda^{-\nu_{\senk}}
) \, ,\\[2mm]
\label{eq:scal_ansatz_Fluc_HS_fss}
a_{\ssstyle \Delta} \,
\Delta \rhoa(\deltap, h, L) 
\; & \sim & \; 
\lambda^{\gamma^{\prime}}\, \, {\tilde D}_{\ssstyle \mathrm{pbc},\Box}
(a_{\ssstyle p} \deltap \; \lambda, 
a_{\ssstyle h} h \, \lambda^{\sigma},
a_{\ssstyle L} L \;\lambda^{-\nu_{\senk}}
)  \, .
\end{eqnarray}
The choice of the corresponding scaling power~$\nu_{\senk}$ is
motivated by the phenomenological finite-size scaling
theory which rests on the assumption
that finite-size effects are controlled within the scaling regime
by the ratio $L/\xi_{\senk}$~\cite{FISHER_6,FISHER_7}.
Approaching the transition point, finite-size effects are 
expected to occur when ${\mathcal{O}}(\xi_{\senk})=L$.
If both lengths, $L$ and $\xi_{\senk}$, are significantly
larger than all other length scales of a given system
(e.g.~the lattice spacing) the above finite-size scaling forms
are universal.
Thus, they may be used to identify a 
system's universality class, additionally to the so far 
considered scaling forms associated to the thermodynamic limit
$L\to \infty$.
For example, it was shown by the use of universal finite-size
scaling functions that the liquid-gas phase transition of a Lennard-Jones
system belongs to the Ising universality 
class~\cite{BRUCE_1,BRUCE_2,NICOLAIDES_1,BRUCE_3}.

The index $\mathrm{pbc},\Box$ in
\refeqs2name{eq:scal_ansatz_EqoS_HS_fss}{eq:scal_ansatz_Fluc_HS_fss}
indicates that the universal
finite-size scaling functions depend on the 
particular choice of the boundary conditions as well as on 
the system shape (see e.g.~\cite{HU_1,KANEDA_1,KANEDA_2,HUCHT_1,ANTAL_1}).
But different lattice structures are contained in the
metric factor~$a_{\ssstyle L}$.
Throughout this work we focus on hypercubic lattices
of size $L^D$ (aspect ratio $1$)
with periodic boundary conditions (pbc).
Of course, the universal scaling 
functions \refeqsname{eq:scal_ansatz_EqoS_HS}{eq:scal_ansatz_Fluc_HS}
are recovered in the thermodynamic limit, e.g.
\begin{equation}
{\tilde R}_{\ssstyle \mathrm {pbc},\Box}(x,y,\infty)
\; = \; {\tilde R}(x,y)\, .
\label{eq:td_limit_uni_fkt}
\end{equation}

Additionally to the order parameter and its fluctuations
we consider the fourth-order cumulant~$Q$ which
is defined\index{cumulant}\index{fourth order cumulant} 
as (see e.g.~\cite{BINDER_2,BINDER_1})
\begin{equation}
Q \; = \;
1 \, - \, \frac{\langle \rhoa^4 \rangle}
{\, 3\,\langle \rhoa^2 \rangle^2\,} \, .
\label{eq:def_binder_cum_op}
\end{equation}
Other combinations of order parameter moments can be
constructed and were investigated for several systems
exhibiting absorbing phase 
transitions~\cite{DICKMAN_4,SILVA_4,DICKMAN_10,HENKEL_2}.
Here, we restrict our attention to~$Q$ since
the cumulant behavior is sufficient for 
our investigation~\cite{LUEB_23}.
For non-vanishing order-parameter 
the cumulant tends to $Q=2/3$ in the thermodynamic limit.
In case of a zero order parameter 
the cumulant vanishes if the 
order parameter is characterized by a Gaussian distribution
symmetrically distributed around zero.
The latter case is observed in equilibrium systems,
e.g.~the Ising model for $T>\Tc$.
But for absorbing phase transitions the order parameter 
is non-negative per definition.
Thus the order parameter is characterized by a non-trivial
distribution and the above scenario does
not apply.

Nevertheless, it is expected that the cumulant obeys the
scaling form 
\begin{equation}
Q(\deltap, h, L) 
\; \sim \; 
{\tilde Q}_{\ssstyle \mathrm{pbc},\Box}
(a_{\ssstyle p} \deltap \; \lambda, 
a_{\ssstyle h} h \;\lambda^{\sigma}, 
a_{\ssstyle L} L \;\lambda^{-\nu_{\senk}}
) \, .
\label{eq:scal_ansatz_Cumu_HS_fss}
\end{equation}
Notice that no metric factor $a_{\ssstyle Q}$
is introduced since the cumulant is already dimensionless.
Choosing $a_{\ssstyle L} L \,\lambda^{-\nu_{\perp}}=1$
we find for zero field 
\begin{eqnarray}
Q(0, 0, L)  & = &
\left . {Q} \vphantom{\tilde Q}
(\deltap, 0, L) \right |_{\ssstyle \deltap=0} \nonumber \\ 
& \sim  &
\left .
{\tilde Q}_{\ssstyle \mathrm{pbc},\Box}
(a_{\ssstyle p} \deltap \; 
(a_{\ssstyle L} L)^{-\nu_{\perp}},  0, 1) \right 
|_{\ssstyle\deltap=0} \nonumber \\  & = &
{\tilde Q}_{\ssstyle \mathrm{pbc},\Box}(0, 0, 1) 
\label{eq:cum_op_inter}
\end{eqnarray}
which is universal despite of the boundary condition and 
the system shape.
The universal value ${\tilde Q}_{\ssstyle \mathrm{pbc},\Box}(0, 0, 1)$
corresponds to an intersection point if one plots
$Q$ as a function of the control parameter~$p$ 
for various system sizes~$L$.
Thus it is possible to determine the 
critical value~$\pc$ 
from the common intersection point.
This cumulant intersection method is very useful and was applied
in numerous works~(see for instance \cite{BINDER_1} and references
therein).
But as pointed out in section~\ref{subsec:apt_steady_state_scal},
it is a characteristic feature of absorbing phase
transitions that the steady state order parameter 
moments~$\langle \rhoa^k\rangle$ vanish as soon as
${\mathcal O}(\xi_{\senk})=L$ even in the active phase.
Thus the powerful cumulant intersection method
can not be applied for absorbing phase transitions.

In order to bypass this problem we consider the 
cumulant as a function of the conjugated field
for $\deltap=0$.
It is convenient to norm the universal
scaling function ${\tilde Q}_{\ssstyle \mathrm{pbc},\Box}$
by the condition
\begin{equation}
{\tilde Q}_{\ssstyle \mathrm{pbc},\Box}(0,1,1) \; = \;0.
\label{eq:metric_factors_a_L}
\end{equation}
Since the metric factor $a_{\ssstyle h}$ 
is already known, the condition above can be used to determine 
the metric factor~$a_{\ssstyle L}$.
Taking into account that the spatial correlation length scales
at criticality as [\refeqname{eq:scal_ansatz_xi_senk_HS}]
\begin{equation}
a_{\senk} \, \xi_{\senk}
\; \sim \; (a_{\ssstyle h}\, h)^{-\nu_{\senk}/\sigma} \, ,
\label{eq:xi_h_at_criticality}
\end{equation}
\refeqname{eq:metric_factors_a_L} implies that
the universal function ${\tilde Q}_{\ssstyle \mathrm{pbc},\Box}$ 
is positive if $a_{\ssstyle L} L > 
a_{\senk} \, \xi_{\senk}$ and negative if
$a_{\ssstyle L} L < a_{\senk} \, \xi_{\senk}$.
Note that in case of equilibrium phase transitions
\refeqname{eq:metric_factors_a_L} is useless since 
the cumulant is usually positive.

\begin{figure}[t] %%%%%%%%%%%%%%%%%%%%%%%%%%%%%%%%%%%%%%%%%%%%%%%%%%%%%%%%%%%%%%
\centering
%\leavevmode 
\includegraphics[clip,width=13cm,angle=0]{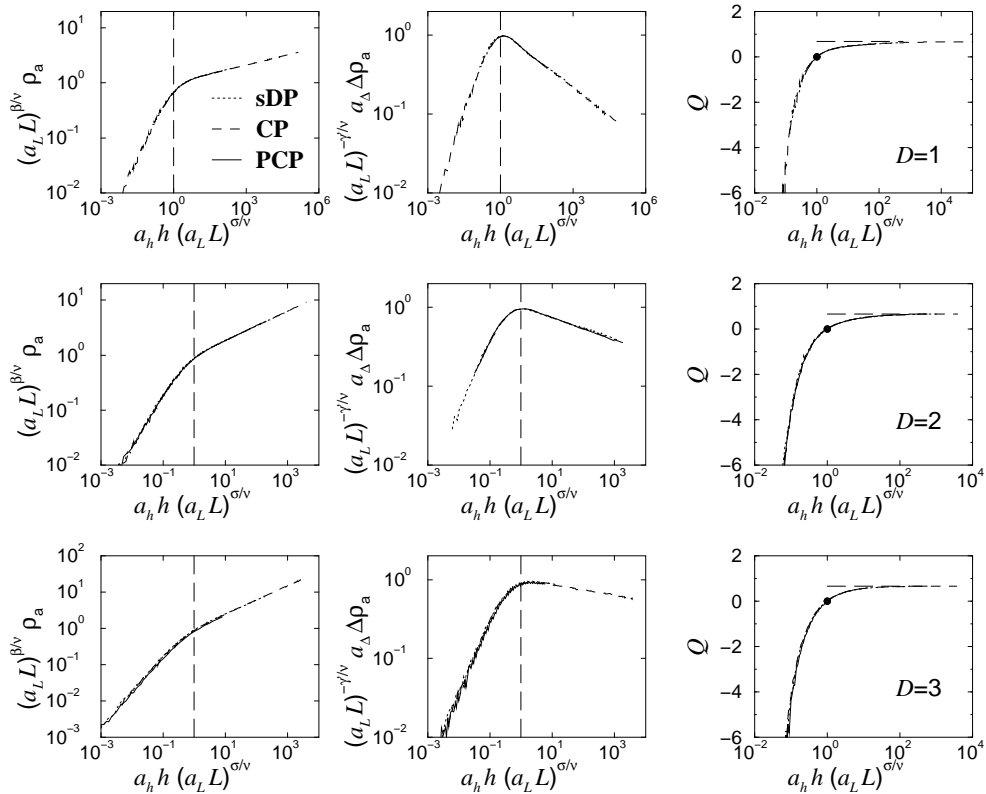}
\caption[Universal finite-size scaling functions, DP class, $D=1,2,3$] 
{The universal finite-size scaling analysis for the
directed percolation universality class in various
dimensions.
The filled circles mark the condition \refeqname{eq:metric_factors_a_L}.
The horizontal lines correspond to the cumulant limit $2/3$.
The vertical lines separate the finite-size regime with 
$a_{\ssstyle L} L < a_{\senk} \, \xi_{\senk}$ (left)
from the regime with $a_{\ssstyle L} L > a_{\senk} \, \xi_{\senk}$ (right).
The data are obtained from simulations of systems sizes
$L=64,128,256,512$ for $D=1$,
$L=64,128,256$ for $D=2$, and 
$L=32,64,128$ for $D=3$.
\label{fig:uni_dp_fss_d123}} 
\end{figure} %%%%%%%%%%%%%%%%%%%%%%%%%%%%%%%%%%%%%%%%%%%%%%%%%%%%%%%%%%%%

As mentioned above we focus on the scaling behavior
at $p=\pc$ and investigate the universal scaling functions
${\tilde R}_{\ssstyle \mathrm{pbc},\Box}(0,x,1)$,
${\tilde D}_{\ssstyle \mathrm{pbc},\Box}(0,x,1)$, 
${\tilde Q}_{\ssstyle \mathrm{pbc},\Box}(0,x,1)$,
by plotting the rescaled quantities
\begin{equation}
\rhoa(0,h,L) \; (a_{\ssstyle L}\, L)^{\beta/\nu_{\senk}} \, ,
\quad\quad
a_{\ssstyle \Delta} \; \Delta \rhoa(0,h,L) 
\; (a_{\ssstyle L}\, L)^{-\gammap/\nu_{\senk}} \,
\label{eq:rescal_quantities_steady_state_fss}
\end{equation}
and $Q(0,h,L)$
as a function of the rescaled field 
$x=a_{\ssstyle h} h (a_{\ssstyle L} L)^{\sigma/\nu_{\senk}}$.
Obviously, only a horizontal rescaling is needed in order
to produce a data-collapse of the cumulant data.
Thus the determination of the critical exponent~$\nu_{\senk}$
via the fourth order cumulant is more accurate
than a corresponding analysis of the order
parameter or of the fluctuations.
Therefore cumulants are widely used in order
to estimate the correlation length exponent.
The finite-size scaling analysis of site directed
percolation, of the contact process, and of the
pair contact process is presented 
in \reffigname{fig:uni_dp_fss_d123}.
Using the best known values for the critical 
exponent~$\nu_{\senk}$ (see \reftablename{table:dp_exponents})
perfect data-collapses are observed.
Note that different lattice structures do not affect the
universal scaling functions.
Whereas bcc lattice types are used for site directed percolation,
simples cubic lattices are used for the contact process and 
pair contact process.
As can be seen the order parameter and its fluctuations 
obey the expected algebraic field dependence for $x\gg 1$, i.e.,
as long as $a_{\senk} \xi_{\senk} \ll a_{\ssstyle L} L$.
The fluctuation curves are characterized by a 
clear maximum signalling its singular behavior
in the thermodynamic limit.
Approaching the transition point
the fourth-order cumulant tends to minus infinity
for all models in all dimensions
\begin{equation}
{\tilde Q}_{\ssstyle {\mathrm {pbc},\Box}}
(0,x,1) \; \longrightarrow \; -\infty
\quad\quad {\mathrm {for}}
\quad\quad
x\; \to \; 0 \,.
\label{eq:uni_value_Q_APT}
\end{equation}
This behavior is caused by the vanishing 
steady state fluctuations, and it is 
assumed that the divergent fourth-order cumulant
is a characteristic feature of all absorbing 
phase transitions~\cite{LUEB_23}. 
A ratio that remains finite at criticality is obtained via
$U=({\langle\rhoa^{2}\rangle\langle\rhoa^{3}\rangle
-\langle\rhoa\rangle\langle\rhoa^{2}\rangle^{2}})/
({\langle\rhoa\rangle\langle\rhoa^{4}\rangle-\langle\rhoa\rangle
\langle\rhoa^{2}\rangle^{2}})$~\cite{LUEB_33}.
The value of $U$ for $x\to 0$
characterizes the universality class.
Numerical investigations yield
$U_{d=1}=0.833\plmi0.011$, $U_{d=2}=0.704\plmi 0.013$, 
and $U_{d=3}=0.61 \plmi 0.02$.

\begin{figure}[t] %%%%%%%%%%%%%%%%%%%%%%%%%%%%%%%%%%%%%%%%%%%%%%%%%%%%%%%%%%%%%%
\centering
%\leavevmode 
\includegraphics[clip,width=13cm,angle=0]{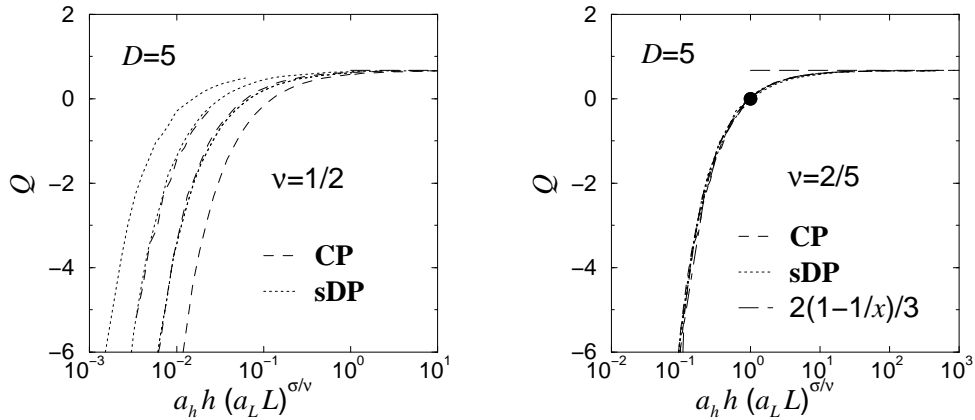}
\caption[Universal finite-size scaling, DP class, $D=5$] 
{The universal finite-size scaling analysis for the
directed percolation universality class above the
upper critical dimension.
Using the standard finite-size scaling form with 
$\betaMF=1$, $\sigmaMF=2$, and $\nuperpMF=1/2$
no data-collapse occurs (left).
Instead the finite-size scaling behavior is described
by a modified scaling form using 
$\betaMF=1$, $\sigmaMF=2$, and $\nuperpMF^{\ast}=2/D$.
The filled circles mark the condition \refeqname{eq:metric_factors_a_L}.
The vertical lines correspond to the cumulant limit $2/3$.
The data are obtained from simulations of system sizes
$L=8,16,32$ for site directed percolation and 
$L=4,8,16$ for the contact process.
As can be seen in the right figure, the numerical data
agree perfectly with the analytical 
result $Q=2/3\, (1-1/x)$~\cite{JANSSEN_P2004,LUEB_33}.
\label{fig:uni_dp_fss_d5}} 
\end{figure}

So far we considered the finite-size scaling behavior
below the upper critical dimension~$\Dc$.
The situation becomes more complicated above~$\Dc$
where the mean field theory applies.
Naively, one would expect that the finite-size scaling behavior 
is described by the mean field exponents $\betaMF$,
$\sigmaMF$, and $\nuperpMF$ and e.g.~the order parameter obeys
\begin{equation}
\label{eq:scal_ansatz_EqoS_HS_fss_MF_falsch}
%a_{\ssstyle a} \,
\rhoa(\deltap, h, L) 
\;  \sim \; 
\lambda^{-\betaMF}\, \, {\tilde R}_{\ssstyle \mathrm{pbc},\Box}
(a_{\ssstyle p}  
\deltap \; \lambda, 
a_{\ssstyle h} h \;\lambda^{\sigmaMF}, 
a_{\ssstyle L} L \;\lambda^{-\nuperpMF}
) \, .
\end{equation}
But as can be seen in \reffigname{fig:uni_dp_fss_d5} for the
order parameter cumulant no data-collapse 
occurs for $\nuperpMF=1/2$
reflecting the failure of phenomenological
finite scaling within the mean field 
regime (see \cite{BRANKOV_2} for a readable review).
This breakdown of common finite-size scaling is well
established in equilibrium although details are 
discussed controversially in the 
literature~\cite{LUIJTEN_4,BINDER_3,LUIJTEN_5,BLOETE_1,LUIJTEN_6,CHEN_1,CHEN_2,CHEN_3,STAUFER_5,BRANKOV_2}.
As pointed out by Br{\'e}zin on the basis of exact
calculations in the limit $n\to \infty$, the usual
finite-size scaling assumption breaks for $D\ggl \Dc$
due to the existence of a dangerous\index{dangerous irrelevant variable} 
irrelevant variable~\cite{BREZIN_5}.
In that case, the above scaling forms are replaced 
by~\cite{BINDER_3}
\begin{equation}
\label{eq:scal_ansatz_EqoS_HS_fss_MF}
%a_{\ssstyle a} \,
\rhoa(\deltap, h, L) 
\;  \sim \; 
\lambda^{-\betaMF}\, \, {\tilde R}_{\ssstyle \mathrm{pbc},\Box}
(a_{\ssstyle p}  
\deltap \; \lambda, 
a_{\ssstyle h} h \;\lambda^{\sigmaMF}, 
a_{\ssstyle L} L \;\lambda^{-\nuperpstar}
) \, .
\end{equation}
%with $\nuperpstar=2/D$.
The modified scaling form reflects that finite-size scaling
above~$\Dc$ is no longer controlled by the ratio 
$L/\xi \asympprop L |\deltap|^{\ssstyle \nuperpMF}$
but by the ratio $L/l_{\ssstyle 0}$.
In equilibrium, the so-called thermodynamic 
length~$l_{\ssstyle 0}$
scales as~\cite{BINDER_3}
\begin{equation}
\label{eq:thermodynamic_length_01}
l_{\ssstyle 0} \, \asympprop \, 
|\deltap|^{-(2 \betaMF \, + \, \gammaMF)/D} \, .
\end{equation}
Below the upper critical dimension, 
the hyperscaling law $2\, \beta + \gamma = 2 - \alpha = \nu_{\senk} D$
holds and~$l_{\ssstyle 0}$ 
coincides with the correlation length~$\xi_{\senk}$.
It is instructive to express the divergence of the 
thermodynamic length in terms of the value of the
upper critical dimension~$\Dc$.
 %%%%%%%%%%%%%%%%%%%%%%%%%%%%%%%%%%%%%%%%%%%%%%%%%%%%%%%%%%%%
According to the Ginzburg criterion (see chapter \ref{chapter:crossover}),
its value is given in equilibrium 
by $\Dc=(2 \betaMF  + \gammaMF)/\nuperpMF$ leading to
\begin{equation}
\left .
\label{eq:thermodynamic_length_02}
l_{\ssstyle 0}^{\vphantom{X}} \right |_{\ssstyle D > \Dc}
 \, \asympprop \, 
|\deltap|^{-\nuperpMF \Dc/D} \, .
\end{equation}
Thus, the scaling power of the finite-size scaling forms
is given by
\begin{equation}
\label{eq:nu_perp_star_Dc}
\nuperpstar \; = \; \nuperpMF \, \frac{\,\Dc}{\,D_{\phantom{\ssstyle c}}} \, .
\end{equation}
For example, the short range interaction Ising model is
characterized by $\nuperpMF=1/2$ and $\Dc=4$
yielding $\nuperpstar=2/D$.
This value is derived by a renormalization group 
analysis~\cite{PRIVMAN_4} and is confirmed by
numerical simulations~\cite{BINDER_3,BLOETE_1}.
Furthermore, \refeqname{eq:nu_perp_star_Dc} 
also describes the modified finite-size scaling
behavior of long-range interacting systems~\cite{LUIJTEN_4,LUIJTEN_5}.
In that case, the interactions decay as $r^{\ssstyle -(D+\sigma)}$ 
($\sigma>2$ corresponds to short-range interacting systems),
yielding the upper critical dimension $\Dc=2 \sigma$ as well as 
$\nuperpMF=1/\sigma$~\cite{JOYCE_2,STELL_1,FISHER_8,AIZENMAN_1}.

Analogous to the equilibrium scenario considered above, 
directed percolation exhibits a dangerous
irrelevant variable within the mean field regime
(the coupling constant $u$ in \refeqname{eq:action_reggeon_field_theory}),
explaining the
breakdown of usual finite-size scaling above $\Dc$.
Following \refeqname{eq:nu_perp_star_Dc},
the modified scaling form is characterized by
the exponent
\begin{equation}
\label{eq:nu_perp_star_DP}
\nuperpstar \; = \; \frac{2}{\,D\,} \, .
\end{equation}
since $\Dc=4$ and $\nuperpMF=1/2$.
As can be seen in \reffigname{fig:uni_dp_fss_d5},
a perfect universal data-collapse is obtained
if the data of the five-dimensional models are 
scaled with $\nuperpstar=2/5$.
%So far, field theoretical analysis of finite-size
%scaling for the directed percolation universality 
%class are restricted to $D< \Dc$~\cite{JANSSEN_9}. 
%It is interesting to see if \refeqname{eq:nu_perp_star_DP} 
%could be confirm by a field theory.
Preliminary results show that corresponding data
of two-dimensional absorbing phase transitions with 
infinite particle hopping collapse onto the same
universal scaling curves if $\nuperpstar=1$ is used.
This indicates that \refeqname{eq:nu_perp_star_DP} holds 
independently of the range of interactions.
Furthermore, the exponent $\nuperpstar$ is $D$-dependent
and therefore non-universal, whereas the finite-size
scaling functions are universal.

A recently performed field theoretical approach of the
mean field scaling behavior of directed percolation
confirms the analysis above.
Similar to equilibrium~\cite{BREZIN_5,BRANKOV_2}, 
a zero-momentum mode treatment
allows to consider finite-size effects of periodic 
systems~\cite{JANSSEN_9}.
Within the mean field regime it is possible to calculate analytically
the critical exponent $\sigma/\nuperpstar=D$ 
[confirming \refeqname{eq:nu_perp_star_DP}]
as well as the scaling functions~\cite{JANSSEN_P2004,LUEB_33}
\begin{eqnarray}
\label{eq:dp_mf_fss_op}
{\tilde R}_{\ssstyle \mathrm{pbc},\Box,\mathrm{mf}}(0,x,1)  
& = &  \sqrt{2} \, \frac{\,\Gamma(\frac{x+1}{2})\,}
{\Gamma(\frac{x}{2})} \, , \\[2mm]
%\label{eq:dp_mf_fss_fluc}
%{\tilde D}_{\ssstyle \mathrm{pbc},\Box,\mathrm{mf}}(0,x,1)  
%& = &  x \, - \, 2 ß, 
%\frac{\,\Gamma(\frac{x+1}{2})^2\,}{\Gamma(\frac{x}{2})^2} \, , \\[2mm]
\label{eq:dp_mf_fss_cum}
{\tilde Q}_{\ssstyle \mathrm{pbc},\Box,\mathrm{mf}}(0,x,1)  
& = & 
%\frac{\,\Gamma(x+2)\,\Gamma(x)}{\Gamma(x+1)^2} \; = \; 
\frac{2}{3} \, \left (1 \, - \, \frac{1}{x} \right ) \, .
\end{eqnarray}
As can be seen in \reffigname{fig:uni_dp_fss_d5}
the numerical data agree perfectly with the 
analytical results.
%{\bf jetzt vielleicht noch offene Randbedingungen}

In conclusion, universal finite-scaling functions can be
used additionally to the scaling behavior of the equation of 
state, the susceptibility, etc.~to identify a system's
universality class.
Analogous to equilibrium, simple finite-size scaling is
valid below~$\Dc$ but has to be modified within the
mean field regime.

\section{Dynamical scaling behavior}
\label{sec:dp_dynamical_scaling}

In this section we discuss the dynamical scaling
behavior close to the transition point.
First we consider how the order parameter decays
starting from a fully occupied lattice, i.e.,
starting from a homogenous particle source.
Second we investigate the activity spreading 
generated from a single active seed.
Interpreting the initial particle density
as another scaling field 
both scaling regimes are connected by
a crossover scenario.

\subsection{Homogeneous particle source}
\label{subsec:dp_dynam_scal_homo_source}

\begin{figure}[t] 
\centering
%\leavevmode 
\includegraphics[clip,width=13cm,angle=0]{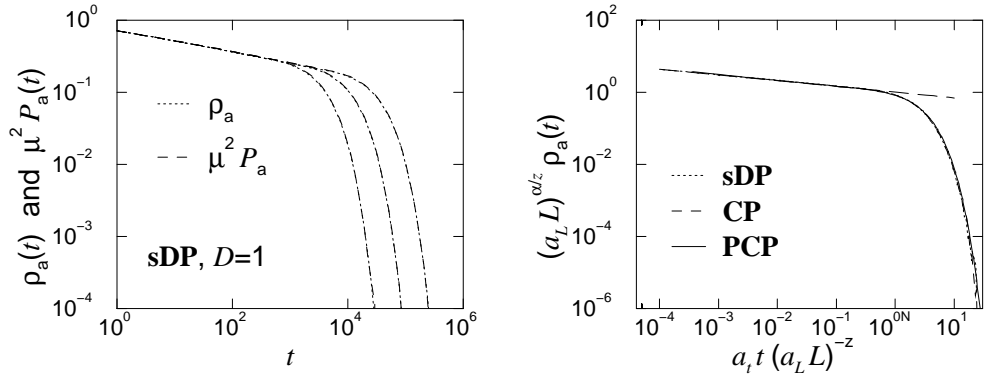}
\caption[Rapidity reversal and universal 
function ${\tilde R (0,0,1,x)}$, DP class, D=1]  
{The order parameter decay at the critical point
for $D=1$.
The left figure shows the unscaled data of 
one-dimensional site directed percolation 
for $L=64,128,256$ (from left to right).
Additionally to the order parameter $\rhoa(t)$ 
the survival probability $\Pa(t)$ is presented
in order to display the rapidity reversal 
symmetry ($\mu=\sqrt{2-\pc\,}\,$).
The rescaled data, i.e., the universal scaling
function ${\tilde R (0,0,1,x)}$ 
[see \refeqname{eq:scal_ansatz_EqoS_HS_dyn_fss_data_collapse}]
is presented in the right figure for site
directed percolation, the contact process
and the pair contact process.
The long-dashed line corresponds to the power-law
behavior of the infinite system [\refeqname{eq:ord_decay_alpha}].
\label{fig:uni_dp_rho_t_1d}} 
\end{figure}

In section \ref{subsec:apt_dynamical_scal} we have 
seen that away from criticality the order parameter 
approaches its steady state value exponentially.
At the critical point the order parameter decays
algebraically
\begin{equation}
\rhoa(t) \; \sim \; (a_{\ssstyle t} \, t)^{-\alpha} \, 
\label{eq:ord_decay_alpha}   
\end{equation}
where $a_{\ssstyle t}$ denotes the associated
non-universal metric factor.
The dynamical behavior of the order parameter can be
incorporated in the scaling forms by an additional 
scaling field
\begin{equation}
%a_{\ssstyle a} \,
\rhoa(\deltap, h, L, t) 
\; \sim  \; 
\lambda^{-\beta}\, \, {\tilde R}_{\ssstyle \mathrm{pbc},\Box,\mathrm{full}}
(a_{\ssstyle p}  
\deltap \; \lambda, 
a_{\ssstyle h} h \;\lambda^{\sigma}, 
a_{\ssstyle L} L \;\lambda^{-\nu_{\senk}},
a_{\ssstyle t} t \;\lambda^{-\nu_{\para}}     ) \, .
\label{eq:scal_ansatz_EqoS_HS_dyn}
\end{equation}
The scaling power of~$t$ has to equal the
scaling power of the correlation time $\xi_{\para}$,
and the index full indicates the initial conditions.
Setting $a_{\ssstyle t} t \;\lambda^{-\nu_{\para}}=1$ leads
in the thermodynamic limit at criticality to
\begin{equation}
\rhoa(\deltap=0, h=0, L\to\infty, t) \; \sim \; 
(a_{\ssstyle t} \, t)^{-\beta/\nu_{\para}} \; \, 
{\tilde R}_{\ssstyle \mathrm{pbc},\Box,\mathrm{full}}(0,0,\infty,1) \, .
\end{equation}
Thus \refeqname{eq:ord_decay_alpha} is recovered for
${\tilde R}_{\ssstyle \mathrm{pbc},\Box,\mathrm{full}}(0,0,\infty,1)=1$
and the exponents $\alpha$, $\beta$, and $\nu_{\para}$ 
fulfill the scaling law
\begin{equation}
\alpha \; = \; \frac{\,\beta\,}{\nu_{\para}} \, .
\label{eq:alpha_beta_nu_para}
\end{equation}
A finite system size limits the power-law behavior above.
The corresponding finite-size scaling form is obtained
by choosing $a_{\ssstyle L} L \lambda^{-\nu_{\senk}}=1$
\begin{equation}
%a_{\ssstyle a} \,
\rhoa(\deltap=0, h=0, L, t) 
\; \sim  \; 
(a_{\ssstyle L} L)^{-\beta/\nu_{\senk}}
\, \, {\tilde R}_{\ssstyle \mathrm{pbc},\Box,\mathrm{full}}
(0,0,1,
a_{\ssstyle t} t \; (a_{\ssstyle L} L)^{-z}     ) \, ,
\label{eq:scal_ansatz_EqoS_HS_dyn_fss}
\end{equation}
where $z=\nu_{\para}/\nu_{\senk}$ denotes the 
dynamical exponent.
Finite-size effects have to be taken into account for
\begin{equation}
{\mathcal O}(t)=t_{\ssstyle \mathrm{FSS}}
\quad\quad\quad
\mathrm{with}
\quad\quad\quad
t_{\ssstyle \mathrm{FSS}} = a_{\ssstyle t}^{-1} \,
(a_{\ssstyle L} \, L)^{z} \, .
\label{eq:scal_ansatz_EqoS_HS_dyn_fss_data_collapse}
\end{equation}
For $t \ll t_{\ssstyle \mathrm{FSS}}$ the scaling 
function obeys the power-law 
${\tilde R}_{\ssstyle \mathrm{pbc},\Box,\mathrm{full}} 
(0,0,1,x) \sim x^{-\alpha}$, whereas
${\tilde R}_{\ssstyle \mathrm{pbc},\Box,\mathrm{full}}(0,0,1,x)$
decays exponentially for $x \gg 1$, i.e., 
$t \gg t_{\ssstyle \mathrm{FSS}}$.
This is shown in \reffigname{fig:uni_dp_rho_t_1d} where the
raw data as well as the rescaled data are presented 
for $D=1$ and various system sizes.
Again the data of all considered models collapse
onto the single universal 
curve ${\tilde R_{\ssstyle \mathrm{pbc},\Box,\mathrm{full}}(0,0,1,x)}$.
It is worth mentioning that the models are simulated
with different update schemes.
Whereas a parallel\index{parallel update}
or synchronous\index{synchronous update scheme}
update scheme is
applied for site directed percolation, the contact
process and the pair contact process are 
simulated with an asynchronous 
update\index{synchronous update scheme},
implemented by a random sequential 
update\index{random sequential update}.
The observed data-collapse confirms 
that different update schemes are contained 
in the non-universal metric factors~$a_{\ssstyle t}$
and do not affect the universal scaling functions.

As already mentioned, the rapidity 
reversal\index{rapidity reversal symmetry}\index{symmetry, rapidity reversal}
of directed percolation implies the asymptotic
equivalence of the survival probability and the 
order parameter decay at criticality 
[\refeqname{eq:p_sur_alpha_rho}]
\begin{equation}
\Pa(t) \; \sim \; \mu^2 \, \rhoa(t) \, .
\label{eq:rapidity_reversal}
\end{equation}
The extension of the above dynamical 
scaling analysis to higher dimensions is redundant
because a detailed analysis of the survival 
probability will be presented in the next section.
But a comment about the rapidity reversal symmetry for
site directed percolation is worth making.
The order parameter $\rhoa(t)$ as well as the 
survival probability $\Pa(t)$ for site directed
percolation are plotted in \reffigname{fig:uni_dp_rho_t_1d}.
The survival probability is vertically shifted 
in order to check \refeqname{eq:rapidity_reversal}.
As can be seen both quantities are identical for all~$t$.
Thus the asymptotic equivalence 
(see the contact process data in \reffigname{fig:cp_5d_duality})
holds in case of site directed
percolation for all values of $t$, i.e.,
\begin{equation}
\Pa(t) \; = \; \mu^2 \, \rhoa(t) \, .
\label{eq:rapidity_reversal_sdp}
\end{equation}
The factor $\mu$ depends on certain non-universal
parameters and can be determined easily.
Starting from a fully occupied lattice, the density
of active sites after one parallel update step is
given by $\rhoa(t=1)=p$,
independently of the dimension and the lattice
structure.
On the other hand, the probability that the activity generated 
by a single seed survived one update step is 
$\Pa(t=1)=1-(1-p)^{z_{\ssstyle \mathrm{fnn}}}$
where $z_{\ssstyle \mathrm{fnn}}$ denotes the number
of forward next neighbors on the lattice.
In case of $1+1$-dimensional systems on a 
square lattice ($z_{\ssstyle \mathrm{fnn}}=2$)
we find at criticality $\mu=\sqrt{2-\pc\,}$.
For the sake of completeness we just mention that
\refeqname{eq:rapidity_reversal_sdp} holds 
with $\mu=1$ for bond directed\index{bond directed percolation} 
percolation~\cite{HINRICHSEN_1}.

%%%%%%%%%%%%%%%%%%%%%%%%%%%%%%%%%%%%%%%%%%%%%%%%%%%%%%%%%%%%%%%%%%%%%%%%%%%
%%%%%%%%%%%%%%%%%%%%%%%%%%%%%%%%%%%%%%%%%%%%%%%%%%%%%%%%%%%%%%%%%%%%%%%%%%%
%%%%%%%%%%%%%%%%%%%%%%%%%%%%%%%%%%%%%%%%%%%%%%%%%%%%%%%%%%%%%%%%%%%%%%%%%%%

\subsection{Localized particle source}
\label{subsec:dp_dynam_scal_located_source}

As pointed out in section~\ref{subsec:apt_dynamical_scal} 
measurements of activity spreading generated from a
single seed have been widely applied in the last two 
decades.
In particular they provide very accurate estimates of the
critical value~$\pc$ as well as of the exponents $\delta$, $\theta$, 
and~$z$.
Here, we will focus on the scaling functions
of the survival probability~$\Pa$ and the
average number of active sites~$\Na$.
At criticality both quantities obey the 
power-laws [\refeqname{eq:def_delta_theta_z}]
\begin{equation}
a_{\ssstyle P} \,    
\Pa
\; \sim  \;
(a_{\ssstyle t}  t)^{-\delta} \, , 
\quad\quad\quad\quad
a_{\ssstyle N} \, 
\Na
\;  \sim  \; 
(a_{\ssstyle t}  t)^{\theta} \, ,
\label{eq:Pa_Na_act_spread}
\end{equation}
where $\theta$ is often termed the critical initial 
slip\index{critical initial slip exponent} exponent~\cite{JANSSEN_10}.
In this notation the rapidity\index{symmetry, rapidity reversal}
reversal [\refeqname{eq:p_sur_alpha_rho}]
implies\index{rapidity reversal symmetry}
the scaling law $\alpha=\delta$.
Sufficiently close to the critical point, $\Pa$ and
$\Na$ are expected to obey the scaling forms
\begin{eqnarray}
\label{eq:scal_ansatz_Pa_dyn}   
a_{\ssstyle P} \,
\Pa(\deltap,h,L,t) 
& \sim  & 
\lambda^{-\delta\nu_{\para}}\, \, {\tilde P}_{\ssstyle \mathrm{pbc},\Box,\mathrm{seed}}
(a_{\ssstyle p}  
\deltap \; \lambda, 
a_{\ssstyle h} h \;\lambda^{\sigma}, 
a_{\ssstyle L} L \;\lambda^{-\nu_{\senk}},
a_{\ssstyle t} t \;\lambda^{-\nu_{\para}}     ) \, , \\[2mm]
\label{eq:scal_ansatz_Na_dyn}   
a_{\ssstyle N} \,
\Na(\deltap,h,L,t) 
& \sim  & 
\lambda^{\theta\nu_{\para}}\, \, {\tilde N}_{\ssstyle \mathrm{pbc},\Box,\mathrm{seed}}
(a_{\ssstyle p}  
\deltap \; \lambda, 
a_{\ssstyle h} h \;\lambda^{\sigma}, 
a_{\ssstyle L} L \;\lambda^{-\nu_{\senk}},
a_{\ssstyle t} t \;\lambda^{-\nu_{\para}}     ) \, .
\end{eqnarray}
Choosing $a_{\ssstyle t} t\lambda^{-\nu_{\para}}=1$
the power-laws [\refeqname{eq:Pa_Na_act_spread}] are 
recovered
for ${\tilde P}_{\ssstyle \mathrm{pbc},\Box}(0,0,\infty,1)=1$ 
as well as ${\tilde N}_{\ssstyle \mathrm{pbc},\Box}(0,0,\infty,1)=1$.
Again the finite-size scaling forms are obtained by 
setting $a_{\ssstyle L} L \;\lambda^{-\nu_{\senk}}=1$
\begin{eqnarray}
\label{eq:scal_ansatz_Pa_dyn_fss}      
a_{\ssstyle P} \,
\Pa(0,0,L,t) 
& \sim  & 
(a_{\ssstyle L} L)^{-\delta z}\, \, {\tilde P}_{\ssstyle \mathrm{pbc},\Box,\mathrm{seed}}
(0,0,1,
a_{\ssstyle t} t \;
(a_{\ssstyle L} L)^{-z}     ) \, , \\[2mm]
\label{eq:scal_ansatz_Na_dyn_fss}   
a_{\ssstyle N} \,
\Na(0,0,L,t) 
& \sim  & 
(a_{\ssstyle L} L)^{\theta z}\, \, {\tilde N}_{\ssstyle \mathrm{pbc},\Box,\mathrm{seed}}
(0,0,1,
a_{\ssstyle t} t \;
(a_{\ssstyle L} L)^{-z}     ) \, .
\end{eqnarray}
Similar to the order parameter behavior 
the scaling functions ${\tilde P_{\ssstyle \mathrm{pbc},\Box,\mathrm{seed}}}(0,0,1,x)$ 
and ${\tilde N_{\ssstyle \mathrm{pbc},\Box,\mathrm{seed}}}(0,0,1,x)$ 
decay exponentially for $t \gg t_{\ssstyle \mathrm{FSS}}$
whereas they exhibit an algebraic behavior
for $t \ll t_{\ssstyle \mathrm {FSS}}$.

\begin{figure}[t] 
\centering
%\leavevmode 
\includegraphics[clip,width=14.3cm,angle=0]{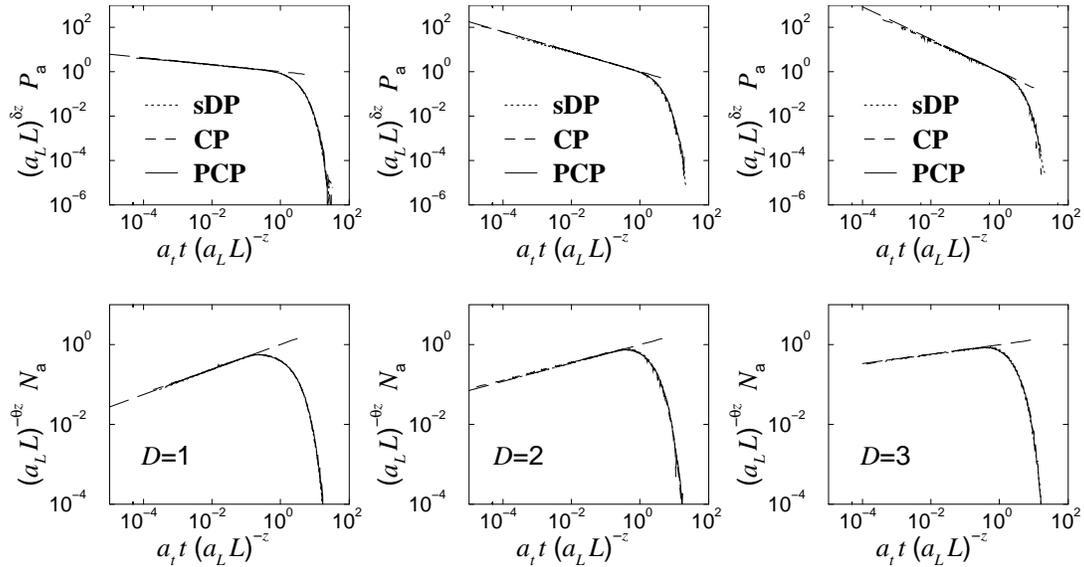}
\caption[Universal scaling functions ${\tilde P(0,0,1,x)}$,
${\tilde N(0,0,1,x)}$, DP class, D=1,2,3]  
{The universal scaling functions 
${\tilde P}_{\ssstyle \mathrm{pbc},\Box,\mathrm{seed}}$ and 
${\tilde N}_{\ssstyle \mathrm{pbc},\Box,\mathrm{seed}}$
of activity spreading for various dimensions.
In case of the pair contact process the simulations
are started from a natural configuration of inactive
particles.
System sizes $L=64,128,256,512$ are considered for $D=1$,
$L=64,128,256,512$ for $D=2$, and 
$L=16,32,64,128$ for $D=3$. 
The dashed lines corresponds to the power-law
behavior of the infinite system $x^{-\delta}$ and $x^{\theta}$,
respectively.
\label{fig:uni_dp_act_spread_d123}} 
\end{figure}

Performing activity spreading simulations of 
site directed percolation and of the contact process
the initial seed is implemented by a single particle on 
an empty lattice.
For absorbing phase transitions with
non-trivial absorbing states, like the pair contact process,
the scaling behavior depends upon the nature
of the initial configuration~\cite{JENSEN_3}.
In that case spreading activity simulations have to be 
performed at the 
so-called natural density of inactive\index{natural density} 
particles~\cite{JENSEN_3,MENDES_1,LUEB_23}.
Starting with a random configuration, an absorbing state at 
criticality is prepared by the natural dynamics. 
Then an active seed is created by adding or moving a particle. 
Thus the numerical effort is significantly increased
for systems exhibiting non-trivial absorbing states
and only small system sizes are available by simulations.
But nevertheless convincing data-collapses, including the pair contact 
process data, are obtained
and the corresponding universal scaling functions
are presented in \reffigname{fig:uni_dp_act_spread_d123}.
The values of the exponents used are listed 
in \reftablename{table:dp_exponents}.
Thus activity spreading from a localized seed
is characterized 
by the same universal scaling functions 
${\tilde P}_{\ssstyle \mathrm{pbc},\Box,\mathrm{seed}}$ and 
${\tilde N}_{\ssstyle \mathrm{pbc},\Box,\mathrm{seed}}$
for all considered models.

The spreading exponents $\delta$ and $\theta$ are related
to the critical exponents $\beta$, $\nu_{\senk}$, and $\nu_{\para}$.
For example, the percolation probability~$\Pperc$ is related to the 
ultimate survival probability~\refeqname{eq:ulti_sur_prob}
\begin{eqnarray}
\Pperc(\deltap) & = & \lim_{t \to \infty} \, \lim_{L \to \infty}  
\Pa(\deltap,h=0,L,t) \nonumber \\[2mm]
& \sim & 
a_{\ssstyle P}^{-1} \; 
\left . \lambda^{-\delta\nu_{\para}}\; \, 
{\tilde P}_{\ssstyle \mathrm{pbc},\Box,\mathrm{seed}}
(a_{\ssstyle p}  
\deltap \; \lambda,0,\infty,\infty )
\right |_{a_{\ssstyle p}  \deltap \lambda=1} \nonumber \\[2mm]
& \sim & 
a_{\ssstyle P}^{-1} \; 
(a_{\ssstyle p} \, \deltap)^{\delta\nu_{\para}}\;
{\tilde P}_{\ssstyle \mathrm{pbc},\Box,\mathrm{seed}}
(1,0,\infty,\infty ).
\label{eq:ulti_sur_prob_scal}
\end{eqnarray}
Comparing this result to \refeqname{eq:both_ord_par} reveals
the scaling law
\begin{equation}
\delta \; = \; \frac{\,\betap \,}{\nu_{\para}} 
\label{eq:delta_betap_nu_para}
\end{equation}   
In case of directed percolation, $\alpha=\delta$ is implied
by the rapidity reversal [\refeqname{eq:p_sur_alpha_rho_dual}] 
leading via \refeqname{eq:alpha_beta_nu_para} to 
\begin{equation}
\beta \; = \; \betap 
\label{eq:delta_beta_nu_para}
\end{equation}   
as well as $\delta=\beta/\nu_{\para}$.

The critical initial slip exponent~$\theta$ is connected to the other
exponents by the hyperscaling\index{hyperscaling} 
relation~\cite{MENDES_1}
\begin{equation}
\theta \, + \, \frac{\,\beta\,}{\nu_{\para}} \, 
+ \, \frac{\,\betap\,}{\nu_{\para}} \; \; = \; \;
\frac{\,D\,}{z}  \, .
\label{eq:gen_hyperscaling_law}
\end{equation}
This scaling law is fulfilled for directed percolation 
as well as other universality classes like
compact directed percolation~\cite{DICKMAN_13},
branching annihilation random walks with an
even number of offsprings~\cite{JENSEN_15}, 
as well as absorbing phase transitions with a conserved 
field~\cite{LUEB_23}.
%{\bf neues Janssen Paper mit GEP dazu,falls ebenfalls erfüllt}
Using \refeqs2name{eq:alpha_beta_nu_para}{eq:delta_betap_nu_para} 
the hyperscaling law is often written as
\begin{equation}
\theta \, + \, \alpha \, + \, \delta \;\;  = \; \;
\frac{\,D\,}{z}  \, .
\label{eq:theta_alpha_delta_D_z}
\end{equation}
In case of directed percolation it 
reduces due to the rapidity reversal ($\alpha=\delta$)
to [\refeqname{eq:hyperscal_dp_theta_delta_z}]
\begin{equation}
\theta \, + \, 2 \, \delta \;\;  = \; \;
\theta \, + \, \frac{\,2\, \beta\,}{\nu_{\para}} \, 
\; \; = \; \;
\frac{\,D\,}{z}  \, .
\label{eq:theta_2delta_D_z}
\end{equation}
Usually the hyperscaling law \refeqname{eq:gen_hyperscaling_law} is 
derived via the pair connectness function 
which is defined as the probability that two 
sites are connected by a path of activity
(see e.g.~\cite{GRASSBERGER_2,KINZEL_1,MENDES_1,MARRO_1,HINRICHSEN_1}).
Here, we present an alternative and instructive
way~\cite{WIJLAND_2,LUEB_23} which
allows to connect the universal scaling functions
${\tilde R}_{\ssstyle \mathrm{full}}$ and 
${\tilde N}_{\ssstyle \mathrm{seed}}$.
Therefore we assume 
that the initial particle density
$\rhoa(t=0)=\rhoanull$ appears as an additional
scaling field 
\begin{eqnarray}
\label{eq:scal_ansatz_EqoS_HS_dyn_initial}   
%a_{\ssstyle a} \,
& & \rhoa(\deltap, h, L, t, \rhoanull) 
\; \sim  \; \\
& & 
\quad \quad \quad \lambda^{-\beta}\, \, {\tilde R}_{\ssstyle \mathrm{pbc},\Box}
(a_{\ssstyle p}  
\deltap \; \lambda, 
a_{\ssstyle h} h \;\lambda^{\sigma}, 
a_{\ssstyle L} L \;\lambda^{-\nu_{\senk}},
a_{\ssstyle t} t \;\lambda^{-\nu_{\para}},
a_{\ssstyle 0} \rhoanull \; \lambda^{D \nu_{\perp}-\nu_{\para} \delta} 
) \, . \nonumber
\end{eqnarray}
and the scaling function behaves asymptotically 
as (see~\cite{HINRICHSEN_1} and references therein)
\begin{equation}
{\tilde R}_{\ssstyle \mathrm{pbc},\Box}
(0,0,\infty,1,x) \; \sim \;
\left \{
\begin{array}{ll}
x & \mathrm{for} \quad x \ll 1 \\
\mathrm{const} & \mathrm{for} \quad x \gg 1  \, .
\end{array}
\right .
\label{eq:R_pbc_asymptotic}
\end{equation}
For example, within the mean field theory 
[see \refeqname{eq:mf_cp_one_site_dyn_at_crit}]
the universal scaling function is given by
\begin{equation}
{\tilde R}_{\ssstyle \mathrm{pbc},\Box}
(0,0,\infty,1,x) \; = \;
\frac{x}{\,1\,+\,x\,}\, .
\label{eq:R_pbc_mf}
\end{equation}
Obviously the scaling function ${\tilde R}_{\ssstyle \mathrm{pbc},\Box}$
is related to ${\tilde R}_{\ssstyle \mathrm{pbc},\Box,\mathrm{full}}$ via  
\begin{equation}
{\tilde R}_{\ssstyle \mathrm{pbc},\Box,\mathrm{full}} 
\; = \; 
\left . 
{\tilde R}_{\ssstyle \mathrm{pbc},\Box} \, \right |_{\rhoanull=1}
\label{eq:R_full_R}
\end{equation}
and due to $\Na=L^D \rhoa$ related 
to ${\tilde N}_{\ssstyle \mathrm{pbc},\Box,\mathrm{seed}}$ by
\begin{equation}
{\tilde N}_{\ssstyle \mathrm{pbc},\Box,\mathrm{seed}} 
\; = \; L^D\,
\left . 
{\tilde R}_{\ssstyle \mathrm{pbc},\Box} \, \right |_{\rhoanull=1/L^{D}} \, .
\label{eq:N_seed_R}
\end{equation}
The scaling power of the initial density is explained by
the following crossover scenario:~Starting 
at criticality from a low density of active 
sites (e.g.~several seeds)
the number of active sites increases as 
$\Na \asympprop \rhoanull t^{\theta}$
until it reaches a maximum 
and crosses over to the expected asymptotic decay 
$\rhoa \asympprop t^{-\alpha}$.
The crossover\index{crossover} time is determined by 
\begin{equation}
{\mathcal O}( a_{\ssstyle 0} \rhoanull \,
( a_{\ssstyle t} t_{\ssstyle \mathrm{co}})^{D/z-\delta} )
\; = \; 1 \, ,
\label{eq:cross_over_initial_slip}
\end{equation}
corresponding to a merging of the 
survived (and former separated) clusters 
of activity~\cite{HINRICHSEN_1}.
Despite of metric factors the crossover takes place
at $t_{\ssstyle \mathrm{co}} \asympprop \rhoanull^{-1/(D/z-\delta)}$
and the scaling form \refeqname{eq:scal_ansatz_EqoS_HS_dyn_initial}
has to recover $\Na \asympprop t^{\theta}$ for 
$t\ll t_{\ssstyle \mathrm{co}}$ whereas 
$\rhoa \asympprop t^{-\alpha}$ has to be obtained for
$t\gg t_{\ssstyle \mathrm{co}}$.

Focusing on criticality we find in the 
thermodynamic limit for 
$a_{\ssstyle t} t \, \lambda^{-\nu_{\para}}=1$
\begin{equation}
%a_{\ssstyle a} \,
\rhoa(0, 0, \infty, t, \rhoanull) 
\; \sim  \; 
(a_{\ssstyle t} t)^{-\beta/\nu_{\para}}\; \; {\tilde R}_{\ssstyle \mathrm{pbc},\Box}
(0, 0, \infty,
%a_{\ssstyle L} L \;(a_{\ssstyle t} t)^{-1/z},
1,
%a_{\ssstyle 0} \rhoanull \; (a_{\ssstyle t} t)^{D/z-\delta} 
(t/t_{\ssstyle \mathrm{co}})^{D/z-\delta} 
) \, . 
\label{eq:scal_ansatz_EqoS_HS_dyn_initial_02}
\end{equation}
Starting with a fully occupied lattice, i.e., with  
$a_{\ssstyle 0} (a_{\ssstyle t} t_{\ssstyle \mathrm{co}})^{D/z-\delta}=1$, 
the order parameter decays as
\begin{eqnarray}
%a_{\ssstyle a} \,
\rhoa(0, 0, \infty, t, 1) 
& \sim  &
(a_{\ssstyle t} t)^{-\beta/\nu_{\para}}\; \; {\tilde R}_{\ssstyle \mathrm{pbc},\Box}
(0, 0, \infty,
%a_{\ssstyle L} L \;(a_{\ssstyle t} t)^{-1/z},
1,
a_{\ssstyle 0} \; (a_{\ssstyle t} t)^{D/z-\delta} 
)  \nonumber \\[2mm]
& \mathop{\approx} \limits_{t \to \infty}   &
(a_{\ssstyle t} t)^{-\alpha}\; \; {\tilde R}_{\ssstyle \mathrm{pbc},\Box}
(0, 0, \infty,
%a_{\ssstyle L} L \;(a_{\ssstyle t} t)^{-1/z},
1, \infty)  \, , \nonumber 
\label{eq:scal_ansatz_EqoS_HS_dyn_initial_03}
\end{eqnarray}
since $D/z-\delta>0$ in all dimensions and where the scaling
law $\alpha=\beta/\nu_{\para}$ is used.
Comparing this results to \refeqname{eq:ord_decay_alpha}
leads to ${\tilde R}_{\ssstyle \mathrm{pbc},\Box}(0,0,\infty,1,\infty)=1$.
On the other hand, single seed initial configurations
$\rhoanull = L^{-D}$ are characterized by 
crossover times $t_{\ssstyle \mathrm{co}} \asympprop L^{D/(D/z-\delta)}$
and we find for $t\ll t_{\ssstyle \mathrm{co}}$
\begin{eqnarray}
%a_{\ssstyle a} \,
\rhoa(0, 0, \infty, t\ll t_{\ssstyle \mathrm{co}}, 1) 
& \sim  &
(a_{\ssstyle t} t)^{-\beta/\nu_{\para}}\; \; {\tilde R}_{\ssstyle \mathrm{pbc},\Box}
(0, 0, \infty,
%a_{\ssstyle L} L \;(a_{\ssstyle t} t)^{-1/z},
1,
(t/t_{\ssstyle \mathrm{co}})^{D/z-\delta} 
)  \nonumber \\[2mm]
& \sim   &
(a_{\ssstyle t} t)^{-\beta/\nu_{\para}}\; \; 
(t/t_{\ssstyle \mathrm{co}})^{D/z-\delta} \nonumber \\[2mm]
& \sim   &
(a_{\ssstyle t} t)^{-\beta/\nu_{\para}+D/z-\delta}\; \; 
a_{\ssstyle 0}\, L^{-D} \, .
\label{eq:scal_ansatz_EqoS_HS_dyn_initial_04}
\end{eqnarray}
Thus $a_{\ssstyle 0} \Na = \rhoa L^{D} \sim (a_{\ssstyle t} t)^{\theta}$
is recovered for $t\ll t_{\ssstyle \mathrm{co}}$, confirming the
hyperscaling relation \refeqname{eq:gen_hyperscaling_law}.
Furthermore the metric factor of the initial
configuration is given by $a_{\ssstyle 0} = 1/a_{\ssstyle N}$.

In this way the asymptotic order parameter 
decay [\refeqname{eq:ord_decay_alpha}]
and the initial particle growth [\refeqname{eq:Pa_Na_act_spread}]
emerges as different scaling regimes of the generalized
scaling function [\refeqname{eq:scal_ansatz_EqoS_HS_dyn_initial}].
The full crossover\index{scaling, crossover}\index{crossover scaling}
can be observed in simulations if 
the particular value of the initial density of active
sites leads to 
$1\ll t_{\ssstyle \mathrm{co}} \ll t_{\ssstyle \mathrm{FSS}}$
(see \cite{LUEB_23} for a detailed discussion).
For example the scaling regime $\rhoa\asympprop t^{-\alpha}$ can
not be observed in simulations starting from single seed
since $t_{\ssstyle \mathrm{FSS}}<t_{\ssstyle \mathrm{co}}$,
i.e., finite-size effects take place before the algebraic
particle decay starts.
On the other hand, too large initial densities $\rhoanull\approx 1$ 
cause too small crossover times 
$\mathcal{O}(t_{\ssstyle \mathrm{co}})=1$ and the short time 
scaling regime 
($\Na\asympprop t^{\theta}$ for $1 \ll\ t \ll t_{\ssstyle \mathrm{co}}$)
does not occur.
But intermediate initial densities allow to investigate the
crossover behavior.
%%%%%%%%%%%%%%%%%%%%%%%%%%%%%%%%%%%%%%%%%%%%%%%%%%%%%%%%%%%%%%%%%%%%%%%%%%%%%%%%%%%%%%
\begin{figure}[t] 
\centering
%\leavevmode 
\includegraphics[clip,width=13cm,angle=0]{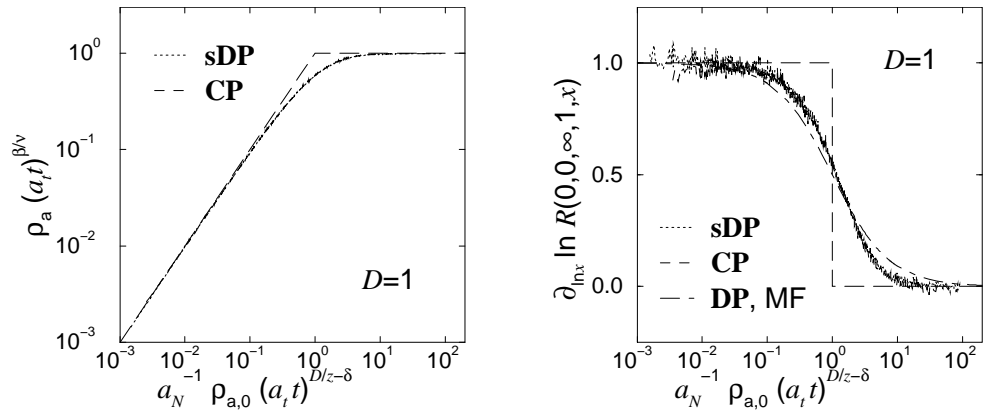}
\caption[Universal crossover scaling function
${\tilde R_{\ssstyle \mathrm{pbc},\Box}(0,0,\infty,1,x)}$, DP class, D=1]  
{The universal scaling function 
${\tilde R}_{\ssstyle \mathrm{pbc},\Box}(0,0,\infty,1,x)$
describes the crossover from initial particle growth 
$\Na\asympprop t^{\theta}$ to the asymptotic order
parameter decay $\rhoa\asympprop t^{-\alpha}$.
The data are obtained from simulations of 
site directed percolation and the contact process.
Different initial particle densities 
$\rhoanull = n / L$ with $n=1,2,4,\ldots,512$
and system sizes $L=1024,2048$ are considered.
Data that are affected by finite-size effects 
(${\mathcal O}(t)\ggl t_{\ssstyle \mathrm{FSS}}$) 
are skipped for a better visualization.
The long-dashed lines correspond to the 
asymptotic scaling behavior of the 
universal scaling function \refeqname{eq:R_pbc_asymptotic}.
Neglecting the metric factors each model would be
characterized by its own scaling function.
In order to scrutinize the data which covers
several decades it is usual to examine the
crossover via the effective exponent (left)
$\partial_{\ln{x}} \, \ln{{\tilde R(0,0,\infty,1,x)}}$.
For the sake of completeness, the effective
exponent of the $1+1$-dimensional models is compared to the
mean field behavior [dot-dashed line, see~\refeqname{eq:R_pbc_mf}].
Extended simulations reveal that the crossover scaling function
depends on the dimension and approaches the mean field 
curve with increasing dimension (not shown).
\label{fig:uni_dp_crossover_1d}} 
\end{figure} 
%%%%%%%%%%%%%%%%%%%%%%%%%%%%%%%%%%%%%%%%%%%%%%%%%%%%%%%%%%%%%%%%%%%%%%%%%%%%%%%%%%%%%%
An alternative way to observe the full crossover is to 
perform simulations with various initial conditions.
According to
\refeqs2name{eq:cross_over_initial_slip}{eq:scal_ansatz_EqoS_HS_dyn_initial_02}
all curves corresponding to different initial configurations
collapse onto the universal 
function ${\tilde R}_{\ssstyle \mathrm{pbc},\Box}(0,0,\infty,1,x)$
by plotting
$\rhoa \, ({a_{\ssstyle t} t})^{\beta/\nu_{\para}}$
as a function of 
$x=a_{\ssstyle N}^{-1} \rhoanull \, (a_{\ssstyle t} t)^{D/z-\delta}$.
This is shown in \reffigname{fig:uni_dp_crossover_1d} 
for $1+1$-dimensional site directed
percolation and the contact process.
Note that the
assumed asymptotic behavior [\refeqname{eq:R_pbc_asymptotic}] 
of the universal scaling 
function ${\tilde R}_{\ssstyle \mathrm{pbc},\Box}(0,0,\infty,1,x)$
is perfectly recovered confirming the scaling
form \refeqname{eq:scal_ansatz_EqoS_HS_dyn_initial}.
Furthermore, the data-collapse reveals that the crossover 
between two scaling regimes
of a given universality class can be described in terms
of universal scaling functions.
We revisit crossover phenomena in a more generalized
context in section \ref{chapter:crossover}.
There we address the question if the crossover
between two different universality classes can be described
in terms of universal functions.

The hyperscaling law [\refeqname{eq:gen_hyperscaling_law}] still
implies more.
Like all hyperscaling laws\index{hyperscaling}
it is fulfilled for the mean field
exponents at the upper critical dimension
\begin{equation}
\thetaMF \, + \, \frac{\,\betaMF\,}{\nuparaMF} \, 
+ \, \deltaMF \; \; = \; \;
\frac{\,\Dc\,}{\zMF}  \, .
\label{eq:gen_hyperscaling_law_mf_dc}
\end{equation}
Within the mean field theory, the activity spreading 
of most absorbing phase transitions
is described by a branching\index{branching process}
process.
The mean field values $\thetaMF=0$ and $\deltaMF=1$ 
lead to 
\begin{equation}
\Dc \; = \; \zMF \, 
\left ( 1 \, + \, \frac{\betaMF}{\nuparaMF} \, \right ) \, .
\label{eq:dc_z_beta_nupara}
\end{equation}
We have already seen that 
the exponents $\betaMF$ and $\nuparaMF$ can be 
derived within a simple mean field approach basing on the
transition rates.
Therefore, the above scaling law allows to determine the
upper critical dimension.
Furthermore a lower bound on~$\Dc$ is 
obtained by taking into account that the order parameter
exponent is positive definite, yielding
\begin{equation}
\Dc \; \ggl \; \zMF \, .
\label{eq:dc_z}
\end{equation}
In case of diffusion-like mean field processes the latter 
result reduces to $\Dc \ggl 2$.

\section{Conclusions and comments}
\label{sec:dp_conclusions}

%Closing this chapter it is worth to remark some points.
The scaling behavior of directed percolation is 
recognized as the paradigm of non-equilibrium phase
transitions from an active to an absorbing state.
The widespread occurrence of such models is reflected by
the universality hypothesis of Janssen\index{universality hypothesis} 
and Grassberger.
%The generic universality class of directed percolation
%is often refered to be a non-equilibrium counterpart
%to the Ising model in equilibrium.
Despite its ubiquity in theoretical models no experiments
observing directed percolation critical behavior are known 
so far.
The lacking experimental realization is explained
by the influence of quenched disorder or the impossibility
to realize a perfect non-fluctuating 
absorbing state~\cite{HINRICHSEN_1}.
For a summary of experimental efforts, including catalytic
reactions, flowing granular matter~\cite{DAERR_1,HINRICHSEN_7},
the onset of turbulence~\cite{RUPP_1},
we refer the interested reader to~\cite{HINRICHSEN_1,HINRICHSEN_8}.
But due to its simplicity and robustness it is still
believed that experimental counterparts to directed percolation exists.
Thus an experimental realization of directed percolation
remains an outstanding problem of top priority~\cite{GRASSBERGER_10}.

Additionally to the experimental situation several 
theoretical aspects are also still open.
For example directed percolation is not solved analytically in
$1+1$-dimensions.
Thus, only approximative estimates (see e.g.~\cite{SIRE_1}) exist 
of the still unknown critical exponents.
Furthermore, the universality hypothesis still awaits a
rigorous proof.
It is worth mentioning that
the universality class of directed percolation
is even larger than expected from the hypothesis.
In particular, the pair contact process and the threshold
transfer process belong to the directed percolation
universality class although they are not characterized
by a unique absorbing state.
Thus the hypothesis defines only sufficient conditions
but fails to describe the DP universality class in full
generality (see~\cite{HINRICHSEN_1} for detailed
discussion).
A refinement of the universality hypothesis is part
of the larger task to provide a classification scheme
for non-equilibrium phase transition.
Compared to equilibrium a full classification 
of non-equilibrium phase transitions is still
open.

%Despite of several attempts (e.g.~for reaction-diffusion like
%systems~\cite{TAUBER_1,KOCKELKOREN_1}) no convincing classification 
%scheme of absorbing phase transitions is known so far.

\chapter{Absorbing phase transitions with a conserved field} 
\setcounter{figure}{37}
\label{chapter:aptcf}

Directed percolation is considered as a paradigm 
for non-equilibrium phase transitions into absorbing states.
Different universality classes are expected to 
occur in the presence of additional symmetries.
For example, a particle-hole symmetry is associated
to the universality class of compact directed 
percolation (see~\ref{subsec:domany_kinzel}).
Branching annihilating random walks with an even
number of offsprings obey\index{parity conserving}\index{PC} 
parity\index{parity symmetry}\index{symmetry, parity} 
conservation (PC)\index{universality class, PC}
and define another universality class~\cite{ZHONG_1,CARDY_2}.
In the following we consider a class of absorbing
phase transitions which is characterized by 
particle conservation.
According to Rossi {\textit{et al.}}~the
additional conservation law leads to the
universality class of stochastic absorbing phase transitions
with a conserved field~\cite{ROSSI_1}.
Similar to the universality hypothesis of Janssen and Grassberger
the authors conjecture that all stochastic models
with an infinite number of absorbing states in which the
order parameter is coupled to a non-diffusive conserved 
field define a unique\index{universality hypothesis} 
universality\index{universality class}
class.
This universality class is of particular interest since
it is related to the concept of 
self-organized~\index{self-organized criticality}
criticality (SOC)~\cite{BAK_1,BAK_2,BAK_3,TURCOTTE_1}.
Compared to directed percolation this universality
class is not well established so far.
Besides field theories~\cite{DICKMAN_1,PASTOR_2,WIJLAND_1},
series\index{series expansion} expansions~\cite{STILCK_1} 
and path integral\index{path integral}
representations~\cite{DICKMAN_15} 
most quantitative results are obtained from simulations.
In particular, a systematic $\epsilon$-expansion is still
lacking. 
In the following we review the numerical results and 
focus again on the universal scaling behavior.
Several lattice models are introduced and the corresponding steady
state and dynamical scaling behavior is investigated.
The determination of the value of the of upper critical
dimensions as well as the relation to self-organized criticality
are discussed.

\section{Lattice models of absorbing phase transitions with a 
conserved field}
\label{sec:aptcf_lattice_models}

In the following we consider various lattice models
obeying particle conservation.
The first one is a modification of Manna's stochastic
sandpile automaton~\cite{MANNA_2}.
This model was intensively investigated in the context
of self-organized criticality 
(see e.g.~\cite{BENHUR_1,DHAR_7,CHESSA_2,CHESSA_3,LUEB_9,LUEB_13,DICKMAN_2,BIHAM_1,SHILO_1}).
The second model is a modification of the threshold
transfer process that obeys particle conservation.
This model is well suited for analytical treatments
at the mean field level.
Furthermore it allows to investigate the crossover between the
universality class of directed percolation and 
the universality class of stochastic absorbing phase
transitions with a conserved field.
The third model is the so-called conserved lattice gas.
Compared to the bosonic Manna model, no multiple particle
occupation is allowed in the conserved lattice gas. 
But more than the other models, the Manna model 
can be considered as the paradigm of the universality
class of stochastic absorbing phase transitions with a 
conserved field.
Since universality classes are often labeled by the simplest 
model belonging to them it is termed  
the\index{Manna universality class} 
Manna\index{universality class, Manna}  
universality class in the following.

Additionally to these lattice models a conserved 
reaction-diffusion model is discussed in the 
literature~\cite{KREE_1,WIJLAND_2,PASTOR_2,PASTOR_5}.
The model is defined by the reaction scheme
\begin{equation}
A \, \longrightarrow \, B\, , 
\quad\quad A \, + \, B \, \longrightarrow \, 2\,A\, , 
\label{eq:reac_diff_aptcf}
\end{equation}
and where $A$-particles diffuse with diffusion rate~$D_{\ssstyle A}$
and $B$-particles with rate~$D_{\ssstyle B}$, respectively. 
The reaction scheme describes systems where a conserved field 
is coupled to a non-conserved order parameter ($A$-particles).
Obviously, the total number of particles 
$N=N_{\ssstyle A}+N_{\ssstyle B}$ is conserved and a 
continuous phase transition occurs by varying $N$.
In the limit $D_{\ssstyle B} \to 0$ the reaction-diffusion
system belongs to the universality class of interest, i.e.,
of absorbing phase transitions where the order parameter is
coupled to a non-diffusive field~\cite{ROSSI_1,PASTOR_5}.
Field theories are derived from the reaction-diffusion
model but the corresponding renormalization group approaches
run into difficulties and the results are controversially
discussed in the 
literature~\cite{ODOR_1,PASTOR_2,WIJLAND_1,JANSSEN_4}.

%%%%%%%%%%%%%%%%%%%%%%%%%%%%%%%%%%%%%%%%%%%%%%%%%%%%%%%%%%%%%%%%%%%%%%%%%%%%%%%%%%%%%%%%%
%%%%%%%%%%%%%%%%%%%%%%%%%%%%%%%%%%%%%%%%%%%%%%%%%%%%%%%%%%%%%%%%%%%%%%%%%%%%%%%%%%%%%%%%%
%%%%%%%%%%%%%%%%%%%%%%%%%%%%%%%%%%%%%%%%%%%%%%%%%%%%%%%%%%%%%%%%%%%%%%%%%%%%%%%%%%%%%%%%%

\subsection{Manna model}
\label{subsec:manna}

\begin{figure}[t] 
\centering
%\leavevmode 
\includegraphics[clip,width=10cm,angle=0]{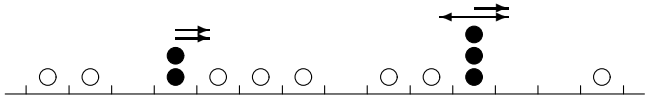}
\caption[Manna model] 
{Sketch of the dynamics of the $1+1$-dimensional Manna
model with $\Nc=2$ according to the original rules 
($n\to 0$ for $n \ggl \Nc$).
Filled circles mark active particles 
whereas non-active particles are marked by open circles.  
The arrows denote possible particle movements 
in the next update step according to a particular
representation of the stochastic rules. 
\label{fig:manna_model_01}} 
\end{figure}

The Manna model\index{Manna model}
was introduced in~\cite{MANNA_2} as 
a stochastic sandpile model in which integer values
represent local energies, number of sand-grains or
particles.
The Manna model is a bosonic lattice
model, i.e., it allows unlimited particle occupation 
of lattice sites ($n=0,1,2,\ldots$).
Lattice sites are considered as inactive if the particle
occupation is below a certain threshold $n<\Nc$.
For $n \ggl \Nc$ the lattice site is active and 
the particles are moved to nearest neighbors,
selected at random for each particle, i.e.,
\begin{equation}
n \; \longrightarrow \; 0 \phantom{\;\, - \, \Nc}
\quad\quad {\mathrm{for\;\;all\;\;sites\;\;with}}
\quad\quad n \ggl \Nc \, .
\label{eq:manna_rule_original}
\end{equation}
The dynamics is sketched in \reffigname{fig:manna_model_01}.
Additionally to this original Manna model~\cite{MANNA_2} a
modified version is considered in the literature
where the occupation number of active lattice sites is not reduced
to zero but 
\begin{equation}
n \; \longrightarrow \; n \, - \, \Nc 
\quad\quad {\mathrm{for\;\;all\;\;sites\;\;with}}
\quad\quad n \ggl \Nc \, , %\phantom{.}
\label{eq:manna_rule_modified}
\end{equation}
and $\Nc$ particles are randomly transferred to the 
nearest neighbors.
%%%%%%%%%%%%%%%%%%%%%%%%%%%%%%%%%%%%%%%%%%%%%%%%%%%%%%%%%%%%%%%%%%%%%%%%%%%%%%%%%%%%
%%%%%%%%%%%%%%%%%%%%%%%%%%%%%%%%%%%%%%%%%%%%%%%%%%%%%%%%%%%%%%%%%%%%%%%%%%%%%%%%%%%%
In the latter case the Manna model is analytically tractable
if the active particles are distributed in a particular 
way~\cite{DHAR_7}.
In both versions the particle transfer conserves the 
particle number, i.e., no particle
creation or annihilation processes take place.
Applying periodic boundary conditions the 
total number of particles~$N$ is conserved. 
The particle density~$\rho=N/L^{D}$ is the control parameter
of a continuous phase transition from an inactive phase
to an active phase.
The active phase is characterized by a non-zero
density~$\rhoa$ of lattice sites occupied by at least
$\Nc$~particles.
The density~$\rhoa$ corresponds to the order
parameter~\cite{VESPIGNANI_4}.
A configuration is absorbing if it contains only empty and 
single occupied lattice sites.
%%%%%%%%%%%%%%%%%%%%%%%%%%%%%%%%%%%%%%%%%%%%%%%%%%%%%%%%%%%%%%%%%%%%%%%%%%%%%%%%%
\begin{figure}[b] 
\centering
%\leavevmode 
\includegraphics[clip,width=13cm,angle=0]{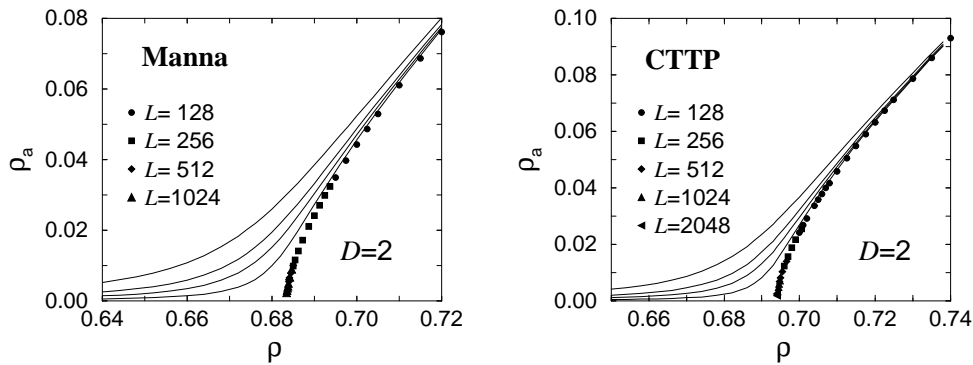}
\caption[Manna model and CTTP, order parameter behavior] 
{The order parameter behavior of the $2+1$-dimensional
Manna model and of the conserved threshold transfer process (CTTP)
for various field values (from $h=5\,10^{-5}$ to $h=10^{-4}$). 
For non-zero field (lines) $\rhoa$ exhibits an analytical behavior.
For zero field (symbols) the order parameter 
vanishes algebraically by approaching the transition point.
The data are obtained from simulations of various system
sizes $L \kgl 2048$ with periodic boundary conditions.
\label{fig:manna_cttp_2d_op}} 
\end{figure} 
%%%%%%%%%%%%%%%%%%%%%%%%%%%%%%%%%%%%%%%%%%%%%%%%%%%%%%%%%%%%%%%%%%%%%%%%%%%%%%%%%
Thus the Manna model is characterized by an infinite 
number of absorbing
configurations in the thermodynamic limit.
Similar to the universality class of directed percolation
it is possible to apply an external field~$h$ which is
conjugated to the order parameter.
But considering absorbing phase transitions with a conserved
field one has to take care that the external field 
does not change the particle number.
A possible realization of the external field was developed
in~\cite{LUEB_22} where a diffusion-like field is implemented.
The field triggers movements of inactive particles which may 
be activated in this way.
The corresponding order parameter behavior of the two-dimensional Manna
model is shown in~\reffigname{fig:manna_cttp_2d_op}.

The driven-dissipative version of the 
Manna model was investigated intensively 
in the context of self-organized criticality 
(see e.g.~\cite{BENHUR_1,DHAR_7,CHESSA_2,CHESSA_3,LUEB_9,LUEB_13,DICKMAN_2,BIHAM_1}).
Following Ben-Hur and Biham~\cite{BENHUR_1} the universality 
class of self-organized critical systems is determined by the 
way how active particles are distributed to the next
neighbors (deterministic, stochastic, directed, undirected, etc.).
At first glance, the Manna model seems to be characterized 
by a stochastic particle transfer.
But this is correct only for moderate, i.e., not to large values of 
the threshold $\Nc$.
In the limit $\Nc \to \infty$ the active particles are
equally distributed to the nearest neighbors, i.e., the 
stochastic character of the particle transfer is lost.
In that case the Manna model exhibits a different 
scaling behavior.
The crossover between the two universality classes 
was observed in numerical simulations and confirms 
the conjecture of~\cite{BENHUR_1}.
Here, we focus our attention on the scaling
behavior of the original Manna model [\refeqname{eq:manna_rule_original}]
for $\Nc=2$ in
various dimensions.
The modified version [\refeqname{eq:manna_rule_modified}]
is considered exemplarily for $D=2$ in order to show the 
identical scaling behavior.

%%%%%%%%%%%%%%%%%%%%%%%%%%%%%%%%%%%%%%%%%%%%%%%%%%%%%%%%%%%%%%%%%%%%%%%%%%%%%%%%%%%%%%%%%
%%%%%%%%%%%%%%%%%%%%%%%%%%%%%%%%%%%%%%%%%%%%%%%%%%%%%%%%%%%%%%%%%%%%%%%%%%%%%%%%%%%%%%%%%
%%%%%%%%%%%%%%%%%%%%%%%%%%%%%%%%%%%%%%%%%%%%%%%%%%%%%%%%%%%%%%%%%%%%%%%%%%%%%%%%%%%%%%%%%

\subsection{Conserved threshold transfer process}
\label{subsec:cttp}

\begin{figure}[b] 
\centering
%\leavevmode 
\includegraphics[clip,width=10cm,angle=0]{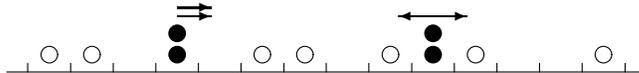}
\caption[Conserved threshold transfer process] 
{Sketch of the dynamics of the $1+1$-dimensional 
conserved transfer threshold process (CTTP).
Filled circles mark active particles 
whereas non-active particles are marked by open circles.  
The arrows denote possible particle movements 
in the next update step according to a particular
representation of the stochastic rules. 
Note that the relaxation event presented on the right site
is deterministic.
\label{fig:cttp_model_01}} 
\end{figure}

The conserved\index{conserved threshold transfer process}
threshold transfer process (CTTP)~\cite{ROSSI_1}\index{CTTP}
is a modification of the threshold transfer process (see 
section~\ref{subsec:dp_other_models}).
Again lattice sites may be empty, occupied by one particle,
or occupied by two particles.
Empty and single occupied lattice sites are considered as
inactive, whereas double occupied lattice sites are considered
as active.
In the latter case one tries to transfer both particles of
a given active site to randomly chosen empty or
single occupied sites (see~\reffigname{fig:cttp_model_01}).
Thus in contrast to the threshold transfer process no
creation and annihilation processes of isolated particles 
take place and the particle number is conserved.
Similar to the Manna model the conjugated field is implemented
by a diffusion-like field~\cite{LUEB_24}.
The order parameter behavior is shown in~\reffigname{fig:manna_cttp_2d_op}
for various external fields.
Numerical simulations turn out that the conserved threshold
transfer process exhibits the same scaling behavior as the
Manna model for $D\ggl 2$~\cite{LUEB_29}.
The situation is more complicated in one-dimensional
systems where a splitting of the universality class
occurs~\cite{LUEB_23}.
The dynamical reduction causes a change of the stochastic 
character of the dynamics of the CTTP.
Whereas all particle movements are stochastic in the Manna model, 
roughly $40\%$ of the relaxation events of the one-dimensional
CTTP are deterministic (see \reffigname{fig:cttp_model_01}).
Furthermore, a perturbation triggered by the external field
inside a cluster of inactive sites performs a simple random 
walk.
This pathologic behavior differs from that of the Manna 
model~\cite{LUEB_23}.
Therefore we focus in the following on the conserved 
threshold transfer process in $D\ggl 2$.

The CTTP can be considered as a modification of the Manna model 
exhibiting a restriction of particle occupation.
This occupation restriction simplifies an analytical treatment
significantly.
For example, the CTTP can be analyzed easily within a 
mean field\index{mean field theory, CTTP}
approximation whereas the multiple particle occupation of 
the Manna model leads to a system of coupled equations~\cite{HEGER_1}.
Therefore, we focus on the CTTP in order to derive 
the mean field behavior of the universality class~\cite{LUEB_25}.
Considering uncorrelated particle behavior, the associated probabilities
for a change of the number of active 
sites by $\Delta n_{\ssstyle \mathrm a}$ are 
\begin{eqnarray}
p(\Delta n_{\ssstyle \mathrm a}=-1) 
& = & \rhoa \, (\rhoe^2 \,+ \,2 \,\rhoe \, \rhoa ) 
\, , \nonumber \\
p(\Delta n_{\ssstyle \mathrm a}=\phantom{-}0) 
& = & \rhoa \, (2 \, \rhoe \,\rhoi \, + \, 2 \, \rhoi \, \rhoa \, + \, \rhoa^2 )
\, + \, \rhoi \, h \, (\rhoe\, +\, \rhoa)  
\, ,\nonumber \\
p(\Delta n_{\ssstyle \mathrm a}=\phantom{-}1) 
& = & \rhoa \, \rhoi^2 \, + \, \rhoi^2 \, h \, . 
\end{eqnarray}   
Here, $\rhoa$ denotes the order parameter, $\rhoi$ corresponds
to the density of single occupied sites and $\rhoe$ is the
density of empty lattice site.
This reaction scheme leads to the differential equation
\begin{equation}
\partial_{\ssstyle t} \, \rhoa(\rho, h) 
\; = \; \rhoa \, (-1 \, + \, 2 \, \rho \, - \,
4\, \rhoa \, + \, \rhoa^2) 
\, + \, h \, (\rho\,-\, 2 \, \rhoa)^2 \, ,
\label{eq:cttp_mean_Langevin}
\end{equation}
where the normalization $\rhoe + \rhoi + \rhoa =1$ as well
as the particle 
conservation $\rho=\rhoi + 2 \rhoa$ are used~\cite{LUEB_25}.
For zero field, the steady state 
condition $\partial_{\ssstyle t} \, \rhoa =0$ leads to
\begin{equation}
\rhoa\;=0\;  \quad \vee \quad 
-1 \, + \, 2 \, \rho \,  -4 \, \rhoa \, +  \, \rhoa^2 \; = \; 0 \, .
\end{equation}   
The first equation corresponds to the absorbing
state and the second equation yields the order parameter
as a function of the particle density~\cite{LUEB_25,DICKMAN_3} 
\begin{equation}
\rhoa \; = \; 2 \, - \, \sqrt{5-2\rho\,} \quad
\quad {\rm for} \quad \rho \ggl 1/2 \, .
\label{eq:cttp_mean_field_ord_par_01}
\end{equation}
The order parameter vanishes in leading order as
%\begin{equation}
%\rho_{\scriptscriptstyle \text a} \; = \; \frac{1}{4} \,\delta\rho^\beta \, , 
%%+ \, {\cal O} (\delta\rho^2),
%\label{eq:cttp_ord_par_02}
%\end{equation}
$\rhoa \;  =  \; \deltarho^\beta / 4$
with the reduced control 
parameter $\deltarho=(\rho-\rhoc)/ \rhoc$ and $\beta=1$.
For non-zero conjugated field the order parameter
scales at the critical density $\rhoc=1/2$ as
$\rhoa = h^{\beta/\sigma} / 4$
with $\sigma=2$.
Close to the critical point, the asymptotic behavior of the 
equation of state is given by~\cite{LUEB_25}
\begin{equation}
\rhoa \; \sim \;  \frac{\, \deltarho \,}{8}
\; + \; \sqrt{ \frac{h}{16} \, + \, 
\left ( \frac{\, \deltarho \, }{8}
\right )^2 \;}
\, .
\label{eq:cttp_mean_field_eqos_01}
\end{equation}

We focus now on the dynamical behavior.
The zero field behavior is determined by the differential equation
\begin{equation}
\partial_t \rhoa \; = \; \rhoa \,
(\deltarho \, - \, 4 \rhoa + \rhoa^2).
\label{eq:cttp_mf_ord_dyn_01}
\end{equation}
For sufficiently small order parameter 
higher order terms ${\cal O}(\rhoa^3)$
can be neglected,
yielding for $\deltarho \neq 0$
\begin{equation}
\rhoa(t) \; = \; 
\frac{\deltarho}
{\; 4-(4-{\deltarho}/{\rhoanull})\;  \exp{(-\deltarho \; t)}\;} \, .
\label{eq:cttp_mf_ord_dyn_02}
\end{equation}
Asymptotically ($t \to \infty$) the order parameter behaves as
\begin{eqnarray}
\label{eq:cttp_mf_ord_dyn_scal_below}
\left . 
\rhoa(t)^{\vphantom X}  
\right |_{\ssstyle \deltarho < 0}
& \sim &  - \, \deltarho \,
\left (4- \frac{\deltarho} { \rhoanull} \right )^{-1}
%\frac{\, \delta\rho\,}{4} 
\; {\rm e}^{\deltarho \; t}  \, , \\
\label{eq:cttp_mf_ord_dyn_scal_above}
\left . 
\rhoa(t)^{\vphantom X} 
\right |_{\scriptscriptstyle \deltarho > 0}
& \sim & \frac{\,\deltarho\,}{4} \; + \;
\frac{\deltarho}{16}
\left (4- \frac{\deltarho} { \rhoanull} \right )
{\rm e}^{-\deltarho \; t} \, .
%\quad \quad {\rm for} \quad \deltarho>0 \, .
\end{eqnarray}
Apart from criticality,
the steady state solutions ($\rhoa =\deltarho/4$ and 
$\rhoa =0$) are approached exponentially, independent of the
initial condition $\rhoanull$.
The corresponding correlation time is given by
$\xi_{\para}=|\deltarho|^{-\nu_{\para}}$,  
with $\nu_{\para}=1$.
At the critical density ($\deltarho=0$) 
the order parameter exhibits an algebraic decay, i.e.,
\begin{equation}
\rhoa(t) \; = \; 
\frac{1}{\, \rhoanull^{-1} \, + \, 4 t \, }
\; \mathop{\longrightarrow}\limits_{t\to \infty}
\; \frac{1}{\, 4 t \, } \, .
\label{eq:cttp_mf_ord_dyn_scal_crit}
\end{equation}

Thus the mean field behavior of the conserved transfer threshold 
process agrees with the mean field solution of directed 
percolation, i.e., the same critical exponents 
($\beta=1$, $\sigma=2$, $\nu_{\para}=1$), and the same universal
scaling functions are obtained. 	
The corresponding non-universal metric factors are given
by $a_{\ssstyle \rho}=1/4$, $a_{\ssstyle h}=1/16$, 
as well as $a_{\ssstyle t}=4$.
Although the scaling behavior of the CTTP and of directed percolation
differ in low dimensions~\cite{LUEB_24} they coincide on the mean field
level.
We will address this point in detail below.

%%%%%%%%%%%%%%%%%%%%%%%%%%%%%%%%%%%%%%%%%%%%%%%%%%%%%%%%%%%%%%%%%%%%%%%%%%%%%%%%%%%%%%%%%
%%%%%%%%%%%%%%%%%%%%%%%%%%%%%%%%%%%%%%%%%%%%%%%%%%%%%%%%%%%%%%%%%%%%%%%%%%%%%%%%%%%%%%%%%
%%%%%%%%%%%%%%%%%%%%%%%%%%%%%%%%%%%%%%%%%%%%%%%%%%%%%%%%%%%%%%%%%%%%%%%%%%%%%%%%%%%%%%%%%

\subsection{Conserved lattice gas}
\label{subsec:clg}

\begin{figure}[t] 
\centering
%\leavevmode 
\includegraphics[clip,width=10cm,angle=0]{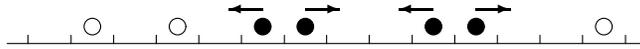}
\caption[Conserved lattice gas] 
{Sketch of the dynamics of the $1+1$-dimensional 
conserved lattice gas (CLG).
Neighboring particles are considered as active (filled
circles) whereas isolated particles remains inactive (open circles).
The arrows denote how the active particles are
moved in the next update step.
\label{fig:clg_model_01}} 
\end{figure}

The\index{conserved lattice gas}\index{CLG} 
conserved lattice gas (CLG)~\cite{ROSSI_1} 
is a stochastic variant
of a model introduced by Jensen~\cite{JENSEN_7}.
In both cases, lattice sites may be empty or occupied
by a particle.
The restriction to single particle occupation differs
from the bosonic Manna model.
Motivated by experiments on flux flow in type-II superconductors
a repulsive interaction is assumed~\cite{JENSEN_7}, 
i.e., a particle is considered as active if at least
one of its neighboring sites on the lattice is occupied
by another particle (see~\reffigname{fig:clg_model_01}).
If all neighboring sites are empty, the particle
remains inactive.
Active particles are moved in the next update step to one of
their empty nearest neighbor sites, selected at random.
Obviously, the dynamics obey particle conservation and 
all configurations consisting only of isolated particles
(without occupied nearest neighbors) are absorbing.
Increasing the particle density, a continuous phase transition
from a non-active to an active phase takes place~\cite{ROSSI_1,LUEB_19}.
Considering simple cubic lattices an upper bound on the 
critical density is $\rhoc=1/2$.
It is known that the one-dimensional system exhibits a trivial
behavior with $\rhoc=1/2$ (see also Ref.\,22  in~\cite{ROSSI_1}) 
whereas non-trivial behavior with $\rhoc<1/2$ occurs for $D\ggl 2$.
It is worth mentioning that the one-dimensional CLG is 
characterized by a deterministic dynamics, i.e., the
one-dimensional CLG does not belong to the universality class
of stochastic absorbing phase transitions with a conserved field.
Therefore, we will consider the CLG only in
higher dimensions in this work.
As for the Manna model and the CTTP, the conjugated field
is implemented by particle diffusion~\cite{LUEB_22}.

%%%%%%%%%%%%%%%%%%%%%%%%%%%%%%%%%%%%%%%%%%%%%%%%%%%%%%%%%%%%%%%%%%%%%%%%%%%%%%%%%%%%%%%%%
%%%%%%%%%%%%%%%%%%%%%%%%%%%%%%%%%%%%%%%%%%%%%%%%%%%%%%%%%%%%%%%%%%%%%%%%%%%%%%%%%%%%%%%%%
%%%%%%%%%%%%%%%%%%%%%%%%%%%%%%%%%%%%%%%%%%%%%%%%%%%%%%%%%%%%%%%%%%%%%%%%%%%%%%%%%%%%%%%%%

\section{Steady state scaling behavior}
\label{sec:aptcf_steady_state_scaling}

In the following section we examine the steady state scaling 
behavior of three different models in various dimensions.
We focus our attention on the scaling behavior
of the order parameter, 
its fluctuations, as well as on the scaling
behavior of the susceptibility.
Again, the Hankey-Stanley scaling 
forms \refeqsname{eq:scal_ansatz_EqoS_HS}{eq:scal_ansatz_Susc_HS}
normed by \refeqname{eq:uni_scal_norm_R_D} are used 
in order to show that the three models of interest 
belong to the same universality class.
In all cases the conjugated field~$h$ is implemented by
particle diffusion and $\deltap$ denotes again the
distance to the critical point, i.e., 
$\deltap = (\rho-\rhoc)/\rhoc$.
A comparison of the scaling behavior of high dimensional systems
with the mean field functions derived above
leads to a reliable determination of the upper critical
dimension.
Additionally, we consider briefly universal amplitude 
combinations as well as steady state finite-size
scaling.

\subsection{Equation of state and fluctuations}
\label{subsec:manna_eqos}

\begin{figure}[b] 
\centering
%\leavevmode 
\includegraphics[clip,width=13cm,angle=0]{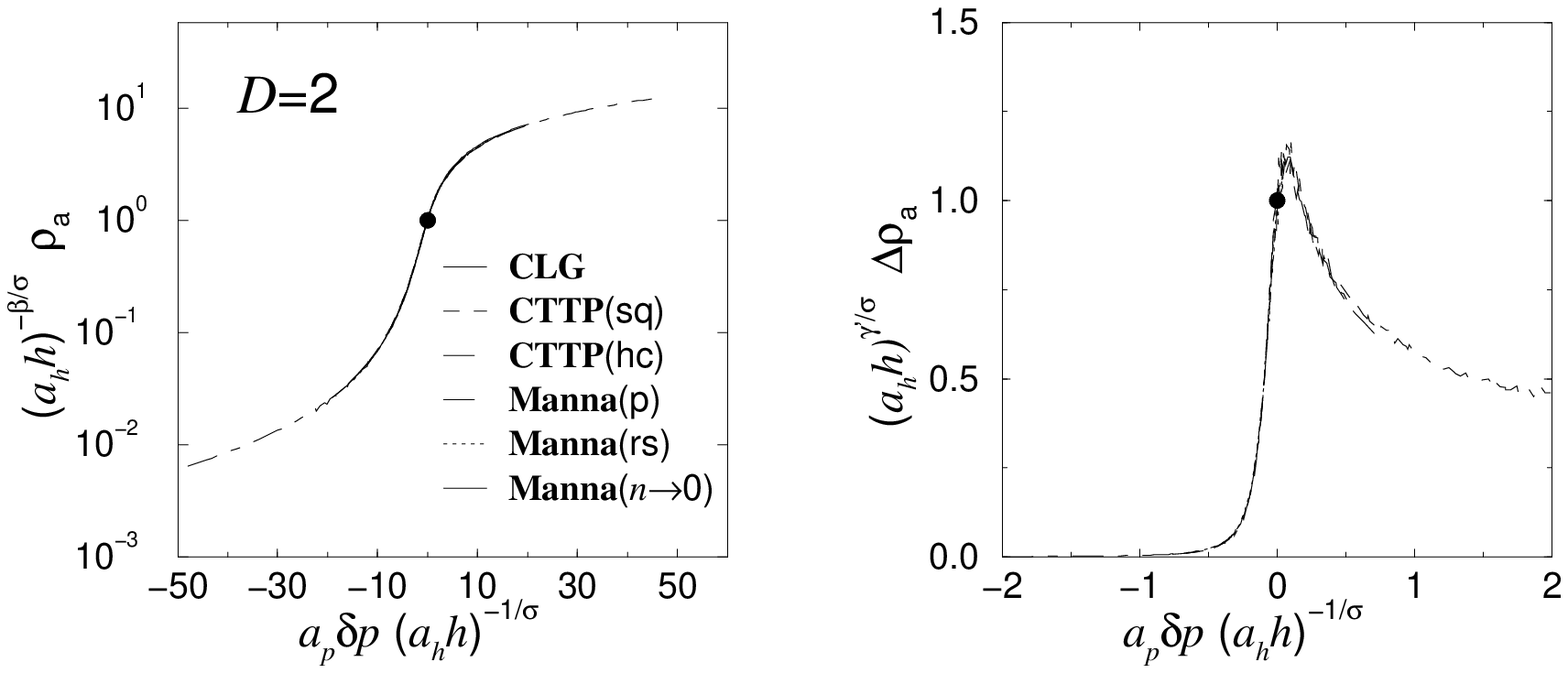}
\caption[Universal scaling functions ${\tilde R}(x,1)$ and ${\tilde D}(x,1)$, 
Manna class, $D=2$] 
{The universal scaling functions ${\tilde R}(x,1)$ and 
${\tilde D}(x,1)$ of the
Manna universality class for $D=2$.
The data are plotted according 
to \refeqname{eq:scal_ansatz_EqoS_HS_collapse}.
The CTTP data are obtained from simulations on a square (sq)
and honeycomb (hc) lattice.
In case of the Manna model, data of the original 
model [\refeqname{eq:manna_rule_original}]
using random sequential (rs) and parallel (p) update 
as well as data of the modified model ($n\to 0$) 
[\refeqname{eq:manna_rule_modified}] are presented.
All considered models are characterized by the same universal
scaling functions demonstrating the robustness of the
Manna universality class with respect to 
variations of the microscopic interactions.
For all considered models the scaling plots contain at least four
different curves corresponding to four different
field values (see e.g.~\reffigname{fig:manna_cttp_2d_op}).
The circles mark the conditions ${\tilde R}(0,1)=1$
and ${\tilde D}(0,1)=1$, respectively.
\label{fig:uni_manna_eqos_HS_2d}} 
\end{figure}

In~\cite{ROSSI_1}, Rossi~{\textit{et al.}} conjecture
that all stochastic models with an infinite number
of absorbing states and an order parameter coupling to a
non-diffusive conserved field define a unique universality
class that is different from directed percolation. 
At the beginning we verify that the Manna model,
the CTTP and the CLG belong to the same
universality class~\cite{ROSSI_1}.
The scaling analysis of the equation of state and of the 
fluctuations are plotted in \reffigname{fig:uni_manna_eqos_HS_2d}.
The corresponding exponents are listed 
in \reftablename{table:manna_exponents}.
The data of all models collapse onto the same scaling 
functions ${\tilde R}(x,1)$ and ${\tilde D}(x,1)$
despite the fact that bosonic and non-bosonic models
on different lattice types using different update schemes
are considered.
This result is a strong manifestation of the robustness of the
Manna universality class with respect to 
variations of the microscopic interactions.
Furthermore, the data-collapse confirms the accuracy of the
numerically estimated values of the critical 
exponents (see \reftablename{table:manna_exponents}).

\begin{figure}[t] 
\centering
%\leavevmode 
\includegraphics[clip,width=13cm,angle=0]{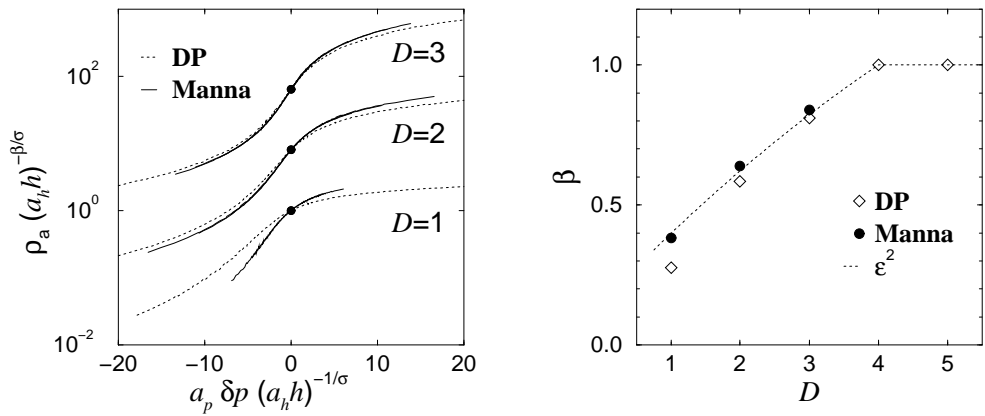}
\caption[Universal scaling functions ${\tilde R}(x,1)$, 
Manna and DP class, $D=1,2,3$] 
{The universal scaling functions ${\tilde R}(x,1)$ of the
Manna and directed percolation (DP) universality class
for $D=1,2,3$ (left).
Data of the CLG ($D=2,3$), the CTTP ($D=2,3$) as well as of 
the Manna model ($D=1,2,3$ with parallel and random sequential update)
are shown.
The two- and three-dimensional data are vertically shifted by 
a factor $8$ and $64$ in order to avoid overlaps.
The circles mark the condition ${\tilde R}(0,1)=1$.
At least four different field values are plotted for each
model and dimension.
As can be seen the universal scaling functions of the Manna
and DP class differs significantly, reflecting the different 
universality.
The order parameter exponents~$\beta$ of both classes are
presented in the right figure.
In all cases the error-bars are smaller than the symbol
sizes and no overlap within the error-bars occurs 
(see \reftables2name{table:dp_exponents}{table:manna_exponents}).
Additionally to the exponent values the renormalization 
group estimates of the DP class 
[\refeqname{eq:dp_exp_epsilon2_beta}] are shown.
\label{fig:uni_manna_eqos_HS_123d_02}} 
\end{figure}

\begin{figure}[t] 
\centering
%\leavevmode 
\includegraphics[clip,width=13cm,angle=0]{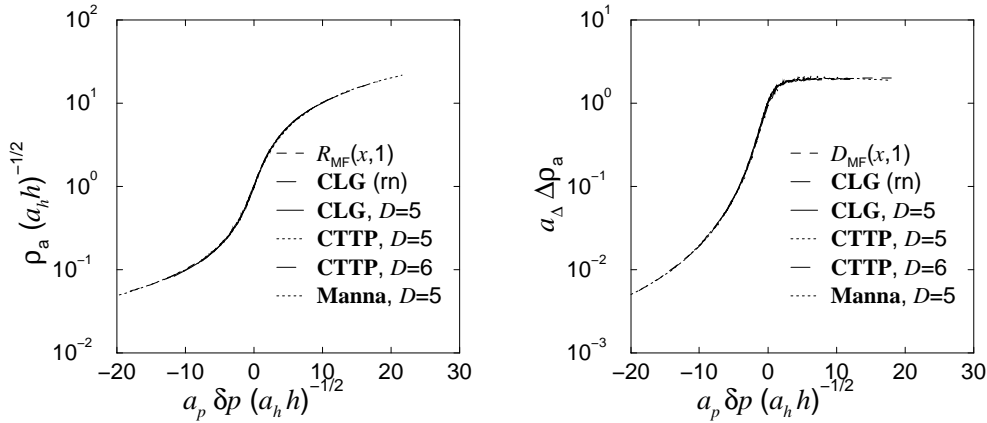}
\caption[Universal scaling functions ${\tilde R}(x,1)$ 
and ${\tilde D}(x,1)$, Manna class, $D=5,6$]
{The universal scaling function of the order parameter
${\tilde R}(x,1)$ (left) and the fluctuations
${\tilde D}(x,1)$ (right) for $D=5,6$.
Additionally to the five- and six-dimensional 
models we present data of a two-dimensional CLG model
with random neighbors (rn)~\protect\cite{LUEB_25}.
At least four different field values are plotted for each
model.
All presented data agree perfectly with the corresponding
mean field functions \refeqname{eq:uni_scal_mf_EqoS_Rx1}
and \refeqname{eq:uni_scal_mf_Fluc_Dx1}, respectively.
\label{fig:uni_manna_eqos_fluc_5d}} 
\end{figure}

After verifying that the Manna model, the CLG as well as the
CTTP share the same scaling behavior we show that 
the universality class differs from that of directed percolation.
Therefore we compare in \reffigname{fig:uni_manna_eqos_HS_123d_02}
the order parameter exponents~$\beta$ as well as the
scaling functions ${\tilde R}(x,1)$ for the DP class 
and the Manna class. 
As can be seen both classes are characterized by significantly
different scaling functions and critical exponents.
Thus as conjectured in~\cite{ROSSI_1}, stochastic absorbing
phase transitions with a conserved field define a unique
universality class different from directed percolation.
Evidently, the difference between the scaling functions and the
exponents reduces with increasing dimension.
Since both classes are characterized by the same mean field 
equation ${\tilde R}(x,y)$, the 
question of the value of the upper critical dimension
of the Manna universality class arises immediately.
%%%%%%%%%%%%%%%%%%%%%%%%%%%%%%%%%%%%%%%%%%%%%%%%%%%%%%%%%%%%%%%%%%%%%%%%%%%%%%%%%%
%%%%%%%%%%%%%%%%%%%%%%%%%%%%%%%%%%%%%%%%%%%%%%%%%%%%%%%%%%%%%%%%%%%%%%%%%%%%%%%%%%
Unfortunately, no convincing answer is provided by field
theoretical approaches so far.
The contrary values $\Dc=4$~\cite{PASTOR_2} 
and $\Dc=6$~\cite{WIJLAND_1} are conjectured
from field theories.
In order to obtain numerically a reliable estimate of $\Dc$ the same
method as described in section~\ref{subsec:dp_at_dc} is applied,
i.e., the data of high-dimensional models are compared
to the analytically known mean field equations.
The result is presented
in \reffigname{fig:uni_manna_eqos_fluc_5d} where the order parameter
as well as fluctuation data of five- and six-dimensional
models are shown.
Furthermore, the data of a modified version of the two-dimensional
CLG with random neighbor hopping of active particles are presented~\cite{LUEB_25}.
There, the unrestricted particle hopping breaks long-range
correlations and the system exhibits mean field scaling behavior.
In all cases the numerical data are in a perfect agreement
with the mean field scaling functions ${\tilde R}(x,1)$
and ${\tilde D}(x,1)$, respectively.
Thus the data-collapse leads to the upper  
bound on the critical dimension $\Dc<5$,
clearly excluding $\Dc=6$ as predicted in~\cite{WIJLAND_1}.

\begin{figure}[t] 
\centering
%\leavevmode 
\includegraphics[clip,width=13cm,angle=0]{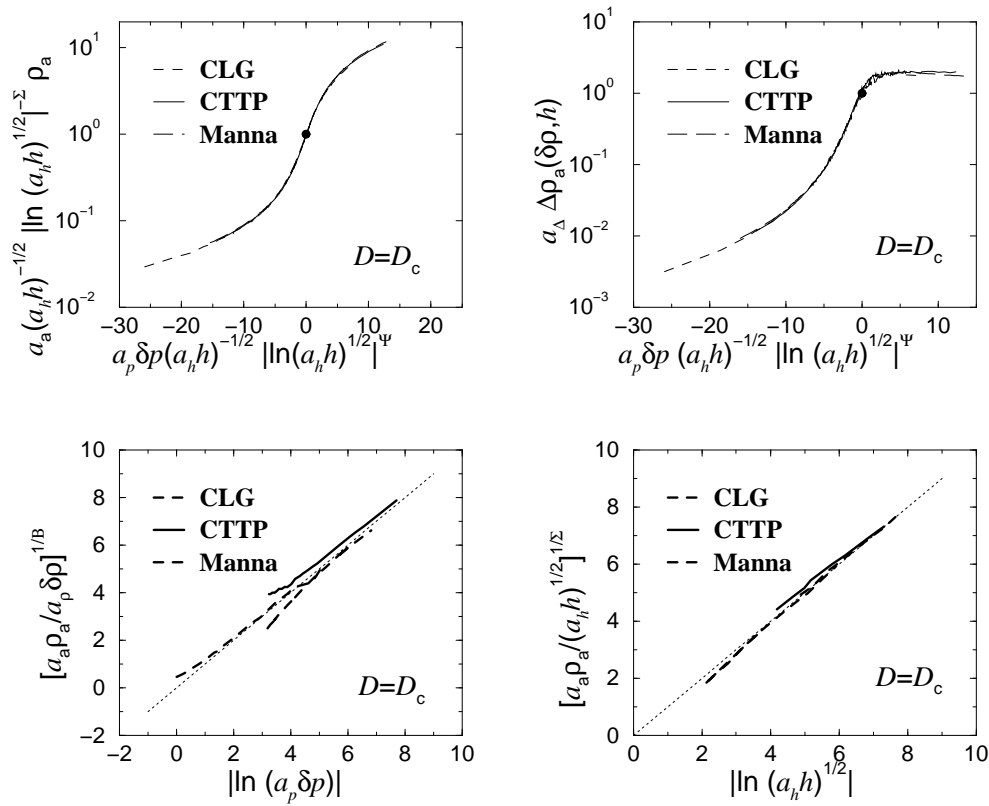}
\caption[Universal scaling functions ${\tilde R}(x,1)$
and ${\tilde D}(x,1)$, Manna class, $D=\Dc$]
{The universal scaling functions
of the order parameter (upper left)
and its fluctuations (upper right) 
at the upper critical dimension $\Dc=4$.
The logarithmic correction exponents are given by
${\mathrm B}=0.20$, $\Sigma = 0.35$, 
and $\Psi=-0.15$~\protect\cite{LUEB_25}.
The lower figures show the order parameter 
for zero field (left) and at the 
critical control parameter (right), respectively.
The order parameter is rescaled according to 
\refeqname{eq:uni_scal_OPzf_dc} and 
\refeqname{eq:uni_scal_OPcp_dc}.
Approaching the transition point 
($h\to 0$ and $\deltap \to 0$) the data tend to the function
$f(x)=x$ (dashed lines) as required (see text).
The circles mark the condition ${\tilde R}(0,1)=1$
and ${\tilde D}(0,1)=1$, respectively.
\label{fig:uni_manna_eqos_fluc_4d}} 
\end{figure}

The observed mean field scaling behavior for $D\ggl 5$ and the
observed non-mean field scaling behavior for $D\kgl 3$ 
restricts the value of the\index{upper critical dimension}
upper critical dimension to
$3< \Dc <5$.
Neglecting non-integer values (which may occur for example
in systems with long-range interactions)
we conclude $\Dc=4$.
This value is also supported by the Ginzburg\index{Ginzburg criterion} 
criterion (see chapter \ref{chapter:crossover})
\begin{equation}
\betaMF=1 \, , \quad 
\nuperpMF=1/2 \, , \quad 
\gammapMF=0  \quad \quad
\Longrightarrow \quad \quad
\Dc \; = \;
\frac{\, \gammapMF + 2 \beta^{\phantom \prime}_{\ssstyle \mathrm{MF}}}
{\nuperpMF}
\; = \; 4 \, .
\end{equation}
Furthermore, $\Dc=4$ is confirmed by direct numerical 
investigations~\cite{LUEB_26}.
Again, the scaling behavior at $D=4$ is characterized by
logarithmic corrections (see \reffigname{fig:uni_manna_eqos_fluc_4d}).
The performed numerical analysis reveals the estimates
for the logarithmic correction exponents 
${\mathrm B}=0.20\plmi 0.06$, $\Sigma = 0.35 \plmi 0.05$, 
and $\Psi={\mathrm B}-\Sigma=-0.15\plmi 0.08$~\protect\cite{LUEB_25}.
As expected these values differ
significantly from the corresponding values of 
directed percolation ${\mathrm B}=\Sigma = 1/3$~\cite{JANSSEN_3}.
Since the logarithmic corrections make the analysis
of the numerical data notoriously difficult it is
desirable to calculate the exponents 
${\mathrm B}$ and $\Sigma$ within a renormalization group
approach.
Unfortunately, these calculations are unavailable so far.

In summary, the investigations of the equation of state
reveals that all considered models belong to the
same universality class which differs from the
directed percolation behavior for $D \kgl \Dc$.
The upper critical dimension is four, which is also
supported by simulations of the Manna model in the
context of self-organized criticality~\cite{LUEB_4,LUEB_PHD}.
It is worth mentioning that the directed percolation 
and the Manna universality class share the same
mean field scaling behavior above $\Dc=4$.
A similar scenario is known from equilibrium
phase transitions, where the Ising model,\index{universality class, Ising}
the XY-model\index{universality class, XY-model}
as well as the Heisenberg model\index{universality class, Heisenberg} 
exhibit different scaling behavior below~$\Dc$
but share the same mean field behavior above~$\Dc$.
Another example is known from restricted random 
walks, like self-avoiding\index{random walk, self-avoiding} 
random walks~\cite{MADRAS_1}
or loop-erased\index{random walk, loop-erased} 
random walks~\cite{LAWLER_1,LAWLER_2,CONIGLIO_1},
which differ below~$\Dc=4$ 
whereas they exhibit the scaling behavior 
of a simple random walk above~$\Dc$~\cite{LAWLER_BOOK}.

\subsection{Universal amplitude combinations}
\label{subsec:aptcf_ampl_comb}

\begin{figure}[t] 
\centering
%\leavevmode 
\includegraphics[clip,width=8cm,angle=0]{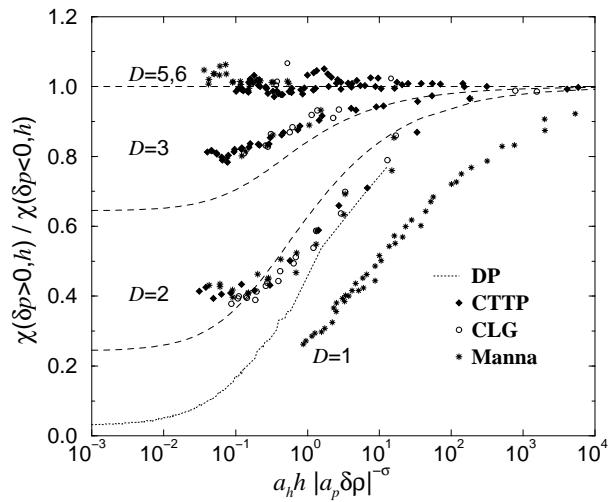}
\caption[Universal scaling function
${\tilde \mathrm{X}}(1,x)/{\tilde \mathrm{X}}(-1,x)$, Manna class]
{The universal scaling function 
${\tilde \mathrm{X}}(1,x)/{\tilde \mathrm{X}}(-1,x)$ 
for various dimensions and models.
The dashed lines mark the corresponding behavior 
${\tilde \mathrm{X}}(1,x)/{\tilde \mathrm{X}}(-1,x)$
for the directed percolation universality class.
The universal amplitude 
${\tilde \mathrm{X}}(1,0)/{\tilde \mathrm{X}}(-1,0)$ 
is obtained from the extrapolation 
$a_{\ssstyle h} h |a_{\ssstyle p} \, \delta p|^{-\sigma} \to 0$
leading to $0.41\pm 0.05$ ($D=2$) and $0.80\pm 0.05$ ($D=3$).
\label{fig:uni_manna_mom_susc}} 
\end{figure}

Universal amplitude combinations are related to
particular values of the universal scaling functions.
Therefore we expect that the amplitude combinations
of the Manna class differ from the corresponding
values of the DP class, as the scaling functions ${\tilde R}$
and ${\tilde D}$ do.
Here, we focus on the susceptibility ratio 
${\tilde \mathrm{X}}(+1,x)/{\tilde \mathrm{X}}(-1,x)$ 
[see \refeqname{eq:uni_ampl_comb_sus_X_x1}].
The susceptibility is obtained by performing 
the numerical derivative of the order parameter
with respect to the conjugated field.
The data of the Manna model, the conserved lattice gas,
and of the conserved threshold transfer process are
shown in \reffigname{fig:uni_manna_mom_susc} 
for various dimensions.
For each dimension the data of the different models collapse onto a 
single universal curve.
Similar to directed percolation,
the systems are well described by the mean field solution
far away from the transition point 
because the critical fluctuations are suppressed.
Approaching criticality ($h\to 0$) the critical fluctuations
increase and a crossover to the $D$-dimensional behavior
takes place.
Below the upper critical dimension the data of the
susceptibility ratio differ from the DP behavior.
In particular, the zero field universal
amplitude ratio ${\tilde \mathrm{X}}(+1,0)/{\tilde \mathrm{X}}(-1,0)$
reflects the different universality classes.
Numerically this ratio is obtained by 
an extrapolation of the susceptibility ratio 
${\tilde \mathrm{X}}(+1,x)/{\tilde \mathrm{X}}(-1,x)$ 
to $x\to 0$.
The estimated values for $D=2,3$ are listed in 
\reftablename{table:manna_exponents}.
In order to estimate the value of the
one-dimensional Manna universality class additional simulations are
required since no saturation occurs so far.

\subsection{Finite-size scaling behavior}
\label{subsec:aptcf_fss_below_Dc}

\begin{figure}[h] %%%%%%%%%%%%%%%%%%%%%%%%%%%%%%%%%%%%%%%%%%%%%%%%%%%%%%%%%%%%%%
\centering
%\leavevmode 
\includegraphics[clip,width=13cm,angle=0]{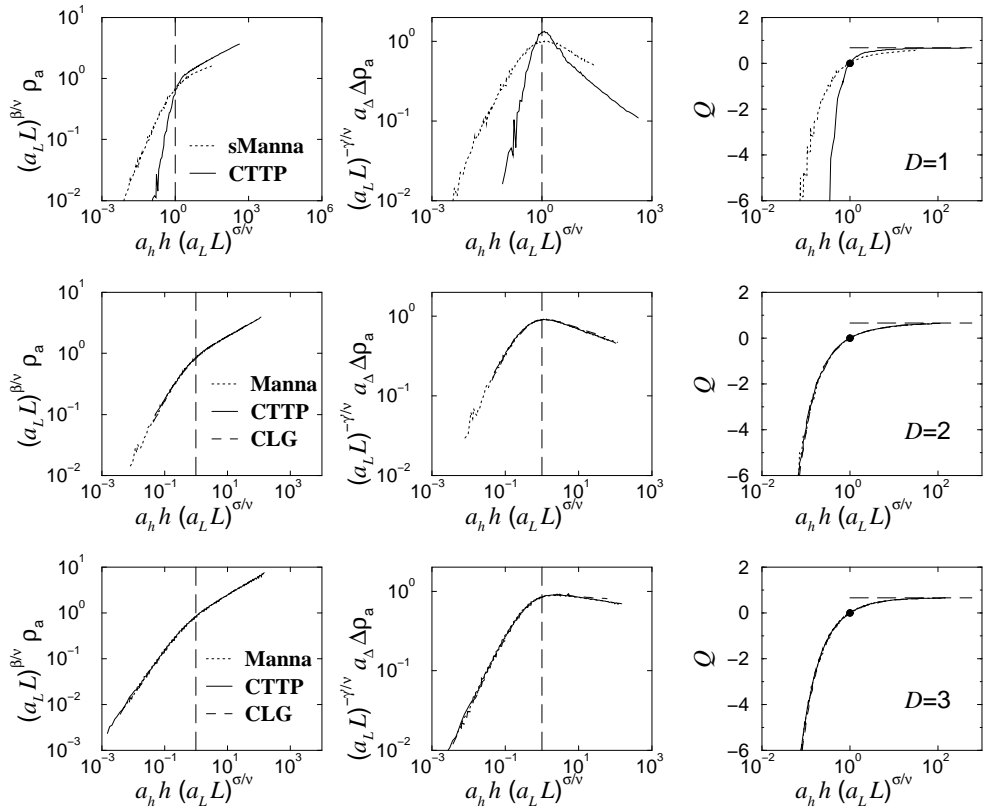}
\caption[Universal finite-size scaling functions, Manna class, $D<\Dc$] 
{The universal finite-size scaling analysis of the
Manna universality class in various
dimensions.
In case of the one-dimensional models, the expected splitting of the 
universality class is observed clearly.
For $D\ggl 2$ all models considered belong to the same 
universality class.
The filled circles mark the condition \refeqname{eq:metric_factors_a_L}.
The vertical lines correspond to the cumulant limit $2/3$.
The data are obtained from simulations of systems sizes
$L=2048,4096,8192$ for $D=1$,
$L=64,128,256$ for $D=2$, and 
$L=16,32,64$ for $D=3$.
\label{fig:uni_manna_fss_d123}} 
\end{figure}

So\index{finite-size scaling}\index{universal finite-size scaling} 
far we considered systems of the Manna universality
class where the correlation length~$\xi_{\senk}$ is small compared 
to the system size~$L$.
Additionally to these simulations of {\textit{infinite}} systems, 
finite-size effects are analysed in the literature~\cite{LUEB_23}.
Analogous to the method described in section~\ref{subsec:dp_fss_below_Dc}
the field dependence of the order parameter~$\rhoa$, 
of its fluctuations~$\Delta\rhoa$ as well as of the
fourth order cumulant~$Q$ at the critical point are 
examined.
Beside of the provided universal scaling functions,
the finite-size scaling analysis of the cumulant
provides accurate estimates of the correlation length
exponent~$\nu_{\senk}$.
The results are listed in \reftablename{table:manna_exponents}.
The finite-size scaling analysis is shown 
in \reffigname{fig:uni_manna_fss_d123} for various models
and $D<\Dc$.
 %%%%%%%%%%%%%%%%%%%%%%%%%%%%%%%%%%%%%%%%%%%%%%%%%%%%%%%%%%%%
As can be seen, the order parameter and its fluctuations 
obey the expected algebraic field dependence for 
$a_{\senk} \xi_{\senk} \ll a_{\ssstyle L} L$.
The fluctuation curves are characterized by a 
clear maximum, signalling its singular behavior
in the thermodynamic limit.
Approaching the transition point,
the fourth-order cumulant tends to infinity
for all models in all dimensions.
Good data-collapses are observed in case of the two-
and three-dimensional models.
In case of the one-dimensional models the expected
splitting of the universality class, i.e., the anomalous
scaling behavior of the $1+1$-dimensional CTTP is 
seen clearly.
Thus additionally to the universal scaling functions of 
infinite systems, 
universal finite-scaling functions are very useful
in order to identify a system's universality class.

\section{Dynamical scaling behavior and hyperscaling relations}
\label{sec:aptcf_dyn_scaling}

In the following we consider the dynamical scaling 
behavior of the Manna model, the conserved lattice gas, and
the conserved threshold transfer process.
Similar to directed percolation we investigate both
homogeneous and localized particle sources.
The obtained estimates of the critical exponents
allow to check certain hyperscaling relations.
It turns out that hyperscaling holds, although the Manna 
universality class is characterized by a scaling anomaly 
which is not understood so far.

\subsection{Homogenous and localized particle source}
\label{subsec:manna_dyn_scal_hom_loc}

\begin{figure}[t] 
\centering
%\leavevmode 
\includegraphics[clip,width=14.3cm,angle=0]{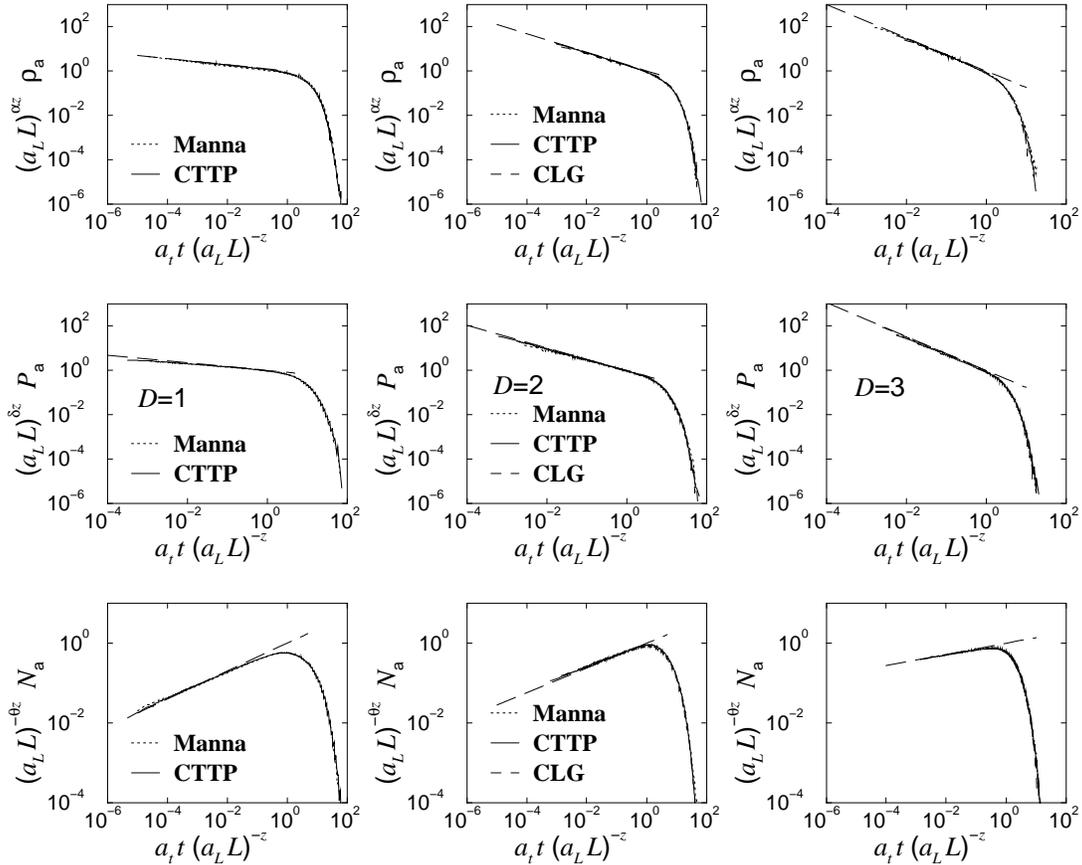}
\caption[Dynamical universal scaling functions, Manna class, $D< \Dc$]  
{The universal scaling functions 
${\tilde R}_{\ssstyle \mathrm{pbc},\Box,\mathrm{full}}$, 
${\tilde P}_{\ssstyle \mathrm{pbc},\Box,\mathrm{seed}}$, and 
${\tilde N}_{\ssstyle \mathrm{pbc},\Box,\mathrm{seed}}$
of the Manna universality class.
In case of activity spreading, the simulations
are started from a natural configuration of inactive
particles.
System sizes $L=512,1024,\ldots,8192$ are considered for $D=1$,
$L=64,128,256,512$ for $D=2$, and 
$L=16,32,64,128$ for $D=3$. 
The dashed lines correspond to the power-law
behavior of the infinite system $x^{-\alpha}$, 
$x^{-\delta}$, and $x^{\theta}$,
respectively.
\label{fig:uni_manna_act_spread_d123}} 
\end{figure}

Starting from a homogeneous particle source
the order parameter is expected to
obey an algebraic decay at criticality described by the exponent $\alpha$,
analogous to directed percolation [\refeqname{eq:ord_decay_alpha}].
Performing simulations, $N=\rhoc L^D$ particles are randomly distributed
on the lattice.
In this way an initial course grained homogeneous particle density is
created which differs significantly from the corresponding initial
density of directed percolation with $\rhoanull =1$.
For example, a random distribution of particles leads for the two-dimensional
CTTP to an initial density $\rhoanull \approx 0.1703$.
Of course, different values are obtained for the Manna model
and the conserved lattice gas.
Nevertheless, asymptotic universal scaling behavior is
observed for the order parameter decay. 
The corresponding scaling plots [see \refeqname{eq:scal_ansatz_EqoS_HS_dyn_fss}]
are shown in \reffigname{fig:uni_manna_act_spread_d123}.
The obtained estimates of the exponents~$\alpha$, $z$ 
are listed in \reftablename{table:manna_exponents}.
The values of the metric factors can be found 
in~\cite{LUEB_23}.

The spreading exponents $\delta$ and $\theta$
are obtained by simulations starting from a localized particle
source, i.e., from a single seed.
Due to the non-trivial absorbing state for the models considered 
the simulations have to be performed at the\index{natural density} 
natural density~\cite{JENSEN_3}.
Therefore, an absorbing state at criticality is prepared by the 
dynamics~\cite{LUEB_23}.
The scaling analysis of the survival probability as well as
of the average number of active particles are presented
in \reffigname{fig:uni_manna_act_spread_d123}.
The obtained values of the dynamical exponent~$z$
agree with those obtained from the order parameter decay measurements.
The results are listed in \reftablename{table:manna_exponents}.
It is worth mentioning that the Manna universality class is
characterized by the inequality $\alpha \neq \delta$.
For example, the two-dimensional systems yield the
estimates $\alpha=0.419\pm0.015$ and $\delta=0.510\pm0.020$~\cite{LUEB_23}.
This indicates the expected violation\index{symmetry, rapidity reversal}
of\index{rapidity reversal symmetry}
the rapidity reversal [that implies \refeqname{eq:p_sur_alpha_rho}].
This point is revisited in detail in the following section.

%%%%%%%%%%%%%%%%%%%%%%%%%%%%%%%%%%%%%%%%%%%%%%%%%%%%%%%%%%%%%%%%%%%%%%%%%%%%%%%%%%%
%%%%%%%%%%%%%%%%%%%%%%%%%%%%%%%%%%%%%%%%%%%%%%%%%%%%%%%%%%%%%%%%%%%%%%%%%%%%%%%%%%%
%%%%%%%%%%%%%%%%%%%%%%%%%%%%%%%%%%%%%%%%%%%%%%%%%%%%%%%%%%%%%%%%%%%%%%%%%%%%%%%%%%%

\subsection{Hyperscaling relations}
\label{subsec:manna_dyn_scal_hyp_scal}

\begin{figure}[t] 
\centering
%\leavevmode 
\includegraphics[clip,width=10.0cm,angle=0]{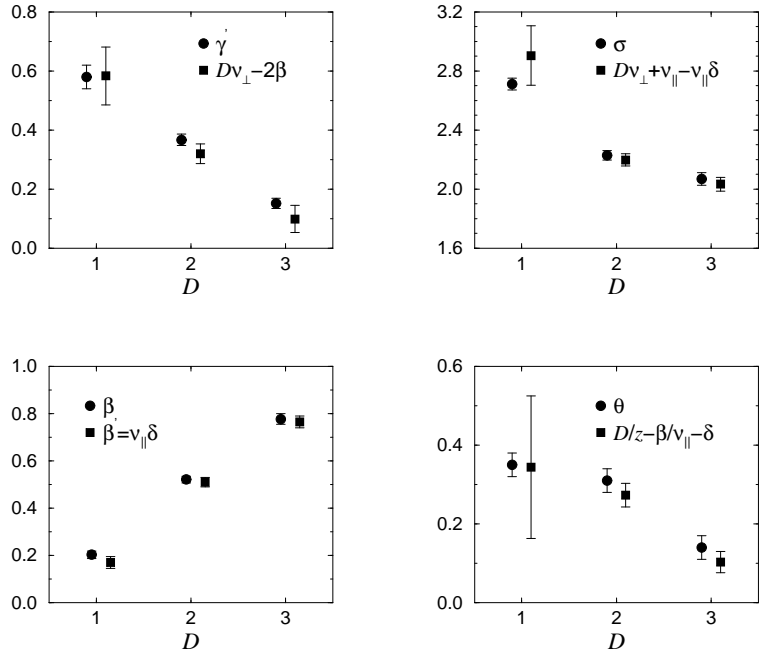}
\caption[Validation of hyperscaling relations, Manna class]  
{Test of certain scaling laws for the Manna universality class.
The hyperscaling laws $\gammap=D\nu_{\senk} -2 \beta$ (upper left),
$\sigma=D\nu_{\senk}+\nu_{\para}-\nu_{\para}\delta$ (upper right),
$\theta=D/z-\beta/\nu_{\para}-\delta$ (lower right), as well as the
equality $\beta=\betap$ (lower left) are checked.
All scaling laws are fulfilled within the error bars.
\label{fig:uni_manna_hyp_scal}} 
\end{figure}

In this section we check several hyperscaling laws.
The validity of the scaling law is immediately related
to the number of independent exponents.
Compared to directed percolation the number of independent
exponents is expected to be larger than three due 
to the violation of the rapidity reversal.
At the beginning we consider the scaling relation
$\gammap = D \nu_{\senk} - 2\beta$ [\refeqname{eq:gammap_nu_D_2beta}].
The corresponding data are plotted in \reffigname{fig:uni_manna_hyp_scal}.
As can be seen, the above scaling law is fulfilled within the 
error bars.

The next hyperscaling law connects the field exponent~$\sigma$
with the correlation exponents.
Taking into consideration that a weak field may trigger
spreading events the field exponent is
given by (see e.g.~\cite{HINRICHSEN_1} or \refeqname{eq:dim_analysis_field})
\begin{equation}
\sigma \; = \; D \, \nu_{\senk} \, + \, \nu_{\para} \,
- \nu_{\para} \, \delta \, .
\label{eq:sigma_D_nu_nu_nu_delta}
\end{equation}
Again, this scaling law is fulfilled within the error bars
(see \reffigname{fig:uni_manna_hyp_scal}).

In case of directed percolation the rapidity 
reversal [\refeqname{eq:p_sur_alpha_rho}]
implies $\alpha=\delta$ and 
therefore $\beta=\betap$ [\refeqname{eq:delta_beta_nu_para}]. 
In a more general context one expects that 
both exponents differ~\cite{HINRICHSEN_1}, 
for example in our case where
a conserved field couples to the order parameter.
The number of independent critical exponents
is therefore expected to be four ($\beta$, $\betap$, 
$\nu_{\para}$, $\nu_{\senk}$) 
instead of three independent exponents for directed percolation 
($\beta$, $\nu_{\para}$, $\nu_{\senk}$).
In order to check this scenario we compare 
the order parameter exponent~$\beta$ with $\nu_{\para} \delta$
in \reffigname{fig:uni_manna_hyp_scal}.
Surprisingly both values agree 
within the error-bars for all dimensions, suggesting 
$\beta = \betap$~\cite{LUEB_23}.
It is worth mentioning that $\beta=\betap$ is not restricted
to systems obeying 
the rapidity\index{rapidity reversal symmetry}\index{symmetry, rapidity reversal}
reversal.
For example, the universality class of 
dynamical\index{dynamical percolation}
percolation\index{universality class, dynamical percolation}
is characterized by $\beta=\betap$ although it 
violates \refeqname{eq:rapidity_trans} 
(see~\cite{ODOR_1,MUNOZ_3,JANSSEN_4} and references therein).
% dies gilt genau genommen nur, wenn man die Dichte nur
% innerhalb der debris misst, Janssen schreibt es schoen auf
In the following $\beta=\betap$ is assumed but we note
that an doubtless verification of that
equality requires direct measurements of the percolation 
probability.
This remains the topic of future research.

Finally, we consider the hyperscaling law\index{hyperscaling}
\begin{equation}
\theta \, + \, \frac{\,\beta\,}{\nu_{\para}} \, 
+ \, \delta \; \; = \; \;
\frac{\,D\,}{z}  \, .
\label{eq:gen_hyperscaling_law_2}
\end{equation}
Violations as well as validations of the 
hyperscaling law are reported in the literature (e.g.~\cite{ROSSI_1,LUEB_23}).
This confusion is caused by the fact that different forms of 
the hyperscaling law [\refeqsname{eq:gen_hyperscaling_law}{eq:theta_2delta_D_z}]
are checked in different studies. 
Here, the numerically obtained exponents are plotted
in \reffigname{fig:uni_manna_hyp_scal} according to 
\refeqname{eq:gen_hyperscaling_law_2}.
As can be seen, the above scaling law is fulfilled within the error bars.

%%%%%%%%%%%%%%%%%%%%%%%%%%%%%%%%%%%%%%%%%%%%%%%%%%%%%%%%%%%%%%%%%%%%%%%%%%%%%%%%%%%%%%
\begin{figure}[t] 
\centering
%\leavevmode 
\includegraphics[clip,width=13cm,angle=0]{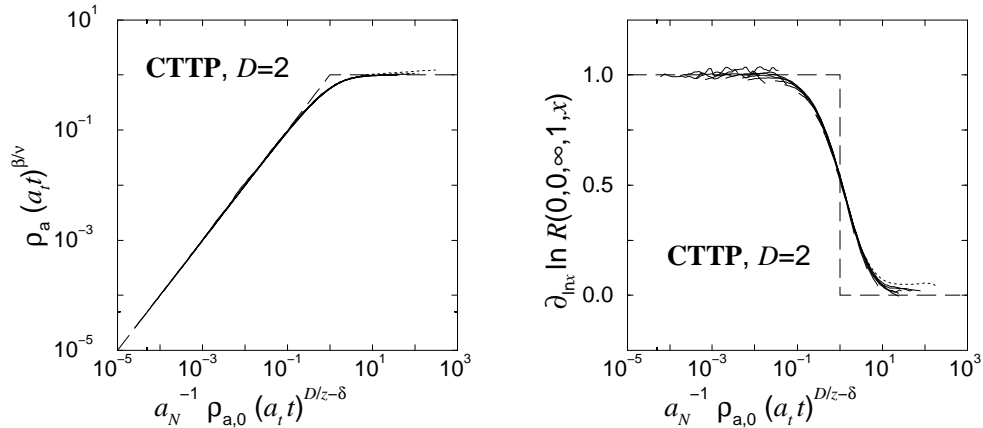}
\caption[Universal crossover function 
${\tilde R_{\ssstyle \mathrm{pbc},\Box}(0,0,\infty,1,x)}$, Manna class, $D=2$]  
{The universal scaling function 
${\tilde R}_{\ssstyle \mathrm{pbc},\Box}(0,0,\infty,1,x)$
describes the crossover from initial particle growth 
$\Na\asympprop t^{\theta}$ to the asymptotic order
parameter decay $\rhoa\asympprop t^{-\alpha}$.
The data are obtained from simulations of 
the two-dimensional conserved threshold transfer process (CTTP). 
Different initial particle densities 
$\rhoanull = n / L^2$ with $n=1,2,5,10,20,\ldots,20000$
for $L=1024, 2048$ and $n=160000$ for $L=4096$ 
are used.
Data that are affected by finite-size effects 
(${\mathcal O}(t)\ggl t_{\ssstyle \mathrm{FSS}}$) 
are skipped for a better visualization.
The long-dashed lines correspond to the 
asymptotic scaling behavior of the 
universal scaling function \refeqname{eq:R_pbc_asymptotic}.
The data are scrutinized using the effective exponent (left)
$\partial_{\ln{x}} \, \ln{{\tilde R(0,0,\infty,1,x)}}$.
For $\rhoanull \to 1$ systematic deviations from the
expected scaling behavior (long dashed line, see \refeqname{eq:R_pbc_asymptotic})
occur.
The largest deviations occur for $\rhoanull \approx 0.02$
and the corresponding curve is marked by a dashed line.
\label{fig:uni_manna_crossover_2d}} 
\end{figure} 
%%%%%%%%%%%%%%%%%%%%%%%%%%%%%%%%%%%%%%%%%%%%%%%%%%%%%%%%%%%%%%%%%%%%%%%%%%%%%%%%%%%%%%

The latter result is confirmed by independent measurements 
of the crossover from the initial particle growth to the asymptotic 
order parameter decay.
In section~\ref{subsec:dp_dynam_scal_located_source} we have seen
that \refeqname{eq:gen_hyperscaling_law_2} is related to the
scaling form 
\begin{eqnarray}
\label{eq:scal_ansatz_EqoS_HS_dyn_initial_2}   
%a_{\ssstyle a} \,
& & \rhoa(\deltap, h, L, t, \rhoanull) 
\; \sim  \; \\
& & 
\quad \quad \quad \lambda^{-\beta}\, \, {\tilde R}_{\ssstyle \mathrm{pbc},\Box}
(a_{\ssstyle p}  
\deltap \; \lambda, 
a_{\ssstyle h} h \;\lambda^{\sigma}, 
a_{\ssstyle L} L \;\lambda^{-\nu_{\senk}},
a_{\ssstyle t} t \;\lambda^{-\nu_{\para}},
a_{\ssstyle N}^{-1} \rhoanull \; \lambda^{D \nu_{\perp}-\nu_{\para} \delta} 
) \, . \nonumber
\end{eqnarray}
The hyperscaling law is valid if the scaling form is fulfilled
for $\rhoanull \to 0$.
On the other hand $\alpha=\beta/\nu_{\para}$ is
obtained if the scaling form holds for $\rhoanull \to 1$.
The scaling form can be checked by performing 
simulations with various initial densities $\rhoanull$.
Therefore, an absorbing state is prepared at criticality
by the natural dynamics.
Then $n$ different active seeds are created by $n$ randomly
selected particle movements on the lattice.
Results of  simulations of the two-dimensional
conserved threshold transfer process are 
presented in \reffigname{fig:uni_manna_crossover_2d}.
As can be seen, the above scaling form is fulfilled
for $\rhoanull \to 0$, confirming that the hyperscaling
law is fulfilled.
But significant deviations from the expected scaling
behavior of the universal scaling function
${\tilde R}_{\ssstyle \mathrm{pbc},\Box}(0,0,\infty,1,x)$
[\refeqname{eq:R_pbc_asymptotic}] occur for $x\gg 1$.
In particular, deviations are observed for $\rhoanull>0.001$.
Due to the violation of the scaling form for increasing initial
densities, the scaling law $\alpha=\beta/\nu_{\para}$ does 
not hold.
This implies the violation of \refeqname{eq:theta_alpha_delta_D_z},
as observed in~\cite{ROSSI_1}.

In conclusion, the hyperscaling law \ref{eq:gen_hyperscaling_law_2}
holds for directed percolation 
as well as for the Manna universality class.
But the universality class of directed percolation fulfills
\begin{equation}
\alpha \; = \; \frac{\,\beta\,}{\nu_{\para}}
\; = \; \frac{\betap}{\nu_{\para}}
\; = \; \delta
\label{eq:dp_uni_class_exp}
\end{equation}
whereas the Manna universality class is characterized by
\begin{equation}
\alpha \; \neq \; \frac{\,\beta\,}{\nu_{\para}}
\; = \; \frac{\betap}{\nu_{\para}}
\; = \; \delta \, .
\label{eq:manna_uni_class_exp}
\end{equation}
So far this scaling anomaly of the Manna class is not understood
(see also~\cite{RAMASCO_1}),
i.e., no scaling law is known involving the order parameter
decay exponent~$\alpha$.
Thus the number of independent exponents of the Manna universality
class is four if one assumes that $\beta=\betap$~\cite{LUEB_23}.

\section{Self-organized criticality}
\label{sec:apt_soc}

The concept of self-organized\index{self-organized criticality} 
criticality (SOC) refers\index{SOC}
to driven-dissipative systems that naturally evolve 
to a critical state, characterized by power-law 
distributions of relaxation events~\cite{BAK_1,BAK_2,BAK_3,TURCOTTE_1}.
The self-organization to the critical
point distinguishes SOC from ordinary 
critical phenomena, e.g.~continuous phase transitions
where a temperature-like control parameter has to be fine tuned
to the critical value.
At criticality the systems jumps among absorbing configurations
via avalanche-like relaxation processes.
In this case the scale invariance of the system is reflected by
an algebraic behavior of certain distribution functions,
characterizing e.g.~the avalanche size and duration. 
Besides of systems characterized by an extremal 
dynamics (see e.g.~\cite{BAK_4}),
the paradigms of SOC are the well-known sandpile\index{sandpile models}
models as the Bak-Tang-Wiesenfeld\index{Bak-Tang-Wiesenfeld model}\index{BTW} 
model (BTW)~\cite{BAK_1} or the
Manna sandpile\index{Manna model} 
model~\cite{MANNA_2}.
In particular the BTW model is analytically tractable
and its theoretical understanding is related 
to its Abelian structure~\cite{DHAR_2}.
This allows to classify stable configurations as
either transient or recurrent.
The recurrent configurations can be represented by\index{spanning tree}
spanning trees~\cite{MAJUMDAR_3} and
many steady state properties are calculated exactly, such as
the height probabilities or 
correlations~\cite{MAJUMDAR_1,PRIEZ_1,IVASH_1}.
On the other hand, the dynamical properties, i.e., the relaxation
events which are often called avalanches, are not fully understood.
A substantial progress was achieved by decomposing the avalanches
into a sequence of relaxation events called waves~\cite{IVASH_3,PRIEZ_2,IVASH_4}.
Due to a spanning tree representation of waves and due to 
the equivalence between spanning trees and 
loop-erased\index{random walk, loop-erased} random
walks exact results of the wave exponents are available~\cite{LUEB_10}.
Furthermore, the intersection probabilities
of random walks and loop-erased random walks 
allow to determine the upper critical 
dimension $\Dc=4$~\cite{PRIEZ_3}, confirming 
a numerical estimate~\cite{LUEB_5}.
But while the behavior of single waves is simple and well understood
the behavior of avalanches is complex and badly understood.
The avalanche exponents are still not known analytically for $D=2$
and reliable numerical estimates are difficult to obtain due to a lack
of simple finite-size scaling.
Thus the interpretation of the numerical data is highly 
controversial (see e.g.~\cite{MANNA_1,GRASSBERGER_1,LUEB_2,DEMENECH_1,TEBALDI_1,CHESSA_2}).

In contrast to the BTW model the Manna model was investigated
successfully by numerical simulations (see for 
example \cite{MANNA_2,BENHUR_1,LUEB_2,CHESSA_2,CHESSA_3,TEBALDI_1,LUEB_9,DICKMAN_2,BIHAM_1})
but no exact results are known so far.
A complete analytical solution of sandpile models is desirable
because our understanding of SOC is still limited.
In particular, such a solution would present a deeper insight into the 
fundamental mechanism of the self-organization to
the critical point.
So far it is recognized phenomenologically that the
critical stationary state is reached through the balance
of driving and dissipation (e.g.~\cite{GRINSTEIN_1,VERGELES_1,VESPIGNANI_1}).
As pointed out by Grinstein, 
the critical behavior is approached in the slow driving 
limit, the so-called separation of time scales~\cite{GRINSTEIN_1}.
An important advance in the understanding of
sandpile models was achieved by a mapping of sandpile models to absorbing phase 
transitions, i.e., to a conventional critical phenomenon. 
This mapping was first discussed within a mean field approach
in~\cite{TANG_1} and elaborated 
in a series of works~\cite{VESPIGNANI_5,DICKMAN_1,VESPIGNANI_1,VESPIGNANI_2}.
Both, the Manna and the BTW model exhibit absorbing phase 
transitions in closed ensembles where the particle density
is strictly conserved (often called fixed-energy\index{fixed-energy sandpiles}
sandpiles~\cite{DICKMAN_1}).
But numerical investigations of the BTW model are notoriously difficult
since it is trapped in limit-cycles due to the deterministic 
dynamics. 
For example, it is still open whether the BTW model exhibits an
absorbing phase transition of first 
or second order~\cite{BAGNOLI_1}.
On the other hand, the various representatives
of the Manna universality are class suitable for numerical 
investigations due to the stochastic dynamics.
Therefore, we focus again on the Manna universality class.
In particular, we consider the dynamical mean field theory
of the conserved threshold transfer process in order to
elucidate the mechanism of the self-organization to criticality.
Numerous mean field treatments of SOC systems
were published, 
e.g.~\cite{TANG_1,ALSTROM_1,DHAR_3,JANOWSKY_1,GARCIA_1,ZAPPERI_1,IVASH_2,KATORI_1,VERGELES_1,CALDARELLI_1,VESPIGNANI_1}.
But most studies focus on the determination of the 
avalanche exponents 
(e.g.~by a mapping to a branching process)\index{branching process}
and do not explain the self-organization to criticality.
The following mean field theory presents a complete picture
of SOC in sandpile models, explaining how the critical state
is attracted by the dynamics as well as the criticality itself.
Due to the mapping of sandpile models to absorbing phase transitions,
the SOC avalanche exponents can be expressed in terms of ordinary 
critical exponents of second order phase transitions~\cite{LUEB_23},
like $\beta$, $\nu_{\para}$, and $\nu_{\senk}$.
This will be discussed in section \ref{subsec:soc_apt_beyond_mf}.

%%%%%%%%%%%%%%%%%%%%%%%%%%%%%%%%%%%%%%%%%%%%%%%%%%%%%%%%%%%%%%%%%%%%%%%%%%%%%%%%%%%%

\subsection{Sandpile models and absorbing phase transitions}
\label{subsec:apt_soc_mapping}

The\index{mean field theory, SOC}
mean field behavior of the conserved threshold transfer process 
with global particle conservation 
is considered in section \ref{subsec:cttp}.
In the following we consider a driven-dissipative modification.
In addition to the dynamical rules above we introduce first
an external drive:~A particle is added from 
outside with probability~$p$ at 
an empty or single occupied lattice site.
Obviously this perturbation breaks the global particle
conservation.
Second we incorporate particle 
dissipation which counteracts the external drive.
For the sake of simplicity we assume that 
a given target site belongs to the (dissipative)
boundary and the particle will leave the system ($\Delta n = -1$).
The corresponding rate equations lead 
to the coupled differential equations
\begin{eqnarray}
\label{eq:cttp_ord_par_langevin_p_e}
\partial_{t} \rhoa & = &
p \, (\rho - 2 \, \rhoa) \ +  \,
\rhoa \, [ -1 + 2 (1-\epsilon) \rho - 4 (1-\epsilon) \rhoa + 
(1-\epsilon)^2\rhoa^2]     \\[2mm]
& + & h \, (\rho - \rhoa)^2 \, ,  \nonumber\\[2mm]
\label{eq:cttp_par_den_langevin_p_e}
\partial_{t} \rho & = & p (1-\rhoa) \, - \, 2 \, \epsilon \, \rhoa \, .
\end{eqnarray}
The latter equation reflects the competition between the 
particle flux in [$p(1-\rhoa)$] and 
out ($2\epsilon \rhoa$) of the system.
Global particle conservation $\rho={\mathrm{const}}$
as well as \refeqname{eq:cttp_mean_Langevin} are recovered
for $p=\epsilon=0$.
For $p=0$, $\epsilon>0$, and $h>0$ the system tends to 
the empty lattice ($\rho=0$ and $\rhoa=0$)
whereas it tends to the fully occupied lattice 
($\rho=2$ and $\rhoa=1$) for $p>0$, $\epsilon=0$.

The steady state solution of the driven-dissipative system
at zero field is given by
\begin{equation}
\label{eq:cttp_steady_state_rho_a_rho}   
\rhoa \; = \;
\frac{p}{\, 2\, \epsilon \, + \, p \,} \, , \quad  \quad  \quad 
\rho \; = \; \frac{\, 2 \, p^2  \, + \, 6 \, \epsilon \, p  \, + \, \epsilon^2 \, (2-p)\, }
{(2\, \epsilon\, +\, p)^2} \, .
\end{equation}
Thus the absorbing phase ($\rhoa=0$) is 
not a steady state solution of the driven-dissipative system
and the control parameter 
is now lost as an independent quantity.
%is now a function of the driving and 
%dissipation rate.
A linear stability analysis reveals that the above
solution is an attractor of the dynamics for all 
values of $\epsilon$ and $p$.
In the slow driving limit ($p \ll 1$) 
the steady state solution
\refeqname{eq:cttp_steady_state_rho_a_rho}
can be expressed in terms of the drive-dissipation ratio
$\kappa=p/ \epsilon$
\begin{equation}
\rhoa \; = \;
\frac{\kappa}{\, 2 + \kappa \,} \, , \quad \quad  \quad
\rho   = 
\frac{2\,  + \, 6 \, \kappa \, + \, 2 \, \kappa^2}{(2 + \kappa)^2} \, .
\label{eq:cttp_steady_state_kappa}
\end{equation}
It turns out that the parameter~$\kappa$, which was 
phenomenologically introduced by Grinstein~\cite{GRINSTEIN_1}, 
is the appropriate parameter to describe the self-organization 
to the critical point.
Eliminating the parameter $\kappa$ 
%in Eq.\,(\ref{eq:steady_state_kappa}) 
%by $\kappa(\rhoa)$
one recovers \refeqname{eq:cttp_mean_field_ord_par_01}, i.e., 
the order parameter behavior in the active phase of the closed
systems
\begin{equation}
%\rhoa( \rho_{\ssstyle \kappa}) \; = \; 
\rhoa(\rho) \; = \; 2 \, - \, \sqrt{\, 5\, -\, 2\, \rho \, }
\quad \quad \mathrm{for} \quad\quad \rho \ggl 1/2 \, .
\label{eq:ord_par_closed_open}
\end{equation}
In this way, the control parameter in the active phase
$\rho \in [ \rhoc,2 ]$ is mapped to the interval of the drive-dissipative 
ratio $\kappa\in [0,\infty [$.
In the strong driving limit $p \gg \epsilon$ ($\kappa\to \infty$) 
the systems tends to the fully occupied lattice ($\rhoa=1$, $\rho = 2 $), 
%\begin{equation}
%\rhoa \to 1 \, , \quad \quad
%\rho  \to 2 \, , .
%\label{eq:steady_state_fast_driving}
%\end{equation}
whereas we find in the weak driving limit
$p \ll \epsilon $ ($\kappa\to 0$) in leading order 
\begin{equation}
\rhoa = \frac{\kappa}{2} %-\kappa^2/4 
\, , \quad \quad
\rho  = \rhoc \, + \, \kappa % \, -5 \kappa^2/8
\, ,\quad \quad 
{\mathrm{for}} \quad \kappa \to 0\, .
\label{eq:cttp_steady_state_slow_driving}
\end{equation}
In the limit $\kappa\to 0$, the driven-dissipative system tends 
to the critical point of the absorbing phase transition 
($\rho = \rhoc$, $\rhoa = 0$).

%%%%%%%%%%%%%%%%%%%%%%%%%%%%%%%%%%%%%%%%%%%%%%%%%%%%%%%%%%%%%%%%%%%%%%%%%%%%%%%%%%%%%%
\begin{figure}[t] 
\centering
%\leavevmode 
\includegraphics[clip,width=13cm,angle=0]{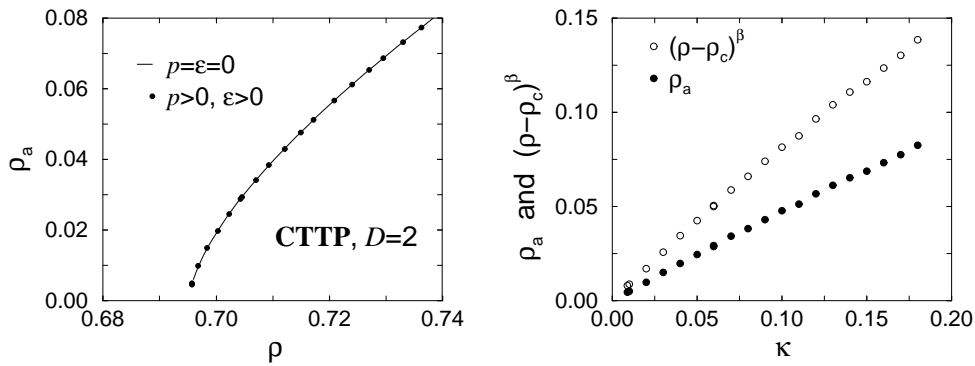}
\caption[Driven-dissipative conserved threshold transfer process, $D=2$]  
{The driven-dissipative conserved threshold transfer process for $D=2$.
The order parameter behavior $\rhoa(\rho)$ of the closed 
($p=\epsilon=0$) 
and 
open ($p>0$,$\epsilon>0$) ensemble are identical (left figure).
Data for $\epsilon=10^{-5}$ and $p=n\,10^{-8}$ with $n=9,10,20,30,40,\ldots\,180$
are shown, obtained from simulations of $L=128,256,\ldots,2048$.
The right figure shows the order parameter and the
rescaled control parameter as a function of the Grinstein
parameter~$\kappa=p/\epsilon$.
As in the mean field approximation, 
the two-dimensional order parameter~$\rhoa$
vanishes linearly if one approaches the transition 
point ($\kappa \to 0$).
As can be seen, the control parameter 
behaves as $(\rho-\rhoc)^{\beta}\asympprop \kappa$.
\label{fig:cttp_soc_2d}} 
\end{figure} 
%%%%%%%%%%%%%%%%%%%%%%%%%%%%%%%%%%%%%%%%%%%%%%%%%%%%%%%%%%%%%%%%%%%%%%%%%%%%%%%%%%%%%%

The crucial role of the Grinstein parameter~$\kappa$ is even
valid beyond the mean field approach.
Simulation data of the driven-dissipative CTTP
in $D=2$ are presented in \reffigname{fig:cttp_soc_2d}.
As can be seen, the data of the driven-dissipative CTTP 
recovers the active phase solution of the closed
CTTP, i.e., for $p=\epsilon=0$.
Furthermore, the critical point is approached for
$\kappa\to 0$.
In leading order, the two-dimensional system behaves as
\begin{equation}
\rhoa = \frac{\kappa}{2} %-\kappa^2/4 
\, , \quad \quad
\rho  = \rhoc \, + \, \kappa^{1/\beta} 
\, ,\quad \quad 
{\mathrm{for}} \quad \kappa \to 0\, .
\label{eq:cttp_steady_state_slow_driving_2d}
\end{equation}
Surprisingly, the two-dimensional order parameter
and the mean field order parameter behave in the same 
way, i.e., they vanish linearly for $\kappa \to 0$.
For the sake of completeness we just mention that
the corresponding fluctuations scale as 
$\Delta\rhoa\asympprop \kappa^{-\gammap}$ and 
$\Delta\rho\asympprop \kappa$.

Now, let us focus on the dynamical behavior close to the steady state
solution for $\kappa\to 0$.
Using \refeqs2name{eq:cttp_par_den_langevin_p_e}{eq:cttp_steady_state_slow_driving}) 
we find
\begin{equation}
\partial_{pt} \rho \; = \; {\mathcal O}(\kappa) \, ,
\quad\quad\quad \partial_{\epsilon t} \rho \; = \; {\mathcal O}(\kappa^2) \, ,
\label{eq:cttp_cont_par_slow_driving}
\end{equation}
%$\partial_{pt} \rho  =  {\mathcal O}(\kappa)$ 
%and $\partial_{\epsilon t} \rho = {\mathcal O}(\kappa^2)$
i.e, the control parameter varies for $\kappa\to 0$ only on the
perturbation scale $\tau_{\ssstyle \mathrm{perp}}=1/p$ whereas
it can be considered approximately as constant on the 
dissipation scale $\tau_{\ssstyle \mathrm{diss}}=1/\epsilon$.
Therefore we set $\rho=\mathrm{const}$ in the following.
Furthermore, we assume that the system displays an avalanche-like
response for infinitesimal driving ($p \to 0$), i.e., no
further perturbations take place until 
an absorbing configuration ($\rhoa=0$) is reached.
This extreme separation of time scales ($\kappa \to 0$)
is necessary in order to identify individual 
relaxation events (avalanches).
Only in this case it is justified to characterize the
scale invariance via avalanche distribution 
functions~\cite{GRINSTEIN_1}.
Otherwise the scale invariance is manifested in terms
of algebraic decaying correlation functions similar 
to usual phase transitions.

Within the approximations discussed above, 
the avalanche processes are determined by 
\refeqname{eq:cttp_ord_par_langevin_p_e} with $p=0$.
At criticality ($\rho=\rhoc$) we find
(neglecting again ${\mathcal O}(\rhoa^3)$)
\begin{equation}
\rhoa(t) \; = \; 
\frac{\rhoanull}
{  \;
\left ( 1 + 4\, \frac{1-\epsilon}{\epsilon} \, 
\rhoanull
\right ) {\mathrm e}^{\epsilon t}
\, - \, 4\, \frac{1-\epsilon}{\epsilon} \, 
\rhoanull \; } \, .
\label{eq:cttp_aval_dyn_01}
\end{equation}
For sufficiently small values of $\epsilon$ the
order parameter decays as
\begin{eqnarray}
\label{eq:cttp_ord_par_aval_long_time}
\rhoa(t) & \sim &  
\frac{1}{\; \rhoanull^{-1} \, + \, 4 \, (1-\epsilon) \, t \; } 
\quad\quad\quad {\mathrm{for}} \quad \quad \epsilon\,t \ll 1 \, ,
\\[2mm]
\label{eq:cttp_ord_par_aval_short_time}
\rhoa(t) 
& \sim &  
\frac{\epsilon}{4(1-\epsilon)} \, {\mathrm e}^{- \epsilon \, t}
\;\;\quad\quad\quad\quad\quad {\mathrm{for}} \quad \quad \epsilon\,t \gg 1 \, .
\end{eqnarray}
Thus close to the critical point ($\kappa \to 0$)
an avalanche, caused by the 
perturbation $\rhoanull$,
decays algebraically before an exponential cutoff occurs.
This cutoff is caused by the particle dissipation and 
takes place at $\tau_{\ssstyle \mathrm{cutoff}} =1/\epsilon$.
A pure power-law behavior is obtained in the limit
$\epsilon \to 0$ only.
Therefore the driven-dissipative CTTP self-organizes itself to
the critical point and exhibits scale invariant avalanches
in the SOC limit $\kappa\to 0$ and $\epsilon \to 0$~\cite{VERGELES_1,VESPIGNANI_1}.

It is customary in the literature to consider the response function 
\begin{equation}
\chi_{\ssstyle p} \; = \; 
\left .
\frac{\partial \rhoa}{\partial p} 
\right |_{p\to 0} \; = \;
\left .
\frac{2 \, \epsilon}{\, (2 \, \epsilon +p)^2} 
\right |_{p\to 0} 
%\; = \; 
%\left .
%\frac{\partial \rho_{\scriptscriptstyle \text a}}{\partial p} 
%\right |_{p\to 0} 
\; = \; \frac{1}{2 \epsilon} %\, ,
\label{eq:chi_p}
\end{equation}
that should be distinguished from the
susceptibility $\chi$ [\refeqname{eq:lin_resp_apt}].
It is often interpreted that the 
singular behavior of $\chi_{\ssstyle p}$ for 
$\epsilon\to 0$ 
signals that criticality is approached in
the SOC limit ($\kappa\to 0$, $\epsilon \to 0$).
But this is a misleading conclusion as can be seen
if one considers the more appropriate form 
\begin{equation}
\chi_{\ssstyle p} \; = \; 
\frac{\partial \rhoa}{\partial p} 
\; = \; \frac{2}{\, \epsilon \, (2+\kappa)^2} \, .
%\; \longrightarrow \; \infty 
\label{eq:chi_p_kappa}
\end{equation}
The response function $\chi_{\ssstyle p}$
exhibits for $\epsilon \to 0$ a singular behavior
in the whole active phase ($\kappa \ge 0$), not only 
at the critical point ($\kappa=0$). 
Thus the singular behavior of $\chi_{\ssstyle p}$
does not reflect the critical behavior at $\rhoc$.
It just reflects that an infinite response is obtained 
if one adds particles to a system with particle 
conservation ($\epsilon=0$).

It is worth to consider how the conjugated field
affects the driven-dissipative system.
%The steady state solution of Eq.\,(\ref{eq:cttp_ord_par_langevin_p_e})
%for non-zero field is given by
%\begin{equation}
%\rho \; = \;
%\frac{-\kappa + \kappa \sqrt{1+4 h \frac{1+\kappa(3+\kappa)}{\kappa\,(2+\kappa)}}}
%{h \, (2+\kappa)} \, .
%\label{eq:rho_dri_diss_non_zero_field}
%\end{equation}
%The behavior for vanishing drive-dissipation ratio~$\kappa$
%is therefore given by
The steady state solution of Eq.\,(\ref{eq:cttp_ord_par_langevin_p_e})
yields 
%$\rho(\kappa\to 0,h=0) =  \rhoc$,
%whereas $\rho(\kappa\to 0,h>0) = 0$.
\begin{equation}
\lim_{\kappa\to 0}{\; \rho(\kappa,h=0)} \,  = \, \rhoc \, , \quad\quad\quad
\lim_{\kappa\to 0}{\; \rho(\kappa,h>0)} \,  = \, 0 \, . 
\label{eq:rho_limit_kappa_to_0_field}
\end{equation}
%In other words, 
The external field drives the system
away from the critical point, i.e., 
SOC is destroyed if one applies an external field
that is conjugated to the order parameter.

In summary,
we have seen that the CTTP undergoes for $p=\epsilon=0$
an absorbing phase transition %at $\rho=\rho_{\scriptscriptstyle \text c}$
from an active to an absorbing phase.
The situation is completely changed if one considers
an open ensemble where fluctuations of the 
particle density occur.
In contrast to the closed ensemble ($p=\epsilon=0$)
the absorbing phase is no longer a solution of the
driven-dissipative system ($\kappa >0$).
The remaining solution of the open ensemble 
is exactly the active phase solution of the closed ensemble. 
It is the key to the understanding of SOC that the 
limit $\kappa \to 0$ tunes the control 
parameter to the boundary of the active phase, i.e.,
to the critical point.
This explains why the {\textit{trivial}\,} limit $\kappa \to 0$
in the open system
corresponds, in the closed system, to the fine tuning of the 
control parameter to the critical point of the 
absorbing phase transition
($\rho\to \rhoc$).
Remarkably, this limit occurs frequently in driven-dissipative
systems in nature, i.e., 
many physical processes are characterized by a large
separation of times scales that makes $\kappa$ extremely 
small~\cite{BAK_3,GRINSTEIN_1}.
For instance the motion of the tectonic plates
that causes earthquakes is extremely slow  
compared to the time scale on which earthquakes 
proceed (see~\cite{GRINSTEIN_1}).
Thus, the dissipation and driving rate are separated by
many orders of magnitude.
Although the drive-dissipation rate remains finite the 
system is very close to the critical point ($\kappa=0$) 
and displays a power-law behavior over several decades,
namely the well known\index{Gutenberg-Richter law}
Gutenberg-Richter law~\cite{GUTENBERG_1}.

%%%%%%%%%%%%%%%%%%%%%%%%%%%%%%%%%%%%%%%%%%%%%%%%%%%%%%%%%%%%%%%%%%%%%%%%%%%%%%%

\subsection{Avalanche exponents}
\label{subsec:soc_apt_beyond_mf}

In the previous section we discussed how SOC is mapped 
to an absorbing phase transition.
Usually self-organized critical systems are 
characterized by avalanche exponents whereas 
absorbing phase transitions are described by the
critical exponents $\beta$, $\betap$, $\nu_{\senk}$, and 
$\nu_{\perp}$.
Thus it remains to show how the avalanche exponents
are connected to the exponents of the second
order phase transition~\cite{LUEB_23}.

In the critical state of SOC systems the external driving~$p$ triggers
scale invariant avalanche-like relaxation events.
%these avalanche processes are described by certain
%critical exponents which can be derived from the spreading
%exponents $\delta$, $\theta$, and $z$~\cite{VESPIGNANI_4,MUNOZ_3}.
In particular the avalanches are characterized by several
quantities (see for example~\cite{BENHUR_1,LUEB_4}), e.g.~the
size~$s$ (number of elementary relaxation events),
the area~$a$ (number of distinct relaxed sites),
the time~$t$ (number of parallel updates until 
the configuration is stable),
as well as the radius exponent~$r$ (radius of gyration).
In the critical steady state the corresponding
probability distributions decay algebraically 
\begin{equation}
P_{\ssstyle x} \; \asympprop \;
x^{-\tau_{\ssstyle x}} \, ,
\label{eq:soc_prob_tau_x}
\end{equation}
characterized by the avalanche exponents 
$\tau_{\ssstyle x}$ with
$x\in \{s,a,t,r \}$.
Assuming that the size, area, etc.~scale as
a power of each other,
\begin{equation}
x \; \asympprop \;
{x^{\prime}}^{\gamma_{\ssstyle xx^{\prime}}} \,
\label{eq:soc_gamma_xx}
\end{equation}
the scaling laws
\begin{equation}
{\gamma_{\ssstyle xx^{\prime}}}
\; = \;
\frac{\,\tau_{\ssstyle x^\prime}-1\,}{\,\tau_{\ssstyle x}-1\,} %\; .
\label{eq:soc_gamma_xx_taux_taux}
\end{equation}
are obtained.
The exponent~$\gamma_{\ssstyle tr}$ equals 
the dynamical exponent~$z$,
the exponent~$\gamma_{\ssstyle ar}$ corresponds to
the fractal dimension of the avalanches and the 
exponent~$\gamma_{\ssstyle sa}$ indicates
whether multiple relaxations of a lattice site are relevant 
($\gamma_{\ssstyle sa}>1$) 
or irrelevant \mbox{($\gamma_{\ssstyle sa}=1$)}.

These avalanche exponents are connected to the
spreading exponents $\delta$, $\theta$, and 
$z$ (see e.g.~\cite{VESPIGNANI_4,MUNOZ_3,LUEB_23}).
First the survival 
probability~$\Pa(t)$ is simply
given by the integrated avalanche duration
\begin{equation}
\Pa(t) \; = \;
\sum_{t^{\prime}=t}^{\infty} \, P_{\ssstyle t}(t^{\prime})
\label{eq:soc_prob_a_prob_t}
\end{equation}
yielding
\begin{equation}
\tau_{\scriptscriptstyle t} \; = \;
1 \, + \, \delta \, .
\label{eq:soc_tau_t}
\end{equation}
Since $\gamma_{\ssstyle tr}=z$ the radius exponent
is given by
\begin{equation}
\tau_{\ssstyle r} \; = \;
1 \, + \, z\, \delta \, .
\label{eq:soc_tau_r}
\end{equation}
Taking into account that the avalanches of the Manna model
are compact ($\gamma_{\ssstyle ar}=D$) 
below the upper critical dimension 
$\Dc=4$~\cite{LUEB_4,LUEB_PHD} we find 
\begin{equation}
\tau_{\scriptscriptstyle a} \; = \;
1 \, + \, \frac{\,z\, \delta\,}{D} \, .
\label{eq:soc_tau_a}
\end{equation}
Finally the number of topplings~$s_{\ssstyle t}$ 
for an avalanche that is active at time~$t$ equals the 
integrated numbers of active sites, i.e.,
\begin{equation}
s_{\ssstyle t} \, \Pa(t)\; = \;
\sum_{{t^\prime} =0}^{t} \, \Na(t^{\prime})
\label{eq:soc_size_Na}
\end{equation}
leading to~\cite{MUNOZ_3}
\begin{equation}
\tau_{\ssstyle s} \; = \;
1 \, + \, \frac{\delta }{\, 1 + \theta + \delta \,} \, .
\label{eq:soc_tau_s}
\end{equation}
Thus the avalanche exponents of sandpile models are 
naturally related to the spreading exponents of absorbing
phase transitions.
%%%%%%%%%%%%%%%%%%%%%%%%%%%%%%%%%%%%%%%%%%%%%%%%%%%%%%%%%%%%%%%%%%%%%%%%%%%%%%%%%%%%%%
\begin{figure}[t] 
\centering
%\leavevmode 
\includegraphics[clip,width=8cm,angle=0]{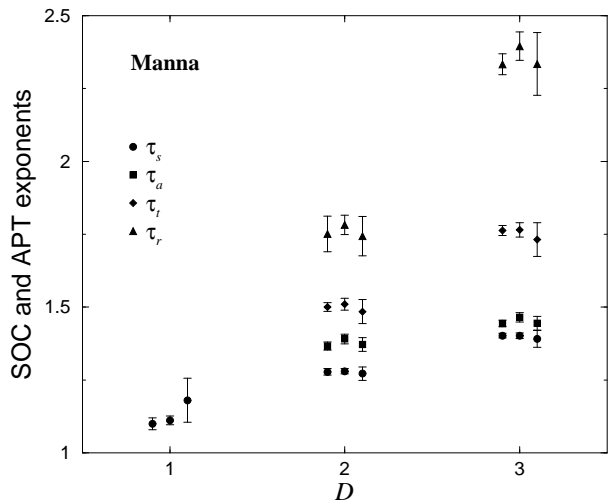}
\caption[Avalanche exponents of the self-organized critical Manna model, $D=2$]  
{The avalanche exponents of the Manna model
in various dimensions.
In order to avoid overlaps the exponents are slightly shifted.
The avalanche exponents of the SOC version of the Manna model
(left) are obtained from~\protect\cite{DICKMAN_6} for $D=1$
and from~\protect\cite{LUEB_PHD,LUEB_9} for $D=2,3$.
Using the Eqs.\,(\protect\ref{eq:soc_tau_t}-\protect\ref{eq:soc_tau_s})
the avalanche exponents (middle) are obtained from the spreading
exponents $\delta$, $\theta$, and $z$.
Using certain hyperscaling relations 
it is possible to express the avalanche exponents (right)
in terms of the exponents of the continuous
absorbing phase transition 
($\beta$, $\betap$, $\nu_{\senk}$, $\nu_{\para}$), 
see Eqs.\,(\protect\ref{eq:soc_exp_apt_exp_tau_r}-\protect\ref{eq:soc_exp_apt_exp_tau_s}).
\label{fig:soc_exp_manna}} 
\end{figure} 
%%%%%%%%%%%%%%%%%%%%%%%%%%%%%%%%%%%%%%%%%%%%%%%%%%%%%%%%%%%%%%%%%%%%%%%%%%%%%%%%%%%%%%
Due to the valid hyperscaling law 
\refeqname{eq:gen_hyperscaling_law} and \refeqname{eq:delta_betap_nu_para}
the SOC avalanche exponents can be expressed in terms of the ordinary
critical exponents of the corresponding absorbing phase 
transition~\cite{LUEB_23}
\begin{eqnarray}
\label{eq:soc_exp_apt_exp_tau_r}
\tau_{\ssstyle r} 
% & = & 
% 1 + \,z\, + D \, - \, \frac{\sigma}{\,\nu_{\ssstyle \perp}\,} \nonumber \\
& = & 1 \, + \, \frac{\beta^{\prime}}{\nu_{\ssstyle \perp}}, \\[2mm]
\label{eq:soc_exp_apt_exp_tau_t}
\tau_{\scriptscriptstyle t} 
%& = & 2 + \, \frac{\,D\,}{z} 
%\, - \, \frac{\sigma}{\, \nu_{\ssstyle \parallel}\,} \nonumber \\
& = & 1 \, + \, \frac{\beta^{\prime}}{\nu_{\ssstyle \parallel}}, \\[2mm]
\label{eq:soc_exp_apt_exp_tau_a}
\tau_{\ssstyle a} 
%& = & 2 + \, \frac{\,z\,}{D} 
%\, - \, \frac{\sigma}{\, D \nu_{\ssstyle \perp}\,} \nonumber \\
& = & 1 \, + \, \frac{\beta^{\prime}}{D \nu_{\ssstyle \perp}}, \\[2mm]
\label{eq:soc_exp_apt_exp_tau_s}
\tau_{\ssstyle s} 
% & = & 1 \, + \,  \frac{ \,\nu_{\ssstyle \parallel} 
% + \nu_{\ssstyle \perp} D - \sigma \,}
% {\nu_{\ssstyle \parallel} 
% + \nu_{\ssstyle \perp} D - \beta
% } \nonumber \\ 
& = & 1 \, + \, \frac{   \beta^{\prime} }
{\, \nu_{\ssstyle \parallel} 
+ \nu_{\ssstyle \perp} D - \beta \,
}\, . 
\end{eqnarray}
In Fig.\,\ref{fig:soc_exp_manna} we compare these values
to the avalanche exponents obtained from 
SOC simulations of the Manna model~\cite{LUEB_PHD,LUEB_9,DICKMAN_6}.
All SOC exponents agree within the error-bars with 
the avalanche exponents derived via the above scaling laws.
Thus it is possible to express the avalanche exponents
($\tau_{\ssstyle s}$, $\tau_{\ssstyle a}$,$\ldots$) 
of SOC systems in terms of the usual critical exponents
of a second order phase transition
($\beta$, $\betap$, $\nu_{\ssstyle \perp}$, $\nu_{\ssstyle \parallel}$).
In this way, the critical state of SOC sandpile models
is closely related to the critical state of an 
ordinary second order phase transition.

\chapter{Universal crossover scaling} 
\setcounter{figure}{52}

\label{chapter:crossover}

Crossovers\index{crossover} 
between different universality classes are well
known from equilibrium phase transitions.
Classical examples are ferromagnetic systems exhibiting 
weak uniaxial spin aniso\-tro\-py (see for example~\cite{PFEUTY_1,AHARONY_4}).
Approaching the transition point~$\Tc$ Heisenberg critical
behavior is observed.
But very close to~$\Tc$ a crossover to Ising
critical behavior occurs.
In the topological language of renormalization
group theory crossover effects may occur if 
more than one fixed point is embedded in the 
critical surface.
In case of the example above,
the crossover corresponds to a trajectory which passes 
close to the Heisenberg fixed point but is eventually 
driven away to the Ising fixed point.
In terms of scaling forms, crossover phenomena are described
by additional relevant scaling fields, characterized by 
a so-called\index{crossover exponent}
crossover exponent~$\phi$~\cite{RIEDEL_2}.

Although crossover phenomena are well understood 
in terms of competing fixed points, 
several aspects are still open and are discussed in 
the literature.
For example, the question whether the
so-called effective\index{effective exponent} 
exponents~\cite{RIEDEL_1} fulfill 
certain scaling laws over the entire crossover 
region was revisited several times
(see e.g.~\cite{CHANG_1,LUIJTEN_1,LUIJTEN_2,MARQUES_1,LUEB_30}
and  references therein).
This question is related closely to the general and 
important question whether effective exponents
obey the scaling laws at all.
For example, it is known experimentally~\cite{GREYWALL_1} 
as well as theoretically~\cite{WEGNER_3}
that the asymptotic scaling behavior
is often masked by corrections to scaling, so-called 
confluent\index{confluent singularities} 
singularities (see section~\ref{sec:rg_theory}).
In this case it is useful to analyze the data
in terms of effective exponents and the 
question above arises naturally~\cite{AHLERS_1}.
The validity of the scaling laws
for effective exponents was often concluded from
experimental (see e.g.~\cite{CHANG_1} and references therein)
and numerical investigations (see for example~\cite{MARQUES_1}).
In contrast, renormalization group approaches predicted 
a violation of the scaling laws for effective exponents~\cite{CHANG_1}.
This renormalization group result is confirmed by numerical investigations
of the Ising model indicating a violation of the 
Rushbrook scaling law~\cite{LUIJTEN_2}.

Another question of interest concerns the 
universality of the crossover scaling functions.
The range where the universal critical scaling
behavior applies is usually restricted to a small vicinity 
around the critical point.
Therefore it is questioned whether the full crossover
region that spans usually several decades 
in temperature or conjugated field, 
can be described in terms of universal scaling
functions~\cite{ANISIMOV_1,BELYAKOV_1,LUIJTEN_1,LUIJTEN_3,MON_1,LUIJTEN_2,PELISSETTO_1,PELISSETTO_2,CARACCIOLO_1,LUEB_29,KIM_1,LUEB_30}.
Renormalization group approaches 
predicted a non-universal behavior if one
uses finite cutoff lengths, whereas 
infinite cutoff lengths (which corresponds to 
unphysical vanishing molecular sizes) lead to a universal 
crossover scaling (see e.g.~\cite{ANISIMOV_1,BELYAKOV_1}).
On the other hand, also the experimental situation is  
unclear since measurements over the whole
crossover region are difficult, and accurate
results are rare (see~\cite{LUIJTEN_1} for a discussion).
Thus several attempts were performed in order
to address this question via numerical 
simulations~\cite{LUIJTEN_1,LUIJTEN_2,MON_1,LUIJTEN_3,LUEB_29,CARACCIOLO_1,LUEB_30}.
But only one model, i.e., only one representative of a
given universality class was considered in previous studies.
Analyses that incorporate different models belonging
to the same universality class were performed 
recently~\cite{LUEB_29,LUEB_30}.
These studies are reviewed in the following.

%\section{Activity spreading with different number of seeds}
%\label{sec:crossover_N_seeds}
%\input{_crossover/__N_seeds.tex}

\section{Crossover to mean field scaling behavior}
\label{sec:crossover_mean_field}

The critical behavior of a system exhibiting a second
order phase transition with non-mean field scaling 
behavior is strongly affected by the
range of interactions.
The longer the range of interactions the stronger
will be the reduction of the critical fluctuations, and in the
limit of infinite interactions the system is 
characterized\index{long-range interactions}
by the mean field scaling behavior.
As stated by the well known\index{Ginzburg criterion} 
Ginzburg criterion~\cite{GINZBURG_1,ALSNIELSEN_1}, 
mean field-like behavior occurs even
for finite interaction ranges sufficiently far 
away from the critical point.
A crossover to the non-mean field scaling
behavior takes place if one approaches the transition
point.
%From the renormalization group point of view,
%the crossover reflects the instability of the 
%mean field fixed point, resulting from an 
%exchange of stabilities between this fixed point and the 
%non-trivial fixed point.
This crossover is described by a crossover
exponent~$\phi$ which is known exactly due to the
Ginzburg criterion.
Therefore the crossover from mean field to non-mean
scaling is theoretically and numerically 
(see e.g.~\cite{LUIJTEN_1,LUIJTEN_3,MON_1,LUIJTEN_2,PELISSETTO_1,PELISSETTO_2,CARACCIOLO_1,LUEB_29,KIM_1,LUEB_30})
well suited to examine if universality holds for crossovers.
Following the spirit of this work, 
we focus on non-equilibrium crossover phenomena 
and consider the crossover scenario in the Manna
universality class (in the so-called critical crossover 
limit~\cite{LUIJTEN_1,PELISSETTO_1}).
% d.h. fuer grosse Korr-laengen, also
% crossover von einem kritischen Verhalten zum naechsten
But it is worth mentioning that the results can be 
applied to continuous phase transitions in general,
including equilibrium crossover phenomena.

According to the definitions of the CLG model, 
the CTTP, and the Manna model (see section \ref{sec:aptcf_lattice_models}), 
particles of active sites are moved to nearest neighbors only, 
i.e., the range of interactions is $R=1$.
It is straightforward to implement various ranges of interactions
into these models~\cite{LUEB_29}.
In these modified models particles of active sites are moved
(according to the rules of each model) to sites randomly 
selected within a radius~$R$.
The dynamics of the models considered are characterized
by simple particle hopping processes, i.e., various interaction
ranges can be implemented easily without affecting the 
simulation performance.
This is an advantage compared to e.g.~equilibrium systems
like the Ising model where increasing interaction ranges
slow down the dynamics.
A disadvantage is that for zero field the crossover occurs 
only in the active phase, whereas equilibrium systems can be
investigated below and above the phase 
transition~\cite{LUIJTEN_1,PELISSETTO_1}.

\begin{figure}[t] 
\centering
%\leavevmode 
\includegraphics[clip,width=13cm,angle=0]{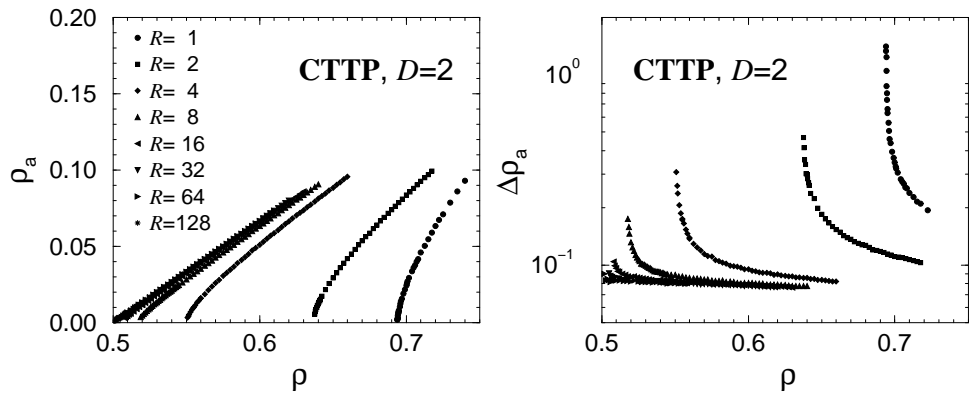}
\caption[Conserved transfer threshold process, various interaction ranges, $D=2$]
{The order parameter (left) and the order parameter fluctuations
(right) of the two-dimensional conserved threshold transfer process
(CTTP) for various values of the interaction range~$R$.
With increasing range the critical density tends to the mean field
value $\rho_{\ssstyle \mathrm{c},R\to\infty}=1/2$.
As can be seen, the power-law behavior of $\rhoa$ and $\Delta\rhoa$
changes with increasing range of interactions. 
The data are obtained from simulations of various
systems sizes $L\kgl 2048$.
\label{fig:cross_over_01}} 
\end{figure}

Notice that the simulations have to be performed in the 
so-called critical crossover limit,
i.e., the corresponding correlation length is sufficiently
large in the whole crossover 
region~\cite{LUIJTEN_3,LUIJTEN_1,PELISSETTO_1}.
For any finite interaction range~$R$ the phase transition
is characterized by non-mean field scaling behavior,
whereas mean field scaling occurs for infinite 
interactions.
The order parameter~$\rhoa$ as well as the order parameter 
fluctuations~$\Delta\rhoa$ of the two-dimensional CTTP are plotted 
in \reffigname{fig:cross_over_01} for various interaction 
ranges at zero field.
As can be seen, the transition point~$\rhocR$ depends on the
range of the interactions and the power-law behavior of 
$\rhoa$ and $\Delta\rhoa$ changes with increasing~$R$.

Before examining how the increasing range of interactions
affects the scaling behavior it is inevitable to mention 
that the parameter~$R$ is not an 
appropriate quantity to describe the scaling behavior since
it describes the maximum range of interactions.
Thus the actual strength of the interactions may 
depend on the lattice structure. 
To avoid these lattice effects one
uses the effective interaction range~\cite{MON_1}
\begin{equation}
\Reff^2 \; = \;
\frac{1}{z_{\ssstyle n}} \, \sum_{i \neq j}  
| {\underline r}_{\scriptscriptstyle i} \, - \,
{\underline r}_{\scriptscriptstyle j}|^2 \, , 
\quad \quad \mathrm{for} \quad  \quad
| {\underline r}_{\scriptscriptstyle i} \, - \,
{\underline r}_{\scriptscriptstyle j}| \kgl R
\label{eq:def_R_eff}
\end{equation}
where $z_{\ssstyle n}$ denotes the number of lattice sites within
a radius~$R$.
The values of the effective interaction ranges
are listed for two- and three dimensional
simple cubic lattices 
in \reftables2name{table:radius_2d}{table:radius_3d}.

Approaching the transition point from the active phase
non-mean field critical behavior is observed 
for all finite values of~$R$.
But mean field behavior occurs away from the critical
point if the long range interactions reduce the 
critical fluctuations sufficiently.
The crossover between these two scaling regimes
is described by the well known
Ginzburg criterion~\cite{GINZBURG_1,ALSNIELSEN_1}
which states that the mean field approximation
is self-consistent
in the active phase as long as the fluctuations within a 
correlation volume are small compared 
to the order parameter itself
\begin{equation}
\; \xi^{\ssstyle -D} \,\Delta\rhoa 
\; \ll \;  \rhoa^2 .
\label{eq:ginzburg_crit}
\end{equation}
In the steady state of absorbing phase transitions~$\xi$
corresponds to the spatial correlation length~$\xi_{\senk}$
that diverges at the critical point according to 
\begin{equation}
\xi_{\ssstyle \senk} \asympprop \Reff \;
(\rho-\rhocR )^{-\nuperpMF} \, .
\label{eq:ginz_corr_length_perp}
\end{equation}
Thus, despite of metric factors the mean field theory applies
at zero field if 
\begin{equation}
1\; \ll \; \Reff^D \;
(\rho-\rhocR)^{\gammapMF+2\beta^{\phantom \prime}_{\ssstyle \mathrm{MF}}
 -\nu^{\phantom \prime}_{\senk,\ssstyle \mathrm{MF}} D} \, ,
\label{eq:ginzburg_crit_zf}
\end{equation}
where the usual power-laws 
\begin{equation}
\rhoa \, \asympprop \, (\rho-\rhocR)^{\betaMF} \, , \quad \quad \quad
\Delta\rhoa \, \asympprop \, (\rho-\rhocR)^{-\gammapMF} \, , 
%\xi_{\senk} \, \asympprop \, \Reff (\rho-\rhocR)^{\nuperpMF} \, , \quad \quad
\label{eq:crossover_mf_power_laws}
\end{equation}
and \refeqname{eq:ginz_corr_length_perp} are used.
Obviously, \refeqname{eq:ginzburg_crit_zf} is fulfilled for
$\rho \to \rhocR$ if 
\begin{equation}
D \; > \; \frac{\, \gammapMF  +  2  
\beta^{\phantom \prime}_{\ssstyle \mathrm{MF}} }{\nuperpMF} \, ,
\label{eq:D_larger_Dc}
\end{equation}
otherwise mean field theory fails to describe the critical behavior.
The marginal case corresponds to the upper critical dimension
\begin{equation}
\Dc \; = \; 
\frac{\, \gammapMF + 2 \beta^{\phantom \prime}_{\ssstyle \mathrm{MF}}} 
{\nuperpMF} \, .
\label{eq:ginzburg_dc}
\end{equation}
In this way the concept of an upper critical dimension
is introduced phenomenologically by the Ginzburg criterion,
independent of the renormalization group scenario. 
In case of directed percolation and the Manna universality class
the mean field values $\betaMF=1$, $\nuperpMF=1/2$, and
$\gammaMF=0$ lead to $\Dc=4$ as expected.
It is convenient to introduce the exponent
\begin{equation}
\phi \; = \; \nuperpMF \, \frac{\, \Dc\, - \, D \, }{D}
\; = \; \frac{\, 4\, - \, D \, }{2\, D} \, .
\label{eq:exp_phi}
\end{equation}
Thus mean field scaling behavior occurs as long as
$1 \ll \Reff \, (\rho-\rhocR)^{\phi}$
whereas non-mean field scaling behavior is observed for 
$1 \gg \Reff \,(\rho-\rhocR)^{\phi}$.
The point at which 
\begin{equation}
{\mathcal O} \left ( \Reff \,(\rho-\rhocR)^{\phi} \right ) \; = \; 1
\label{eq:crossover_point}
\end{equation}
is usually considered as the crossover point and~$\phi$
is termed the\index{crossover exponent} 
crossover exponent~\cite{RIEDEL_2}.

\begin{figure}[t] 
\centering
%\leavevmode 
\includegraphics[clip,width=8cm,angle=0]{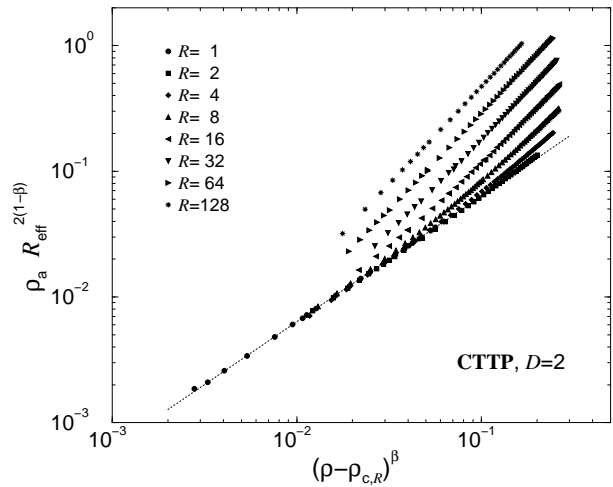}
\caption[Critical amplitude behavior, conserved threshold transfer process, $D=2$]
{The order parameter data of the two-dimensional conserved threshold 
transfer process (CTTP).
The data are rescaled according to \refeqname{eq:cross_op_ampl}.
The dashed line corresponds to a power-law behavior
of the two-dimensional order parameter with exponent 
$\beta=\beta_{\ssstyle D=2}$.
\label{fig:cross_over_ampl}} 
\end{figure}

In order to describe the crossover behavior the 
scaling forms have to incorporate the 
range of interactions as an additional scaling field.
For example, the non-universal scaling form of the 
order parameter is given by
\begin{equation}
\rhoa(\rho,h,\Reff) \; \sim \;
\lambda^{-\betaMF} \; {\tilde \EuFrak r}( \, \lambda\, (\rho-\rhocR),
h \, \lambda^{\sigmaMF}, \Reff^{-1} \, \lambda^{\phi}\, ) \, ,
\label{eq:cross_op_non_uni}
\end{equation}
where the scaling power of $\Reff$ is 
motivated by \refeqname{eq:crossover_point}.
In the following, ${\tilde\EuFrak r}$ and ${\tilde\EuFrak R}$ are used
to denote the non-universal and universal scaling
functions, describing the particular crossover.
Obviously, the added scaling field is relevant ($\phi>0$) 
below the upper critical dimension whereas it is 
irrelevant ($\phi<0$) above~$\Dc$. 
The mean field scaling behavior should be recovered for
$R\to\infty$
\begin{equation}
{\tilde \EuFrak r}(x,0,0) \; \asympprop \;
x^{\betaMF} 
\quad\quad {\mathrm{for}} \quad\quad
x\to 0 \, . 
\label{eq:r_asym_mf}
\end{equation}
On the other hand, the $D$-dimensional order parameter behavior 
is observed for finite interaction ranges
\begin{equation}
{\tilde \EuFrak r}(x,0,\Reff^{-1}) \; \asympprop \;
x^{\beta_{\ssstyle D}} 
\quad\quad {\mathrm{for}} \quad\quad
x\to 0 \, . 
\label{eq:r_asym_D}
\end{equation}
It is instructive to consider the amplitudes of the
corresponding power-laws because the amplitudes display
a singular dependence on the range of interactions.
Setting $\Reff^{-1} \lambda^{\phi}=1$ in \refeqname{eq:cross_op_non_uni}
yields sufficiently close to the critical point 
\begin{eqnarray}
\rhoa(\rho,0,\Reff) & \sim &
\Reff^{-\betaMF/\phi} \;\; {\tilde \EuFrak r}( \, \Reff^{1/\phi}\, 
(\rho-\rhocR),0,1 ) \nonumber \\[2mm]
& \asympprop & 
\Reff^{(\beta_{\ssstyle D}-\betaMF)/\phi} \; \;
(\rho-\rhocR)^{\beta_{\ssstyle D}} \, .
\label{eq:cross_op_ampl}
\end{eqnarray}
Thus this power-law behavior and the corresponding 
power-laws of the
fluctuations, the susceptibility, etc.~are valid only for
finite interaction ranges whereas they become useless for
infinite~$R$, signalling the change in the universality
class. 
In this way the singular amplitudes reflect the breakdown of
a scaling regime if a crossover is approached.
The critical amplitude behavior can be observed in simulations.
The corresponding data of the two-dimensional conserved threshold
transfer process is presented in \reffigname{fig:cross_over_ampl}.
The rescaled order parameter data tend to the same power-law
behavior if one approaches the transition point.

\begin{figure}[t] 
\centering
%\leavevmode 
\includegraphics[clip,width=13cm,angle=0]{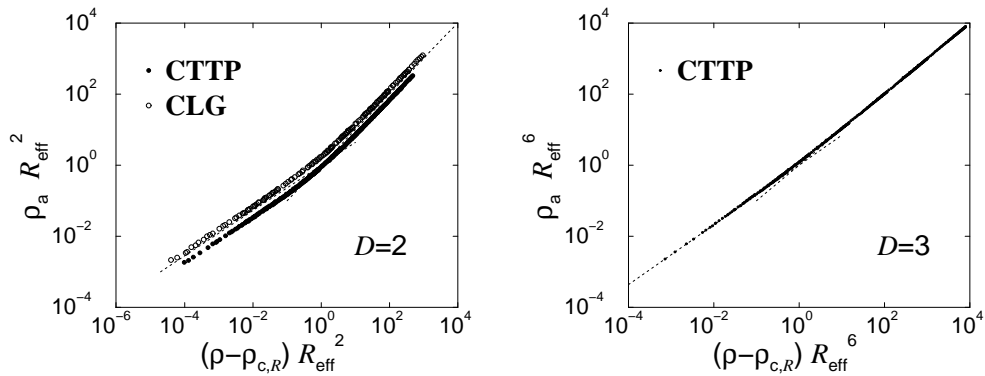}
\caption[Crossover scaling function, order parameter, $D=2,3$]
{The order parameter of the two-dimensional (left, $\phi=1/2$) 
and three dimensional (right, $\phi=1/6$) 
conserved threshold transfer process (CTTP) and conserved lattice gas model
for various values of the interaction range~$R$.
The data are rescaled according to \refeqname{eq:cross_op_non_uni}.
The dashed line corresponds to the power-law behavior
of the $D$-dimensional models and to the mean field 
scaling behavior, respectively.
The data are obtained from simulations of various system
sizes ($L\kgl 2048$ for $D=2$ and $L\kgl 128$
for $D=3$).
\label{fig:cross_over_scal}} 
\end{figure}

Furthermore, data obtained from
simulations at zero field and various interaction ranges
have to collapse onto a single curve by plotting 
$\rhoa \, \Reff^{\betaMF/\phi}$ as a function of
$(\rho-\rhocR)\, \Reff^{1/\phi}$.
The respective plots of the two- and three-dimensional CTTP and 
the CLG model are shown in \reffigname{fig:cross_over_scal}.
A good collapse is observed over the entire crossover which 
spans roughly $8$~decades.
The asymptotic power-law behavior of the $D$-dimensional
scaling behavior
as well as of the mean field scaling behavior are recovered well.
But as can be seen for $D=2$, each model is characterized 
by its own scaling function~${\EuFrak r}$.
In order to obtain the universal crossover
scaling function non-universal metric factors have to be introduced.
This will be done in the next section where we focus
on the control parameter driven crossover 
(temperature driven crossover in equilibrium).
The subsequent section is devoted to the field driven
crossover.
In particular, the data of the field driven crossover allows to
check the validity of the Widom scaling law over the entire
crossover region.

%%%%%%%%%%%%%%%%%%%%%%%%%%%%%%%%%%%%%%%%%%%%%%%%%%%%%%%%%%%%%%%%%%%%%%%%%%%%%%%%%%%%%%

\subsection{Control parameter driven crossover}
\label{subsec:crossover_temp_driven}

In the following we address the question whether the 
crossover scaling functions, which spans several decades,
can be described in terms of universal scaling functions.
Therefore, non-universal metric factors have to be incorporated
in the scaling forms, i.e., we make the 
phenomenological ansatz~\cite{LUEB_29}
\begin{equation}
\rhoa(\rho,h,\Reff) \; \sim \;
\lambda^{-\betaMF} \;\; {\tilde \EuFrak R}( 
{\EuFrak a}_{\ssstyle \rho} \,(\rho-\rhocR)\, \lambda, 
{\EuFrak a}_{\ssstyle h} \, h \, \lambda^{\sigmaMF}, 
{\EuFrak a}_{\ssstyle R} \, \Reff^{-1} \, \lambda^{\phi}\, ) \, ,
\label{eq:scal_ansatz_EqoS_co}
\end{equation}
The mean field scaling function should be recovered
for $R\to \infty$, thus
\begin{equation}
{\tilde {\EuFrak R}}(x,y,0) \; = \; 
{\tilde R}_{\ssstyle {\mathrm{MF}}}(x,y) 
\label{eq:mf_limit_R}
\end{equation}
and, therefore,
\begin{equation}
{\tilde {\EuFrak R}}(1,0,0) \; = \; 
{\tilde R}_{\ssstyle {\mathrm{MF}}}(1,0) \, = \, 1 \, , \quad \quad\quad
{\tilde {\EuFrak R}}(0,1,0) \, = \,
{\tilde R}_{\ssstyle {\mathrm{MF}}}(0,1) \, = \, 1 \, ,
\label{eq:mf_limit_R_norm}
\end{equation}
which implies
\begin{equation}
{\EuFrak a}_{\ssstyle \rho} \, = \,
\frac{a_{\ssstyle \rho, R\to\infty}}
{\rho_{\ssstyle {\mathrm{c}}, R\to\infty}} \, , \quad\quad\quad
{\EuFrak a}_{\ssstyle h} \, = \,
{a_{\ssstyle h, R\to\infty}} \, .
\label{eq:metric_factors_fractur_rho_h}
\end{equation}
These metric factors were already determined in independent
studies where absorbing phase transitions with infinite 
particle hopping were investigated~\cite{LUEB_20,LUEB_25}.

\begin{figure}[t] 
\centering
%\leavevmode 
\includegraphics[clip,width=13cm,angle=0]{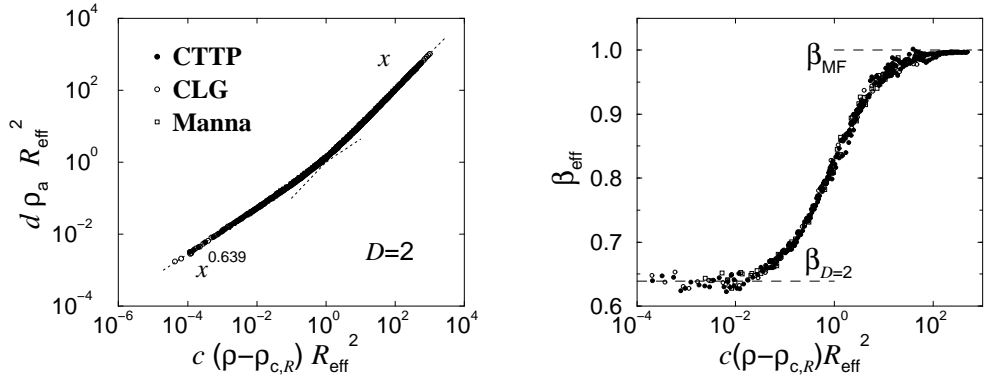}
\caption[Universal crossover scaling function, order parameter, 
Manna class, control parameter driven crossover, $D=2$]
{The universal crossover scaling function of order parameter
at zero field for $D=2$ (left).
The metric factors are given by 
$c={\EuFrak a}_{\ssstyle \rho} {\EuFrak a}_{\ssstyle R}^2$ 
and $d={\EuFrak a}_{\ssstyle R}^2$.
The data of all models display an excellent 
collapse onto the universal crossover scaling function 
${\tilde {\EuFrak R}}(x,0,1)$.
The dashed lines correspond to the asymptotic behavior
of the two-dimensional system 
($\beta_{\ssstyle D=2}=0.639$)
and of the mean field behavior
($\betaMF=1$).
The right figure displays the corresponding effective 
exponent $\beta_{\ssstyle \mathrm{eff}}$.
\label{fig:uni_co_opzf_2d}} 
\end{figure}

The non-universal metric factor~${\EuFrak a}_{\ssstyle R}$
has to be determined by a third condition.
Several ways are possible (e.g.~${\tilde {\EuFrak R}}(0,0,1)=1$)
but it is convenient to enforce that ${\tilde {\EuFrak R}}$ 
scales as
\begin{equation}
{\tilde {\EuFrak R}}(x,0,1) \; \sim \;
x^{\beta_{\ssstyle D}} \, ,
\quad\quad {\mathrm{for}} \quad\quad x\to 0 \, .
\label{eq:fractur_R_small_x}
\end{equation}
Setting ${\EuFrak a}_{\ssstyle R}^{-1} \Reff^{-1} 
\, \lambda^{\phi} =1$ in \refeqname{eq:scal_ansatz_EqoS_co}
yields at zero field 
\begin{equation}
\rhoa
(\rho,0,\Reff)
\;  \sim \;  
({\EuFrak a}_{\ssstyle R} \, \Reff)^{-\betaMF/\phi} 
\; \; {\tilde {\EuFrak R}}
({\EuFrak a}_{\ssstyle \rho} (\rho-\rhocR) 
\, {\EuFrak a}_{\ssstyle R}^{1/\phi} \; 
\Reff^{1/\phi},0,1) \, . 
\label{eq:scal_plot_EqoS_co}   
\end{equation}
Taking into account that the $D$-dimensional scaling
behavior is recovered for $R=1$ we find
\begin{equation}
{\EuFrak a}_{\ssstyle R} \; = \;
\left ( \frac{\rho_{\ssstyle {\mathrm{c}},R=1}}
{a_{\ssstyle \rho, R=1}} \;
\frac{a_{\ssstyle \rho,R\to \infty}}
{\rho_{\ssstyle {\mathrm{c}},R\to \infty}}
\right )^{{\phi \beta_{\ssstyle D}}/
{(\betaMF-\beta_{\ssstyle D}})}  \, .
\label{eq:metric_factors_fractur_R}
\end{equation}
According to the scaling form above, 
we plot in 
\reffigname{fig:uni_co_opzf_2d} the rescaled order parameter 
as a function of 
the rescaled control parameter 
for the two-dimensional ($\phi=1/2$) models.
The data of the metric factors are obtained from
directed measurements of the corresponding 
amplitudes (see~\cite{LUEB_26}).
An excellent data-collapse is observed over the entire
range of the crossover confirming the phenomenological
ansatz.

Since the full crossover region covers several 
decades it could be difficult to observe small but systematic
differences between the scaling functions of both models.
Therefore, it is customary to scrutinize the crossover
via the so-called effective\index{effective exponent} 
exponents~\cite{RIEDEL_1,LUIJTEN_2,CARACCIOLO_1,LUEB_29}
\begin{equation}
\beta_{\ssstyle \mathrm{eff}} \; = \;
\frac{\partial\hphantom{\ln{x}}}{\partial \ln{x}}
\, \ln{{\tilde{\EuFrak R}}(x,0,1)} .
\label{eq:def_eff_exp_beta}
\end{equation}
The corresponding data are shown in \reffigname{fig:uni_co_opzf_2d}.
The effective exponent changes monotonically from its 
mean field value to its two-dimensional value.
Non-monotonic crossover behavior may also occur, e.g.~in
equilibrium systems~\cite{LUIJTEN_1} and even in the
Manna universality class~\cite{LUEB_13}.
The excellent collapse 
over more than 7 decades supports strongly the hypothesis
that the crossover function is a universal function.
It is worth mentioning that the collapse includes
the data for small interaction ranges, in particular $R=1$ and $R=2$.
Renormalization group analyses of the
same crossover scenario in spin models 
reveal non-universal corrections to the universal scaling
functions~\cite{PELISSETTO_2}.
These corrections scales as $R^{-D}$ (in two loop order), 
i.e., significant corrections are expected to occur 
for small interaction ranges~$R$.

We now consider the order parameter fluctuations.
Analogous to the order parameter we use the scaling
form
\begin{equation}
{\EuFrak a}_{\ssstyle \Delta} \,
\Delta\rhoa
(\rho, h, \Reff) 
\; \sim \; 
\lambda^{\gammapMF}\; \; {\tilde {\EuFrak D}}
({\EuFrak a}_{\ssstyle \rho} \, (\rho-\rhocR) \,
\; \lambda, 
{\EuFrak a}_{\ssstyle h} \, h \, \lambda^{\sigmaMF},
{\EuFrak a}_{\ssstyle R}^{-1} \, \Reff^{-1} 
\; \lambda^{\phi} \, ) \, . 
\label{eq:scal_ansatz_Fluc_co}
\end{equation}
Again the mean field behavior should be recovered for
$R\to\infty$, implying
\begin{equation}
{\tilde {\EuFrak D}}(x,y,0) \; = \; 
{\tilde D}_{\ssstyle {\mathrm{MF}}}(x,y) 
\label{eq:mf_limit_D}
\end{equation}
and thus
\begin{equation}
{\tilde {\EuFrak D}}(1,0,0) \, = \,
{\tilde D}_{\ssstyle {\mathrm{MF}}}(1,0) \, = \, 2 \, , \quad\quad\quad
{\tilde {\EuFrak D}}(0,1,0) \, = \, 
{\tilde D}_{\ssstyle {\mathrm{MF}}}(0,1) \, = \, 1 \, ,
\label{eq:mf_limit_D_norm}
\end{equation}
as well as 
\begin{equation}
{\EuFrak a}_{\ssstyle \Delta} \; = \;
{a_{\ssstyle \Delta, R\to\infty}} \, .
\label{eq:metric_factors_fractur_a_delta}
\end{equation}
Setting ${\EuFrak a}_{\ssstyle R}^{-1} 
\Reff^{-1} \, \lambda^{\phi} =1$ in \refeqname{eq:scal_ansatz_Fluc_co}
yields at zero field and for $\gammapMF=0$
\begin{equation}
{\EuFrak a}_{\ssstyle \Delta} \,
\Delta\rhoa
(\rho,0,\Reff) 
\; \sim \; 
{\tilde {\EuFrak D}} ({\EuFrak a}_{\ssstyle \rho} \, 
(\rho-\rhocR) \; {\EuFrak a}_{\ssstyle R}^{1/\phi} 
\Reff^{1/\phi}, 0, 1) . 
\label{eq:scal_plot_Fluc_co}
\end{equation}
The fluctuations diverge at the critical point, i.e.,
the universal function ${\tilde {\EuFrak D}}$ scales as
\begin{equation}
{\tilde {\EuFrak D}}(x,0,1) \; \sim \; m_{\ssstyle\Delta,\rho} \,
x^{-\gammap_{\ssstyle D}} \, ,
\quad\quad {\mathrm{for}} \quad\quad x\to 0 \, . 
\label{eq:fractur_D_small_x}
\end{equation}
The power-law amplitude~$m_{\ssstyle\Delta,\rho}$
is determined in the following way~\cite{LUEB_29}:~The scaling form
\refeqname{eq:scal_ansatz_Fluc_co} has to equal for $R=1$
the $D$-dimensional scaling behavior 
\begin{equation}
\Delta \rhoa \; \sim \;
a_{\ssstyle \Delta,R=1}^{-1} \; {\tilde D}_{\ssstyle D}(1,0) \;
\left ( a_{\ssstyle \rho,R=1} \,
\frac{(\rho-\rho_{\ssstyle {\mathrm{c}},R=1})}
{\rho_{\ssstyle {\mathrm{c}},R=1}}
\right )^{-\gammap_{D}} \, .
\label{eq:scaling_fluc_R1}
\end{equation}
The value of the universal scaling function ${\tilde D}_{\ssstyle D}(1,0)$ can be 
%%%%%%%%%%%%%%%%%%%%%%%%%%%%%%%%%%%%%%%%%%%%%%%%%%%%%%%%%%%%%%%%%%%%%%%%%%%%%%%%%%%%%%%%%%
\begin{figure}[t] 
\centering
%\leavevmode 
\includegraphics[clip,width=13cm,angle=0]{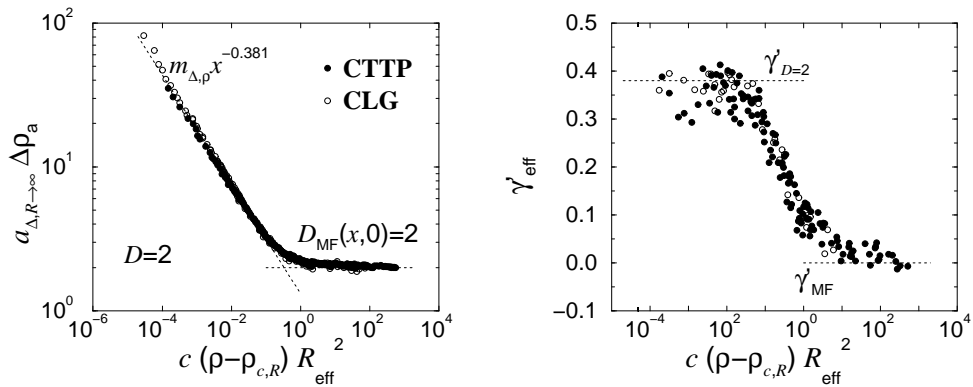}
\caption[Universal crossover scaling function, fluctuations, 
Manna class, control parameter driven crossover, $D=2$]
{The universal crossover scaling function of the order parameter
fluctuations at zero field for $D=2$ (left).
The metric factor is given by 
$c={\EuFrak a}_{\ssstyle \rho} {\EuFrak a}_{\ssstyle R}^2$.
The data of both models display a good 
collapse to the universal crossover scaling function 
${\tilde {\EuFrak D}}(x,0,1)$.
The dashed lines correspond to the asymptotic behavior
of the two-dimensional system 
($\gammap_{\ssstyle D=2}=0.381$)
and of the mean field behavior
($\gammapMF=0$).
The universal amplitude is given 
by $m_{\ssstyle \Delta,\rho} =1.28$.
The right figure displays the corresponding effective
exponent~$\gammap_{\ssstyle \mathrm{eff}}$.
\label{fig:uni_co_flzf_2d}} 
\end{figure} 
%%%%%%%%%%%%%%%%%%%%%%%%%%%%%%%%%%%%%%%%%%%%%%%%%%%%%%%%%%%%%%%%%%%%%%%%%%%%%%%%%%%%%%%%
obtained from a direct measurement of the corresponding
$D$-dimensional system (see \reftablename{table:manna_exponents})
and the universal amplitude of the
crossover fluctuation function is therefore given by
\begin{equation}
m_{\ssstyle \Delta,\rho} \; \sim \;
{\tilde D}_{\ssstyle D}(1,0) \; \frac{a_{\ssstyle \Delta, R\to \infty}}
{a_{\ssstyle \Delta, R=1}} \;
\left ( 
\frac{\rho_{\ssstyle {\mathrm{c}},R=1}}
{a_{\ssstyle \rho, R=1}}\,
\frac{a_{\ssstyle \rho,R\to \infty}}
{\rho_{\ssstyle {\mathrm{c}},R\to \infty}}
\right )^{\gammap_{\ssstyle D}\betaMF/(\betaMF-\beta_{\ssstyle D})} 
\;  . 
\label{eq:uni_ampl_m_delta}  
\end{equation}
According to the scaling form \refeqname{eq:scal_plot_Fluc_co}
we plot in \reffigname{fig:uni_co_flzf_2d} the rescaled
fluctuations as a function of the rescaled control
parameter for the two-dimensional CLG model as well as
for the CTTP.
%In order to see the crossover only behavior 
% those data are plotted 
%which do not suffer by finite-size effects of 
%which lie not outside the critical region.
We observe again a good data-collapse of the data over 
the entire region of the crossover.
Furthermore, the asymptotic behavior is recovered for
both power-laws,
confirming the scaling ansatz \refeqname{eq:scal_ansatz_Fluc_co}.
The corresponding effective exponent
\begin{equation}
\gammap_{\ssstyle \mathrm{eff}} \; = \; - \, 
\frac{\partial\hphantom{\ln{x}}}{\partial \ln{x}}
\, \ln{{\tilde{\EuFrak D}}(x,0,1)} 
\label{eq:def_eff_exp_gammap}
\end{equation}
is displayed in \reffigname{fig:uni_co_flzf_2d}.
Although the data of the effective exponents 
are suffering from statistical fluctuations
one can see that both models are characterized by the 
same universal behavior.

\subsection{Field driven crossover:~Violation of Widom scaling law}
\label{subsec:crossover_field_driven}

\begin{figure}[t] 
\centering
%\leavevmode 
\includegraphics[clip,width=13cm,angle=0]{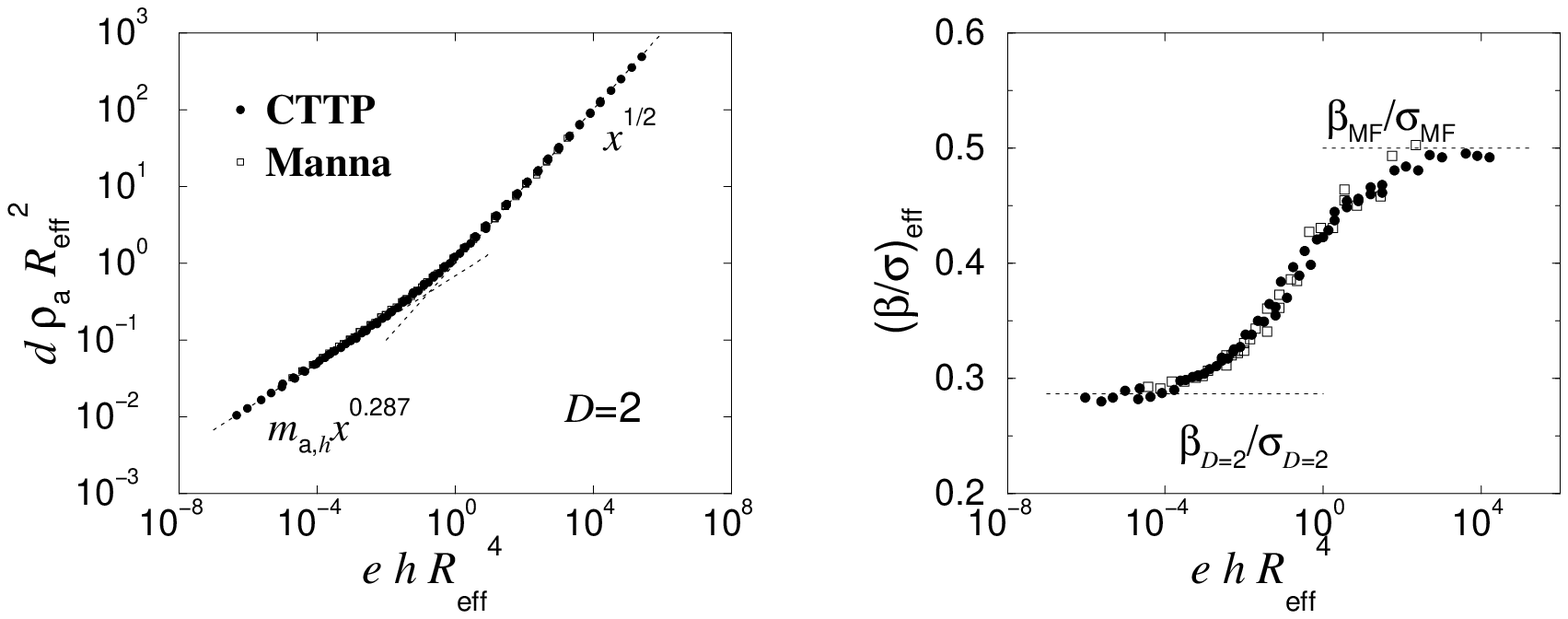}
\caption[Universal crossover scaling function, order parameter, Manna class, 
field driven crossover, $D=2$]
{The universal crossover scaling function 
of the order parameter at the critical density for $D=2$ (left).
The metric factors are given by 
$e={\EuFrak a}_{\ssstyle h} {\EuFrak a}_{\ssstyle R}^4$
and $d={\EuFrak a}_{\ssstyle R}^2$.    
The dashed lines correspond to the asymptotic behavior
of the two-dimensional system 
($\beta_{\ssstyle D=2}/\sigma_{\ssstyle D=2}=0.287$)
and of the mean field behavior
($\betaMF/\sigmaMF=1/2$).
The universal amplitude is given 
by $m_{\ssstyle {\mathrm{a}} , h} =0.681$.
The right figure displays the corresponding effective
exponent~$(\beta/\sigma)_{\ssstyle \mathrm{eff}}$.  
\label{fig:uni_co_opcp_2d}} 
\end{figure}

In this section we focus our attention on the field driven 
crossover, i.e., we consider the scaling behavior at the
critical density~$\rhocR$ as a function of the conjugated field.
The scaling form at the critical point is given by~\cite{LUEB_30}
\begin{equation}
\rhoa
(\rhocR,h,\Reff) \;  \sim \;  
({\EuFrak a}_{\ssstyle R} \,
\Reff)^{-\betaMF/\phi} 
\; \; {\tilde {\EuFrak R}}
(0,{\EuFrak a}_{\ssstyle h} \, h \;
{\EuFrak a}_{\ssstyle R}^{\sigmaMF/\phi} 
\Reff^{\sigmaMF/\phi},1) \,    . 
\label{eq:scal_plot_OPcp_co}
\end{equation}
For vanishing field the universal function scales as
\begin{equation}
{\tilde {\EuFrak R}}(0,x,1) \; \sim \; m_{\ssstyle {\mathrm{a}},h} \,
x^{\beta_{\ssstyle D}/\sigma_{\ssstyle D}} \, ,
\quad\quad {\mathrm{for}} \quad \quad x\to 0 \, . 
\label{eq:fractur_R_small_x_field}
\end{equation}
The universal amplitude~$m_{\ssstyle {\mathrm{a}},h}$ can be determined
similar to the fluctuation amplitude~$m_{\ssstyle \Delta,\rho}$, yielding
\begin{equation}
\label{eq:uni_ampl_m_h}
m_{\ssstyle {\mathrm{a}},h} \; = \;
\left ( \frac{a_{\scriptscriptstyle h, R=1}}
{a_{\ssstyle h,R\to \infty}} \right )^{\beta_{\ssstyle D}/\sigma_{\ssstyle D}} \;
%\left ( 
%\frac{\rho_{\scriptscriptstyle {\mathrm{c}},R=1}}
%{a_{\scriptscriptstyle \rho, R=1}}\,
%\frac{a_{\scriptscriptstyle \rho,R\to \infty}}
%{\rho_{\scriptscriptstyle {\mathrm{c}},R\to \infty}}\right )
{\EuFrak a}_{\ssstyle R}^{\betaMF/\phi
- \sigmaMF \beta_{\ssstyle D}/
\sigma_{\ssstyle D} \phi}\, .
\end{equation}
We plot in \reffigname{fig:uni_co_opcp_2d} the
rescaled order parameter $\rhoa\,({\EuFrak a}_{\ssstyle R} 
\Reff)^{2}$ 
as a function of the rescaled field 
${\EuFrak a}_{\ssstyle h} h ({\EuFrak a}_{\ssstyle R} 
\Reff)^{4}$.
Again the data cover the entire crossover region and 
both asymptotic power-laws are clearly recovered.
The data-collapse of the effective exponent
\begin{equation}
\left (
\frac{\beta}{\sigma} 
\right )_{\ssstyle \mathrm{eff}} \; = \;
\frac{\partial\hphantom{\ln{x}}}{\partial \ln{x}}
\, \ln{{\tilde{\EuFrak R}}(0,x,1)} 
\label{eq:def_eff_exp_sigma}
\end{equation}
confirms again the universality of the crossover scaling
function~${\tilde {\EuFrak R}}$.

Next the order parameter susceptibility is considered.
The scaling form of the susceptibility is given by~\cite{LUEB_30}
\begin{equation}
{\EuFrak a}_{\ssstyle h}^{-1} \,
\chi(\rho,h,\Reff) 
\; \sim \; 
\lambda^{\gammaMF}\; \; {\tilde {\EuFrak X}}
({\EuFrak a}_{\ssstyle \rho} (\rho-\rhocR)
\; \lambda, 
{\EuFrak a}_{\ssstyle h} \, h \; \lambda^{\sigmaMF},
{\EuFrak a}_{\ssstyle R}^{-1} \Reff^{-1} 
\; \lambda^{\phi} ) \, .
\label{eq:scal_ansatz_Sucp_co}   
\end{equation}
The mean field behavior is
recovered for $R\to \infty$, i.e.,
\begin{equation}
{\tilde {\EuFrak X}}(x,y,0) \; = \; 
{\tilde {\mathrm{X}}}_{\ssstyle {\mathrm{MF}}}(x,y) \, , 
\label{eq:mf_limit_X}
\end{equation}
implying
\begin{equation}
{\tilde {\EuFrak X}}(1,0,0) \, = \,
{\tilde {\mathrm{X}}}_{\ssstyle {\mathrm{MF}}}(1,0) \, = \, 1 \, ,
\quad\quad\quad
{\tilde {\EuFrak X}}(0,1,0) \, = \,
{\tilde {\mathrm{X}}}_{\ssstyle {\mathrm{MF}}}(0,1) \, = \, \frac{1}{2} \, .
\label{eq:mf_limit_X_norm}
\end{equation}
In order to obtain a data-collapse we set 
${\EuFrak a}_{\ssstyle R}^{-1} \Reff^{-1} \, \lambda^{\phi} =1$ 
in \refeqname{eq:scal_ansatz_Sucp_co},
yielding at the critical density
\begin{equation} 
{\EuFrak a}_{\ssstyle h}^{-1} \,
\chi
(\rhocR,h,\Reff) 
\; \sim \; 
({\EuFrak a}_{\ssstyle R} \Reff)^{\gammaMF/\phi} 
\;\; {\tilde {\EuFrak X}}
(0,{\EuFrak a}_{\ssstyle h} \, h \; 
({\EuFrak a}_{\ssstyle R} 
\Reff)^{\sigmaMF/\phi},1) \, . 
\label{eq:scal_plot_Sucp_co}   
\end{equation}
%%%%%%%%%%%%%%%%%%%%%%%%%%%%%%%%%%%%%%%%%%%%%%%%%%%%%%%%%%%%%%%%%%%%%%%%%%%%%%%%%%%%%
\begin{figure}[t] 
\centering
%\leavevmode 
\includegraphics[clip,width=13cm,angle=0]{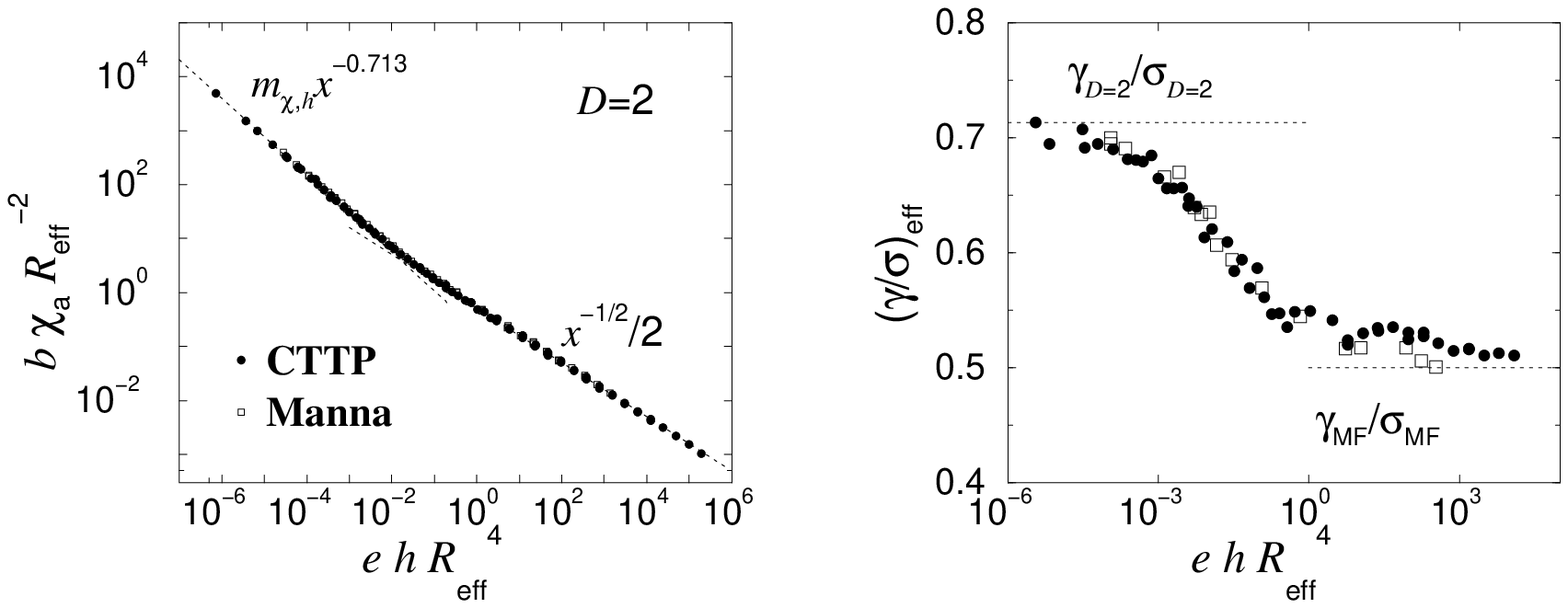}
\caption[Universal crossover scaling function, susceptibility, Manna class, 
field driven crossover,  $D=2$]
{The universal crossover scaling function 
of the susceptibility at the critical density for $D=2$ (left).
The metric factors are given by 
$e={\EuFrak a}_{\ssstyle h} {\EuFrak a}_{\ssstyle R}^4$
and $b={\EuFrak a}_{\ssstyle h}^{-1} {\EuFrak a}_{\ssstyle R}^2$.    
The dashed lines correspond to the asymptotic behavior
of the two-dimensional system 
($\gamma_{\ssstyle D=2}/\sigma_{\ssstyle D=2}=0.713$)
and of the mean field behavior ($\gammaMF/\sigmaMF=1/2$).
The universal amplitude is given 
by $m_{\ssstyle \chi,h} =0.208$ which agrees within the 
error-bars with 
$m_{\ssstyle a,h} \beta_{\ssstyle D}/\sigma_{\ssstyle D}=0.195 \plmi 0.013$
[see \refeqname{eq:uni_ampl_m_chi_h}].
The right figure displays the corresponding effective
exponent~$(\gamma/\sigma)_{\ssstyle \mathrm{eff}}$.  
\label{fig:uni_co_susc_2d}} 
\end{figure} 
%%%%%%%%%%%%%%%%%%%%%%%%%%%%%%%%%%%%%%%%%%%%%%%%%%%%%%%%%%%%%%%%%%%%%%%%%%%%%%%%%%%%%%%%%%
Approaching the transition point the susceptibility is expected
to scale as
\begin{equation}
{\tilde {\EuFrak X}}(0,x,1) \; \sim \; 
m_{\ssstyle \chi,h} \,
x^{-\gamma_{\ssstyle D}/\sigma_{\ssstyle D}} \, ,
\quad\quad {\mathrm{for}} \quad \quad x\to 0 \, . 
\label{eq:fractur_X_small_x_field}
\end{equation}
On the other hand, the susceptibility obeys for $R=1$
\begin{equation}
\chi \; \sim \;
a_{\ssstyle h,R=1} \; \; {\tilde {\mathrm{X}}}_{\ssstyle D}(0,1) \; \;
\left ( a_{\ssstyle h,R=1} h
\right )^{-\gamma_{\ssstyle D}/\sigma_{\ssstyle D}} .
\label{eq:scaling_sucp_R1}
\end{equation}
Thus the universal power-law amplitude is given by
\begin{equation}
m_{\ssstyle \chi,h} \; = \;
\left ( \frac{a_{\ssstyle h, R=1}}
{a_{\ssstyle h, R\to \infty}} 
\right )^{1-\gamma_{\ssstyle D}/\sigma_{\ssstyle D}} \;
{\EuFrak a}_{\ssstyle R}^{-\gammaMF/\phi
 +\sigmaMF \gamma_{\ssstyle D}/
\sigma_{\ssstyle D} \phi}
\; \frac{\, \beta_{\ssstyle D} \, }{\sigma_{\ssstyle D}} 
\; = \; m_{\ssstyle a,h} \,
\frac{\, \beta_{\ssstyle D} \, }{\sigma_{\ssstyle D}} 
\, ,
%{\tilde {\mathrm{X}}}_{\ssstyle D}(0,1) \, .
\label{eq:uni_ampl_m_chi_h}   
\end{equation}
where \refeqname{eq:susc_Chi_01} is used.
The rescaled susceptibility is 
shown in \reffigname{fig:uni_co_susc_2d}.
Over the entire crossover region we got an excellent
data-collapse including both asymptotic scaling regimes.
The right figure displays the effective exponent
\begin{equation}
\left (
\frac{\gamma}{\sigma} 
\right )_{\ssstyle \mathrm{eff}} \; = \; - \,
\frac{\partial\hphantom{\ln{x}}}{\partial \ln{x}}
\, \ln{{\tilde{\EuFrak X}}(0,x,1)} 
\label{eq:def_eff_exp_gamma_sigma}
\end{equation}
which exhibits again a monotonic crossover from the
two-dimensional scaling regime to the mean field
scaling behavior.

\begin{figure}[t] 
\centering
%\leavevmode 
\includegraphics[clip,width=8cm,angle=0]{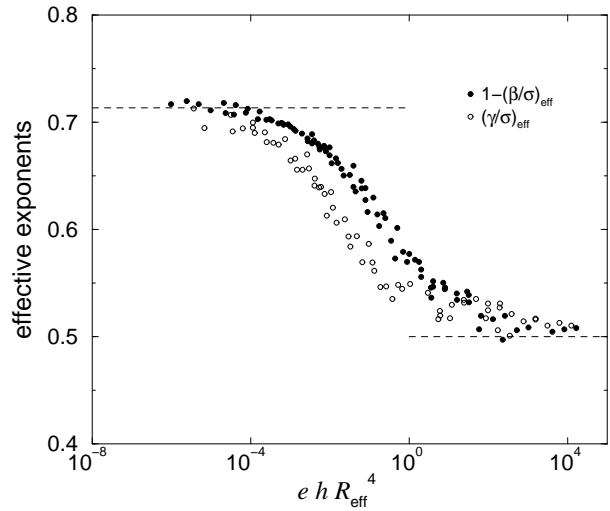}
\caption[Violation of Widom law in the crossover regime, Manna class, $D=2$]
{The violation of the Widom scaling law 
\refeqname{eq:widom_eff} in the crossover regime.
Data of the Manna model and the conserved threshold transfer
process are plotted.
The dashed lines correspond to the asymptotic regimes
(two-dimensional behavior and mean field scaling behavior).
\label{fig:widom_01}} 
\end{figure}

In this way we have obtained the effective
exponents 
$(\beta / \sigma )_{\ssstyle \mathrm{eff}}$
and $(\gamma / \sigma )_{\ssstyle \mathrm{eff}}$
for the field driven crossover from 
mean field to non-mean field behavior.
Thus we are able to check the corresponding 
Widom scaling law [\refeqname{eq:widom_apt}]
\begin{equation}
\left (
\frac{\gamma}{\sigma}
\right )_{\ssstyle \mathrm{eff}} \; = \;
1 \, - \, \left (
\frac{\beta}{\sigma}
\right )_{\ssstyle \mathrm{eff}} \, ,
\label{eq:widom_eff}
\end{equation}
for the whole crossover region.
The corresponding data are shown in \reffigname{fig:widom_01}.
As can be seen the Widom scaling law is 
fulfilled for the asymptotic regimes ($D=2$
and mean field scaling behavior) but it is clearly
violated for the intermediate crossover region.
This result is not surprising if one notices that the
above Widom law \refeqname{eq:widom_eff}
corresponds to the differential 
equation [see \refeqs2name{eq:def_eff_exp_sigma}{eq:def_eff_exp_gamma_sigma}]
\begin{equation}
- \frac{\partial\ln{\hphantom{x}}}{\partial \ln{x}}
\, \frac{\partial\hphantom{x}}{\partial x}
\, {\tilde{\EuFrak R}}(0,x,1)
\; = \;
1 \, - \,
\frac{\partial\ln{\hphantom{x}}}{\partial \ln{x}} \,
{\tilde{\EuFrak R}}(0,x,1) \, .
\label{eq:widom_dlg}
\end{equation}
Using $1= \partial\ln{a x} / \partial\ln{x}$ we get
\begin{equation}
- \ln{\partial_{\ssstyle x} {\tilde{\EuFrak R}}(0,x,1)}
\; = \; \ln{ax} \, - \,
\ln{{\tilde{\EuFrak R}}(0,x,1)} \, + \, c \, ,
\label{eq:widom_dlg_2}
\end{equation}
%$- \ln{\partial_{\scriptscriptstyle x} {\tilde{\EuFrak R}}(0,x,1)}
%= \ln{ax} -
%\ln{{\tilde{\EuFrak R}}(0,x,1)} + c $
where $c$ is some constant.
It is straightforward to show that this differential
equation is solved by simple power-laws 
[${\tilde{\EuFrak R}}(0,x,1)= c_{\ssstyle 0}
x^{c_{\ssstyle 1}}$
with $c_{\ssstyle 1}=1/a \exp{c}$].
Thus the Widom scaling law is fulfilled in the asymptotic regimes only.
In the case that the scaling behavior is affected by
crossovers, confluent singularities, etc.~no pure power-laws
occur and the scaling laws 
do\index{effective exponent}\index{confluent singularities} 
not hold for the corresponding
effective exponents.

%%%%%%%%%%%%%%%%%%%%%%%%%%%%%%%%%%%%%%%%%%%%%%%%%%%%%%%%%%%%%%%%%%%%%%%%%%%%%%%%%%%%%%%%
%%%%%%%%%%%%%%%%%%%%%%%%%%%%%%%%%%%%%%%%%%%%%%%%%%%%%%%%%%%%%%%%%%%%%%%%%%%%%%%%%%%%%%%%

$~$ \\

In conclusion, the numerically investigated crossover from mean field 
to non-mean field scaling behavior within the
Manna universality class reveals that the 
corresponding crossover scaling functions are universal.
Although the systems considered exhibit non-equilibrium 
phase transitions, the results can be applied to 
continuous phase transitions in general, including
equilibrium systems.
All effective exponents considered change monotonically
from their mean field values to their $D$-dimensional values.
In contrast, the crossover from the stochastic Manna universality 
class to the universality class of deterministic particle
distribution is characterized by a non-monotonic behavior~\cite{LUEB_13}.
Regardless of a monotonic or non-monotonic behavior,
the crossover scaling functions are universal.
This is of experimental relevance 
since measurements of critical behavior are 
performed often within a crossover region.

\chapter{Concluding remarks and outlook} 
\label{chapter:outlook}

Similar to equilibrium critical phenomena, 
the great variety of 
non-equilibrium phase transitions
can be grouped into different universality classes.
Each universality class is characterized by a 
certain symmetry which is often masked 
within the Langevin equation approach, but it is
expressed clearly within the corresponding 
path integral formulation
(see e.g.~\cite{ODOR_1,JANSSEN_4} and references therein).
For example, non-equilibrium critical systems
belong to the directed percolation
universality class if the associated 
absorbing phase transition is 
described by a single component order parameter
and if the corresponding course grained system 
obeys the rapidity reversal symmetry asymptotically.
Different universality classes occur if the 
rapidity reversal is broken, e.g.~by quenched disorder or
by a coupling of the
order parameter to a non-diffusive conserved field
as in the case of the Manna universality class.
Since no complete classification scheme is known so far,
a phenomenological identification
of non-equilibrium universality classes is needed.
Beyond the values of the critical exponents, universal 
scaling functions are a very sensitive and accurate
tool to verify a system's universality class.
Therefore, universal scaling plots reflect
the robustness of a given universality class impressively.
For example, a striking demonstration of the universality class
of directed percolation is presented 
in \reffigname{fig:uni_dp_eqos_HS_2d}
where a universal data-collapse of five different models 
is shown.
%the data-collapse of the 
%equation of state of five different models 
%is a striking demonstration of the universality class
%of directed percolation (see~\reffigname{fig:uni_dp_eqos_HS_2d}).

Additionally to the manifestation of universality classes, universal
scaling functions are useful in order to check
renormalization group results quantitatively.
Since the renormalization group theory is the basis of
our understanding of critical phenomena it is of fundamental
interest to examine the obtained results.
Due to the continuing improvement of computer hardware,
accurate numerical data have become available in the last years,
resulting in a fruitful and instructive interplay
between numerical investigations and renormalization
group analyzes.
Incorporating the conjugated 
field, a series of opportunities is offered to compare
renormalization group results to those of numerical
investigations.
Simulations performed for non-zero field
include the measurements of the equation of state,
of the susceptibility, as well as a modified 
finite-size scaling analysis 
appropriate for absorbing phase transitions.

In this work, two non-equilibrium universality
classes are considered in detail, namely the
directed percolation and the Manna universality class.
%A picture gallery of universal scaling functions
%is presented and we hope that it will be useful for
%future research.
Although steady state, dynamical, as well as finite-size scaling 
functions are determined future work is needed to complete the gallery of 
universal scaling functions.
For example, the universal scaling behavior of correlation
functions has not been considered so far.
Another example is the scaling behavior of persistence distributions 
which has attracted a lot of research interests in the recent years
(see 
e.g.~\cite{DERRIDA_1,BRAY_1,STAUFFER_1,MAJUMDAR_4,CUEILLE_1,MAJUMDAR_5,DERRIDA_2,LEE_1,OERDING_1,HINRICHSEN_2,ALBANO_1,HINRICHSEN_3,LUEB_21}).
Persistence distributions are defined as the probability that a certain
physical quantity does not change its state during the
stochastic evolution.
%Investigations of both reversible and irreversible phase transitions
%are reported.
Various distributions are investigated, associated with the
persistence behavior of the local as well as global order
parameter.
In particular, the global distributions are expected to be
universal, i.e., they are suitable to reflect the corresponding
universality class.
So far, most work has focused on the determination of the
exponents, neglecting the corresponding universal scaling 
functions.

Furthermore, it is interesting to consider crossover
phenomena between the different universality classes.
The threshold transfer process provides\index{threshold transfer process}
an opportunity to examine the crossover from 
directed percolation to the Manna universality class.
Assuming that particle creation/annihilation processes occur
with rate~$\kappa$, a crossover to the conserved
threshold\index{conserved threshold transfer process}
transfer process takes place for $\kappa\to 0$.

Within this work, we focus our attention on short range interacting
systems.
Long-range interactions may affect the scaling behavior
significantly.
It is known from equilibrium that slowly decaying
long-range interactions reduce the value of the upper
critical dimension~$\Dc$~\cite{JOYCE_2,STELL_1,FISHER_8,AIZENMAN_1}.
Since lower values of the upper critical dimension allow
to simulate larger system sizes, it should be possible 
to investigate the logarithmic corrections at $\Dc$
with a higher accuracy in that way.

Additionally to the considered directed percolation and 
Manna universality class, it is desirable
to determine the universal scaling functions of 
other non-equilibrium phase transitions.
According to field theories as well as numerical
investigations, different universality classes emerge
from disorder\index{disorder effects} 
effects~\cite{HINRICHSEN_1,JANSSEN_11,JENSEN_10,MOREIRA_1,CAFIERO_1,HOOYBERGHS_1,VOJTA_1}.
The different universal behavior is of particular interest since
disorder effects are suspected to destroy possible experimental 
realizations of directed percolation~(see e.g.~\cite{HINRICHSEN_8}).
If this explanation is valid, the disorder associated
scaling behavior will also occur experimentally and universal 
scaling functions will be useful for an 
accurate identification. 
Another established universality class
is related to the parity\index{parity conserving} 
conservation process~\cite{ZHONG_1,CARDY_2}.
So far, no systematic analysis of the corresponding
universal scaling functions has been performed.

Pinning transitions in driven-disordered media
are another well known\index{depinning transition}
non-equi\-li\-brium critical 
phenomenon (see e.g.~\cite{KARDAR_2,BRAZOVSKII_1}).
These transitions occur at zero temperature 
due to the competition between an external driving force
and an internal quenched disorder.
Taking thermal fluctuations into account,
universal scaling functions become available in
numerical simulations~\cite{NOWAK_1,LUEB_11,LUEB_15,LUEB_17} 
and can be compared to 
experimental data (see e.g.~\cite{REPAIN_1}).
Further numerical, 
% various disorder distributions
theoretical 
% RG analysis including thermal fluctuations
as well as experimental
% various temperatures
research is needed to provide a complete 
picture of the scaling behavior of depinning transitions.

Eventually, another open point and perhaps the most important 
one is the
formulation of a universality hypothesis of non-equilibrium
phase transitions.
Due to a lack of a unifying theoretical framework,
a classification scheme similar to Kadanoff's hypothesis
in equilibrium is still unknown.
%and remains therefore an exciting challenge for future research.
Thus, the concept of universality
remains the major tool to order 
the great variety of critical phenomena.
Therefore, we hope that the presented picture gallery
of universal scaling functions will be useful for future
research, because 
universal scaling functions are the most impressive and
most beautiful manifestations of universality at all.

%\backmatter

\chapter{Appendix}

\appendix

\section{Branching process}

In probability theory (see e.g.~\cite{HARRIS_1,GRIMMETT_1}), 
a branching process\index{branching process} models 
the evolution of a population.
Each individual in generation $n$ (e.g. particle at
time $t$)
creates in the next generation (time $t+1$)
$k$~offsprings (particles) 
with probability $p_{\ssstyle k}$.
It is assumed that the probabilities $p_{\ssstyle k}$ do not
vary from individual to individual.
A central question is whether the branching process
leads to ultimate extinction, i.e., no individuals
exist after some finite number of generations.
The expected number of offsprings of each
generation is denoted by $\mu$.
Starting in generation zero with one individual,
the expected size of generation~$n$ is given by
$\mu^n$.
Thus three cases are distinguished:~the subcritical
case $\mu <1$,
the critical case $\mu=1$, and the supercritical
case $\mu>1$.
Ultimate extinction is certain for $\mu \kgl 1$
whereas the probability of ultimate extinction is 
less than one for $\mu>1$.
At criticality the survival probability~$P(n)$,
i.e., the probability that the branching process
is still alive after~$n$ generations 
is known to scale as $P(n)\asympprop n^{-1}$. 
Furthermore, the probability $P(s)$ that a critical 
branching process creates exactly~$s$ individuals scales 
for large~$s$ 
as $P(s)\asympprop s^{-3/2}$~\cite{OTTER_1}.

Considering activity spreading simulations, the 
mean field scaling behavior 
at criticality equals that of a critical branching process.
Thus the mean field values of the spreading exponents 
are given by $\thetaMF=0$, $\deltaMF=1$, 
as well as $\tau_{\ssstyle s,{\mathrm{MF}}}=3/2$.

\section{Scaling laws}
\label{subsec_app:scaling_laws}

In\index{scaling law} 
general, absorbing phase transitions are characterized
by four independent exponents, for example
$\beta$, $\betap$, $\nu_{\senk}$,
and $\nu_{\para}$ (see e.g.~\cite{HINRICHSEN_1}).
All other exponents can be expressed in terms of 
these exponents.
The spreading exponents $\delta$, $\theta$, 
and the dynamical exponent~$z$
are given by 
[see \refeqs2name{eq:delta_betap_nu_para}{eq:gen_hyperscaling_law}]
\begin{equation}
\delta \; = \; \frac{\,\betap \,}{\nu_{\para}} \, ,
\quad\quad\quad
\theta \, = \, \frac{\,D\,}{z} \, - \, \frac{\,\beta\,}{\nu_{\para}} \, 
 -\, \frac{\,\betap\,}{\nu_{\para}}   \, ,
\quad\quad\quad z \; = \; \frac{\,\nu_{\para}\,}{\nu_{\senk}} \, .
\label{eq:spreading_exponents}
\end{equation}   
The steady state critical exponents $\gamma$, $\gammap$ and $\sigma$
are related to ($\beta$, $\betap$, $\nu_{\senk}$, $\nu_{\para}$) via 
\begin{equation}
\gamma \; = \; \sigma-\beta \, ,
\quad\quad\quad
\sigma \; = \; D \, \nu_{\senk} \, + \, \nu_{\para}\, - \, \betap \, ,
\quad\quad\quad
\gammap \; = \; D \, \nu_{\senk} \, - \, 2\, \beta \, .
\label{eq:steady_state_exponents}
\end{equation}   
Furthermore, the fractal dimension 
of the spreading clusters [see \refeqname{eq:def_frac_dim_cluster}]
is given by 
\begin{equation}
D_{\ssstyle f}\; = \; z \, (\theta + \delta) 
\; = \; D  -  z \, \delta 
\; = \; D - \frac{\beta}{\nu_{\senk}} \, .
\label{eq:frac_dim}
\end{equation}
Simple dimensional analysis offers a convenient way to derive
these scaling laws.
Therefore, we remind that the various quantities of interest 
enter the scaling forms in combinations with their 
scaling powers, e.g.,
\begin{eqnarray}
& &\lambda \, \deltap, \quad
\lambda^{\beta} \rhoa, \quad
\lambda^{\sigma} h,  \quad
\lambda^{-\gamma} \chia,  \quad
\lambda^{-\gammap} \Delta\rhoa,  \quad
\lambda^{-\nu_{\senk}} {\underline x},  \quad
\lambda^{-\nu_{\senk}} L,  \quad \nonumber \\
& & \lambda^{-\nu_{\senk}} R,  \quad 
\lambda^{-\nu_{\para}} t,  \quad
\lambda^{D\nu_{\senk}-\nu_{\para}\delta} \rhoanull,  \quad 
\lambda^{-\theta \nu_{\para}} \Na,  \quad
\lambda^{\delta \nu_{\para}} \Pa,  \quad
\ldots \, . 
\label{eq:scaling_powers}  
\end{eqnarray}
In the language of real-space renormalization, 
the rescaling is usually related to the transformation 
${\underline x} \mapsto b {\underline x}$,
corresponding to $\lambda=b^{1/\nu_{\senk}}$.
Enforcing scale invariance, the definition of
e.g. the susceptibility leads to 
\begin{equation}
\label{eq:dim_analysis_widom}   
\chia = \frac{\partial \, \rhoa}{\partial \, h}
\quad \Longrightarrow
\quad \gamma=\sigma-\beta \, .
\end{equation}
Similar, the hyperscaling law of the fluctuation exponent
is obtained from
\begin{equation}
\label{eq:dim_analysis_fluc}   
\Delta\rhoa = L^D \; (\langle \rhoa^2 \rangle - \langle \rhoa \rangle ^2) 
\quad \Longrightarrow
\quad \gammap=D\nu_{\senk} - 2 \beta \, .
\end{equation}
Taking into account that an initial homogenous particle
density $\rhoanull$ may be represented by the external
field we find
\begin{equation}
\label{eq:dim_analysis_field}   
h = \rhoanull \, \delta(t)
\quad \Longrightarrow
\quad 
\sigma =  D \nu_{\senk}  +  \nu_{\para} -  \nu_{\para}\delta \, .
\end{equation}
Considering activity spreading, the average number of active sites 
(usually averaged over all runs) scales as %~\cite{GRASSBERGER_4}
\begin{equation}
\label{eq:dim_analysis_Na}   
\Na = \mathrm{const} \, \rhoa \, \Pa \, R^{\ssstyle D} 
\quad \Longrightarrow
\quad 
\nu_{\para} \theta =  D \nu_{\senk}  - {\beta}
-  \nu_{\para} \delta \, .
\end{equation}
Eventually, we mention that the correlation function of 
active sites
\begin{equation}
\Gamma({\underline r}_{\ssstyle i},{\underline r}_{\ssstyle j})
\; = \; \left \langle \, \left (\rhoa({\underline r}_{\ssstyle i})^{\vphantom{X}} 
\, - \, \langle \rhoa({\underline r}_{\ssstyle i})\rangle  \right )\,
\left (\rhoa({\underline r}_{\ssstyle j}) \, - \,
\langle \rhoa({\underline r}_{\ssstyle j}) \rangle^{\vphantom{X}}  \right) \,
\right \rangle \, 
\label{eq:def_corr_function}
\end{equation}
is expected to decay for translationary invariant systems 
(${\underline r}={\underline r}_{\ssstyle i}-{\underline r}_{\ssstyle j}$)
as
\begin{equation}
\Gamma({\underline r},0,0) \, \asympprop \, r^{-D+2-\eta_{\senk}} \, 
\label{eq:def_eta}.
\end{equation}
The correlation function is related to the order
parameter fluctuations 
\begin{equation}
\Delta\rhoa(\deltap, h) \; = \; \sum_{{\underline r}} \, 
\Gamma({\underline r}, \deltap, h) \, ,
%\; \sim \;
%\int  {\mathrm d}r \; r^{D-1} \, \Gamma(r, \deltap, h)  
\label{eq:fluc_corr_func}
\end{equation}
yielding the scaling power of the correlation
function~${2\beta}$.
Thus, the correlation function exponent $\eta_{\senk}$ obeys
the scaling law
\begin{equation}
(2\, -\, \eta_{\senk})\; \nu_{\senk} \; = \;
D \, \nu_{\senk} \, - \, 2 \, \beta \; = \; \gammap
\label{eq:non_equil_fisher}
\end{equation}
which corresponds to the 
Fisher\index{scaling law, Fisher}\index{Fisher scaling law}
scaling law
of equilibrium phase transitions.

\clearpage
   
\section{Critical exponents, universal amplitude combinations
and critical parameters}
 
\subsection{Directed percolation}

\vspace{-5mm}
\begin{table}[h]
\centering
\caption{
The critical exponents and various universal amplitude combinations
of directed percolation for various dimensions~$D$.
In $D=1$, the exponents $\gamma$, $\nu_{\protect\senk}$, and
$\nu_{\protect\para}$ are obtained from a 
series expansion by Jensen~\protect\cite{JENSEN_5}.
For $D=2$ and $D=3$ activity spreading simulations
are performed yielding $\delta$, $\theta$, as well 
as $z$~\protect\cite{VOIGT_1,JENSEN_6}.
Additionally, the exponent $\nu_{\protect\para}$ is
determined~\cite{GRASSBERGER_3,JENSEN_6} in order to 
estimate the full set of exponents via scaling laws.
}
\vspace{0.3cm}
\label{table:dp_exponents}
\begin{tabular}{|c|l|l|l|c|}
\hline
$$       
& $D=1$ {\protect{\cite{JENSEN_5}}}
& $D=2$ {\protect\cite{VOIGT_1,GRASSBERGER_3}}
& $D=3$ {\protect\cite{JENSEN_6}}
& Mean field\\  
\hline
$\beta=\betap$    &  $0.276486(8)$	& $0.5834\pm0.0030\quad$ & $0.813\pm0.009$ & $1$\\   
$\nu_{\senk}$     &  $1.096854(4)$	
& $0.7333\pm0.0075$ &  $0.584\pm0.005$& $1/2$\\    
$\nu_{\protect\para}$  &  $1.733847(6)$	
& $1.2950\pm0.0060$ &  $1.110\pm0.010$ & $1$\\    
$\sigma$     &  $2.554216(13)$	& $2.1782\pm0.0171$ &
$2.049\pm0.026$ & $2$\\    
$\protect\gammap$     &  $0.543882(16)$	&
$0.2998\pm0.0162$ & $0.126\pm0.023$& $0$\\   
$\gamma$        & $2.277730(5)$	&
$1.5948\pm0.0184$ & $1.237\pm 0.023$ & $1$\\  
$\eta_{\senk}$&  $1.504144(19)$    & $1.5912\pm 0.0148$      & $1.783\pm 0.016$ & $2$\\        
\hline
$\delta=\alpha$ &  $0.159464(6)$	    & $0.4505\pm0.0010$   
& $0.732\pm 0.004$ & $1$\\ 
$\theta$      &  $0.313686(8)$    & $0.2295\pm0.0010$  & $0.114\pm 0.004$ & $0$\\
$z$ 	      &  $1.580745(10)$   & $1.7660\pm0.0016$  & $1.901\pm0.005$ & $2$\\ 
\hline
${\tilde D}(1,0)$ {\protect{\cite{LUEB_28}}}  
&  $1.46\pm0.12$	    & $1.65\pm0.09$   & $1.83\pm0.11$ & $2$\\      
$\frac{{\tilde \mathrm{X}}(+1,0)}{{\tilde \mathrm{X}}(-1,0)}$ {\protect{\cite{LUEB_28}}} 	     
&  $0.033\pm0.004$	    & $0.25\pm0.01$   & $0.65\pm0.03$ & $1$\\      
$R_{\chi}$ {\protect{\cite{LUEB_28}}} 	     
&  $0.60\pm0.04$	    & $0.72\pm0.04$   & $0.86\pm0.08$ & $1$\\ 
$U$ {\protect{\cite{LUEB_33}}} 	     
&  $0.833\pm0.011$	    & $0.704\pm0.013$   & $0.61\pm0.02$ & $1/2$\\ 
\hline
\end{tabular}
\end{table}

\subsection{Compact directed percolation}
   \begin{table}[h]
\centering
\caption{The critical exponents of the compact directed percolation
universality class ($D=1$).
The critical behavior equals that of the $1+1$-dimensional 
voter model~\protect\cite{LIGGETT_1} and is tractable analytically
due to the mapping to random 
walks~\protect\cite{DOMANY_1,ESSAM_1}.}
\vspace{0.5cm}
\label{table:cdp_exponents}
\begin{tabular}{|c|c|c|c|c|c|c|}
\hline
\phantom{X}$\betap$ \phantom{X}
& \phantom{X}$\beta$\phantom{X} 	
& \phantom{X}$\nu_{\senk}$\phantom{X}
& \phantom{X}$\nu_{\para}$\phantom{X} 
& \phantom{X}$z$\phantom{X}   
& \phantom{X}$\delta$\phantom{X} 
& \phantom{X}$\theta$\phantom{X} \\  
\hline
$1$ 		& $0$ 	& $1$ 	& $2$ & $2$ & $1/2$ & $0$\\  
\hline
\end{tabular}
\end{table}

\clearpage

\subsection{Manna universality class}

\begin{table}[h]
\centering
\caption{The critical exponents and various universal amplitude combinations
of the Manna universality class 
below the upper critical dimension $\Dc=4$.
The data of the exponents $\beta$, $\sigma$, 
$\nu_{\protect\senk}$, and $\protect\gammap$  
are obtained from steady state 
simulations~\protect\cite{LUEB_24,LUEB_23},
whereas activity spreading reveals the values 
of $\alpha$, $\delta$, $\theta$, and $z$~\protect\cite{LUEB_23}.
The exponents $\protect\betap$ and $\nu_{\protect\para}$
are determined via scaling laws.
In particular, the values of $\nu_{\protect\para}$ are 
in good agreement with those of direct measurements of the order 
parameter persistence distribution~\protect\cite{LUEB_21}. 
In case of the one-dimensional models we observe a 
splitting of the universality class.}
\vspace{0.5cm}
\label{table:manna_exponents}
\begin{tabular}{|c|l|l|l|c|}
\hline
$$       &  $D=1$	& $D=2$	& $D=3$ & Mean field\\  
\hline
$\beta$      &  $0.382\pm0.019$	& $0.639\pm0.009\quad$ & $0.840\pm0.012$ & $1$\\   
$\beta^{\prime}$  &  $0.319\pm0.052_{\ssstyle\mathrm{Manna}}$	
& $0.624\pm0.029\quad$ & $0.827\pm0.034$ & $1$\\ 
$\nu_{\senk}$     &  $1.347\pm0.091_{\ssstyle\mathrm
{Manna}}$	& $0.799\pm0.014$ & $0.593\pm0.013$& $1/2$\\    
$$           &  $1.760\pm0.060_{\ssstyle\mathrm{CTTP}}$	& $$ & $$ & $$\\   
$\nu_{\para}$     &  $1.876\pm 0.135_{\ssstyle\mathrm{Manna}}$	
& $1.225\pm0.029$ & $1.081\pm0.027$ & $1$\\    
$$           &  $2.452\pm0.106_{\ssstyle\mathrm {CTTP}}$	& $$ & $$ & $$\\   
$\sigma$     &  $2.710\pm0.040_{\ssstyle\mathrm {Manna}}\quad$	& $2.229\pm0.032$ &
$2.069\pm0.043$ & $2$\\    
$$           &  $1.770\pm0.058_{\ssstyle\mathrm {CTTP}}\quad$	& $$ & $$ & $$\\   
$\protect\gammap$     &  $0.550\pm0.040_{\ssstyle\mathrm {Manna}}$	&
$0.367\pm0.019$ & $0.152\pm0.017$& $0$\\    
$$           &  $0.670\pm0.040_{\ssstyle\mathrm {CTTP}}$	& $$ & $$ & $$ \\
$\gamma$     &  $2.328\pm0.044_{\ssstyle\mathrm {Manna}}$	&
$1.590\pm0.033$ & $1.229\pm0.045$ & $1$\\    
$$           &  $1.388\pm0.040_{\ssstyle\mathrm {CTTP}}$	& $$ & $$ & $$ \\
$\eta_{\senk}$     &  $1.592\pm0.040_{\ssstyle\mathrm {Manna}}$	&
$1.541\pm0.025$ & $1.744\pm0.029$& $2$\\    
$$           &  $1.619\pm0.028_{\ssstyle\mathrm {CTTP}}$	& $$ & $$ & $$ \\
\hline
\phantom{XX}$\alpha$\phantom{XX}&  $0.141\pm0.024$	    & $0.419\pm0.015$   
& $0.745\pm0.017$ & $1$\\ 
$\delta$     &  $0.170\pm0.025$     & $0.510\pm0.020$   & $0.765\pm0.025$ & $1$\\ 
$\theta$     &  $0.350\pm0.030$     & $0.310\pm0.030$   & $0.140\pm0.030$ & $0$\\
$z$ 	     &  $1.393\pm0.037$	    & $1.533\pm0.024$   & $1.823\pm0.023$ & $2$\\      
\hline
${\tilde D}(1,0)$  &  $$	    & $1.81\pm0.03$   & $1.91\pm0.08$ & $2$\\      
$\frac{{\tilde \mathrm{X}}(+1,0)}{{\tilde \mathrm{X}}(-1,0)}$ 	     
&  $$	    & $0.41\pm0.05$   & $0.8\pm0.05$ & $1$\\      
\hline
\end{tabular}
\end{table}

\clearpage

\subsection{Critical parameters}

\begin{table}[h]
\centering
\caption{Estimates of the critical probabilities ($\pc$ and $\lambda_{\ssstyle \mathrm{c}}$) 
for the directed percolation universality class.}
\vspace{0.5cm}
\label{table:pc_values_dp}
\begin{tabular}{|c|c|c|c|}
\hline
& ${\mathrm{DP}}_{\ssstyle {\mathrm{bcc}}}$ 
& ${\mathrm{CP}}_{\ssstyle {\mathrm{sc}}}$
& ${\mathrm{PCP}}_{\ssstyle {\mathrm{sc}}}$\\
\hline
$D=1$
& 0.70548515(20)\protect\cite{JENSEN_16}        & 3.297848(22)\protect\cite{JENSEN_11}   & 0.077093(3)\protect\cite{LUEB_27}   \\
$D=2$
& 0.34457(1)\protect\cite{GRASSBERGER_3}       & 1.64877(3)\protect\cite{DICKMAN_7}   & 0.20053(9)\protect\cite{LUEB_32}   \\
$D=3$
& 0.160958(6)\protect\cite{GRASSBERGER_P2004}  & 1.31686(1)\protect\cite{DICKMAN_7}   & 0.25803(13)\protect\cite{LUEB_32}   \\
$D=4$
& 0.0755850(3)\protect\cite{GRASSBERGER_12}    & 1.19505(15)\protect\cite{LUEB_32}   &    \\
$D=5$
& 0.0359725(2)\protect\cite{GRASSBERGER_P2004} & 1.13846(11)\protect\cite{LUEB_32}   &  0.29874(15)\protect\cite{LUEB_32}  \\
\hline
\end{tabular}
\end{table}

\begin{table}[h]
\centering
\caption{Estimates of the critical densities $\rhoc$ 
for the Manna universality class.}
\vspace{0.5cm}
\label{table:pc_values_manna}
\begin{tabular}{|c|c|c|c|}
\hline
& ${\mathrm{Manna}}_{\ssstyle {\mathrm{sc}}}$ 
& ${\mathrm{CLG}}_{\ssstyle {\mathrm{sc}}}$
& ${\mathrm{CTTP}}_{\ssstyle {\mathrm{sc}}}$\\
\hline
$D=1$
& 0.89199(5)\protect\cite{LUEB_26}   & 1/2 & 0.96929(3)\protect\cite{LUEB_24} \\
$D=2$
& 0.68333(3)\protect\cite{LUEB_26}   & 0.34494(3)\protect\cite{LUEB_19} & 0.69392(1)\protect\cite{LUEB_24} \\
$D=3$
& 0.60018(4)\protect\cite{LUEB_26}   & 0.21791(9)\protect\cite{LUEB_26} & 0.60489(2)\protect\cite{LUEB_26}    \\
$D=4$
& 0.56451(7)\protect\cite{LUEB_26}   & 0.15705(10)\protect\cite{LUEB_26} & 0.56705(3)\protect\cite{LUEB_26} \\
$D=5$
& 0.54704(9)\protect\cite{LUEB_26}   & 0.12298(15)\protect\cite{LUEB_26} & 0.54864(5)\protect\cite{LUEB_26} \\
\hline
\end{tabular}
\end{table}

\clearpage

\subsection{Crossover parameters}

\begin{table}[h]
\centering
\caption{The range of interactions $R$, the corresponding
number of interacting neighbors~$z_{\ssstyle {\mathrm{nn}}}$ on a 
two-dimensional square lattice and 
the effective range of 
interactions~$R_{\ssstyle {\mathrm{eff}}}$ for which 
simulations have been performed.
Additionally, the values of the critical densities are listed.}
\vspace{0.5cm}
\label{table:radius_2d}
\begin{tabular}{|c|c|c|l|l|l|}
\hline
$R$ & $z_{\ssstyle {\mathrm{nn}}}$ 
& $R_{\ssstyle {\mathrm{eff}}}^2$
& $\rho_{\ssstyle {\mathrm c}, R}^{\ssstyle  {\mathrm{Manna}}}\quad\quad\quad\quad$ 
& $\rho_{\ssstyle {\mathrm c}, R}^{\ssstyle  {\mathrm{CLG}}}\quad\quad\quad\quad$ 
& $\rho_{\ssstyle {\mathrm c}, R}^{\ssstyle  {\mathrm{CTTP}}}\quad$\\
\hline
1 	   & 4 	    & 1                         & 0.68333(3) & 0.34494(3)     & 0.69392(1)   \\
2 	   & 12     & $\frac{7}{3}$             & 0.60595(4)  & 0.22432(4)     & 0.63649(2)   \\
4 	   & 48     & 8                         & 0.54378(6)  & 0.16802(7)     & 0.55005(3)   \\
8 	   & 196    & $\frac{1546}{49}$         & 0.51549(8)  & 0.14050(9)     & 0.51688(4)   \\
16 	   & 796    & $\frac{25274}{199}$       & 0.50481(9)  & 0.12977(10)     & 0.50552(6)   \\
32 	   & 3208   & $\frac{204875}{401}$      & 0.50146(10)  & 0.12598(11)     & 0.50161(7)   \\
64  	   & 12852  & $\frac{13146247}{6426}$   & 0.50045(12)  & 0.12499(16)     & 0.50046(8)   \\
$\quad$128$\quad$ & 51432  & $\quad\frac{105255421}{12858}
\quad$ & & 0.12465(19) & 0.50019(9)\\
\hline
\end{tabular}
\end{table}

\begin{table}[h]
\centering
\caption{The range of interactions $R$, the corresponding
number of interacting neighbors~$z_{\ssstyle {\mathrm{nn}}}$ on a 
three-dimensional square lattice and 
the effective range of 
interactions~$R_{\ssstyle {\mathrm{eff}}}$ for which 
simulations have been performed.
Additionally, the values of the critical densities are listed.}
\vspace{0.5cm}
\label{table:radius_3d}
\begin{tabular}{|c|c|c|l|}
\hline
$R$ & $z_{\ssstyle {\mathrm{nn}}}$ 
& $R_{\ssstyle {\mathrm{eff}}}^2$
& $\rho_{\ssstyle {\mathrm c}, R}^{\ssstyle  {\mathrm{CTTP}}}\quad$\\
\hline
$1$ 	    & 6       & 1                         & 0.60483(1)   \\
$\sqrt{2}$  & 18      & $\frac{5}{3}$             & 0.56770(2)   \\
$2$ 	    & 32      & $\frac{39}{16}$           & 0.54406(3)   \\
$2\sqrt{2}$ & 92      & $\frac{219}{46}$          & 0.51720(4)   \\
$4$ 	    & 256     & $\frac{597}{64}$          & 0.50694(6)   \\
$4\sqrt{2}$ & 750     & $\frac{2383}{125}$        & 0.50260(7)   \\
$8$  	    & 2108    & $\frac{20016}{527}$       & 0.50136(8)   \\
$16$  	    & 17076   & $\frac{217886}{1423}$     & 0.50044(8)   \\
\hline
\end{tabular}
\end{table}

%\chapter*{Acknowledgements} 
%\input{_sonstiges/_acknow.tex}

\bibliographystyle{sven_prsty}
\bibliography{/users/sven/Paper/_library}

\begin{thebibliography}{100} \renewcommand{\baselinestretch}{1.1} \normalsize
  \footnotesize \setlength{\itemsep}{0.2ex}

\bibitem{STANLEY_1}
{H.\,E.~Stanley}, {\em Scaling, universality, and renormalization:~three
  pillars of modern critical phenomena}, Rev.~Mod.~Phys. {\bf 71},  S358
  (1999).

\bibitem{KADANOFF_2}
{L.\,P.~Kadanoff}, {\em Critical behavior, universality and scaling,
  {$\mathrm{proceedings\; of\; the}$ {\textit 1970 Varenna summer school on
  critical phenomena}}}, edited by {M.\,S.~Green} (Academic Press, New York,
  1971).

\bibitem{GUGGENHEIM_1}
{E.\,A.~Guggenheim}, {\em The principle of corresponding states},
  J.~Chem.~Phys. {\bf 13},  253  (1945).

\bibitem{MILOSEVIC_2}
S. Milo{\u {s}}evi{\'{c}} and {H.\,E.~Stanley}, {\em Equation of state near the
  critical point:~II.~Comparison with experiment and possible universality with
  respect to lattice structure and spin quantum number}, Phys.~Rev.~B {\bf 6},
  1002  (1972).

\bibitem{PELISSETTO_3}
A. Pelissetto and E. Vicari, {\em Critical phenomena and renormalization-group
  theory}, Phys.~Rep. {\bf 368},  549  (2002).

\bibitem{HAKEN_2}
H. Haken, {\em Cooperative phenomena in systems far from thermal equilibrium
  and in nonphysical systems}, Rev.~Mod.~Phys. {\bf 47},  67  (1975).

\bibitem{HINRICHSEN_1}
H. Hinrichsen, {\em Nonequilibrium critical phenomena and phase transitions
  into absorbing states}, Adv.~Phys. {\bf 49},  815  (2000).

\bibitem{ODOR_1}
G. {\'O}dor, {\em Universality classes in nonequilibrium lattice systems},
  Rev.~Mod.~Phys. {\bf 76},  663  (2004).

\bibitem{WEIDLICH_1}
W. Weidlich, {\em Physics and social science -- The approach of synergetics},
  Phys.~Rev. {\bf 204},  1  (1991).

\bibitem{HELBING_1}
D. Helbing, {\em Quantitative sociodynamics -- Stochastic methods and models of
  social interaction processes} (Kluwer, Amsterdam, 1995).

\bibitem{KAMPEN_1}
{N.\,G.~van\,Kampen}, {\em Stochastic processes in physics and chemistry}
  (North Holland, Amsterdam, 1992).

\bibitem{RISKEN_1}
H. Risken, {\em The Fokker-Planck equation} (Springer, Berlin, 1989).

\bibitem{WOLFRAM_1}
S. Wolfram, {\em Statistical mechanics of cellular automata}, Rev.~Mod.~Phys.
  {\bf 55},  601  (1983).

\bibitem{SCHUETZ_1}
{G.\,M.~Sch{\"u}tz}, {\em Exactly solvable models for many-body systems far
  from equilibrium {\textit in Phase Transitions and Critical Phenomena},
  Vol.\,19}, edited by C. Domb and {J.\,L.~Lebowitz} (Academic Press, London,
  2001).

\bibitem{WILSON_1}
{K.\,G.~Wilson}, {\em Renormalization group and critical phenomena. I.
  Renormalization group and the Kadanoff scaling picture}, Phys.~Rev.~B {\bf
  4},  3174  (1971).

\bibitem{WILSON_2}
{K.\,G.~Wilson}, {\em Renormalization group and critical phenomena. II.
  Phase-space cell analysis of critical behavior}, Phys.~Rev.~B {\bf 4},  3184
  (1971).

\bibitem{PFEUTY_1}
P. Pfeuty and G. Toulouse, {\em Introduction to the renormalization group and
  critical phenomena} (John Wiley \protect\,\&\,Sons, Chichester, 1994).

\bibitem{PLISCHKE_1}
M. Plischke and B. Bergersen, {\em Equilibrium statistical physics} (World
  Scientific, Singapore, 1994).

\bibitem{YEOMANS_1}
{J.\,M.~Yeomans}, {\em Statistical mechanics of phase transitions} (Clarendon,
  Oxford, 1992).

\bibitem{WILSON_3}
{K.\,G.~Wilson} and J. Kogut, {\em The renormalization group and the
  $\epsilon$~expansion}, Phys.~Rep. {\bf 12C},  75  (1974).

\bibitem{FISHER_4}
{M.\,E.~Fisher}, {\em The renormalization group in the theory of critical
  behavior}, Rev.~Mod.~Phys {\bf 46},  597  (1974).

\bibitem{WEGNER_2}
{F.\,J.~Wegner}, {\em The critical state, general aspects {\textit in Phase
  Transitions and Critical Phenomena}, Vol.\,6}, edited by C. Domb and
  {M.\,S.~Green} (Academic Press, London, 1976).

\bibitem{JANSSEN_1}
{H.\,K.~Janssen}, {\em On the nonequilibrium phase transition in
  reaction-diffusion systems with an absorbing stationary state}, Z.~Phys.~B
  {\bf 42},  151  (1981).

\bibitem{GRASSBERGER_2}
P. Grassberger, {\em On phase transitions in Schl{\protect\"o}gl's second
  model}, Z.~Phys.~B {\bf 47},  365  (1982).

\bibitem{LANDAU_E_1}
{E.~Landau}, {\em Handbuch {\"u}ber die Lehre von der Verteilung der
  Primzahlen} (Teubner, Leizig, 1909).

\bibitem{ABRAMOWITZ_1}
M. Abramowitz and {I.\,A.~Stegun}, {\em Hand Book of Mathematical Functions}
  (Dover, New York, 1973).

\bibitem{ONSAGER_1}
L. Onsager, {\em Crystal statistics.~I.\,A two-dimensional model with an
  order-disorder transition}, Phys.~Rev. {\bf 65},  117  (1944).

\bibitem{YANG_1}
{C.\,N.~Yang} and {T.\,D.~Lee}, {\em Statistical theory of equation of state
  and phase transition:~I.\,Theory of condensation}, Phys.~Rev. {\bf 87},  404
  (1952).

\bibitem{LEE_2}
{T.\,D.~Lee} and {C.\,N.~Yang}, {\em Statistical theory of equation of state
  and phase transition:~II.\,Lattice gas and Ising model}, Phys.~Rev. {\bf 87},
   410  (1952).

\bibitem{LANDAU_5}
{L.\,D.~Landau} and {E.\,M.~Lifschitz}, {\em Lehrbuch der theoretischen Physik,
  {\rm Band~V}} (Akademie, Berlin, 1966).

\bibitem{ESSAM_2}
{J.\,W.~Essam} and {M.\,E.~Fisher}, {\em Pad{\'e} approximant studies of
  lattice gas and Ising ferromagnet below critical point}, J.~Chem.~Phys. {\bf
  38},  802  (1963).

\bibitem{DOMB_2}
C. Domb and {D.\,L.~Hunter}, {\em On critical behavior of ferromagnets},
  P.~Phys.~Soc.~Lond. {\bf 86},  1147  (1965).

\bibitem{WIDOM_1}
B. Widom, {\em Equation of state in the neighborhood of the critical point},
  J.~Chem.~Phys. {\bf 43},  3898  (1965).

\bibitem{KADANOFF_3}
{L.\,P.~Kadanoff}, {\em The introduction of the idea that exponents could be
  derived from real-space scaling arguments}, Physics (NY) {\bf 2},  263
  (1966).

\bibitem{PATASHINSKII_1}
{A.\,Z.~Patashinskii} and {V.\,L.~Pokrovskii}, {\em Behavior of ordered systems
  near transition point}, Sov.~Phys.~JETP {\bf 23},  292  (1966).

\bibitem{GRIFFITHS_2}
{R.\,B.~Griffiths}, {\em Thermodynamic functions for fluids and ferromagnets
  near the critical point}, Phys.~Rev. {\bf 158},  176  (1967).

\bibitem{HANKEY_1}
A. Hankey and {H.\,E.~Stanley}, {\em Systematic application of generalized
  homogeneous functions to static scaling, dynamical scaling, and
  universality}, Phys.~Rev.~B {\bf 6},  3515  (1972).

\bibitem{FISHER_3}
{M.\,E.~Fisher}, {\em Quantum corrections to critical-point behavior},
  Phys.~Rev.~Lett. {\bf 16},  11  (1966).

\bibitem{JASNOW_1}
D. Jasnow and M. Wortis, {\em High-temperature critical indices for the
  classical anisotropic Heisenberg model}, Phys.~Rev. {\bf 176},  739  (1968).

\bibitem{WATSON_1}
{P.\,G.~Watson}, {\em Formation of invariants from critical amplitudes of
  ferromagnets}, J.~Phys.~C {\bf 2},  1883  (1969).

\bibitem{GRIFFITHS_1}
{R.\,B.~Griffiths}, {\em Dependence of critical indices on a parameter},
  Phys.~Rev.~Lett. {\bf 24},  1479  (1970).

\bibitem{BETTS_1}
{D.\,D.~Betts}, {A.\,J.~Guttmann}, and {G.\,S.~Joyce}, {\em Lattice-lattice
  scaling and the generalized law of corresponding states}, J.~Phys.~C {\bf 4},
   1994  (1971).

\bibitem{HUBBARD_1}
J. Hubbard and P. Schofield, {\em Wislon theory of a liquid-vapour critical
  point}, Phys.~Lett.~A {\bf 40},  245  (1972).

\bibitem{BRILLIANTOV_1}
{N.\,V.~Brilliantov}, {\em Effective magnetic Hamiltonian and Ginzburg
  criterion for fluids}, Phys.~Rev.~E {\bf 58},  2628  (1998).

\bibitem{HOCKEN_1}
R. Hocken and {M.\,R.~Moldover}, {\em Ising critical exponents in real
  fluids:~An experiment}, Phys.~Rev.~Lett. {\bf 37},  29  (1976).

\bibitem{LEVELT_1}
{A.~Levelt\,Sengers}, R. Hocken, and {J\,V.~Sengers}, {\em Critical-point
  universality and fluids}, Physics Today {\bf 28},  42 (December)  (1977).

\bibitem{SENGERS_1}
{J.\,V.~Sengers} and {J.\,M.\,H.~Levelt\,Sengers}, {\em Critical phenomena in
  classical fluids {\textit in Progress in liquid physics}}, edited by
  {C.\,A.~Croxton} (John Wiley\protect\&Sons, New York, 1978).

\bibitem{LEVELT_2}
{J.\,M.\,H.~Levelt\,Sengers}, {\em From van\,der\,Waals' equation to the
  scaling law}, Physica~A {\bf 73},  73  (1974).

\bibitem{BALZARINI_1}
D. Balzarini and K. Ohrn, {\em Coexistence curve of sulfur hexafluoride},
  Phys.~Rev.~Lett. {\bf 28},  840  (1972).

\bibitem{ALSNIELSEN_2}
J. Als-Nielsen, {\em Neutron scattering and spatial correlation near the
  critical point {\textit in Phase Transitions and Critical Phenomena},
  Vol.\,5a}, edited by C. Domb and {M.\,S.~Green} (Academic Press, London,
  1976).

\bibitem{KAJANTIE_1}
K. Katajantie, M. Laine, R. Rummukainen, and M. Shaposhnikov, {\em Is there a
  hot electroweak phase transition at $m_{\scriptscriptstyle H} \ge
  m_{\scriptscriptstyle W}$}, Phys.~Rev.~Lett. {\bf 77},  2887  (1996).

\bibitem{HALPERIN_1}
{B.\,I.~Halperin}, {T.\,C.~Lubensky}, and {S.-K.~Ma}, {\em First-order phase
  transitions in superconductors and smectic-A liquid crystals},
  Phys.~Rev.~Lett. {\bf 32},  292  (1974).

\bibitem{HALPERIN_2}
{B.\,I.~Halperin} and {T.\,C.~Lubensky}, {\em On the analogy between smetic~A
  liquid crystals and superconductors}, Solid State Comm. {\bf 14},  997
  (1974).

\bibitem{CHAIKIN_1}
{P.\,M.~Chaikin} and {T.\,C.~Lubensky}, {\em Principles of condensed matter
  physics} (Cambridge University, Cambridge, 1995).

\bibitem{AHLERS_1}
G. Ahlers, {\em Critical phenomena and the superfluid transition in
  $^4\mathrm{He}$ {\textit in } Quantum Liquids}, edited by J. Ruvalds and T.
  Regge (North-Holland, Amsterdam, 1978).

\bibitem{LIPA_1}
{J.\,A.~Lipa} {\it et~al.}, {\em Heat capacity and thermal relaxation of bulk
  Helium very near the lambda point}, Phys.~Rev.~Lett. {\bf 76},  944  (1996).

\bibitem{LIPA_2}
{J.\,A.~Lipa} {\it et~al.}, {\em Specific heat of Helium confined to a
  $57\mathrm{µm}$ planar geometry near the lambda point}, Phys.~Rev.~Lett. {\bf
  84},  4894  (2000).

\bibitem{DOHM_1}
V. Dohm, {\em Renormalization-group flow equations of model F}, Phys.~Rev.~B
  {\bf 44},  2697  (1991).

\bibitem{ZHANG_2}
{S.-C.~Zhang}, {\em A unified theory based on ${\mathrm{SO}}(5)$ symmetry of
  superconductivity and antiferromagnetism}, Science {\bf 275},  1089  (1997).

\bibitem{DEMLER_1}
E. Demler, W. Hanke, and {S.-C.~Zhang}, {\em SO(5) theory of antiferromagnetism
  and superconductivity}, Rev.~Mod.~Phys. {\bf 76},  909  (2004).

\bibitem{JONES_1}
D. Jones, A. Love, and M. Moore, {\em Phase transitions in superfluid
  $^3{\mathrm He}$}, J.~Phys.~C {\bf 9},  743  (1976).

\bibitem{GENNES_1}
{G.\,P.~de\,Gennes}, {\em Scaling concepts on polymer physics} (Cornell
  University, Ithaca, 1979).

\bibitem{BERLIN_1}
{T.\,H.~Berlin} and M. Kac, {\em The spherical model of a ferromagnet},
  Phys.~Rev. {\bf 86},  821  (1952).

\bibitem{STANLEY_3}
{H.\,E.~Stanley}, {\em Spherical model as the limit of infinite spin
  dimensionality}, Phys.~Rev. {\bf 176},  718  (1968).

\bibitem{JOYCE_1}
{G.\,S.~Joyce}, {\em Critical properties of the spherical model {\textit in
  Phase Transitions and Critical Phenomena}, Vol.\,2}, edited by C. Domb and
  {M.\,S.~Green} (Academic Press, London, 1972).

\bibitem{PRIVMAN_3}
V. Privman and {M.\,E.~Fisher}, {\em Universal critical amplitudes in
  finite-size scaling}, Phys.~Rev.~B {\bf 30},  322  (1984).

\bibitem{MILOSEVIC_1}
S. Milo{\u {s}}evi{\'{c}} and {H.\,E.~Stanley}, {\em Equation of state near the
  critical point:~I.~Calculation of the scaling function for $S=1/2$ and
  $S=\protect\infty$ Heisenberg models using high-temperature
  series-expansion}, Phys.~Rev.~B {\bf 6},  986  (1972).

\bibitem{PRIVMAN_2}
V. Privman, {P.\,C.~Hohenberg}, and A. Aharony, {\em Universal critical-point
  amplitude relations {\textit in Phase Transitions and Critical Phenomena},
  Vol.\,14}, edited by C. Domb and {J.\,L.~Lebowitz} (Academic Press, London,
  1991).

\bibitem{BERKER_1}
{A.\,N.~Berker} and S. Ostlund, {\em Renormalisation-group calculations of
  finite systems:~order parameter and specific heat for epitaxial ordering},
  J.~Phys.~C {\bf 12},  4961  (1979).

\bibitem{KAUFMANN_1}
M. Kaufmann and {R.\,B.~Griffiths}, {\em Exactly soluble Ising models on
  hierarchical lattices}, Phys.~Rev.~B {\bf 24},  496  (1981).

\bibitem{MIGDAL_1}
{A.\,A.~Migdal}, {\em Phase transitions in gauge and spin-lattice systems},
  Sov.~Phys.~JETP {\bf 42},  743  (1976).

\bibitem{KADANOFF_4}
{L.\,P.~Kadanoff}, {\em Notes on Migdal's recursion formula}, Ann.~Phys.~(NY)
  {\bf 100},  359  (1976).

\bibitem{BREZIN_3}
E. Br{\'e}zin, {J.\,C.~Le\,Guillou}, and J. Zinn-Justin, {\em Field theoretical
  approach to critical phenomena {\textit in Phase Transitions and Critical
  Phenomena}, Vol.\,6}, edited by C. Domb and {M.\,S.~Green} (Academic Press,
  London, 1976).

\bibitem{WEGNER_1}
{F.\,J.~Wegner} and {E.\,K.~Riedel}, {\em Logarithmic corrections to the
  molecular-field behavior of critical and tricritical systems}, Phys.~Rev.~B
  {\bf 7},  248  (1973).

\bibitem{WEGNER_3}
{F.\,J.~Wegner}, {\em Corrections to scaling laws}, Phys.~Rev.~B {\bf 5},  4529
   (1972).

\bibitem{FISHER_5}
{M.\,E.~Fisher}, {\em Renormalization group in critical phenomena and quantum
  field theory}, edited by {D.\,J.~Gunton} and {M.\,S.~Green} (Temple
  University, Philadelphia, 1974).

\bibitem{PRIVMAN_4}
V. Privman and {M.\,E.~Fisher}, {\em Finite-size effects at first-order
  transitions}, J.~Stat.~Phys. {\bf 33},  385  (1983).

\bibitem{BRANKOV_2}
{J.\,G.~Brankov}, {D.\,M.~Danchev}, and {N.\,S.~Tonchev}, {\em Theory of
  critical phenomena in finite-size systems}, in {\textit{Series in Mordern
  Condensed Matter Physics}}, World Scientific {\bf 9},  $\,$  (2000).

\bibitem{BREZIN_5}
E. Br{\'e}zin, {\em An investigation of finite size scaling}, J.~Phys.~(France)
  {\bf 43},  15  (1982).

\bibitem{WILSON_4}
{K.\,G.~Wilson} and {M.\,E.~Fisher}, {\em Critical exponents in 3.99
  dimensions}, Phys.~Rev.~Lett. {\bf 28},  240  (1972).

\bibitem{WALLACE_1}
{D.\,J.~Walace}, {\em The $\epsilon$-expansion for exponents and the equation
  of state in isotropic systems {\textit in Phase Transitions and Critical
  Phenomena}, Vol.\,6}, edited by C. Domb and {M.\,S.~Green} (Academic Press,
  London, 1976).

\bibitem{BREZIN_4}
E. Br{\'e}zin, {D.\,J.~Wallace}, and {K.\,G.~Wilson}, {\em Feynman-graph
  expansion for the equation of state near the critical point}, Phys.~Rev.~B
  {\bf 7},  232  (1973).

\bibitem{BAROUCH_1}
E. Barouch, {B.\,M.~Mc\,Coy}, and {T.\,T.~Wu}, {\em Zero-field susceptibility
  of the two-dimensional Ising model near $T_{\mathrm c}$}, Phys.~Rev.~Lett.
  {\bf 31},  1409  (1973).

\bibitem{NICOLL_1}
{J.\,F.~Nicoll} and {P.\,C.~Albright}, {\em Crossover functions by
  renormalization-group matching:~Three-loop results}, Phys.~Rev.~B {\bf 31},
  4576  (1985).

\bibitem{DELFINO_2}
G. Delfino, {\em Universal amplitude ratios in the two-dimensional Ising
  model}, Phys.~Lett.~B {\bf 419},  291  (1998).

\bibitem{STAUFFER_3}
D. Stauffer, {\em Scaling theory of percolation clusters}, Phys.~Rep. {\bf 54},
   1  (1979).

\bibitem{ESSAM_3}
{J.\,W.~Essam}, {\em Percolation theory}, Rep.~Prog.~Phys. {\bf 43},  833
  (1980).

\bibitem{STAUFFER_2}
D. Stauffer and A. Aharony, {\em Introduction to percolation theory}
  (Taylor\,\&\,Francis, London, 1992).

\bibitem{KARDAR_1}
M. Kadar, G. Parisi, and {Y.-\,C.~Zhang}, {\em Dynamic scaling of growing
  interfaces}, Phys.~Rev.~Letts. {\bf 56},  889  (1986).

\bibitem{KARDAR_2}
M. Kadar, {\em Nonequilibrium dynamics of interfaces and lines}, Phys.~Rep.
  {\bf 301},  85  (1998).

\bibitem{BARABASI_1}
{A.\,L.~Barab{\'a}si} and {H.\,E.~Stanley}, {\em Fractal concepts in surface
  growth} (Cambridge University Press, Cambridge, 1995).

\bibitem{WITTEN_1}
{T.\,A.~Witten} and {L.\,M.~Sander}, {\em Diffusion limited aggregation:~A
  kinetic critical phenomenon}, Phys.~Rev.~Lett. {\bf 47},  1400  (1981).

\bibitem{KOLMOGOROV_1}
{A.\,N.~Kolmogorov}, {\em The local structure of turbulence in incompressible
  viscous fluid for very large Reynolds numbers}, C.~R.~Acad.~Sci.~USSR {\bf
  30},  299  (1941).

\bibitem{FRISCH_1}
U. Frisch, {\em Turbulence:~The legacy of A. N. Kolmogorov} (Cambridge
  University Press, Cambridge, 1995).

\bibitem{GRINSTEIN_2}
G. Grinstein and {M.\,A.~Mu\~{n}oz}, {\em The statistical mechanics of
  absorbing states}, in {\textit{Lecture Notes in Physics}}, Springer {\bf
  493},  223  (1997).

\bibitem{ALBANO_2}
{E.\,V.~Albano}, {\em Critical behavior of a forest fire model with immune
  trees}, J.~Phys.~A {\bf 27},  L881  (1994).

\bibitem{CLAR_2}
S. Clar, B. Drossel, and F. Schwabl, {\em Forest fires and other examples of
  self-organized criticality}, J.~Phys.:~Condens.~Matter {\bf 8},  6803
  (1996).

\bibitem{MOLLISON_1}
D. Mollison, {\em Spatial contact models for ecological and epidemic spread},
  J.~Roy.~Stat.~Soc.~B {\bf 39},  283  (1977).

\bibitem{ZIFF_1}
{R.\,M.~Ziff}, E. Gular{\'{\i}}, and Y. Barshad, {\em Kinetic phase transitions
  in an irreversible surface-reaction model}, Phys.~Rev.~Lett. {\bf 56},  2553
  (1986).

\bibitem{POMEAU_1}
Y. Pomeau, {\em Front motion, metastability and subcritical bifurcations in
  hydrodynamics}, Physica D {\bf 23},  3  (1986).

\bibitem{MENON_1}
{G.\,I.~Menon}, S. Sinha, and P. Ray, {\em Persistence at the onset of
  spatiotemporal intermittency in coupled map lattices}, Europhys.~Lett. {\bf
  61},  40  (2003).

\bibitem{RUPP_1}
P. Rupp, R. Richter, and I. Rehberg, {\em Critical exponents of directed
  percolation measured in spatiotemporal intermittency}, Phys.~Rev.~E {\bf 67},
   036209  (2003).

\bibitem{CARDY_1}
{J.\,L.~Cardy} and {R.\,L.~Sugar}, {\em Directed percolation and Reggeon field
  theory}, J.~Phys.~A {\bf 13},  L423  (1980).

\bibitem{NARAYAN_1}
O. Narayan and {A.\,A.~Middleton}, {\em Avalanches and the renormalization
  group for pinned charge-density waves}, Phys.~Rev.~B {\bf 49},  244  (1994).

\bibitem{BAK_1}
P. Bak, C. Tang, and K. Wiesenfeld, {\em Self-organized criticality:~an
  explanation of $1/f$ noise}, Phys.~Rev.~Lett. {\bf 59},  381  (1987).

\bibitem{BAK_2}
P. Bak, C. Tang, and K. Wiesenfeld, {\em Self-organized criticality},
  Phys.~Rev.~A {\bf 38},  364  (1988).

\bibitem{BAK_3}
P. Bak, {\em How Nature Works} (Springer, New York, 1996).

\bibitem{TURCOTTE_1}
{D.\,L.~Turcotte}, {\em Self-organized criticality}, Rep.~Prog.~Phys. {\bf 62},
   1377  (1999).

\bibitem{KAUFFMAN_1}
{S.\,A.~Kaufmann}, {\em Emergent properties in random complex automata},
  Physica~D {\bf 10},  145  (1984).

\bibitem{HERRMANN_1}
{H.\,J.~Herrmann}, {\em Damage spreading}, Physica~A {\bf 168},  516  (1990).

\bibitem{HINRICHSEN_9}
H. Hinrichsen, R. Livi, D. Mukamel, and A. Politi, {\em A model for
  nonequilibrium wetting transitions in two dimensions}, Phys.~Rev.~Lett. {\bf
  79},  2710  (1997).

\bibitem{BINDER_1}
K. Binder and {D.\,W.~Heermann}, {\em Monte Carlo Simulation in Statistical
  Physics} (Springer, Berlin, 1997).

\bibitem{HARRIS_2}
{T.\,E.~Harris}, {\em Contact interactions on a lattice}, Ann.~Prob. {\bf 2},
  969  (1974).

\bibitem{MARRO_1}
J. Marro and R. Dickman, {\em Nonequilibrium phase transitions in lattice
  models} (Cambridge University Press, Cambridge, 1999).

\bibitem{GARDINER_1}
{C.\,W.~Gardiner}, {K.\,J.~Mc\,Neil}, {D.\,F.~Walls}, and {I.\,S.~Matheson},
  {\em Correlations in stochastic theories of chemical reactions},
  J.~Stat.~Phys. {\bf 14},  307  (1976).

\bibitem{MORI_1}
H. Mori and {K.\,J.~Mc\,Neil}, {\em Critical dimensionality for normal
  fluctuations of macrovariables in nonequilibrium states}, Prog.~Theor.~Phys.
  {\bf 57},  770  (1977).

\bibitem{ELDERFIELD_1}
D. Elderfield and {D.\,D.~Vvedensky}, {\em Non-equilibrium scaling in the
  Schl{\"o}gl model}, J.~Phys.~A {\bf 18},  2591  (1985).

\bibitem{HENKEL_5}
M. Henkel, {\em Local scale invariance and strongly anisotropic equilibrium
  critical systems}, Phys.~Rev.~Lett. {\bf 78},  1940  (1997).

\bibitem{HORNREICH_1}
{R.\,M.~Hornreich}, M. Luban, and S. Shtrikman, {\em Critical behavior at the
  onset of $\vec{k}$-space instability on the $lambda$ line}, Phys.~Rev.~Lett.
  {\bf 35},  1678  (1975).

\bibitem{SELKE_1}
A. Aharony, {\em Spatially modulated structures in systems with competing
  interactions {\textit in Phase Transitions and Critical Phenomena},
  Vol.\,15}, edited by C. Domb and {J.\,L.~Lebowitz} (Academic Press, New York,
  1992).

\bibitem{HUCHT_1}
A. Hucht, {\em On the symmetry of universal finite-size scaling functions in
  anisotropic systems}, J.~Phys.~A {\bf 35},  L481  (2002).

\bibitem{ESSAM_4}
{J.\,W.~Essam}, {K.~de\,Bell}, {J.~Adler}, and {F.\,M.~Bhatti}, {\em Analysis
  of extended series for bond percolation on the directed square lattice},
  Phys.~Rev.~B {\bf 33},  1982  (1986).

\bibitem{ADLER_1}
J. Adler and {J.\,A.\,M.\,S.~Duarte}, {\em Directed percolation:~field
  exponents and a test of scaling in two and three dimensions}, Phys.~Rev.~B
  {\bf 35},  7046  (1987).

\bibitem{ADLER_2}
M. Adler, J. Berger, {J.\,A.\,M.\,S.~Duarte}, and Y. Meir, {\em Directed
  percolation in $3+1$ dimensions}, Phys.~Rev.~B {\bf 37},  7529  (1988).

\bibitem{ESSAM_5}
{J.\,W.~Essam}, {A.\,J.~Guttmann}, and {K.~de\,Bell}, {\em On two-dimensional
  directed percolation}, J.~Phys.~A {\bf 21},  3815  (1988).

\bibitem{JENSEN_11}
I. Jensen and R. Dickman, {\em Time-dependent perturbation theory for
  nonequilibrium lattice models}, J.~Stat.~Phys. {\bf 71},  89  (1993).

\bibitem{JENSEN_12}
I. Jensen and R. Dickman, {\em Time-dependent perturbation theory for diffisive
  non-equilibrium lattice models}, J.~Phys.~A {\bf 26},  L151  (1993).

\bibitem{JENSEN_13}
I. Jensen and {A.\,J.~Guttmann}, {\em Series expansions of the percolation
  probability for directed square and honeycomb lattices}, J.~Phys.~A {\bf 28},
   4813  (1995).

\bibitem{JENSEN_10}
I. Jensen, {\em Temporally disordered bond percolation on the directed square
  lattice}, Phys.~Rev.~Lett. {\bf 77},  4988  (1996).

\bibitem{JENSEN_14}
I. Jensen and {A.\,J.~Guttmann}, {\em Series expansions for two-dimensional
  directed percolation}, Nucl.~Phys.~B (Proc.~Suppl.) {\bf 47},  835  (1996).

\bibitem{JENSEN_1}
I. Jensen, {\em Low-density series expansions for directed percolation on
  square and triangular lattices}, J.~Phys.~A {\bf 29},  7013  (1996).

\bibitem{JENSEN_5}
I. Jensen, {\em Low-density series expansions for directed percolation, I.~A
  new efficient algorithm with applications to the square lattice}, J.~Phys.~A
  {\bf 32},  5233  (1999).

\bibitem{JENSEN_16}
I. Jensen, {\em Low-density series expansions for directed percolation,
  III.~Some two-dimensional lattices}, J.~Phys.~A {\bf 37},  6899  (2004).

\bibitem{JENSEN_2}
I. Jensen, {\em Critical behavior of the pair contact process},
  Phys.~Rev.~Lett. {\bf 70},  1465  (1993).

\bibitem{JENSEN_3}
I. Jensen and R. Dickman, {\em Nonequilibrium phase transitions in systems with
  infinitely many absorbing states}, Phys.~Rev.~E {\bf 48},  1710  (1993).

\bibitem{MENDES_1}
{J.\,F.\,F.~Mendes}, R. Dickman, M. Henkel, and {M.\,C.~Marques}, {\em
  Generalized scaling for models with multiple absorbing states}, J.~Phys.~A
  {\bf 27},  3019  (1994).

\bibitem{LUEB_19}
S. L{\protect\"u}beck, {\em Scaling behavior of the absorbing phase transition
  in a conserved lattice gas around the upper critical dimension}, Phys.~Rev.~E
  {\bf 64},  016123  (2001).

\bibitem{LUEB_22}
S. L{\protect\"u}beck, {\em Scaling behavior of the order parameter and its
  conjugated field in an absorbing phase transition around the upper critical
  dimension}, Phys.~Rev.~E {\bf 65},  046150  (2002).

\bibitem{LUEB_24}
S. L{\protect\"u}beck, {\em Scaling behavior of the conserved threshold
  transfer process}, Phys.~Rev.~E {\bf 66},  046114  (2002).

\bibitem{LUEB_27}
S. L{\protect\"u}beck and {R.\,D.~Willmann}, {\em Universal scaling behaviour
  of directed percolation and the pair contact process in an external field},
  J.~Phys.~A {\bf 35},  10205  (2002).

\bibitem{LUEB_26}
S. L{\protect\"u}beck and {P.\,C.~Heger}, {\em Universal scaling behavior at
  the upper critical dimension of non-equilibrium continuous phase
  transitions}, Phys.~Rev.~Lett. {\bf 90},  230601  (2003).

\bibitem{LUEB_28}
S. L{\protect\"u}beck and {R.\,D.~Willmann}, {\em Universal scaling behavior of
  directed percolation around the upper critical dimension}, J.~Stat.~Phys.
  {\bf 115},  1231  (2004).

\bibitem{LUEB_32}
S. L{\protect\"u}beck and {R.\,D.~Willmann}, {\em Scaling behavior of the
  directed percolation universality class}, in preparation {\bf ~},  ~  (2005).

\bibitem{GRASSBERGER_4}
P. Grassberger and {A.~de\,la\,Torre}, {\em Reggeon field theory (Sch{\"o}gl's
  first model) on a lattice:~Monte Carlo calculation of critical behavior},
  Ann.~Phys.~(N.Y.) {\bf 122},  373  (1979).

\bibitem{GRASSBERGER_8}
P. Grassberger, {\em Directed percolation in 2+1 dimensions}, J.~Phys.~A {\bf
  22},  3673  (1989).

\bibitem{JENSEN_8}
I. Jensen, {H.\,C.~Fogedby}, and R. Dickman, {\em Critical exponents for an
  irreversible surface reaction model}, Phys.~Rev.~A {\bf 41},  3411  (1990).

\bibitem{DICKMAN_9}
R. Dickman, {\em Nonequilibrium critical behavior of the triplet annihilation
  model}, Phys.~Rev.~A {\bf 42},  6985  (1990).

\bibitem{JENSEN_9}
I. Jensen, {\em Universality class of a one-dimensional cellular automaton},
  Phys.~Rev.~A {\bf 43},  3187  (1991).

\bibitem{JENSEN_6}
I. Jensen, {\em Critical behavior of the three-dimensional contact process},
  Phys.~Rev.~A {\bf 45},  R563  (1992).

\bibitem{ODOR_2}
G. {\'O}dor and A. Szolnoki, {\em Directed percolation conjecture for cellular
  automata}, Phys.~Rev.~E {\bf 53},  2231  (1996).

\bibitem{VOIGT_1}
{C.\,A.~Voigt} and {R.\,M.~Ziff}, {\em Epidemic analysis of the second-order
  transition in the Ziff-Gulari-Barshad surface-reaction model}, Phys.~Rev.~E
  {\bf 56},  R6241  (1997).

\bibitem{LUEB_23}
S. L{\protect\"u}beck and {P.\,C.~Heger}, {\em Universal finite-size scaling
  behavior and universal dynamical scaling behavior of absorbing phase
  transitions with a conserved field}, Phys.~Rev.~E {\bf 68},  056102  (2003).

\bibitem{ODOR_6}
G. {\'O}dor, {\em Critical behavior in reaction-diffusion systems exhibiting
  absorbing phase transition}, Braz.~J.~Phys. {\bf 33},  431  (2003).

\bibitem{JENSEN_15}
I. Jensen, {\em Critical exponents for branching annihilating random walks with
  an even number of offspring}, Phys.~Rev.~E {\bf 50},  3623  (1994).

\bibitem{NOH_1}
{J.\,D.~Noh} and H. Park, {\em Universality class of absorbing phase
  transitions with continuously varying exponents}, Phys.~Rev.~E {\bf 69},
  016122  (2004).

\bibitem{RIEDEL_1}
{E.\,K.~Riedel} and {F.\,J.~Wegner}, {\em Effective critical and tricritical
  exponents}, Phys.~Rev.~B {\bf 9},  294  (1974).

\bibitem{DICKMAN_17}
R. Dickman, {\em Numerical study of a field theory for directed percolation},
  Phys.~Rev.~E {\bf 50},  4404  (1994).

\bibitem{LOPEZ_1}
C. L{\'o}pez and {M.\,A.~Mu{\~{n}}oz}, {\em Numerical analysis of a Langevin
  equation for systems with infinite absorbing states}, Phys.~Rev.~E {\bf 56},
  4864  (1997).

\bibitem{RAMASCO_1}
{J.\,J.~Ramasco}, {M.\,A.~Mu{\~n}oz}, and {C.\,A.~da\,Silva\,Santos}, {\em
  Numerical study of the Langevin theory for fixed-energy sandpiles},
  Phys.~Rev.~E {\bf 69},  045105  (2004).

\bibitem{BROADBENT_1}
{S.\,R.~Broadbent} and {J.\,M.~Hammersley}, {\em Percolation
  processes:~I.~crystals and mazes}, Proc.~Camb.~Phil.~Soc. {\bf 53},  629
  (1957).

\bibitem{POTTS_1}
{R.\,B.~Potts}, {\em Some generalized order-disorder transformations},
  Proc.~Camb.~Phil.~Soc. {\bf 48},  106  (1952).

\bibitem{WU_1}
{F.\,U.~Wu}, {\em The Potts model}, Rev.~Mod.~Phys. {\bf 54},  235  (1982).

\bibitem{FORTUIN_1}
{C.\,M.~Fortuin} and {P.\,W.~Kasteleyn}, {\em On the random cluster
  model.~I.~Introduction and relation to other models}, Physica {\bf 57},  536
  (1972).

\bibitem{FORTUIN_2}
{C.\,M.~Fortuin}, {\em On the random cluster model.~II.~The percolation model},
  Physica {\bf 58},  393  (1972).

\bibitem{NIJS_1}
{M.\,P.\,M.~den\,Nijs}, {\em A relation between the temperature exponents of
  the eight-vertex and $q$-state Potts model}, J.~Phys.~A {\bf 12},  1857
  (1979).

\bibitem{NIENHUIS_1}
B. Nienhuis, {E.\,K.~Riedel}, and M. Schick, {\em Magnetic exponents of the
  two-dimensional $q$-state Potts model}, J.~Phys.~A {\bf 13},  L189  (1980).

\bibitem{OBUKHOV_1}
{S.\,P.~Obukhov}, {\em The upper critical dimension and $\epsilon$-expansion
  for self-organized critical phenomena, {\rm in {\textit Random Fluctuations
  and Pattern Growth}}}, edited by H.~E. Stanley and {N.~Ostrowsky, NATO ASI
  Series E:~Applied Sciences Vol.~157} (Kluwer, Dordrecht, 1988).

\bibitem{ABARBANEL_1}
{H.\,D.\,I.~Abarbanel}, {J.\,B.~Bronzan}, {R.\,L.~Sugar}, and {A.\,R.~White},
  {\em Reggeon field theory:~formulation and use}, Phys.~Rep. {\bf 21},  119
  (1975).

\bibitem{OHTSUKI_1}
T. Ohtsuki and T. Keyes, {\em Nonequilibrium critical phenomena in
  one-component reaction-diffusion systems}, Phys.~Rev.~A {\bf 35},  2697
  (1987).

\bibitem{OHTSUKI_2}
T. Ohtsuki and T. Keyes, {\em Crossover in nonequilibrium multicritical
  phenomena of reaction-diffusion systems}, Phys.~Rev.~A {\bf 36},  4434
  (1987).

\bibitem{CHAICHIAN_1}
M. Chaichian and A. Demichev, {\em Path integrals in physics,
  {$\mathrm{Volume}\;1$}} (Institute of Physics, Bristol, 2001).

\bibitem{JANSSEN_7}
{H.\,K.~Janssen}, {\em Directed percolation with colors and flavors},
  J.~Stat.~Phys. {\bf 103},  801  (2001).

\bibitem{DEDOMINICIS_1}
{C.\,J.~De\,Dominicis}, {\em Techniques de renormalisation de la th{\'e}orie
  des champs et dynamique des ph{\'e}nom{\`e}nes critiques},
  J.~Phys.~C~(France) {\bf 37},  247  (1976).

\bibitem{JANSSEN_8}
{H.\,K.~Janssen}, {\em On a Lagrangean for classical field dynamics and
  renormalization group calculations of dynamical critical properties},
  Z.~Phys.~B {\bf 23},  377  (1976).

\bibitem{MARTIN_1}
{P.\,C.~Martin}, {E.\,D.~Siggia}, and {H.\,A.~Rose}, {\em Statistical dynamics
  of classical systems}, Phys.~Rev.~A {\bf 8},  423  (1973).

\bibitem{JANSSEN_4}
{H.\,K.~Janssen}, {\em Survival and percolation probabilities in the field
  theory of growth models}, cond-mat/0304631  (2004).

\bibitem{LIGGETT_1}
{T.\,M.~Liggett}, {\em Interacting particle systems} (Springer, New York,
  1985).

\bibitem{BRONZAN_1}
{J.\,B.~Bronzan} and {J.\,W.~Dash}, {\em Higher-order epsilon terms in Reggeon
  field theory}, Phys.~Rev.~D {\bf 10},  4208  (1974).

\bibitem{BRONZAN_2}
{J.\,B.~Bronzan} and {J.\,W.~Dash}, {\em Erratum: higher-order epsilon terms in
  Reggeon field theory}, Phys.~Rev.~D {\bf 12},  1850  (1974).

\bibitem{BRONZAN_3}
{J.\,B.~Bronzan} and {J.\,W.~Dash}, {\em Higher-order $\epsilon$-terms in the
  renormalization group approach to Reggeon field theory}, Phys.~Rev.~D {\bf
  12},  1850  (1974).

\bibitem{JANSSEN_2}
{H.\,K.~Janssen}, {\protect\"U}. Kutbay, and K. Oerding, {\em Equation of state
  for directed percolation}, J.~Phys.~A {\bf 32},  1809  (1999).

\bibitem{DOMANY_1}
E. Domany and W. Kinzel, {\em Equivalence of cellular automata to Ising models
  and directed percolation}, Phys.~Rev.~Lett. {\bf 53},  311  (1984).

\bibitem{DURRETT_1}
R. Durrett, {\em On the growth of the one dimensional contact processes},
  Ann.~Prob. {\bf 8},  890  (1980).

\bibitem{DURRETT_2}
R. Durrett and D. Griffeath, {\em On the growth of the one dimensional contact
  processes}, Z.~f{\"u}r~Wahr. {\bf 59},  535  (1982).

\bibitem{BEZUIDENHOUT_1}
C. Brezuidenhout and G. Grimmett, {\em The critical contact process dies out},
  Ann.~Prob. {\bf 18},  1462  (1990).

\bibitem{KINZEL_2}
W. Kinzel, {\em Phase transition of cellular automata}, Z.~Phys.~B {\bf 58},
  229  (1985).

\bibitem{TOME_1}
T. Tom{\'e}, {\em Spreading of the damage in the Domany-Kinzel cellular
  automaton:~a mean-field approach}, Physica~A {\bf 212},  99  (1994).

\bibitem{RIEGER_1}
H. Rieger, A. Schadschneider, and M. Schreckenberg, {\em Re-entrant behaviour
  in the Domany-Kinzel cellular automaton}, J.~Phys.~A {\bf 27},  L423  (1994).

\bibitem{ATMAN_1}
{A.\,P.\,F.~Atman}, R. Dickman, and {J.\,G.~Moreira}, {\em Phase diagram of a
  probabilistic cellular automaton with three-site interactions}, Phys.~Rev.~E
  {\bf 67},  016107  (2003).

\bibitem{ZEBENDE_1}
{G.\,F.~Zebende} and {T.\,J.\,P.~Penna}, {\em The Domany-Kinzel cellular
  automaton phase diagram}, J.~Stat.~Phys. {\bf 74},  1273  (1994).

\bibitem{ESSAM_1}
{J.\,W.~Essam}, {\em Directed compact percolation: cluster size and
  hyperscaling}, J.~Phys.~A {\bf 22},  4927  (1989).

\bibitem{KEARNEY_1}
{M.\,J.~Kearney} and {R.\,J.~Martin}, {\em On the finite-size scaling of
  clusters in compact directed percolation}, J.~Phys.~A: Math.~Gen. {\bf 36},
  6629  (2003).

\bibitem{GRASSBERGER_3}
P. Grassberger and {Y.-C.~Zhang}, {\em Self-organized formulation of standard
  percolation phenomena}, Physica~A {\bf 224},  169  (1996).

\bibitem{CARLON_1}
E. Carlon, M. Henkel, and U. Schollw{\"o}ck, {\em Critical properties of the
  reaction-diffusion model $2A\to 3A$, $2A \to 0$}, Phys.~Rev.~E. {\bf 63},
  036101  (2001).

\bibitem{DICKMAN_11}
R. Dickman, {\em Nonuniversality and critical point shift in systems with
  infinitely many absorbing configurations}, cond-mat/9909347  (1999).

\bibitem{DICKMAN_4}
R. Dickman and {J.~Kamphorst\,Leal\,da\,Silva}, {\em Moment ratios for
  absorbing-state phase transitions}, Phys.~Rev.~E {\bf 58},  4266  (1998).

\bibitem{DICKMAN_5}
R. Dickman, W. Rabelo, and G. {\'O}dor, {\em Pair contact process with a
  particle source}, Phys.~Rev.~E {\bf 65},  016118  (2001).

\bibitem{MUNOZ_1}
{M.\,A.~Mu\~{n}oz}, G. Grinstein, R. Dickman, and R. Livi, {\em Critical
  behavior of systems with many absorbing states}, Phys.~Rev.~Lett. {\bf 76},
  451  (1996).

\bibitem{WIJLAND_1}
{F.~van\,Wijland}, {\em Universality class of nonequilibrium phase transitions
  with infinitely many absorbing states}, Phys. Rev. Lett. {\bf 89},  190602
  (2002).

\bibitem{WIJLAND_3}
{F.~van\,Wijland}, {\em Infinitely-many absorbing-state nonequilibrium phase
  transitions}, Braz.~J.~Phys. {\bf 33},  551  (2003).

\bibitem{GRASSBERGER_9}
P. Grassberger, {\em On the critical behavior of the general epidemic process
  and dynamical percolation}, Math.~Biosci. {\bf 63},  157  (1983).

\bibitem{CARDY_5}
{J.\,L.~Cardy}, {\em Field theoretic formulation of an epidemic process with
  immunisation}, J.~Phys.~A {\bf 16},  L706  (1983).

\bibitem{CARDY_3}
{J.\,L.~Cardy} and P. Grassberger, {\em Epidemic models and percolation},
  J.~Phys.~A {\bf 18},  L267  (1985).

\bibitem{JANSSEN_6}
{H.\,K.~Janssen}, {\em Renormalized field theory of dynamical percolation},
  Z.~Phys.~B {\bf 58},  311  (1985).

\bibitem{SILVA_4}
{J.~Kamphorst\,Leal\,da\,Silva} and R. Dickman, {\em The pair contact process
  in two dimensions}, Phys.~Rev.~E {\bf 60},  5126  (1999).

\bibitem{HOWARD_2}
{M.\,J.~Howard} and {U.\,C.~T{\"a}uber}, {\em Real versus imaginary noise in
  diffusion-limited reactions}, J.~Phys.~A {\bf 30},  7721  (1997).

\bibitem{ODOR_4}
G. {\'O}dor, {\em Critical behaviour of the 1d annihilation fission process
  $2A\to 0$, $2A\to 3A$}, Phys.~Rev.~E {\bf 62},  3027  (2000).

\bibitem{HINRICHSEN_4}
H. Hinrichsen, {\em Pair contact process with diffusion:~A new type of
  nonequilibrium critical phenomena?}, Phys.~Rev.~E {\bf 63},  036102  (2001).

\bibitem{HINRICHSEN_5}
H. Hinrichsen, {\em Cyclically coupled spreading and pair annihilation},
  Physica~A {\bf 291},  275  (2001).

\bibitem{ODOR_3}
G. {\'O}dor, {\em Critical behavior of the one-dimensional diffusive pair
  contact process}, Phys.~Rev.~E {\bf 67},  016111  (2003).

\bibitem{DICKMAN_10}
R. Dickman and {M.\,A.\,F.~de\,Menezes}, {\em Nonuniversality in the pair
  contact process with diffusion}, Phys.~Rev.~E {\bf 66},  045101  (2002).

\bibitem{ODOR_5}
G. {\'O}dor, {M.\,C.~Marques}, and {M.\,A.~Santos}, {\em Phase transition of a
  two-dimensional binary spreading model}, Phys.~Rev.~E {\bf 65},  056113
  (2002).

\bibitem{PARK_1}
K. Park and {I.-M.~Kim}, {\em Well-defined set of exponents for a pair contact
  process with diffusion}, Phys.~Rev.~E {\bf 66},  027106  (2002).

\bibitem{KOCKELKOREN_1}
J. Kockelkoren and H. Chat{\'e}, {\em Absorbing phase transitions of
  branching-annihilating random walks}, Phys.~Rev.~Lett. {\bf 90},  125701
  (2003).

\bibitem{BARKEMA_1}
{G.\,T.~Barkema} and E. Carlon, {\em Universality in the pair contact process
  with diffusion}, Phys.~Rev.~E. {\bf 68},  036113  (2003).

\bibitem{HINRICHSEN_6}
H. Hinrichsen, {\em The diffusive pair contact process and non-equilibrium
  wetting}, cond-mat/0302381  (2003).

\bibitem{PAESSENS_1}
M. Paessens and {G.\,M.~Sch{\"u}tz}, {\em Phase transitions and correlations in
  the bosonic pair contact process with diffusion:~exact results}, J.~Phys.~A
  {\bf 37},  4709  (2004).

\bibitem{JANSSEN_12}
{H.\,K.~Janssen}, {F.~van\,Wijland}, O. Deloubri{\`e}re, and
  {U.\,C.~T{\protect\"a}uber}, {\em Pair contact process with
  diffusion:~Failure of master equation field theory}, Phys.~Rev.~E {\bf 70},
  056114  (2004).

\bibitem{HENKEL_1}
M. Henkel and H. Hinrichsen, {\em The non-equilibrium phase transition of the
  pair-contact process with diffusion}, J.~Phys.~A {\bf 37},  R117  (2004).

\bibitem{GRINSTEIN_3}
G. Grinstein, {Z.\,W.~Lai}, and {R.\,J.~Browne}, {\em Critical phenomena in a
  nonequilibrium model of heterogeneous catalysis}, Phys.~Rev.~A {\bf 40},
  4820  (1989).

\bibitem{LIU_1}
{D.-J.~Liu}, N. Pavlenko, and {J.\,W.~Evans}, {\em Crossover between mean-field
  and Ising critical behavior in a lattice-gas reaction-diffusion model},
  J.~Stat.~Phys. {\bf 114},  101  (2004).

\bibitem{STAUFFER_4}
D. Stauffer, M. Ferer, and M. Wortis, {\em Universality of second-order phase
  transitions:~The scale factor for the correlation length}, Phys.~Rev.~Lett.
  {\bf 29},  345  (1972).

\bibitem{JANSSEN_3}
{H.\,K.~Janssen} and O. Stenull, {\em Logarithmic corrections in directed
  percolation}, Phys.~Rev.~E {\bf 69},  016125  (2004).

\bibitem{SCHOFIELD_1}
P. Schofield, {\em Parametric representation of the equation of state near a
  critical point}, Phys.~Rev.~Lett. {\bf 22},  606  (1969).

\bibitem{JOSEPHSON_1}
{B.\,D.~Josephson}, {\em Equation of state near the critical point}, J.~Phys.~C
  {\bf 2},  1113  (1969).

\bibitem{OBUKHOV_2}
{S.\,P.~Obukhov}, {\em The problem of directed percolation}, Physica~A {\bf
  101},  145  (1980).

\bibitem{LARKIN_1}
{A.\,I.~Larkin} and {D.\,E.~Khmel'nitski{\u \i}}, {\em Phase transition in
  uniaxial ferroelectrics}, JETP {\bf 29},  1123  (1969).

\bibitem{AHARONY_1}
A. Aharony, {\em Critical behavior of magnets with dipolar
  interactions.~V.~Uniaxial magnets in $d$ dimensions}, Phys.~Rev.~B {\bf 8},
  3363  (1973).

\bibitem{GRIFFIN_1}
{J.\,A.~Griffiths}, {J.\,D.~Litster}, and A. Linz, {\em Spontaneous
  magnetization at marginal dimensionality in {$\mathrm{LiTbF}_4$}},
  Phys.~Rev.~Lett. {\bf 38},  251  (1977).

\bibitem{BRINKMANN_1}
J. Bringmann, R. Courths, and {H.\,J.~Guggenheim}, {\em Logarithmic corrections
  to the critical behavior of uniaxial, ferromagnetic $\mathrm{TbF}_3$},
  Phys.~Rev.~Lett. {\bf 40},  1286  (1978).

\bibitem{AKTEKIN_1}
N. Aktekin, {\em The finite-size scaling functions of the four-dimensional
  Ising model}, J.~Stat.~Phys. {\bf 104},  1397  (2001).

\bibitem{GRASSBERGER_6}
P. Grassberger, {\em Pruned-enriched Rosenbluth method: Simulations of theta
  polymers of chain length up to 1000000}, Phys.~Rev.~E {\bf 56},  3682
  (1997).

\bibitem{GRASSBERGER_7}
P. Grassberger, R. Hegger, and L. Sch{\protect\"a}fer, {\em Self-avoiding walks
  in four dimensions: logarithmic corrections}, J.~Phys.~A {\bf 27},  7265
  (1994).

\bibitem{OWCZAREK_1}
{A.\,L.~Owczarek} and T. Prellberg, {\em Scaling of self-avoiding walks in high
  dimensions}, J.~Phys.~A {\bf 34},  5773  (2001).

\bibitem{LUEB_5}
S. L{\protect\"u}beck, {\em Logarithmic corrections of the avalanche
  distributions of sandpile models at the upper critical dimension},
  Phys.~Rev.~E {\bf 58},  2957  (1998).

\bibitem{LUEB_10}
{D.\,V.~Ktitarev}, S. L{\protect\"u}beck, P. Grassberger, and
  {V.\,B.~Priezzhev}, {\em Scaling of waves in the Bak-Tang-Wiesenfeld sandpile
  model}, Phys.~Rev.~E {\bf 61},  81  (2000).

\bibitem{LUEB_17}
L. Roters, S. L{\protect\"u}beck, and {K.\,D.~Usadel}, {\em Depinning
  transition of a driven interface in the random-field Ising model around the
  upper critical dimension}, Phys.~Rev.~E {\bf 66},  069901  (2002).

\bibitem{GRASSBERGER_5}
P. Grassberger, {\em Critical percolation in high dimensions}, Phys.~Rev.~E
  {\bf 67},  036101  (2003).

\bibitem{FEDORENKO_1}
{A.\,A.~Fedorenko} and S. Stepanow, {\em Depinning transition at the upper
  critical dimension}, Phys.~Rev.~E {\bf 67},  057104  (2003).

\bibitem{JANSSEN_5}
{H.\,K.~Janssen} and O. Stenull, {\em Logarithmic corrections in dynamical
  isotropic percolation}, Phys.~Rev.~E {\bf 68},  036131  (2003).

\bibitem{STENULL_1}
O. Stenull and {H.\,K.~Janssen}, {\em Logarithmic corrections to scaling in
  critical percolation and random resistor networks}, Phys.~Rev.~E {\bf 68},
  036129  (2003).

\bibitem{GRUENEBERG_1}
D. Gr{\protect\"u}neberg and A. Hucht, {\em Universal finite-size scaling
  analysis of Ising models with long-range interactions at the upper critical
  dimensionality:~Isotropic case}, Phys.~Rev.~E {\bf 69},  036104  (2004).

\bibitem{KENNA_1}
R. Kenna, {\em Finite size scaling for ${\mathcal{O}}(N)$ $\phi^4$-theory at
  the upper critical dimension}, Nucl.~Phys.~B {\bf 691},  292  (2004).

\bibitem{GRASSBERGER_12}
P. Grassberger, {\em Logarithmic corrections in ($4+1$)-dimensional directed
  percolation},   (2004).

\bibitem{HENKEL_2}
M. Henkel and U. Schollw{\"o}ck, {\em Universal finite-size scaling amplitudes
  in anisotropic scaling}, J.~Phys.~A {\bf 34},  3333  (2001).

\bibitem{CARDY_4}
{}, {\em Finite-size scaling}, edited by {J.\,L.~Cardy} (North-Holland, New
  York, 1988).

\bibitem{FISHER_6}
{M.\,E.~Fisher}, {\em Theory of critical point singularities,
  {$\mathrm{proceedings\; of\; the}$ {\textit 1970 Varenna summer school on
  critical phenomena}}}, edited by {M.\,S.~Green} (Academic Press, New York,
  1971).

\bibitem{FISHER_7}
{M.\,E.~Fisher} and {M.\,N.~Barber}, {\em Scaling theory for finite-size
  effects in the critical region}, Phys.~Rev.~Lett. {\bf 28},  1516  (1972).

\bibitem{BRUCE_1}
{A.\,D.~Bruce}, {\em Probability density functions for collective coordinates
  in Ising-like systems}, J.~Phys.~C {\bf 14},  3667  (1981).

\bibitem{BRUCE_2}
{A.\,D.~Bruce}, {\em Universality in the two-dimensional continuous spin
  model}, J.~Phys.~A {\bf 18},  L873  (1985).

\bibitem{NICOLAIDES_1}
D. Nicolaides and {A.\,D.~Bruce}, {\em Universal configurational structure in
  two-dimensional scalar models}, J.~Phys.~A {\bf 21},  233  (1988).

\bibitem{BRUCE_3}
{A.\,D.~Bruce} and {N.\,B.~Wilding}, {\em Scaling fields and universality of
  the liquid-gas critical point}, Phys.~Rev.~Lett. {\bf 68},  193  (1992).

\bibitem{HU_1}
{C.-K.~Hu}, {C.-Y.~Lin}, and {J.-A.~Chen}, {\em Universal scaling functions in
  critical phenomena}, Phys.~Rev.~Lett. {\bf 75},  193  (1995).

\bibitem{KANEDA_1}
K. Kaneda, Y. Okabe, and M. Kikuchi, {\em Shape effects of finite-size scaling
  functions for anistropic three-dimensional Ising models}, J.~Phys.~A {\bf
  32},  7263  (1999).

\bibitem{KANEDA_2}
K. Kaneda and Y. Okabe, {\em Finite-size scaling for the Ising model on the
  M{\"o}bius strip and the Klein bottle}, Phys.~Rev.~Lett. {\bf 86},  2134
  (2001).

\bibitem{ANTAL_1}
T. Antal, M. Droz, and Z. R{\'a}cz, {\em Probability distribution of
  magnetization in the one-dimensional Ising model:~effects of boundary
  conditions}, J.~Phys.~A {\bf 37},  1465  (2004).

\bibitem{BINDER_2}
K. Binder, {\em Finite size scaling analysis of Ising model block distribution
  functions}, Z.~Phys.~B {\bf 43},  119  (1981).

\bibitem{LUEB_33}
S. L{\protect\"u}beck and {H.\,K.~Janssen}, {\em Finite-size scaling of
  directed percolation above the upper critical dimension}, submitted for
  publication {\bf ~},  ~  (2004).

\bibitem{LUIJTEN_4}
E. Luijten and {H.\,W.\,J.~Bl{\protect\"o}te}, {\em Finite-size scaling and
  universality above the upper critical dimensionality}, Phys.~Rev.~Lett. {\bf
  76},  1557  (1996).

\bibitem{BINDER_3}
K. Binder, M. Nauenberg, V. Privman, and {A.\,P.~Young}, {\em Finite-size tests
  of hyperscaling}, Phys.~Rev.~B {\bf 31},  1498  (1985).

\bibitem{LUIJTEN_5}
E. Luijten and {H.\,W.\,J.~Bl{\protect\"o}te}, {\em Classical critical behavior
  of spin models with long-range interactions}, Phys.~Rev.~B {\bf 56},  8945
  (1997).

\bibitem{BLOETE_1}
{H.\,W.\,J.~Bl{\protect\"o}te} and E. Luijten, {\em Universality and
  five-dimensional Ising model}, Europhys.~Lett. {\bf 38},  565  (1997).

\bibitem{LUIJTEN_6}
E. Luijten, K. Binder, and {H.\,W.\,J.~Bl{\protect\"o}te}, {\em Finite-size
  scaling above the upper critical dimension revisited:~the case of the
  five-dimensional Ising model}, Eur.~Phys.~J.~B {\bf 9},  289  (1999).

\bibitem{CHEN_1}
{X.\,S.~Chen} and V. Dohm, {\em Failure of universal finite-size scaling above
  the upper critical dimension}, Physica~A {\bf 251},  439  (1998).

\bibitem{CHEN_2}
{X.\,S.~Chen} and V. Dohm, {\em Finite-size scaling in the $\phi^4$ theory
  above the upper critical dimension}, Eur.~Phys.~J.~B {\bf 5},  529  (1998).

\bibitem{CHEN_3}
{X.\,S.~Chen} and V. Dohm, {\em Universality and $\phi^4$ theory of finite-size
  scaling above the upper critical dimension}, Phys.~Rev.~E {\bf 63},  016113
  (2000).

\bibitem{STAUFER_5}
D. Stauffer, {\em World records in the size of simulated Ising models},
  Braz.~J.~Phys. {\bf 30},  787  (2000).

\bibitem{JANSSEN_P2004}
{H.\,K.~Janssen}, {\em \,}, private communication {\bf \,},  \,  (2004).

\bibitem{JOYCE_2}
{G.\,S.~Joyce}, {\em Spherical model with long-range ferromagnetic
  interactions}, Phys.~Rev. {\bf 146},  349  (1966).

\bibitem{STELL_1}
G. Stell, {\em Extension of the Ornstein-Zernike theory of the critical
  region.~II}, Phys.~Rev.~B {\bf 1},  2265  (1970).

\bibitem{FISHER_8}
{M.\,E.~Fisher}, {S.-K.~Ma}, and {B.\,G.~Nickel}, {\em Critical exponents for
  long-range interactions}, Phys.~Rev.~Lett. {\bf 29},  917  (1972).

\bibitem{AIZENMAN_1}
M. Aizenman and R. Fern{\'a}ndez, {\em Critical exponents for long-range
  interactions}, Lett.~Math.~Phys. {\bf 16},  39  (1988).

\bibitem{JANSSEN_9}
{H.\,K.~Janssen}, B. Schaub, and B. Schmittmann, {\em Finite size scaling for
  directed percolation and related stochastic evolution processes}, Z.~Phys.~B
  {\bf 71},  377  (1988).

\bibitem{JANSSEN_10}
{H.\,K.~Janssen}, B. Schaub, and B. Schmittmann, {\em New universal short-time
  scaling behaviour of critical relaxation processes}, Z.~Phys.~B {\bf 73},
  539  (1988).

\bibitem{DICKMAN_13}
R. Dickman and {A.\,Yu.~Tretyakov}, {\em Hyperscaling in the Domany-Kinzel
  cellular automaton}, Phys.~Rev.~E {\bf 52},  3218  (1995).

\bibitem{KINZEL_1}
W. Kinzel, {\em Directed percolation}, Ann.~Israel~Phys.~Soc. {\bf 5},  425
  (1983).

\bibitem{WIJLAND_2}
{F.~van\,Wijland}, K. Oerding, and {H.\,J.~Hilhorst}, {\em Wilson
  renormalization of a reaction--diffusion process}, Physica~A {\bf 251},  179
  (1998).

\bibitem{DAERR_1}
A. Daerr and S. Douady, {\em Two types of avalanche behavior in granular
  media}, Nature {\bf 399},  241  (1999).

\bibitem{HINRICHSEN_7}
H. Hinrichsen, {\em Flowing sand:~A physical realization of directed
  percolation}, Phys.~Rev.~Lett. {\bf 83},  4999  (1999).

\bibitem{HINRICHSEN_8}
H. Hinrichsen, {\em On possible experimental realizations of directed
  percolation}, Braz.~J.~Phys. {\bf 30},  69  (2000).

\bibitem{GRASSBERGER_10}
P. Grassberger, {\em Directed percolation:~results and open problems {\textit
  in Nonlinearities in complex systems}}, edited by {S.~Pury \textit{et al.}}
  (Narosa Publishing, New Dehli, 1997).

\bibitem{SIRE_1}
C. Sire, {\em Approximation for directed percolation in d=1+1}, Phys.~Rev.~E
  {\bf 66},  046133  (2002).

\bibitem{ZHONG_1}
D. Zhong and {D.~ben\,Avraham}, {\em Universality class of two-offspring
  branching annihilating random walks}, Phys.~Lett.~A {\bf 209},  333  (1995).

\bibitem{CARDY_2}
{J.\,L.~Cardy} and {U.\,C.~T{\protect\"a}uber}, {\em Theory of branching and
  annihilating random walks}, Phys.~Rev.~Lett. {\bf 13},  4780  (1996).

\bibitem{ROSSI_1}
M. Rossi, R. Pastor-Satorras, and A. Vespignani, {\em The universality class of
  absorbing phase transition with a conserved field}, Phys.\,Rev.\,Lett. {\bf
  85},  1803  (2000).

\bibitem{DICKMAN_1}
R. Dickman, A. Vespignani, and S. Zapperi, {\em Self-organized criticality as
  an absorbing-state phase transition}, Phys.~Rev.~E {\bf 57},  5095  (1998).

\bibitem{PASTOR_2}
R. Pastor-Satorras and A. Vespignani, {\em Field theory of absorbing phase
  transitions with a nondiffusive conserved field}, Phys.~Rev.~E {\bf 62},
  5875  (2000).

\bibitem{STILCK_1}
{J.\,F.~Stilck}, R. Dickman, and {R.\,R.~Vidigal}, {\em Series expansion for a
  stochastic sandpile}, J.~Phys.~A {\bf 37},  1145  (2004).

\bibitem{DICKMAN_15}
R. Dickman and {R.\,R.~Vidigal}, {\em Path-integral representation for a
  stochastic sandpile}, J.~Phys.~A {\bf 35},  7269  (2002).

\bibitem{MANNA_2}
{S.\,S.~Manna}, {\em Two-state model of self-organized criticality}, J.~Phys.~A
  {\bf 24},  L363  (1991).

\bibitem{BENHUR_1}
A. Ben-Hur and O. Biham, {\em Universality in sandpile models}, Phys.~Rev.~E
  {\bf 53},  R1317  (1996).

\bibitem{DHAR_7}
D. Dhar, {\em Some results and a conjecture for Manna's stochastic sandpile
  model}, Physica A {\bf 270},  69  (1999).

\bibitem{CHESSA_2}
A. Chessa, {H.\,E.~Stanley}, A. Vespignani, and S. Zapperi, {\em Universality
  in sandpiles}, Phys.~Rev.~E {\bf 59},  12  (1999).

\bibitem{CHESSA_3}
A. Chessa, A. Vespignani, and S. Zapperi, {\em Critical exponents in stochastic
  sandpile models}, Comp.~Phys.~Com. {\bf 121},  299  (1999).

\bibitem{LUEB_9}
S. L{\protect\"u}beck, {\em Moment analysis of the probability distributions of
  different sandpile models}, Phys.~Rev.~E {\bf 61},  204  (2000).

\bibitem{LUEB_13}
S. L{\protect\"u}beck, {\em Crossover phenomenon in self-organized critical
  sandpile models}, Phys.~Rev.~E {\bf 62},  6149  (2000).

\bibitem{DICKMAN_2}
R. Dickman {\it et~al.}, {\em Critical behavior of a one-dimensional
  fixed-energy stochastic sandpile}, Phys.~Rev.~E {\bf 64},  056104  (2001).

\bibitem{BIHAM_1}
O. Biham, E. Milshtein, and O. Maclai, {\em Evidence for universality within
  the classes of deterministic and stochastic sandpile models}, Phys.~Rev.~E
  {\bf 63},  061309  (2001).

\bibitem{SHILO_1}
Y. Shilo and O. Biham, {\em Sandpile models and random walkers on finite
  lattices}, Phys.~Rev.~E {\bf 67},  066102  (2003).

\bibitem{KREE_1}
R. Kree, B. Schaub, and B. Schmitmann, {\em Effects of pollution on critical
  population dynamics}, Phys.~Rev.~A {\bf 39},  2214  (1989).

\bibitem{PASTOR_5}
R. Pastor-Satorras and A. Vespignani, {\em Reaction-diffusion system with
  self-organized critical behavior}, Eur.~Phys.~J.~B {\bf 19},  583  (2001).

\bibitem{VESPIGNANI_4}
A. Vespignani, R. Dickman, {M.\,A.~Mu\~{n}oz}, and S. Zapperi, {\em Absorbing
  state phase transition in fixed-energy sandpiles}, Phys.~Rev.~E {\bf 62},
  4564  (2000).

\bibitem{LUEB_29}
S. L{\protect\"u}beck, {\em Universal behavior of crossover scaling functions
  for continuous phase transitions}, Phys.~Rev.~Lett. {\bf 90},  210601
  (2003).

\bibitem{HEGER_1}
{P.\,C.~Heger}, {\em Computersimulationen zur Bestimmung des Skalenverhaltens
  absorbierender Phasen{\protect{\"u}}berg{\protect{\"a}}nge}, Diplomarbeit,
  Universit{\protect{\"a}}t Duisburg-Essen, 2003.

\bibitem{LUEB_25}
S. L{\protect\"u}beck and A. Hucht, {\em Mean-field scaling function of the
  universality class of absorbing phase transitions with a conserved field},
  J.~Phys.~A {\bf 35},  4853  (2002).

\bibitem{DICKMAN_3}
R. Dickman, T. Tom{\'{e}}, and {M.\,J.~de\,Oliveira}, {\em Sandpiles with
  height restrictions}, Phys.~Rev.~E {\bf 66},  016111  (2002).

\bibitem{JENSEN_7}
{H.\,J.~Jensen}, {\em Lattice gas as a model of $1/f$ noise}, Phys.~Rev.~Lett.
  {\bf 64},  3103  (1990).

\bibitem{LUEB_4}
S. L{\protect\"u}beck and {K.\,D.~Usadel}, {\em The Bak-Tang-Wiesenfeld
  sandpile model around the upper critical dimension}, Phys.~Rev.~E {\bf 56},
  5138  (1997).

\bibitem{LUEB_PHD}
S. L{\protect\"u}beck, Ph.D. thesis, Gerhard-Mercator-Universit{\protect{\"a}}t
  Duisburg, 1998.

\bibitem{MADRAS_1}
N. Madras and G. Slade, {\em The self-avoiding walk} (Birkh{\"a}user, Boston,
  1996).

\bibitem{LAWLER_1}
{G.\,F.~Lawler}, {\em A self-avoiding random walk}, Duke Math.~J. {\bf 47},
  655  (1980).

\bibitem{LAWLER_2}
{G.\,F.~Lawler}, {\em Gaussian behavior of loop-erased self-avoiding random
  walk in four dimensions}, Duke Math.~J. {\bf 53},  249  (1986).

\bibitem{CONIGLIO_1}
A. Coniglio, {\em Fractal structure of Ising and Potts clusters: exact
  results}, Phys.~Rev.~Lett. {\bf 62},  3054  (1989).

\bibitem{LAWLER_BOOK}
{G.\,F.~Lawler}, {\em Intersections of Random Walks} ({Birkh\"auser}, Boston,
  1991).

\bibitem{MUNOZ_3}
{M.\,A.~Mu\~{n}oz}, R. Dickman, A. Vespignani, and S. Zapperi, {\em Avalanche
  and spreading exponents in systems with absorbing states}, Phys.~Rev.~E {\bf
  59},  6175  (1999).

\bibitem{BAK_4}
P. Bak and K. Sneppen, {\em Punctuated equilibrium and criticality in a simple
  model of evolution}, Phys.~Rev.~Lett. {\bf 71},  4083  (1993).

\bibitem{DHAR_2}
D. Dhar, {\em Self-organized critical state of sandpile automaton models},
  Phys.~Rev.~Lett. {\bf 64},  1613  (1990).

\bibitem{MAJUMDAR_3}
{S.\,N.~Majumdar} and D. Dhar, {\em Equivalence between the Abelian sandpile
  model and the $q\to 0$ limit of the Potts model}, Physica A {\bf 185},  129
  (1992).

\bibitem{MAJUMDAR_1}
{S.\,N.~Majumdar} and D. Dhar, {\em Height correlations in the Abelian sandpile
  model}, J.~Phys.~A {\bf 24},  L357  (1991).

\bibitem{PRIEZ_1}
{V.\,B.~Priezzhev}, {\em Structure of the two-dimensional sandpile},
  J.~Stat.~Phys. {\bf 74},  955  (1994).

\bibitem{IVASH_1}
{E.\,V.~Ivashkevich}, {\em Boundary height correlations in two-dimensional
  Abelian sandpile}, J.~Phys.~A {\bf 27},  3643  (1994).

\bibitem{IVASH_3}
{E.\,V.~Ivashkevich}, {D.\,V.~Ktitarev}, and {V.\,B.~Priezzhev}, {\em Waves of
  topplings in an Abelian sandpile}, Physica A {\bf 209},  347  (1994).

\bibitem{PRIEZ_2}
{V.\,B.~Priezzhev}, {D.\,V.~Ktitarev}, and {E.\,V.~Ivashkevich}, {\em Formation
  of avalanches and critical exponents in an Abelian sandpile model},
  Phys.~Rev.~Lett. {\bf 76},  2093  (1994).

\bibitem{IVASH_4}
{E.\,V.~Ivashkevich}, {D.\,V.~Ktitarev}, and {V.\,B.~Priezzhev}, {\em Critical
  exponents for boundary avalanches in two-dimensional Abelian sandpile},
  J.~Phys.~A {\bf 27},  L585  (1994).

\bibitem{PRIEZ_3}
{V.\,B.~Priezzhev}, {\em The upper critical dimension of the Abelian sandpile
  model}, J.~Stat.~Phys. {\bf 98},  667  (2000).

\bibitem{MANNA_1}
{S.\,S.~Manna}, {\em Large-Scale simulation of avalanche cluster distribution
  in sandpile model}, J.~Stat.~Phys. {\bf 59},  509  (1990).

\bibitem{GRASSBERGER_1}
P. Grassberger and {S.\,S.~Manna}, {\em Some more sandpiles}, J.~Phys.~(France)
  {\bf 51},  1077  (1990).

\bibitem{LUEB_2}
S. L{\protect\"u}beck and {K.\,D.~Usadel}, {\em Numerical determination of the
  avalanche exponents of the Bak-Tang-Wiesenfeld model}, Phys.~Rev.~E {\bf 55},
   4095  (1997).

\bibitem{DEMENECH_1}
M. {De\,Menech}, {A.\,L.~Stella}, and C. Tebaldi, {\em Rare events and
  breakdown of simple scaling in the Abelian sandpile model}, Phys.~Rev.~E {\bf
  58},  2677  (1998).

\bibitem{TEBALDI_1}
C. Tebaldi, M. {De\,Menech}, and {A.\,L.~Stella}, {\em Multifractal scaling in
  the Bak-Tang-Wiesenfeld Sandpile and edge events}, Phys.~Rev.~Lett. {\bf 83},
   3952  (1999).

\bibitem{GRINSTEIN_1}
G. Grinstein, {\em Generic scale invariance and self-organized criticality,
  {${\mathrm{in}}$} Scale Invariance, Interfaces, and Non-Equilibrium
  Dynamics}, edited by {A.~{McKane {\textit{et al.}}}, NATO ASI Series
  B:~Physics Vol.~344} (Plenum Press, London, 1995).

\bibitem{VERGELES_1}
M. Vergeles, A. Maritan, and J. Banavar, {\em Mean-field theory of sandpiles},
  Phys.~Rev.~E {\bf 55},  1998  (1997).

\bibitem{VESPIGNANI_1}
A. Vespignani and S. Zapperi, {\em How self-organized criticality works:~A
  unified mean-field picture}, Phys.~Rev.~E {\bf 57},  6345  (1998).

\bibitem{TANG_1}
C. Tang and P. Bak, {\em Mean field theory of self-organized critical
  phenomena}, J.~Stat.~Phys. {\bf 51},  797  (1988).

\bibitem{VESPIGNANI_5}
A. Vespignani and S. Zapperi, {\em Order parameter and scaling fields in
  self-organized criticality}, Phys.~Rev.~Lett. {\bf 78},  4793  (1997).

\bibitem{VESPIGNANI_2}
A. Vespignani, R. Dickman, {M.\,A.~Mu\~{n}oz}, and S. Zapperi, {\em Driving,
  conservation, and absorbing states in sandpiles}, Phys.~Rev.~Lett. {\bf 81},
  5676  (1998).

\bibitem{BAGNOLI_1}
F. Bagnoli, F. Cecconi, A. Flammini, and A. Vespignani, {\em Short period
  attractors and non-ergodic behavior in the deterministic fixed energy
  sandpile model}, Europhys.~Lett. {\bf 63},  512  (2003).

\bibitem{ALSTROM_1}
P. Alstr{\o}m, {\em Mean-field exponents for self-organized critical
  phenomena}, Phys.~Rev.~A {\bf 38},  4905  (1988).

\bibitem{DHAR_3}
D. Dhar and {S.\,N.~Majumdar}, {\em Abelian sandpile model on the Bethe
  lattice}, J.~Phys.~A {\bf 23},  4333  (1990).

\bibitem{JANOWSKY_1}
{S.\,A.~Janowsky} and {C.\,A.~Laberge}, {\em Exact solutions for a mean-field
  Abelian sandpile}, J.~Phys.~A {\bf 26},  L973  (1993).

\bibitem{GARCIA_1}
{R.~Gar{\'c}ia-Pelayo}, I. Salazar, and {W.\,C.~Schieve}, {\em A branching
  process model for sandpile avalanches}, J.~Stat.~Phys. {\bf 72},  167
  (1993).

\bibitem{ZAPPERI_1}
S. Zapperi, {K.\,B.~Lauritsen}, and {H.\,E.~Stanley}, {\em Self-organized
  branching process:~mean-field theory for avalanches}, Phys.~Rev.~Lett. {\bf
  75},  4071  (1995).

\bibitem{IVASH_2}
{E.\,V.~Ivashkevich}, {\em Critical behavior of the sandpile model as a
  self-organized branch\-ing process}, Phys.~Rev.~Lett. {\bf 76},  3368
  (1996).

\bibitem{KATORI_1}
M. Katori and H. Kobayashi, {\em Mean-field theory of avalanches in
  self-organized critical states}, Physica A {\bf 229},  461  (1996).

\bibitem{CALDARELLI_1}
G. Caldarelli, {\em Mean field theory for ordinary and hot sandpiles},
  Physica~A {\bf 252},  295  (1998).

\bibitem{GUTENBERG_1}
B. Gutenberg and {C.\,F.~Richter}, {\em Seismicity of the earth and associated
  phenomena} (Princeton University Press, Princeton, New Jersey, 1954).

\bibitem{DICKMAN_6}
R. Dickman, {M.\,A.~Mu\~{n}oz}, A. Vespignani, and S. Zapperi, {\em Paths to
  self-organized criticality}, Braz.~J.~Phys. {\bf 30},  27  (2000).

\bibitem{AHARONY_4}
A. Aharony, {\em Dependence of universal critical behavior on symmetry and
  range of interaction {\textit in Phase Transitions and Critical Phenomena},
  Vol.\,6}, edited by C. Domb and {M.\,S.~Green} (Academic Press, London,
  1976).

\bibitem{RIEDEL_2}
{E.\,K.~Riedel} and {F.\,J.~Wegner}, {\em Scaling approach to anisotropic
  magnetic systems statics}, Z.~Phys. {\bf 225},  195  (1969).

\bibitem{CHANG_1}
{M.-C.~Chang} and A. Houghton, {\em Universal ratios among
  correction-to-scaling amplitudes on the coexistence curve}, Phys.~Rev.~Lett.
  {\bf 44},  785  (1980).

\bibitem{LUIJTEN_1}
E. Luijten, {H.\,W.\,J.~Bl{\protect\"o}te}, and K. Binder, {\em Nonmomotonic
  crossover of the effective susceptibility exponent}, Phys.~Rev.~Lett. {\bf
  79},  561  (1997).

\bibitem{LUIJTEN_2}
E. Luijten and K. Binder, {\em Nature of crossover from classical to Ising-like
  critical behavior}, Phys.~Rev.~E {\bf 58},  4060  (1998).

\bibitem{MARQUES_1}
{{M.\,I.~Marqu{\'e}s} and {J.\,A.~Gonzalo}}, {\em Scaling relationship between
  effective critical exponents throughout the crossover region in thin films},
  Eur.~Phys.~J.~B {\bf 14},  317  (2000).

\bibitem{LUEB_30}
S. L{\protect\"u}beck, {\em Violation of the Widom scaling law for effective
  crossover exponents}, Phys.~Rev.~E {\bf 69},  066101  (2004).

\bibitem{GREYWALL_1}
{D.\,S.~Greywall} and G. Ahlers, {\em Second-sound velocity, scaling, and
  universality in {$\mathrm He$} II under pressure near the superfluid
  transition}, Phys.~Rev.~Lett. {\bf 28},  1251  (1972).

\bibitem{ANISIMOV_1}
{M.\,A.~Anisimov}, {S.\,B.~Kiselev}, {J.\,V.~Sengers}, and S. Tang, {\em
  Crossover approach to global critical phenomena in fluids}, Physica~A {\bf
  188},  487  (1992).

\bibitem{BELYAKOV_1}
{M.\,Y.~Belyakov} and {S.\,B.~Kiselev}, {\em Crossover behavior of the
  susceptibility of the susceptibility and the specific heat near a
  second-order phase transition}, Physica~A {\bf 190},  75  (1992).

\bibitem{LUIJTEN_3}
E. Luijten, {H.\,W.\,J.~Bl{\protect\"o}te}, and K. Binder, {\em Medium-range
  interactions and crossover to classical critical behavior}, Phys.~Rev.~E {\bf
  54},  4626  (1996).

\bibitem{MON_1}
{K.\,K.~Mon} and K. Binder, {\em Finite-size scaling and the crossover to
  mean-field critical behavior in the two-dimensional Ising model with
  medium-ranged interactions}, Phys.~Rev.~E {\bf 48},  2498  (1993).

\bibitem{PELISSETTO_1}
A. Pelissetto, P. Rossi, and E. Vicari, {\em Crossover scaling from classical
  to nonclassical critical behavior}, Phys.~Rev.~E {\bf 58},  7146  (1998).

\bibitem{PELISSETTO_2}
A. Pelissetto, P. Rossi, and E. Vicari, {\em Mean-field expansion for spin
  models with medium-range interactions}, Nucl.~Phys.~B {\bf 554},  552
  (1999).

\bibitem{CARACCIOLO_1}
S. Caracciolo {\it et~al.}, {\em Crossover phenomena in spin models with
  medium-range interactions and self-avoiding walks with medium-range jumps},
  Phys.~Rev.~E {\bf 64},  046130  (2001).

\bibitem{KIM_1}
{Y.\,C.~Kim}, {M.\,A.~Anisimov}, {J.\,V.~Sengers}, and E. Luijten, {\em
  Crossover critical behavior in the three-dimensional Ising model},
  J.~Stat.~Phys. {\bf 110},  591  (2003).

\bibitem{GINZBURG_1}
{V.\,L.~Ginzburg}, {\em Some remarks on the phase transitions of the second
  kind and the microscopic theory of ferroelectric materials}, Sov.~Phys.~Solid
  State {\bf 2},  1824  (1960).

\bibitem{ALSNIELSEN_1}
J. Als-Nielsen and {R.\,J.~Birgeneau}, {\em Mean-field theory, the Ginzburg
  criterion, and marginal dimensionality of phase transitions}, Am.~J.~Phys.
  {\bf 45},  554  (1977).

\bibitem{LUEB_20}
S. L{\protect\"u}beck and A. Hucht, {\em Absorbing phase transition in a
  conserved lattice gas with random neighbors}, J.~Phys.~A {\bf 34},  L577
  (2001).

\bibitem{DERRIDA_1}
B. Derrida, {A.\,J.~Bray}, and {C.~Godr\`{e}che}, {\em Non-trivial exponents in
  the zero temperature dynamics of the 1D Ising and Potts model}, J.~Phys.~A
  {\bf 27},  L357  (1994).

\bibitem{BRAY_1}
{A.\,J.~Bray}, B. Derrida, and {C.~Godr\`{e}che}, {\em Non-trivial algebraic
  decay in a soluble model of coarsening}, Europhys.~Lett. {\bf 27},  175
  (1994).

\bibitem{STAUFFER_1}
D. Stauffer, {\em Ising spinodal decomposition at T=0 in one to five
  dimensions}, J.~Phys.~A {\bf 27},  5029  (1994).

\bibitem{MAJUMDAR_4}
{S.\,N.~Majumdar}, {A.\,J.~Bray}, {S.\,J.~Corwell}, and C. Sire, {\em Global
  persistence exponent for nonequilibrium critical dynamics}, Phys.~Rev.~Lett.
  {\bf 77},  3704  (1996).

\bibitem{CUEILLE_1}
S. Cueille and C. Sire, {\em Spin block persistence at finite temperature},
  J.~Phys.~A {\bf 30},  L791  (1997).

\bibitem{MAJUMDAR_5}
{S.\,N.~Majumdar}, {\em Persistence in nonequilibrium systems},
  Curr.~Sci.~India {\bf 77},  370  (1999).

\bibitem{DERRIDA_2}
B. Derrida, V. Hakim, and V. Pasquier, {\em Exact first-passage exponents of 1D
  domain growth: relation to a reaction-diffusion model}, Phys.~Rev.~Lett. {\bf
  75},  751  (1995).

\bibitem{LEE_1}
{B.\,P.~Lee} and {A.\,D.~Rutenberg}, {\em Persistence, poisoning, and
  autocorrelations in dilute coarsening}, Phys.~Rev.~Lett. {\bf 79},  4842
  (1997).

\bibitem{OERDING_1}
K. Oerding and {F.~van\,Wijland}, {\em Global persistence in directed
  percolation}, J.~Phys.~A {\bf 31},  7011  (1998).

\bibitem{HINRICHSEN_2}
H. Hinrichsen and M. Antoni, {\em Identification of domain walls in coarsening
  systems at finite temperature}, Phys.~Rev.~E {\bf 57},  2650  (1998).

\bibitem{ALBANO_1}
{E.\,V.~Albano} and {M.\,A.~Mu\~{n}oz}, {\em Numerical study of persistence in
  models with absorbing states}, Phys.~Rev.~E {\bf 63},  031104  (2001).

\bibitem{HINRICHSEN_3}
H. Hinrichsen and {H.\,M.~Koduvely}, {\em Numerical study of local and global
  persistence in directed percolation}, Eur.~Phys.~J.~B {\bf 5},  257  (1998).

\bibitem{LUEB_21}
S. L{\protect\"u}beck and A. Misra, {\em Persistence distributions in a
  conserved lattice gas with absorbing states}, Eur.~Phys.~J.~B {\bf 26},  75
  (2002).

\bibitem{JANSSEN_11}
{H.\,K.~Janssen}, {\em Renormalization field theory of the Gribov process with
  quenched disorder}, Phys.~Rev.~E {\bf 55},  6253  (1997).

\bibitem{MOREIRA_1}
{A.\,G.~Moreira}, {\em Critical dynamics of the contact process with quenched
  disorder}, Phys.~Rev.~E {\bf 54},  3090  (1996).

\bibitem{CAFIERO_1}
R. Cafiero, A. Gabrielli, and {M.\,A.~Mu\~{n}oz}, {\em Disordered
  one-dimensional contact process}, Phys.~Rev.~E {\bf 57},  5060  (1998).

\bibitem{HOOYBERGHS_1}
J. Hooyberghs, F. Igl{\'o}i, and C. Vanderzande, {\em Absorbing state phase
  transitions with quenched disorder}, Phys.~Rev.~E {\bf 69},  066140  (2004).

\bibitem{VOJTA_1}
T. Vojta, {\em Broadening of a nonequilibrium phase transition by extended
  structural defects}, Phys.~Rev.~E {\bf 70},  026108  (2004).

\bibitem{BRAZOVSKII_1}
S. Brazovskii and T. Nattermann, {\em Pinning and sliding of driven elastic
  systems:~from domain walls to charge density waves}, Adv.~Phys. {\bf 53},
  177  (2004).

\bibitem{NOWAK_1}
U. Nowak and {K.\,D.~Usadel}, {\em Influence of the temperature on the
  depinning transition of driven interfaces}, Europhys.~Lett. {\bf 44},  634
  (1998).

\bibitem{LUEB_11}
L. Roters {\it et~al.}, {\em Depinning transition and thermal fluctuations in
  the random-field Ising model}, Phys.~Rev.~E {\bf 60},  5202  (1999).

\bibitem{LUEB_15}
L. Roters, S. L{\protect\"u}beck, and {K.\,D.~Usadel}, {\em Creep motion in a
  random-field Ising model}, Phys.~Rev.~E {\bf 63},  026113  (2001).

\bibitem{REPAIN_1}
{V.~Repain {\textit{et al.}}}, {\em submitted to}, Europhys.~Lett. {\bf ~},  ~
  (2004).

\bibitem{HARRIS_1}
{T.\,E.~Harris}, {\em The Theory of Branching Processes} (Springer, Berlin,
  1963).

\bibitem{GRIMMETT_1}
{G.\,R.~Grimmet} and {D.\,R.~Stirzaker}, {\em Probability and Random Processes}
  (Clarendon Press, Oxford, 1992).

\bibitem{OTTER_1}
R. Otter, {\em The multiplicative process}, Ann.~Math.~Stat. {\bf 20},  206
  (1949).

\bibitem{DICKMAN_7}
R. Dickman, {\em Reweighting in non-equilibrium simulations}, Phys.~Rev.~E {\bf
  60},  2441  (1999).

\bibitem{GRASSBERGER_P2004}
P. Grassberger, {\em \,}, private communication {\bf \,},  \,  (2004).

\end{thebibliography}

\newpage

\listoffigures
\cleardoublepage

\chapter*{Acknowledgements} 
\thispagestyle{empty}

%This work is based on my Habilitation thesis submitted to the 
%Universit{\"a}t Duisburg-Essen in June 2004.
The author has benefited from discussions with a number
of colleagues, 
in particular
O.~Biham,
E.~Domany,
G.~Foltin,
P.~Grassberger,
P.\,C.~Heger,
H.~Hinrichsen,
A.~Hucht, 
H.\,K.~Janssen,
M.~K{\"u}pper,
O.~Stenull,
K.\,D.~Usadel, and
R.\,D.~Willmann.
Furthermore, I would like to thank E.~Domany for his kind
hospitality at the Weizmann Institute of Science. 
Most of the numerical work presented herein was 
carried out during my postdoctoral stay at the Weizmann
Institute which was supported financially by the 
Minerva Foundation (Max Planck Gesellschaft).
I thank P.~Grassberger and H.\,K.~Janssen
for communicating their results prior to publication.
I am grateful to G.~Foltin, A.~Hucht, A.~Siebke, and
R.\,D.~Willmann for a careful reading of the manuscript and 
useful comments.
Eventually, I am indebted to K.\,D.~Usadel
who made this work possible.

\printindex
\cleardoublepage  

\end{document}